\documentclass[a4paper,onecolumn,11pt,unpublished]{quantumarticle}
\pdfoutput=1

\usepackage[english]{babel}

\usepackage[letterpaper,top=2cm,bottom=2cm,left=3cm,right=3cm,marginparwidth=1.75cm]{geometry}
\usepackage{calc}
\usepackage{fancyhdr} 
\usepackage{amsmath}
\usepackage{amssymb}
\usepackage{amsfonts}
\usepackage{enumitem}
\usepackage{mathtools}
\usepackage{graphicx}
\usepackage{upgreek}

\usepackage{caption}
\usepackage{subcaption}
\usepackage{rotating}
\usepackage{pdflscape}
\usepackage{amsthm}
\usepackage{mathabx}
\usepackage{array}

\usepackage[square,comma,numbers,sort&compress]{natbib}
\usepackage{titletoc}
\usepackage{tikz}
\usetikzlibrary{positioning,shapes.misc,shapes.geometric}
\usepackage[colorlinks=true, allcolors=blue]{hyperref}

\newcommand{\appendixhead}%
{\centering\textbf{Appendices}
\vspace{0.25in}}

\newcommand\etc{etc.\ }

\newcommand\ie{i.e.\ }
\newcommand\eg{e.g.\ }
\newcommand\cf{c.f.\ }

\newcommand\citeR{\cite}
\newcommand\citeRef{Ref.~\cite}
\newcommand\citeRefs{Refs~\cite}

\theoremstyle{definition}

\theoremstyle{remark}

\newcommand\eqR[1]{Eq.~\protect\eqref{eq:#1}}
\newcommand\eqRB[2]{Eqs~\protect\eqref{eq:#1} and \protect\eqref{eq:#2}}
\newcommand\eqRC[3]{Eqs~\protect\eqref{eq:#1}, \protect\eqref{eq:#2}, and \protect\eqref{eq:#3}}
\newcommand\eqRD[4]{Eqs~\protect\eqref{eq:#1}, \protect\eqref{eq:#2}, \protect\eqref{eq:#3}, and (\ref{eq:#4})}

\newcommand\eqAR[1]{Eq.~\protect\eqref{eqA:#1}}
\newcommand\eqARB[2]{Eqs~\protect\eqref{eqA:#1} and \protect\eqref{eqA:#2}}
\newcommand\eqARC[3]{Eqs~\protect\eqref{eqA:#1}, \protect\eqref{eqA:#2}, and \protect\eqref{eqA:#3}}
\newcommand\eqARD[4]{Eqs~\protect\eqref{eqA:#1}, \protect\eqref{eqA:#2}, \protect\eqref{eqA:#3}, and (\ref{eqA:#4})}

\newcommand\eqARF[6]{Eqs~\protect\eqref{eqA:#1}, \protect\eqref{eqA:#2}, \protect\eqref{eqA:#3}, \protect\eqref{eqA:#4}, \protect\eqref{eqA:#5}, and \protect\eqref{eqA:#6}}

\newcommand\eqARList[2]{Eqs~\protect\eqref{eqA:#1}-\protect\eqref{eqA:#2}}

\newcommand\secR[1]{Sec.~\protect\ref{sec:#1}}
\newcommand\secRB[2]{Secs~\protect\ref{sec:#1} and \ref{sec:#2}}
\newcommand\secRC[3]{Secs~\protect\ref{sec:#1}, \protect\ref{sec:#2}, and \protect\ref{sec:#3}}
\newcommand\secRD[4]{Secs~\protect\ref{sec:#1}, \protect\ref{sec:#2}, \protect\ref{sec:#3}, and \ref{sec:#4}}

\newcommand\appR[1]{App.~\protect\ref{app:#1}}

\newcommand\appSR[1]{App.~\protect\ref{appS:#1}}

\newcommand\appSRB[2]{Apps~\protect\ref{appS:#1} and \protect\ref{appS:#2}}

\newcommand\figR[1]{Fig.~\protect\ref{fig:#1}}
\newcommand\figRB[2]{Figs~\protect\ref{fig:#1} and \protect\ref{fig:#2}}

\newcommand\figRList[2]{Figs~\protect\eqref{fig:#1}-\protect\eqref{fig:#2}}
\newcommand\tabR[1]{Table~\protect\ref{tab:#1}}
\newcommand\tabRB[2]{Tables~\protect\ref{tab:#1} and \protect\ref{tab:#2}}

\newcommand\tabAR[1]{Table~\protect\ref{tabA:#1}}

\newcommand\listL[1]{\protect\label{list:#1}}
\newcommand\stepL[1]{\protect\label{step:#1}}
\newcommand\eqL[1]{\protect\label{eq:#1}}
\newcommand\eqAL[1]{\protect\label{eqA:#1}}

\newcommand\secL[1]{\protect\label{sec:#1}}
\newcommand\figL[1]{\protect\label{fig:#1}}

\newcommand\tabL[1]{\protect\label{tab:#1}}
\newcommand\tabAL[1]{\protect\label{tabA:#1}}

\newcommand\appL[1]{\protect\label{app:#1}}
\newcommand\appSL[1]{\protect\label{appS:#1}}

\newcommand\noisyLabelIS[1]{\scalebox{#1}{\(\displaystyle \sim\)}}
\newcommand\gateISa[4]{\scalebox{#1}{\textcolor{black}{\(\displaystyle #2_{#3}^{#4}\)}}}
\newcommand\gateISm[2]{\scalebox{#1}{\textcolor{black}{\(\displaystyle #2\)}}}

\newcommand \xVac{\left| 0\right\rangle}
\newcommand\obsOp{\hat{O}}

\newcommand\obsOpN{\tilde{O}}

\newcommand\xObs{\left\langle\obsOp\right\rangle}
\newcommand\xObsN{\left\langle\obsOpN\right\rangle}

\newcommand\xObsNi[1]{\left\langle\obsOpN\right\rangle_{#1}}
\newcommand\xObsM{\xObsNi{\mitigationLabel}}

\newcommand\gateNameIS{}
\newcommand\tagNameIS{}
\newcommand\noiseNameIS{}

\newcommand\circuitSIS[1]{\begin{matrix}\scalebox{0.1}{\input{figures/inkscape/#1.pdf_tex}}\end{matrix}}

\newcommand\noisyLabelSIS{\noindent\hspace{1.5 cm}\noisyLabelIS{8}}
\newcommand\shiftWidth[1]{\hspace{72pt-\widthof{#1}/2}}
\newcommand\shiftHeight[1]{\vspace{65pt-\heightof{#1}/2}}
\newcommand\shiftGate[1]{\shiftHeight{#1}\shiftWidth{#1}#1}

\newcommand\gateSISa[3]{\shiftGate{\gateISa{8}{#1}{#2}{#3}}}

\newcommand\gateSISm[1]{\gateISm{8}{#1}}

\newcommand\gateSISi[2]{\gateSISa{#1}{#2}{\phantom{\dagger}}}
\newcommand\gateSISdi[2]{\gateSISa{#1}{#2}{\dagger}}

\newcommand\proxy{\scriptscriptstyle\heartsuit}
\newcommand\runtimeScaling{\mathrm{RS}}
\newcommand\noiseBoundary{\mathrm{NB}}
\newcommand\target{\pluscirc}

\newcommand\addnoise[1]{\tilde{#1}}
\newcommand\addnoisew[1]{\widetilde{#1}}
\newcommand\makeQuasi[1]{\breve{#1}}
\newcommand\makeNormQuasi[1]{\check{#1}}

\newcommand\gausMean{\mu_\overrotationNoiseLabel}
\newcommand\gausSD{\sigma_\overrotationNoiseLabel}

\newcommand\identityGlobal{\mathbb{I}}
\newcommand\iOp{\mathrm{I}}

\newcommand\xOp{\mathrm{X}}
\newcommand\yOp{\mathrm{Y}}
\newcommand\zOp{\mathrm{Z}}

\newcommand\rotationOp{R}
\newcommand\pOp{\mathrm{P}}
\newcommand\hermitianOp{P}
\newcommand\hermitianOpi[1]{\hermitianOp_{#1}}

\newcommand\sOp{\mathrm{S}}

\newcommand\hOp{\mathrm{H}}

\newcommand\tOp{\mathrm{T}}

\newcommand\pOpi[1]{\pOp_{#1}}

\newcommand\rOpi[1]{{\rotationOp_{#1}}}

\newcommand\sOpdi[1]{\sOp^\dagger_{#1}}
\newcommand\sOpi[1]{\sOp_{#1}}
\newcommand\sOpdH{\sOpdi{\hermitianOp}}
\newcommand\sOpH{\sOpi{\hermitianOp}}

\newcommand\tOpi[1]{\tOp_{#1}}

\newcommand\circuitLabel{\mathrm{C}}
\newcommand\gateLabel{\mathrm{G}}

\newcommand\globalLabel{\mathrm{Gl}}
\newcommand\hiddenInverseLabel{\mathrm{HI}}
\newcommand\localCancellationLabel{\mathrm{LC}}
\newcommand\localLabel{\mathrm{L}}
\newcommand\mitigationLabel{\mathrm{M}}

\newcommand\pauliNoiseLabel{\mathrm{P}}
\newcommand\qubitLabel{\mathrm{Q}}
\newcommand\rotationNoiseLabel{\mathrm{RE}}
\newcommand\overrotationNoiseLabel{\mathrm{ORE}}
\newcommand\runLabel{\mathrm{R}}
\newcommand\setLabel{\mathrm{S}}
\newcommand\stochasticLabel{\mathrm{SN}}

\newcommand\thresholdLabel{\mathrm{B}}

\newcommand\componentLabel{\mathrm{N}}

\newcommand\acton[1]{\left[#1\right]}
\newcommand\coefGlobali[1]{c_{\globalLabel; #1}}

\newcommand\coefSNi[1]{c_{\stochasticLabel; #1}}
\newcommand\coefREi[1]{c_{\rotationNoiseLabel; #1}}

\newcommand\coefLocali[1]{c_{\localLabel; #1}}
\newcommand\coefLCi[1]{c_{\localCancellationLabel; #1}}
\newcommand{\chiLabel}{\mathrm{CHI}}
\newcommand{\cLabel}{\mathrm{C}}

\newcommand\coefLocalCHIni[1]{\addnoise{c}_{\chiLabel; #1}}
\newcommand\coefLocalCLMni[1]{\addnoise{c}_{\cLabel; #1}}

\newcommand\costGlobal{C_{\globalLabel}}

\newcommand\costSN{C_{\stochasticLabel}}
\newcommand\costRE{C_{\rotationNoiseLabel}}

\newcommand\costLocal{C_{\localLabel}}
\newcommand\costLocali[1]{C_{\localLabel,#1}}

\newcommand\multiplicity[1]{\#\left[#1\right]}

\newcommand\eExpecti[1]{e_{\scriptscriptstyle #1}}
\newcommand\eExpect{\eExpecti{\mathrm{G}}}

\newcommand\noiseLevel{\eExpect}
\newcommand\noiseLeveli[1]{\eExpecti{#1}}
\newcommand\noiseLevelM{\noiseLeveli{\mathrm{G};\mitigationLabel}}

\newcommand\eThreshold{\eExpecti{\thresholdLabel}}
\newcommand\eExpectP{\eExpecti{\stochasticLabel}}
\newcommand\eThresholdP{\eExpecti{\stochasticLabel,\thresholdLabel}}
\newcommand\eExpectR{\eExpecti{\rotationNoiseLabel}}
\newcommand\eExpectSOR{\eExpecti{\overrotationNoiseLabel}}

\newcommand\eigenTransferi[1]{\lambda_{#1}}

\newcommand\eBias{\varepsilon}
\newcommand\eBiasi[1]{\eBias_{#1}}
\newcommand\eBiasO{\eBiasi{\scriptscriptstyle\mathrm{O}}}

\newcommand\eBiasTarget{\eBiasi{\scriptscriptstyle\target}}

\newcommand\eBiasProxyM{\eBiasi{\proxy;\scriptscriptstyle\mitigationLabel}}
\newcommand\eBiasProxyN{\eBiasi{\proxy}}

\newcommand\chiTerm{\chi}
\newcommand\chiTermi[1]{\makeQuasi{\chiTerm}_{\scriptscriptstyle{#1}}}

\newcommand\nAMi[1]{N_{\mathrm{A},#1}}
\newcommand\nCircGlobal{N_{\circuitLabel}}
\newcommand\nCircLocal{N_{\localLabel,\circuitLabel}}
\newcommand\nCircLocali[1]{N_{\localLabel,\circuitLabel,#1}}
\newcommand\nCircLC{N_{\localCancellationLabel}}
\newcommand\nCircLCi[1]{N_{\localCancellationLabel,#1}}

\newcommand\nMit{N_{\mitigationLabel}}

\newcommand\nMitSN{N_{\mitigationLabel,\stochasticLabel} }
\newcommand\nMitRE{N_{\mitigationLabel,\rotationNoiseLabel} }
\newcommand\nGate{N_{\gateLabel}}

\newcommand\nComponent{N_{\componentLabel}}

\newcommand\dIIi[1]{d_{i}}
\newcommand\nQubit{{N_{\qubitLabel}}}
\newcommand\nQubitM{N_{\qubitLabel,\mitigationLabel}}
\newcommand\nRun{N_{\runLabel}}
\newcommand\nRunM{N_{\runLabel,\mitigationLabel}}
\newcommand\nRepeat{N_{\setLabel}}

\newcommand\orderGeneral{\mathcal{O}}
\newcommand\orderI[1]{\orderGeneral_{#1}}

\newcommand\orderZero{\orderI{0}}
\newcommand\orderSmall{\orderI{\mathrm{S}}}
\newcommand\orderLarge{\orderI{\mathrm{L}}}
\newcommand\orderLargeSmall{\orderI{\mathrm{LS}}}
\newcommand\orderLargeorSmall{\orderI{\mathrm{L/S}}}

\newcommand\channelCoefficienti[1]{f_{#1}}

\newcommand\channelCoefficientPrimei[1]{f'_{#1}}

\newcommand\circuitCoeffficientLabel{g}
\newcommand\circuitCoefficienti[1]{\circuitCoeffficientLabel_{\scriptscriptstyle{#1}}}

\newcommand\circuitCoefficientMi[1]{\circuitCoeffficientLabel_{\mitigationLabel,\scriptscriptstyle{#1}}}
\newcommand\noiseAmplitude{x}
\newcommand\noiseAmplitudeY{y}
\newcommand\noiseAmplitudeZ{z}

\newcommand\noiseAmplitudeM{\noiseAmplitude_{\mitigationLabel}}
\newcommand\noiseAmplitudei[1]{\noiseAmplitude_{#1}}

\newcommand\noiseFactorSi[1]{y_{#1}}
\newcommand\noiseFactorRi[1]{z_{#1}}
\newcommand\noiseFactorSP{\noiseFactorSi{\pOp}}
\newcommand\noiseFactorSS{\noiseFactorSi{\sOp}}
\newcommand\noiseFactorSHI{\noiseFactorSi{\hiddenInverseLabel}}
\newcommand\noiseFactorRP{\noiseFactorRi{\pOp}}
\newcommand\noiseFactorRS{\noiseFactorRi{\sOp}}
\newcommand\noiseFactorRHI{\noiseFactorRi{\hiddenInverseLabel}}
\newcommand\eAngle{\phi}
\newcommand\eAnglei[1]{\eAngle_{#1}}
\newcommand\eAngleM{\eAnglei{\mitigationLabel}}
\newcommand\eAngleHI{\eAnglei{\hiddenInverseLabel}}
\newcommand\tAngle{\omega}
\newcommand\tAnglei[1]{\tAngle_{#1}}
\newcommand\tAngleHI{\tAnglei{\hiddenInverseLabel}}
\newcommand\cAngle{\theta}

\newcommand\cAnglei[1]{\cAngle_{#1}}

\newcommand\eStoch{p}
\newcommand\eStochi[1]{\eStoch_{#1}}

\newcommand\eStochMax{\eStochi{\mathrm{Max}}}
\newcommand\eStochM{\eStochi{\mitigationLabel}}

\newcommand\eStochLC{\eStochi{\localCancellationLabel}}

\newcommand\aParam{a}
\newcommand\bValue{b}
\newcommand\dTaylor{d}

\newcommand\dTaylori[1]{\dTaylor_{#1}}

\newcommand\cStochShift{r}
\newcommand\cStochShifti[1]{\cStochShift_{#1}}

\newcommand\cStoch{\cStochShift}
\newcommand\cStochi[1]{\cStoch_{#1}}
\newcommand\pPaulii[1]{\eStoch_{\pauliNoiseLabel,#1}}

\newcommand\mProb{q}
\newcommand\mProbi[1]{\mProb_{#1}}
\newcommand\probGlobali[1]{\mProbi{\globalLabel; #1}}
\newcommand\probLocali[1]{\mProbi{\localLabel; #1}}
\newcommand\gProb{p}
\newcommand\gProbi[1]{\gProb_{#1}}

\newcommand\mSign{s}
\newcommand\mSigni[1]{\mSign_{#1}}

\newcommand\signGlobali[1]{\mSigni{\globalLabel; #1}}
\newcommand\signLocali[1]{\mSigni{\localLabel; #1}}
\newcommand\scalabilityi[1]{S_{#1}}
\newcommand\scalabilityGlobal{\scalabilityi{\globalLabel}}

\newcommand\scalabilityTarget{\scalabilityi{\target}}

\newcommand\transferMatrixi[1]{\hat{T}_{#1}}
\newcommand\transferMatrixD{\hat{D}}
\newcommand\transferMatrixDi[1]{\transferMatrixD_{#1}}
\newcommand\transferMatrixW{\hat{W}}
\newcommand\transferMatrixWd{\transferMatrixW^{-1}}

\newcommand\timeInit{\tau_{0}}
\newcommand\timeExec{\tau_{\mathrm{E}}}
\newcommand\timeExecM{\tau_{\mathrm{E},\mitigationLabel}}
\newcommand\timeExeci[1]{\tau_{\mathrm{E},#1}}
\newcommand\timeExecSNi[1]{\tau_{\mathrm{E},\stochasticLabel,#1}}
\newcommand\timeExecREi[1]{\tau_{\mathrm{E},\rotationNoiseLabel,#1}}
\newcommand\timeExecLocal{\tau_{\mathrm{E},\localLabel}}
\newcommand\timeExecLocali[1]{\tau_{\mathrm{E},\localLabel,#1}}
\newcommand\timeN{t_{0}}
\newcommand\timeM{t_{\mitigationLabel}}

\newcommand\rhoi[1]{\rho_{#1}}
\newcommand\initialState{\rhoi{0}}
\newcommand\rhoN{\addnoise{\rho}}

\newcommand\rhoNi[1]{\rhoN_{#1}}

\newcommand\rhoM{\rhoNi{\mitigationLabel}}

\newcommand\kroneckerDelta[1]{\delta_{#1}}

\newcommand\traceSymbol{\mathrm{Tr}}
\newcommand\traceOp[1]{\traceSymbol\left[#1\right]}
\newcommand\blochVariable{b}

\newcommand\blochOperator{{\vec{\blochVariable}}}
\newcommand\blochOperatorN{\vec{\addnoise{\blochVariable}}}

\newcommand\nomark{\(\times\)}

\newcommand\yesmark{\checkmark}

\newcommand\namedOperator[1]{\mathrm{#1}}
\newcommand\channelOperator[1]{\mathcal{#1}}
\newcommand\identityOperator[1]{\mathbb{#1}}
\newcommand\stateCover[1]{\left|#1\right>}
\newcommand\makeLabel[1]{\mathrm{#1}}
\newcommand\separationSymbolExternal{;}
\newcommand\separationSymbolInternal{,}
\newcommand\internalSeparation[2]{{\makeLabel{#1}
\if\relax\detokenize{#1}\relax
\else 
	\if\relax\detokenize{#2}\relax\else \separationSymbolInternal\fi
\fi #2}}
\newcommand\fullSeparation[4]{{\internalSeparation{#1}{#2}
\if\relax\detokenize{#3}\if\relax\detokenize{#4}\relax\else \separationSymbolExternal\fi
\else \separationSymbolExternal
\fi
\internalSeparation{#3}{#4}}}

\newcommand\gG[6]{
	\begingroup
	\renewcommand\noiseNameIS{}
	\renewcommand\gateNameIS{\gateSISi{{#1}}{\internalSeparation{#2}{#3}}}
	\renewcommand\tagNameIS{\internalSeparation{#4}{#5}}	
	\circuitSIS{gateM#6}
	\endgroup
}
\newcommand\oG[6]{#1_\fullSeparation{#2}{#3}{#4}{#5}}
\newcommand\gGN[6]{
	\begingroup
	\renewcommand\noiseNameIS{\noisyLabelSIS}
	\renewcommand\gateNameIS{\gateSISi{{#1}}{\internalSeparation{#2}{#3}}}
	\renewcommand\tagNameIS{\internalSeparation{#4}{#5}}
	\circuitSIS{gateM#6}
	\endgroup
}
\newcommand\oGN[6]{\addnoise{#1}_\fullSeparation{#2}{#3}{#4}{#5}}
\newcommand\gGD[6]{
	\begingroup
	\renewcommand\noiseNameIS{}
	\renewcommand\gateNameIS{\gateSISdi{{#1}}{\internalSeparation{#2}{#3}}}
	\renewcommand\tagNameIS{\internalSeparation{#4}{#5}}	
	\circuitSIS{gateM#6}
	\endgroup
}
\newcommand\oGD[6]{{#1}^{\dagger}_\fullSeparation{#2}{#3}{#4}{#5}}

\newcommand\gGDN[6]{
	\begingroup
	\renewcommand\noiseNameIS{\noisyLabelSIS}
	\renewcommand\gateNameIS{\gateSISdi{{#1}}{\internalSeparation{#2}{#3}}}
	\renewcommand\tagNameIS{\internalSeparation{#4}{#5}}	
	\circuitSIS{gateM#6}
	\endgroup
}

\newcommand\oGDN[6]{\addnoisew{{#1}^{\dagger}}_\fullSeparation{#2}{#3}{#4}{#5}}

\newcommand\gGm[6]{
	\begingroup
	\renewcommand\noiseNameIS{}
	\renewcommand\gateNameIS{\gateSISm{#1}}
	\renewcommand\tagNameIS{\internalSeparation{#4}{#5}}	
	\circuitSIS{gateM#6}
	\endgroup
}

\newcommand\gO[6]{\gG{\namedOperator{#1}}{#2}{#3}{#4}{#5}{#6}}
\newcommand\oO[6]{\oG{\namedOperator{#1}}{#2}{#3}{#4}{#5}{#6}}

\newcommand\oON[6]{\oGN{\namedOperator{#1}}{#2}{#3}{#4}{#5}{#6}}

\newcommand\gC[6]{\gG{\channelOperator{#1}}{#2}{#3}{#4}{#5}{#6}}
\newcommand\oC[6]{\oG{\channelOperator{#1}}{#2}{#3}{#4}{#5}{#6}}
\newcommand\gCN[6]{\gGN{\channelOperator{#1}}{#2}{#3}{#4}{#5}{#6}}
\newcommand\oCN[6]{\oGN{\channelOperator{#1}}{#2}{#3}{#4}{#5}{#6}}

\newcommand\oI[6]{\oG{\identityOperator{#1}}{#2}{#3}{#4}{#5}{#6}}

\newcommand\gS[6]{\gG{\stateCover{#1}}{#2}{#3}{#4}{#5}{#6}}

\newcommand\quasiChannelOperator[1]{\makeQuasi{\channelOperator{#1}}}

\newcommand\gQN[6]{\gGN{\quasiChannelOperator{#1}}{#2}{#3}{#4}{#5}{#6}}
\newcommand\oQN[6]{\oGN{\quasiChannelOperator{#1}}{#2}{#3}{#4}{#5}{#6}}

\newcommand\normQuasiChannelOperator[1]{\makeNormQuasi{\channelOperator{#1}}}

\newcommand\oN[6]{\oG{\normQuasiChannelOperator{#1}}{#2}{#3}{#4}{#5}{#6}}

\newcommand\variableOperator[1]{#1}

\newcommand\oV[6]{\oG{\variableOperator{#1}}{#2}{#3}{#4}{#5}{#6}}

\newcommand\variantOperator[1]{\mathfrak{#1}}

\newcommand\oM[6]{\oG{\variantOperator{#1}}{#2}{#3}{#4}{#5}{#6}}

\newcommand\rotG[2]{\mathrm{e}^{\mathrm{i}\frac{#1}{2}#2}}
\newcommand\rotGNF[2]{\mathrm{e}^{\mathrm{i}#1#2}}
\newcommand\rotGD[2]{\mathrm{e}^{-\mathrm{i}\frac{#1}{2}#2}}

\graphicspath{{./figures/},{./figures/inkscape/},{./figures/plots/}}

\title{Certification and Classification of Linear Quantum Error Mitigation Methods}
\author{Zach Blunden-Codd}
\email{zblundencodd@gmail.com}
\affiliation{Universit\'{e} Grenoble Alpes, CEA, List, F-38000 Grenoble, France}
\author{Mohamed Tamaazousti}
\affiliation{Universit\'{e} Paris-Saclay, CEA, List, F-91120, Palaiseau, France}

\begin{document}
\maketitle

\begin{abstract}
Numerous mitigation methods exist for quantum noise suppression, making it challenging to identify the optimum approach for a specific application\ \textendash\ especially as ongoing advances in hardware tuning and error correction are expected to reduce logical error rates. In order to facilitate the `future proof' application-dependent comparison of mitigation methods, we develop a set of quantitative metrics that account for continual improvements in logical gate quality. We use these metrics to define qualitative criteria (e.g. scalability, efficiency, and robustness to characterised imperfections in the mitigation implementation), which we combine into application-specific certifications. We then provide a taxonomy of linear mitigation methods, characterising them by their features and requirements. Finally, we use our framework to produce and evaluate a mitigation strategy. A mitigation strategy is a collections of mitigation methods and compilation procedures designed to mitigate relevant errors for a given piece of characterised hardware. Our example mitigation strategy is targeted at mitigating the outputs of hardware suffering from stochastic noise and/or rotational errors. We find the most significant determinant of efficient mitigation is accurate and precise characterisation.

\end{abstract}

\startcontents[sections]
\printcontents[sections]{l}{1}{\section*{Contents}\setcounter{tocdepth}{2}}

\section{Introduction}
In this section we introduce and motivate our study (\secR{intro}), highlight our contributions (\secR{contributions}), discuss the assumptions inherent to our study (\secR{assumptions}), and provide an overview of the remainder of this document (\secR{overview}). This document also has a companion piece: \citeRef{MyLongWork}, which follows the same basic layout but has more extensive cross-referenced appendices, which contain all the derivations required for this work. In \appSR{workedExample} we provide a complete worked example for the characterisation of a mitigation method for uniform dephasing noise, to aid comprehension.

\subsection{Context and Motivation}\secL{intro}
Quantum devices inherently suffer from noise \citeR{Preskill2018Lecture}. While unavoidable at present, one effective way to reduce its impact on computations is to employ error mitigation techniques (see \cite{Cai2023a,Endo2021}). Error mitigation permits one to increase the accuracy of quantum calculations at the cost of a longer runtime (and/or increased numbers of qubits). 

There are a great many mitigation methods available, all with different parameters, assumptions, requirements, trade-offs, \etc Our aim is to provide a more comprehensive framework to facilitate the selection of mitigation methods for particular applications implemented on well-characterised hardware. We build on the existing literature to simplify the assessment of mitigation methods. 

In \citeRef{Pascuzzi2022} the authors discuss a trade-off between the sampling cost (due to the increased variance of mitigated circuits) and the increase in the runtime of an individual circuit (due to additional gates being used to effect the mitigation) for methods using identity insertion. We combine both these factors and any increase in the number of qubits required for the mitigation into a single metric: the runtime scaling, which measures the effective increase in total runtime due to the mitigation.

 In \citeRef{Takagi2022} a lower bound is provided for the sampling cost of a variety of classes of mitigation methods as a function of the maximum bias for any observable measurement. We provide a strategy to calculate an upper bound for the post-mitigation bias without requiring quantum measurements of the observable, or simulation of the circuit.
 
 In \citeRef{Cai2023b} they provide order of magnitude estimates for the sampling costs and fidelities of a variety of classes of mitigation methods. We provide finer grained measures of bias and sampling cost to permit the comparison of mitigation methods of the same class, such as different variants of zero-noise extrapolation or probabilistic error cancellation.

 Error mitigation is only one of three main types of noise suppression:
\begin{enumerate}
\item \textbf{Hardware Tuning} \cite{Ballance2016}: Improving the quality of physical gates and qubits through improved manufacturing techniques, pulse shaping, lowering temperatures, feedback control, \etc
 \item \textbf{Error Correction} \cite{Google2024}: Improving the quality of logical gates and qubits (and therefore outputs) for each run of the circuit through redundancy, encoding, real time intervention, \etc 
 \item \textbf{Error Mitigation}  \cite{Cai2023a,Endo2021}: Improving the quality of observable expectation values, on average, for an ensemble of runs; through the use of symmetry verification, effective quasi-stochastic superpositions of gates, \etc 
\end{enumerate}
Although we focus on error mitigation, our framework is designed to be generalisable. It accounts for future reductions in the gate-wise noise amplitude of quantum circuits, due to hardware tuning and/or error correction, by defining metrics in terms of the overall noise level. The cost of error mitigation has been shown to grow exponentially with the number of gates (for stochastic noise with fixed gate-wise error probability, see \citeRef{Takagi2022}). This means that the only way to mitigate larger circuits efficiently is to reduce the gate-wise error probability, which means we can have the same total noise level with a greater number of gates. The error probability could be reduced using error correction and in this case we should mitigate the logical qubits rather than physical qubits, in order to profit from the noise reduction. 

It should be straightforward to extend this framework to incorporate some aspects of error correction, such as: increased numbers of qubits and increased execution time. In order to generalise the framework further, to incorporate the costs associated with hardware tuning, and the classical computing resources required for error correction, we might have to use energy or resource consumption as our metric of choice to characterise the cost of noise suppression (akin to  \citeRef{Fellous2023}).

\subsection{Contributions}\secL{contributions}
\begin{enumerate}
\item \textbf{Taxonomy (\secR{taxonomy}):} We introduce a taxonomy dedicated to linear mitigation methods for quantum noise suppression, categorising them according to key features such as the use of hidden inverses, custom noise-amplification channels, and identity-insertion techniques. This taxonomy enables a systematic assessment of the utility and limitations of individual components of mitigation, and provides a foundation for generalising and optimising existing methods.
\item \textbf{Certification (\secR{metrics}):} There are myriad mitigation methods in the literature and it can be difficult to determine the most appropriate method for a given application. We provide a selection of novel (proxy bias, noise boundary, runtime scaling, and area scale factor) and existing (sampling cost and length scale factor) quantitative metrics for the circuit-independent evaluation of linear mitigation methods (\ie these metrics do not require numerical simulation of circuits or their implementation on a quantum computer). We then use these metrics to determine the compliance of a method with a set of qualitative criteria (robust, scalable, efficient, precise, \etc\!\!), which can be used to certify the method for a particular task. 
\item \textbf{Case Studies:} We carry out the certification procedure for a selection of mitigation methods (\secR{evaluation}) for stochastic noise (\secR{SN}) and rotational errors (\secR{RE}). We also introduce the idea of a mitigation strategy (combination of mitigation method, gate set, and compilation scheme; see \secR{certificationStrategies}) and certify it to be USEful (\secR{certificationStrategies}) if it is a scalable strategy that is efficacious for the production of the mitigated circuit for any quantum algorithm. Finally, we study the USEM:ORE strategy (\secR{strategy}) for the universal and scalable mitigation of hardware suffering from (stochastic) over-rotation errors (\secR{ORE}) and demonstrate that it complies with the USEful certification.
\end{enumerate}

\subsection{Assumptions}\secL{assumptions}
\begin{itemize}
\item \textbf{Linear Methods:} Our discussion is limited to linear mitigation methods (\secR{LinM}) since they are particularly amenable to circuit-independent analysis\footnote{In this work an analysis is circuit independent if it depends only on coarse properties of the circuit such as the size or number of gates and not on the precise nature of the algorithm/gates being implemented.}. There are many non-linear methods (such as virtual distillation \cite{Koczor2021a,Huggins2021, Czarnik2021V} and symmetry verification \cite{Cai2023b}) that would also be worthy of study but are beyond the scope of this paper. 
 \item \textbf{Noise Models:} Though we provide some relatively general analyses (see \secR{metrics}), our main focus is on stochastic noise (\secR{SN}) and rotational errors (\secR{RE}). So our conclusions should not be applied to other error channels without further verification. We assume perfect characterisation of our noise models but in practice there will always be limitations to the precision of characterisation, which will affect the final bias. 
\item \textbf{Large-Circuit Limit:} One of our core focuses is scalability. So a number of our criteria are only appropriate in the asymptotic regime of large circuits (\secR{LargeCircuitRegime}). Care should be taken when applying our analysis to shorter circuits.
\item \textbf{Small Noise Levels:} Our analysis of the bias assumes the small-noise-level regime (\secR{LowNoiseRegime}). Additional analysis would be required to generalise to larger noise levels.
\item \textbf{Pauli Observables:} We assume that our measured observables are single qubit Pauli operators, this assumption can be imposed without loss of generality, but would require extra gates and post-processing for the measurement of arbitrary observables.
\item \textbf{Gate Errors:} We focus on noise related to the application of gates but one could also apply some of these methods to SPAM (state preparation and measurement) errors, if one assigned the effective error channels to virtual identity gates.
\item  \textbf{Uniform Noise Models:} For simplicity we assume that our noisy circuits have uniform noise. That is, every gate suffers from the same noise (see \secR{fullCircuit}). However, our analysis can be generalised to non-uniform noise models (\eg see \secR{composition}).  
\end{itemize}

\subsection{Overview} \secL{overview}
The structure of this paper is the following:
\begin{enumerate}
\setcounter{enumi}{0}
\item \textbf{Certification Scheme (\secR{criteria}):} We provide quantitative metrics for the evaluation of mitigation methods: the proxy bias, sampling cost, area scale factor, runtime scaling, and noise boundary. Then we use these metrics to define qualitative criteria for certifying mitigation methods (\secR{styles}) and strategies (\secR{certificationStrategies}).
\item \textbf{Characterisation of Noise Models (\secR{characterisation}):} We `characterise' the three types of noise we analyse in this work: stochastic noise (\secR{SN}), rotational errors (\secR{RE}), and (stochastic) over-rotation errors (\secR{ORE}).
\item  \textbf{Taxonomical Classification Scheme (\secR{taxonomy}):} We classify the resources we have available to modify error channels: customised channels (\secR{C}), identity insertions (\secR{II}), hidden inverses (\secR{HI}); the scope of noise amplification (\secR{namingConvention}): synchronous, asynchronous, and local; and the type of pre-tailoring we apply: none or local cancellation (\secR{LC}). Then we produce a taxonomy of mitigation methods derived from various combinations of these components (\figR{Taxonomy}).
\item \textbf{Case Study for Methods (\secR{evaluation}):} We consider the mitigation methods in our taxonomy (\tabR{Taxonomy}) and evaluate them with respect to the metrics and certification criteria developed in \secR{criteria} for stochastic noise (\secR{SN}) and rotational errors (\secR{RE}).
\item \textbf{Case Study for Strategy (\secR{USEM:ORE}):} We use the results of \secR{evaluation} to design USEM:ORE, a mitigation strategy for hardware suffering from (stochastic) over-rotation errors and certify that this strategy is USEful (see \secR{certificationStrategies}).
\item \textbf{Conclusions:} We conclude by summarising our contributions and discussing possible applications of our work (see \secR{conclude}).
\end{enumerate}

\section{Certification Scheme}\secL{criteria}
Here we will work through the quantitative metrics (\secR{metrics}) and qualitative criteria  (\secR{certifications}) that we develop for the certification of mitigation methods. But first, we shall briefly discuss linear mitigation methods and their advantages.

\subsection{Linear Mitigation}\secL{LinM}
The purpose of quantum error mitigation is to suppress the bias (\(\eBiasO\)) of an observable (\(\obsOp\)) estimate generated using a quantum computer. We define the bias by:
\begin{align}\eqL{bias}
\eBiasO = &\left|\xObsM-\xObs\right|,
\end{align}
where \(\xObs\) is the target observable expectation in the absence of noise and \(\xObsM\) is the mitigated estimate.

Linear\footnote{An example of a non-linear mitigation method is virtual distillation (see \citeRefs{Koczor2021a,Huggins2021, Czarnik2021V}), where the effective state after mitigation is the original noisy state raised to some integer power and is not a linear combination of physical states, in general.} mitigation is a way of mitigating an observable using a linear combination of (noisy) observable estimates from noisy circuits:
\begin{align}\eqL{linearMitigationObservables}
\xObsM = \sum_{i=0}^{\nCircGlobal}\coefGlobali{i}\xObsNi{i}\approx \xObs,
\end{align} 
where \(\coefGlobali{i}\) is the (global) mitigation coefficient for the \(i\)th observable; \(\xObsNi{i}\) is the \(i\)th noisy observable estimate, \eg the result of the measurement of observable \(\obsOp\) using the \(i\)th noisy circuit variant (\(\oCN{C}{}{}{}{i}{}\)); \(\nCircGlobal\) is the number of circuits used for the mitigation; and \(\xObs\) is the noise-free observable expectation. We can also define our mitigation in terms of noisy states (\(\rhoNi{i}\)) or noisy circuit variants (\(\oCN{C}{}{}{}{i}{}\)):
\begin{align}\eqL{linearMitigationStatesAndCircuits}
\rhoM =& \sum_{i=0}^{\nCircGlobal}\coefGlobali{i}\rhoNi{i}\approx \rho,&\oCN{C}{}{}{M}{}{} =& \sum_{i=0}^{\nCircGlobal}\coefGlobali{i}\oCN{C}{}{}{}{i}{}\approx \oC{C}{}{}{}{}{},
\end{align}
where we use the same coefficients (\(\coefGlobali{i}\)) for observables (\eqR{linearMitigationObservables}), states, and circuits. Here \(\rhoM\) is the mitigated state, \(\rhoNi{i}\) is the noisy states output by the \(i\)th noisy circuit, and \(\rho\) is the (target) noise-free state. \(\oCN{C}{}{}{M}{}{}\) is the channel corresponding to the mitigated circuit, \(\oCN{C}{}{}{}{i}{}\) is the (noisy) channel corresponding to the \(i\)th  circuit, and \(\oC{C}{}{}{}{}{}\) is the (noise-free) channel corresponding to the circuit we wish to implement. To move between the representations we use:
\begin{align}\eqL{observablesFullSet}
\xObsM=&\traceOp{\obsOp\rhoM}=\traceOp{\obsOp\oCN{C}{}{}{M}{}{}\left(\initialState\right)},&\xObsNi{i}=&\traceOp{\obsOp\rhoNi{i}}=\traceOp{\obsOp\oCN{C}{}{}{}{i}{}\left(\initialState\right)},
\end{align}
where (\(\initialState\)) is the initial state input into the circuit. Some common examples of linear mitigation are probabilistic error cancellation \cite{Endo2018}, subspace expansion \cite{Yoshioka2022}, and Richardson extrapolation \cite{Temme2017}. 

Linear mitigation methods are interesting because it is possible to develop analytic bounds for their bias and costs (see \secR{metrics}). An additional advantage of linear mitigation methods is that they can be composed (see \secRB{composition}{concatenation}). So we can combine multiple mitigation methods in order to mitigate different sections of our circuit or different types of noise. In the next section we will discuss how we can model the effect of noise in a circuit-independent way (\secR{models}) and then use these models to analyse mitigation methods (\secR{metrics}).

\subsection{Circuit-Independent Noise Modelling}\secL{models}
Here we model the effect of noise in a circuit-independent way. The idea is to find a model for error channels that denotes them as linear combinations of (c-number) explicit functions of some noise amplitude (\(x\)) and (q-number) quantum channels, which are independent of the noise amplitude. This allows one to model the effect of multiple error channels within a circuit without requiring precise knowledge of the circuit itself or the precise nature of the quantum channels within our error channels.

\subsubsection{Channels}\secL{noiseModelling}
In this work we focus on noise associated to gates so we model each noisy gate (\eg \(\oGN{U}{}{}{}{}{}\)) as a noise-free gate (\eg \(\oG{U}{}{}{}{}{}\)) followed\footnote{The analysis is analogous when the error channel occurs before the noise-free gate so we will not explicitly consider that case in this work.} by an error channel (\(\oC{E}{}{}{}{}{}\)). Each error channel has a noise amplitude\footnote{In this work we mainly focus on noise models with a single noise parameter but the analysis can be readily generalised to multi-noise-parameter noise models.} (\(\noiseAmplitude\)) where the larger the noise amplitude the large the effect of the noise (in general\footnote{When we have coherent errors, sometimes, larger noise amplitudes can result in smaller errors due to periodicity in the noise parameter. However, we will focus on the small-noise-amplitude regime, \(\noiseAmplitude\ll 1\), in this work so that shouldn't be an issue.}). We separate the full error channel (\(\oC{E}{}{}{}{}{}\)) into \(\nComponent\) sub-components, each with a scalar valued coefficient (\(\channelCoefficienti{k}\left(x\right)\)) that depends on the noise amplitude and a `channel valued' term known as a noise channel (\(\oC{N}{}{k}{}{}{}\)), which is independent of the amplitude (\(x\)):
\begin{align}\eqL{errorChannelDecomposition}
\oC{E}{}{}{}{}{}=\sum_{k=0}^{\nComponent} \oV{f}{}{k}{}{}{}\!\left(\noiseAmplitude\right)\oC{N}{}{k}{}{}{},
\end{align}
where we assume:
\begin{align}\eqL{firstIdentityChannel}
\oC{E}{}{}{}{}{}\underset{\noiseAmplitude\rightarrow 0}{\rightarrow}\oC{I}{}{}{}{}{},
\end{align}
where \(\oC{I}{}{}{}{}{}\) is the identity channel, which has no effect on the quantum state. For the rest of this work we also assume that:
\begin{align}\eqL{firstKroneckerDelta}
\oV{f}{}{k}{}{}{}\!\left(0\right)=&\kroneckerDelta{0k},&\oC{N}{}{0}{}{}{}=&\oC{I}{}{}{}{}{},
\end{align}
where \(\kroneckerDelta{0k}\) is the Kronecker delta (so zero unless \(k=0\), when it is unity) and that \(\oC{N}{}{k}{}{}{}\) is a completely-positive and trace-preserving (CPTP) channel. 

To illustrate this notation we consider the example of a local-dephasing-noise (DPh) channel:
\begin{align}
\oC{E}{DPh}{}{}{}{}\!\left(\rho\right) = \left(1-\eStoch\right) \rho +\eStoch\zOp\rho\zOp,
\end{align}
where \(\rho\) is the density operator representing a quantum state, \(\eStoch\) is the noise probability, and \(\zOp\) is the Pauli Z operator. For this channel we can define:
\begin{align}\eqL{localDephasingDecomposition}
\oC{E}{DPh}{}{}{}{} = &\oV{f}{}{0}{}{}{}\!\left(\noiseAmplitude\right)\oC{N}{}{0}{}{}{}+\oV{f}{}{1}{}{}{}\!\left(\noiseAmplitude\right)\oC{N}{}{1}{}{}{},&\noiseAmplitude=&\eStoch,&\oV{f}{}{0}{}{}{}\!\left(\noiseAmplitude\right)=&1-\noiseAmplitude,&\oC{N}{}{0}{}{}{}=&\oC{I}{}{}{}{}{},\\
&&\nComponent=&2,&\oV{f}{}{1}{}{}{}\!\left(\noiseAmplitude\right)=&\noiseAmplitude,&\oC{N}{}{1}{}{}{}\!\left(\rho\right) =&\zOp\rho\zOp,
\end{align}
which satisfies our requirements since \(\oV{f}{}{0}{}{}{}\!\left(0\right)=1\), \(\oV{f}{}{1}{}{}{}\!\left(0\right)=0\), and \(\oC{N}{}{0}{}{}{}\) and \(\oC{N}{}{1}{}{}{}\) are CPTP channels.

\subsubsection{Circuits}\secL{fullCircuit}
We assume a uniform noise model, meaning each error channel in the circuit shares the same type and noise amplitude. This means all error channels in the circuit have the same functional form for the coefficients (\(\channelCoefficienti{k}\));  and the same noise amplitude  (\(\noiseAmplitude\)); but they could have different associated channels (\(\oC{N}{}{k}{}{}{}\)). The uniformity assumption is limiting, but we can account for non-uniform noise models by sequentially applying this procedure for every error channel that appears in the circuit. We assume there are \(\nGate\) (uniform) noisy gates in the circuit and therefore \(\nGate\) error channels. Between the noisy gates we permit any number of noise-free gates and we place no restriction on the form of the noise-free components of the noisy gates. The result is that our noisy circuit (\(\oCN{C}{}{}{}{}{}\)) can be expressed by the channel:
\begin{align}\eqL{generalUniformErrors}
\oCN{C}{}{}{}{}{}=&\sum_{\left\{l_k\right\}:\sum_{j=0}^{\nComponent}l_j = \nGate}\circuitCoefficienti{\left\{l_k\right\}}\!\left(\noiseAmplitude\right) \chiTermi{\left\{l_k\right\}},&\text{ where } \circuitCoefficienti{\left\{l_k\right\}}\!\left(\noiseAmplitude\right)=&\left(\prod_{k=0}^{\nComponent} \left(\channelCoefficienti{k}\left(\noiseAmplitude\right)\right)^{l_k}\right),
\end{align}
but this expression needs a bit of unpacking. We use the short-hand notation:
\begin{align}\eqL{expansionOfL}
\left\{l_k\right\}=&\left\{l_k\right\}_{k=0}^{\nComponent}=l_0,l_1,\dots,l_{\nComponent}.
\end{align}
The sum in \eqR{generalUniformErrors} includes every list of \(l_k\)'s (\(\left\{l_k\right\}\) where \(0\le l_k\le \nGate\)) for which the sum of the list elements (\(l_k\)'s) is equal to the number of error channels in the circuit (\(\nGate\)):
\begin{align}
\sum_{j=0}^{\nComponent}l_j = \nGate.
\end{align} 
\(\chiTermi{\left\{l_k\right\}}\) is a quasi-channel\footnote{Quasi-channels are linear sums of completely-positive and trace-preserving (CPTP) channels but may not be CPTP themselves. \(\chiTermi{\left\{l_k\right\}}\) is a special type of quasi-channel where all the linear coefficients are equal to unity.} and is the linear combination of all the channels that can be obtained by replacing \(l_k\) of the error channels (\(\oC{E}{}{}{}{}{}\)) in the circuit with \(\oC{N}{}{k}{}{}{}\) for each \(k\) (\(0\le k\le \nComponent\)). \(\chiTermi{\left\{l_k\right\}}\) is independent of the noise amplitude (\(\noiseAmplitude\)) since all noise amplitude dependence is encoded in \(\circuitCoefficienti{\left\{l_k\right\}}\!\left(\noiseAmplitude\right)\). 

As an example we will consider the case of a circuit with \(\nGate\) gates, each of which are followed by a local-dephasing-noise channel (\(\oC{E}{DPh}{}{}{}{}\)), so the noisy circuit is given by:
\begin{align}\eqL{dephasingExampleMetric}
\oCN{C}{}{}{}{}{}=&\bigcirc_{i=1}^{\nGate}\left(\oC{E}{DPh}{}{}{}{}\circ \oC{U}{}{i}{}{}{}\right) =\sum_{l_0, l_1:l_0+l_1 = \nGate}\circuitCoefficienti{l_0,l_1}\!\left(\noiseAmplitude\right) \chiTermi{l_0,l_1} ,
\end{align}
where:
\begin{align}
 \oC{U}{}{i}{}{}{}\!\left(\rho\right) =&\oG{U}{}{i}{}{}{}\rho \oGD{U}{}{i}{}{}{},
 \end{align}
  is the quantum channel associated with the \(i\)th quantum gate (\(\oG{U}{}{i}{}{}{}\)), and (see \eqRB{localDephasingDecomposition}{generalUniformErrors}):
  \begin{align}
\circuitCoefficienti{l_0,\nGate-l_0} =&\left(1-\noiseAmplitude\right)^{l_0}\noiseAmplitude^{\nGate-l_0},&\chiTermi{l_0,\nGate-l_0} =&\sum_{j=1}^{\frac{\nGate!}{l_0!\left(\nGate-l_0\right)!}}\bigcirc_{i=1}^{\nGate}\left(\oC{N}{}{\oV{S}{}{i}{}{l_0,j}{}}{}{}{}\circ \oC{U}{}{i}{}{}{}\right),
 \end{align}
 where \(\oV{S}{}{i}{}{l_0,j}{}\) is either zero or one, \ie \(\oC{N}{}{\oV{S}{}{i}{}{l_0,j}{}}{}{}{}\) is either \(\oC{N}{}{0}{}{}{}\) or \(\oC{N}{}{1}{}{}{}\). More concretely \(\oV{S}{}{i}{}{l_0,j}{}\) is the \(i\)th element of the \(j\)th permutation of a set of \(l_0\) zeros and \(l_1=\nGate-l_0\) ones, we sum over all possible distinct permutations. Since \(l_1 = \nGate-l_0\) (see \eqR{dephasingExampleMetric}) these are the only terms that need to be considered. 
 
Fortunately, we do not need to understand the structure of \(\chiTermi{\left\{l_k\right\}}\) since it is independent of the noise amplitude (\(\noiseAmplitude\)). For the following analysis all we need to know is that a decomposition of the form in \eqR{generalUniformErrors} exists and the definition of \(\circuitCoefficienti{\left\{l_k\right\}}\!\left(\noiseAmplitude\right)\). The nature of \(\chiTermi{\left\{l_k\right\}}\) is irrelevant for our calculations as long as it is independent of the noise amplitude and we know its multiplicity (how many CPTP channels it comprises, see \eqR{multiplicity}).

\subsection{Quantitative Mitigation Method Metrics}\secL{metrics}
Mitigation suppresses the bias (\eqR{bias}) at the cost of an increase in the variance, thus we require a greater number of repeats of an experiment and therefore longer cumulative runtimes to obtain mitigated results to the same precision as unmitigated results. In the following sections, we derive a novel, circuit-independent, upper bound for the bias\ \textendash\ termed the proxy bias (\secRB{proxyBias}{proxyBiasMitigated})\ \textendash\ as well as a new characterisation of runtime scaling, which quantifies the relative increase in runtime resulting from mitigation (\secR{scalability}). We combine the two to determine the noise boundary (\secR{Thresholds}), a new measure of the maximum noise level (see \eqR{NL}) that can be mitigated for a given target bias and tolerated runtime scaling. We  then use these metrics to help us define unambiguous qualitative criteria for certification (\secR{certifications}).

\subsubsection{Proxy Bias}\secL{proxyBias}
In order to determine the true bias (\eqR{bias}) we would have to implement or simulate the full quantum circuit both with and without noise. This is not feasible for (novel) large-scale circuits, where no ground truth is available. As an alternative we provide the proxy bias (\eqR{ungainlyExpression}), which is an upper bound for the true bias that can be determined without any explicit simulation, once we have a model for the noise such as that derived in \secR{fullCircuit}. This facilitates the comparison of mitigation methods.

 It is useful to define the multiplicity\footnote{In general the multiplicity of a quasi-channel is the sum of the magnitudes of the linear coefficients pre-multiplying the CPTP channels that are summed together to give the quasi-channel. Since all the linear coefficients of \(\chiTermi{\left\{l_k\right\}}\) are unity the multiplicity \(\chiTermi{\left\{l_k\right\}}\) is just the number of constituent channels.} of the quasi-channel \(\chiTermi{\left\{l_k\right\}}\) (\ie the number of CPTP channels that make it up):
 \begin{align}\eqL{multiplicity}
 \multiplicity{\chiTermi{\left\{l_k\right\}}}=\frac{\nGate !}{\prod_{j=0}^{\nComponent}\left(l_j!\right)}.
 \end{align}
This means any physical (\ie properly normalised) state (\(\initialState\)) acted on by the quasi-channel \(\chiTermi{\left\{l_k\right\}}\) will have an expectation value that satisfies:
 \begin{align}
0\le \left| \traceOp{\obsOp\chiTermi{\left\{l_k\right\}}\left(\initialState\right)}\right|\le  \multiplicity{\chiTermi{\left\{l_k\right\}}},
 \end{align}
 if \(\obsOp\in \left(\iOp,\xOp,\yOp,\zOp\right)^{\nQubit}\) is a Pauli word operator and \(\nQubit\) is the number of qubits. We have this upper bound because we know that the maximum expectation magnitude for any Pauli observable with respect to a physical state is unity and CPTP channels (\eg \(\oC{N}{}{k}{}{}{}\)) acting on physical states return physical states. Since there are \(\multiplicity{\chiTermi{\left\{l_k\right\}}}\) physical channels that make up \(\chiTermi{\left\{l_k\right\}}\) that is the maximum possible observable expectation after passing through the quasi-channel: \(\chiTermi{\left\{l_k\right\}}\).
 
 So the noisy state will have a bias given by (\eqRC{bias}{observablesFullSet}{generalUniformErrors}):
 \begin{align}\eqL{biasBoundFirst}
\eBiasO = &\left|\traceOp{\obsOp\rho}-\traceOp{\obsOp\oCN{C}{}{}{}{}{}\left(\initialState\right)}\right|\nonumber\\
=&\left|\left(1-\left(\channelCoefficienti{0}\left(\noiseAmplitude\right)\right)^{\nGate}\right)\traceOp{\obsOp\rho}-\sum_{\left\{l_k\right\}:\sum_{j=0}^{\nComponent}l_j = \nGate ; l_0\ne \nGate} \circuitCoefficienti{\left\{l_k\right\}}\!\left(\noiseAmplitude\right)\traceOp{\obsOp\chiTermi{\left\{l_k\right\}}\left(\initialState\right)}\right|\le \eBiasProxyN,
\end{align}
where we have separated out the noise-free component (\(\rho\)), \ie that corresponding to the case where we act with the identity component (\(\oC{N}{}{0}{}{}{}\)) of every error channel (see \eqR{firstKroneckerDelta}). In this case \(l_0=\nGate\) so (see \eqR{expansionOfL}):
\begin{align}
\left\{l_k\right\}=&\nGate,0,\dots,0,
\end{align}
and \(\chiTermi{\nGate,0,\dots,0}\left(\initialState\right)=\rho\) is the noise-free state. \(\obsOp\) is any Pauli operator, and we define:
 \begin{align}\eqL{ungainlyExpression}
\eBiasProxyN = &\left|1-\left(\channelCoefficienti{0}\left(\noiseAmplitude\right)\right)^{\nGate}\right|+\sum_{\left\{l_k\right\}_{k}:\sum_{k=0}^{\nComponent}l_k = \nGate ; l_0\ne \nGate} \left|\circuitCoefficienti{\left\{l_k\right\}}\!\left(\noiseAmplitude\right)\right| \multiplicity{\chiTermi{\left\{l_k\right\}}}\nonumber\\
 =&\left|1-\left(\channelCoefficienti{0}\left(\noiseAmplitude\right)\right)^{\nGate}\right|+\left(\sum_{k=0}^{\nComponent}\left|\channelCoefficienti{k}\left(\noiseAmplitude\right)\right|\right)^{\nGate}-\left|\channelCoefficienti{0}\left(\noiseAmplitude\right)\right|^{\nGate},
 \end{align}
 as the (unmitigated) proxy bias.
 
 If we consider our local-dephasing-noise example (see \eqRB{localDephasingDecomposition}{dephasingExampleMetric}) we find the proxy bias is given by (see \eqRB{localDephasingDecomposition}{ungainlyExpression}):
\begin{align}\eqL{dephasingProxyBias}
\eBiasProxyN =&2\left(1-\left(1-\noiseAmplitude\right)^{\nGate}\right),
\end{align}
 since \(0\le \noiseAmplitude\le 1\). This is upper bounded by 2 (which occurs in the limit \(\nGate\rightarrow \infty\) or \(\noiseAmplitude=1\)). This bound is tight since it is indeed the maximum possible error for a Pauli expectation, which must be in the range \(-1\) to \(1\) (inclusive). However, in general we do not expect this method to yield a tight upper bound on the bias. By choosing the decomposition components (\(\oC{N}{}{k}{}{}{}\)) carefully one can make the bound tighter.

\subsubsection{Noise Level}\secL{NL}
The noise level (\(\noiseLevel\)) is the leading order term of the (unmitigated) proxy bias (\eqR{ungainlyExpression}) in the small-noise regime (\(\nGate \left|\noiseAmplitude\right|\ll 1\)). So if we define:
\begin{align}\eqL{derivativeFfirst}
\channelCoefficienti{k}\left(\noiseAmplitude\right)=&\channelCoefficienti{k}\left(0\right)+\channelCoefficientPrimei{k}\left(\noiseAmplitude\right)+\orderSmall\left(\noiseAmplitude^{2}\right),&\channelCoefficienti{0}\left(0\right)=&1,&\channelCoefficienti{k>0}\left(0\right)=&0,&\channelCoefficientPrimei{k}\left(\noiseAmplitude\right)=&\orderSmall\left(\noiseAmplitude\right),
\end{align}
where \(\orderSmall\left(\noiseAmplitude\right)\) is of linear order in \(\noiseAmplitude\) (see \secR{LowNoiseRegime}), then the noise level is given by (see \eqR{ungainlyExpression}):
\begin{align}\eqL{NL}
\noiseLevel=\nGate\sum_{k=0}^{\nComponent}\left|\channelCoefficientPrimei{k}\left(\noiseAmplitude\right)\right|.
\end{align}
We find this to be a useful parameter to quantify the strength of the noise and it is proportional to the noise level defined in \citeRef{Cai2023b} for stochastic noise (see \secR{SN}). It is independent of the precise structure of the circuit, only depending on the number of error channels the circuit contains. If we have multiple different types of error channel then the noise level of the full circuit is just the sum of the noise levels due to the constituent channels. For lower noise levels we expect smaller biases, for the unmitigated circuits, but owing to the complex interaction between gates and error channels this will not always hold in practice. The noise level is always an upper bound for the bias (to leading order in the small noise regime), but it is not guaranteed to be tight. One should choose the channel decomposition (see \eqR{errorChannelDecomposition}) that results in the smallest noise level, to achieve the tightest bound.

 If we consider our local-dephasing-noise example (see \eqRB{localDephasingDecomposition}{dephasingExampleMetric}) we find the noise level is given by (see \eqR{dephasingProxyBias}):
\begin{align}\eqL{dephasingNoiseLevel}
\noiseLevel =&2\nGate\noiseAmplitude,
\end{align}
 since (\eqR{localDephasingDecomposition}):
 \begin{align}
 \channelCoefficienti{0}\left(0\right)=&1,& \channelCoefficientPrimei{0}\left(\noiseAmplitude\right)=&-\noiseAmplitude,  &\channelCoefficienti{1}\left(0\right)=&0,& \channelCoefficientPrimei{1}\left(\noiseAmplitude\right)=&\noiseAmplitude.
 \end{align}

\subsubsection{Proxy Bias after Mitigation}\secL{proxyBiasMitigated}
The proxy bias after mitigation is simply the proxy bias (see \secR{proxyBias}) of the mitigated channel, which serves as an upper bound for the bias of the mitigated state (for a Pauli observable). So the lower the proxy bias the better the mitigation, in general (the true bias after mitigation will depend on the specific structure of the circuit and could be significantly less than the proxy bias). In this work we consider linear mitigation and assume that all the noisy circuits used for our mitigation have the same uniform noise model but different noise amplitudes. In this case the mitigated circuit has the channel (\eqRB{linearMitigationStatesAndCircuits}{generalUniformErrors}):
\begin{align}
\oCN{C}{}{}{M}{}{} =& \sum_{i=0}^{\nCircGlobal}\coefGlobali{i}\oCN{C}{}{}{}{i}{}=\!\!\!\!\sum_{\left\{l_k\right\}_{k}:\sum_{k=0}^{\nComponent}l_k = \nGate}\!\!\!\!\!\!\!\!\!\!\!\! \circuitCoefficientMi{\left\{l_k\right\}}\!\left(\noiseAmplitude\right)\chiTermi{\left\{l_k\right\}},&\text{ where }&&\circuitCoefficientMi{\left\{l_k\right\}}\!\left(\noiseAmplitude\right)=& \sum_{i=0}^{\nCircGlobal}\coefGlobali{i}
\circuitCoefficienti{\left\{l_k\right\}}\!\left(\noiseAmplitudei{i}\right),
\end{align}
and \(\noiseAmplitudei{i}\) is the noise amplitude of the \(i\)th noisy circuit used for the mitigation (generally we assume \(\noiseAmplitudei{0}=\noiseAmplitude\) is the original noise level). Notice the \(\chiTermi{\left\{l_k\right\}}\) terms are not altered by the mitigation only their coefficients (\(\circuitCoefficientMi{\left\{l_k\right\}}\!\left(\noiseAmplitude\right)\)). The proxy bias is therefore given by (see \eqR{ungainlyExpression}):
 \begin{align}\eqL{proxyBiasDefinition}
\eBiasProxyN = \left|1-\circuitCoefficientMi{\left\{\nGate,0,\dots,0\right\}}\!\left(\noiseAmplitude\right)\right|+\sum_{\left\{l_k\right\}_{k}:\sum_{k=0}^{\nComponent}l_k = \nGate ; l_0\ne \nGate} \left|\circuitCoefficientMi{\left\{l_k\right\}}\!\left(\noiseAmplitude\right)\right| \multiplicity{\chiTermi{\left\{l_k\right\}}}.
 \end{align}
In this work we will only make use of the most significant order approximation to the proxy bias (\ie the noise level after mitigation). We provide example derivations of the proxy bias in \appSRB{proxyBias}{mitigatedProxyBias}.

\subsubsection{Runtime Scaling}\secL{scalability}
Although mitigation reduces the bias, it will increase the variance (\(\sigma^2\)) and may increase the number of gates required within a circuit, and therefore each circuit's runtime. In order to account for both the increase in variance and the increase in the runtime we introduce the runtime scaling (\(\scalabilityGlobal\)), which is the ratio of the total runtime required in the mitigated case to the total runtime in the unmitigated case to obtain the same precision. Naturally, the smaller the runtime scaling the better (for a fixed bias). For the equations below we assume that the observables measured are Pauli word operators\footnote{The application of the derivation in \secR{scalability} to non-Pauli observables is straightforward since any observable can be decomposed as a function of Pauli observables.} \(\obsOp\in \left\{\iOp,\xOp,\yOp, \zOp\right\}^{\otimes \nQubit}\), where \(\nQubit\) is the number of qubits. 

We define the unmitigated total runtime by:
\begin{align}\eqL{totalRuntimeFirst}
\timeN=\nRun\left(\timeExec+\timeInit\right),
\end{align}
where \(\nRun\) is the number of repeated runs (or shots) of a circuit we must make (on average) to obtain our target variance (\(\sigma^2\)). For the worst case observable expectation we have:
\begin{align}\eqL{nrunDefinitionFirst}
\nRun=\frac{1}{\sigma^2}.
\end{align}
\(\timeExec\) is the average execution time of our unmitigated circuit and \(\timeInit\) is the time taken for initialisation and readout (treated as independent of the size of the circuit). For mitigated circuits the total runtime is given by:
\begin{align}\eqL{mitigatedRunTime}
\timeM =\costGlobal^{2}\nRun\left(\oV{F}{Gl,L}{}{}{}{} \timeExec+\timeInit\right),
\end{align}
where \(\oV{F}{Gl,L}{}{}{}{}\) is the length scale factor, \ie the average increase in runtime owing to the mitigation; and \(\costGlobal\) is the global sampling cost for the mitigation, \ie the scale factor for the increase in the number of samples required for the mitigation to obtain the same variance as the unmitigated circuit \cite{Endo2018}:
\begin{align}\eqL{costScale}
\costGlobal=&\sum_{i=0}^{\nCircGlobal}\left|\coefGlobali{i}\right|, &\oV{F}{Gl,L}{}{}{}{} =& \frac{\sum_{i=0}^{\nCircGlobal}\left|\coefGlobali{i}\right|\timeExeci{i}}{\costGlobal\timeExec},
\end{align}
where \(\timeExeci{i}\) and \(\coefGlobali{i}\) are the average execution time and the mitigation coefficient, respectively, of the \(i\)th circuit variant (see \eqR{linearMitigationStatesAndCircuits} and \citeRef{Endo2018}). 

The runtime scaling (\(\scalabilityGlobal\)) is given by the ratio of the mitigated runtime to the unmitigated runtime:
 \begin{align}\eqL{scalability}
\scalabilityGlobal = \frac{\timeM}{\timeN}\approx\begin{cases}
\costGlobal^{2} & \text{ if }\timeExec\ll \timeInit,\\
\costGlobal^{2}\oV{F}{Gl,L}{}{}{}{} & \text{ if }\timeExec\gg \timeInit.
\end{cases}
 \end{align}
Since we will mainly focus on the large-circuit regime (see \secR{LargeCircuitRegime}) we will use the latter limit. 
 
 We can also define the width scale factor:
\begin{align}\eqL{widthDefinition}
\oV{F}{Gl,W}{}{}{}{}=\frac{\nQubitM}{\nQubit},
\end{align}
which is the ratio of qubits required for the mitigation (\(\nQubitM\)) to those required for the original noisy circuit (\(\nQubit\)). This provides a natural definition for the area scale factor:
\begin{align}\eqL{areaDefinition}
\oV{F}{Gl,A}{}{}{}{}=\oV{F}{Gl,L}{}{}{}{}\left\lfloor\oV{F}{Gl,W}{}{}{}{}\right\rfloor,
\end{align}
which expresses the number of times we could run the original circuit using the time and resources required to run the mitigated circuit, by running the original circuit multiple times in parallel on the circuit used for the mitigated results. We take the largest integer smaller than \(\oV{F}{Gl,W}{}{}{}{}\) because we cannot split up the qubits required to implement the original circuit between runs, in general. In this more general case the runtime scaling is given by:
 \begin{align}\eqL{scalabilityGeneral}
\scalabilityGlobal = \frac{\timeM\left\lfloor\oV{F}{Gl,W}{}{}{}{}\right\rfloor}{\timeN}\approx\begin{cases}
\costGlobal^{2}\left\lfloor\oV{F}{Gl,W}{}{}{}{}\right\rfloor & \text{ if }\timeExec\ll \timeInit,\\
\costGlobal^{2}\oV{F}{Gl,A}{}{}{}{} & \text{ if }\timeExec\gg \timeInit.
\end{cases}
 \end{align}
However, for the methods we consider in this work \(\oV{F}{Gl,W}{}{}{}{}=1\).

 \subsubsection{Noise Boundaries}\secL{Thresholds}
It is often possible to reduce the bias at the cost of increasing the runtime scaling, for a given mitigation method (such as zero-noise extrapolation). We call this `increasing the order of the mitigation' and define an order parameter \(\nMit\), where larger \(\nMit\) means, in general, lower bias but longer runtime. For most applications one must solve a Pareto optimisation problem where the relative weights given to the (proxy) bias and the runtime scaling will depend on the application.

In order to permit the comparison of mitigation methods that manage the bias-runtime-scaling (or equivalently bias-variance) trade-off relation differently, we introduce the noise boundary (\(\eThreshold\)). The noise boundary is the maximum noise level (\secR{NL}) that can be mitigated, given that we require a bias smaller than or equal to \(\eBiasTarget\) and a runtime scaling less than or equal to \(\scalabilityTarget\) (where the exact values of \(\eBiasTarget\) and \(\scalabilityTarget\) will be set by the problem we wish to solve and the equipment we have available). Naturally, the larger the noise boundary, the better the mitigation. A general expression for the noise boundary can be difficult to formulate analytically, so we will focus on the large-circuit regime (see \secR{LargeCircuitRegime}), considering only the most significant contributions to the proxy bias and runtime scaling\footnote{Where approximations to the runtime scaling and/or bias must be made one should err on the side of caution (\ie round up the bias and runtime scaling) so that the noise boundary is guaranteed to be an upper bound (to the specified order).}.

In order to calculate the noise boundary we first characterise the proxy bias and the runtime scaling as functions of the number of noisy gates (\(\nGate\)), noise level (\(\eExpect\)), and the order of mitigation (\(\nMit\)) for a given mitigation method and noise model:
\begin{align}\eqL{generalFunctionfDef}
\eBiasProxyM =& f_{\proxy}\left(\nGate,\eExpect,\nMit\right),&\scalabilityGlobal =& f_{\runtimeScaling}\left(\nGate,\eExpect,\nMit\right).
\end{align}
Then we solve these equations simultaneously with \(\eBiasProxyM=\eBiasTarget\) and \(\scalabilityGlobal=\scalabilityTarget\) to obtain \(\eExpect\) as a function of \(\eBiasTarget\) and \(\scalabilityTarget\) (and possibly \(\nGate\)). In this case \(\eExpect=\eThreshold\) (\ie the noise boundary):
\begin{align}\eqL{boundaryEquationFirst}
\eThreshold =  f_{\noiseBoundary}\left(\nGate,\eBiasTarget,\scalabilityTarget\right).
\end{align}
The larger the boundary the higher the noise level we can mitigate, so larger boundaries imply better mitigation methods. However, it is the rate of change with \(\scalabilityTarget\) that is more interesting when we are comparing methods in the large-circuit regime (see \secR{LargeCircuitRegime}).

For an examples of a noise boundary calculation see \appSR{noiseBoundaryWorkedExample}.

\subsection{Regimes}\secL{regimes}
In order to permit the comparison of different mitigation methods in a tractable way we will focus on two main regimes: large circuits and small noise levels. To facilitate our analysis we define our order parameter:
\begin{align}\eqL{OrderParameters}
\orderGeneral\left(f\left(y_0,y_1,\dots\right)\right),
\end{align}
as a function that tends to zero in the limit \(f\rightarrow 0\). \(f\) is an arbitrary function that will indicate the most significant term of the Laurent series of \(\orderGeneral\), up to constant scale factors. We add a subscript (or two) to \(\orderGeneral\) in order to specify the regime that we are in.

\subsubsection{Small-Amplitude Regime}\secL{LowAmplitudeRegime}
In all regimes we assume that the noise amplitude of each individual gate is small, \eg \(\noiseAmplitude, \noiseAmplitude^{2}\nGate \ll 1\), where \(\nGate \) is the number of noisy gates in the circuit. So we can neglect all but the smallest powers of \(\noiseAmplitude\) in our calculations, remaining terms can be replaced by:
\begin{align}\eqL{OrderZero}
\orderZero\left(f\left(\noiseAmplitude\right)\right),
\end{align}
where \(\orderZero\rightarrow 0\) as \(\noiseAmplitude\rightarrow 0\).

\subsubsection{Small-Noise-Level Regime}\secL{LowNoiseRegime}
In the small-noise-level regime, the noise level (a function of the noise amplitude: \(\noiseAmplitude\) and the number of gates: \(\nGate\), see \secR{NL}) in the circuit is small (\eg \(\noiseLevel\ll 1\)) so that we consider only the lowest orders of \(\noiseLevel\). In this regime we denote the neglected component:
\begin{align}\eqL{OrderParameterSmall}
\orderSmall\left(f\left(\noiseLevel\right)\right),
\end{align}
this tends to zero in the limit \(\noiseLevel\rightarrow 0\), which is the small-noise limit. It is in this regime that we calculate the proxy bias but it may be possible to generalise to other regimes.

\subsubsection{Large-Circuit Regime}\secL{LargeCircuitRegime}
In the large-circuit regime we assume that the number of gates in the circuit is large (\(\nGate \gg 1\)) so that we consider only the highest orders in \(\nGate \). In this regime we denote the neglected component:
\begin{align}\eqL{OrderParameterLarge}
\orderLarge\left(f\left(\nGate\right)\right),
\end{align}
this tends to zero in the limit \(\nGate\rightarrow \infty\) (so \(\nGate\) will generally appear in the denominator). The large-circuit limit is when the number of gates tends to infinity but the (full circuit) noise level (\(\noiseLevel\)) stays fixed, so this requires that the (gate-wise) noise level tends to zero, \ie \(\noiseAmplitude\rightarrow 0\) as \(\nGate \rightarrow \infty\) such that \(\noiseLevel\) is independent of \(\nGate\) (at least to leading order). If we are in the regime of large circuits and low noise levels then we use \(\orderLargeSmall\) or, if the behaviour is the same for either large circuits or small noise levels, we use \(\orderLargeorSmall\).

\subsection{Qualitative Certification Criteria}\secL{certifications}
We can use our quantitative metrics (\secR{metrics}) to derive some qualitative criteria for certifying our mitigation methods (\secR{styles}) and strategies (\secR{certificationStrategies}). One can pick and choose the relevant certification criteria depending on the target application.

\subsubsection{Mitigation Methods}\secL{styles}
We certify mitigation methods with respect to a particular noise model (such as stochastic noise (\secR{SN}) or rotational errors (\secR{RE})) as their behaviour can depend strongly on the type of noise they are mitigating. We propose the following certification criteria:
\begin{enumerate}\listL{QC}
\item \stepL{Scalable}\textbf{Scalable (S):} A scalable mitigation method is one for which the runtime scaling (\(\scalabilityGlobal\)) is finite and independent of circuit size (\(\nGate\)) to leading order, in the large-circuit regime (\(\nGate\gg 1\), see \secR{LargeCircuitRegime}) when the noise level (\(\noiseLevel\), \secR{NL}) is independent of \(\nGate\). \ie \(\scalabilityGlobal<\infty\) for \(\nGate\rightarrow \infty\) if \(\noiseLevel<\infty\). This means we can mitigate arbitrarily large circuits without a change in the runtime scaling, provided the noise amplitude (\(\noiseAmplitude\)) is reduced by a proportionate amount, so that the noise level (\(\noiseLevel\)) remains fixed. This will generally mean the length of circuit we can mitigate using our method will be inversely proportional to the noise amplitude. 

\item \textbf{Unbounded (U):} An unbounded (in principle) mitigation method can mitigate arbitrarily large noise levels, assuming we permit arbitrarily large runtime scaling, in the large-circuit limit, \ie \(\eThreshold\rightarrow \infty\) (\secR{Thresholds}) as \(\scalabilityTarget\rightarrow \infty\) for fixed \(\eBiasTarget\) and \(\nGate=\infty\). 
\item \textbf{Precise (P):} A precise mitigation method is either unbiased or has a bias after mitigation that can be made arbitrarily small for suitably large runtime scaling, in the small-noise regime, \ie \(\eBiasProxyM\rightarrow 0\) for \(\scalabilityTarget\rightarrow \infty\), where \(\eExpect\ll 1\).
\item \textbf{Efficient (E):} An efficient mitigation method has a runtime scaling  (\(\scalabilityGlobal\), \secR{scalability}) that tends to unity (\ie cost free) as the noise level tends to zero, for fixed target bias (\(\eBiasTarget\)). \ie  \(\scalabilityGlobal\rightarrow 1\) as \(\noiseLevel\rightarrow 0\) for \(\eBiasTarget\) fixed.
\item \stepL{Robust}\textbf{Robust (R):} A robust mitigation method has metrics (\ie runtime scaling, \secR{scalability}, and proxy bias, \secR{proxyBiasMitigated}) that are independent of deviations from the target noise due to small, characterised, imperfections in variant generation (see \secR{VGS}), to leading order. In general, we expect the process of generating noisy gate variants, \ie noise amplification or transformation (\eg see \secRB{HI}{C}) to be imperfect and to result in noise differing slightly from the target noise. A method is robust if these imperfections, provided they are characterised and of the same order as the original noise amplitude (\(\noiseAmplitude\)), do not change the value of the metrics to leading order in the noise level for the small-noise regime (\secR{LowNoiseRegime}) and to leading order in the number of gates for the large-circuit regime (\secR{LargeCircuitRegime}). If the coefficients, but not the order, of the leading order terms change then it is quasi-robust (qR). A (quasi-)hardware-robust method is (quasi-)robust for the imperfections that occur on a particular hardware, not necessarily for all possible imperfections. Imperfections must be well characterised and taken into account in the coefficient and metric determination. 
\end{enumerate}
Some of the implications of these certifications are give below:
\begin{itemize}
\item \textbf{Scalable} mitigation methods can take advantage of reductions in noise levels (\eg through hardware tuning \cite{Ballance2016} or error correction \cite{Google2024}) to mitigate larger circuits without an increase in relative cost.
\item \textbf{Unbounded} mitigation methods can mitigate, arbitrarily large, arbitrarily noisy circuits with a runtime that depends only on their noise level and (possibly) their target bias (in the large-circuit limit). 
\item \textbf{Precise} mitigation methods make high precision applications, \eg precise characterisation, possible. Since they can make the bias after mitigation arbitrarily small.
\item \textbf{Efficient} mitigation methods make high precision applications practical (in the small-noise regime). Since the cost of mitigation disappears as the (pre-mitigation) noise level decreases.
\item \textbf{Robust} methods maintain their performance in the face of inevitable hardware imperfections. 
\end{itemize}
A mitigation method that is scalable, unbounded, precise, efficient, and robust is certified as a SUPER mitigation method. If the method is only quasi-robust, we certify it SUPEqR. This latter designation is sufficient for inclusion in an efficacious mitigation strategy (see \secR{certificationStrategies}).

\subsubsection{Mitigation Strategies}\secL{certificationStrategies}
A mitigation strategy (\eg see \secR{strategy}) is a complete prescription for producing mitigated observable estimates from a quantum algorithm using a particular piece of noisy quantum hardware. They include: 
\begin{itemize}
\item \textbf{Physical Operations:} The (characterised) physical operations\footnote{Physical operations are a generalisation of gates and consists of all gates implementable on a piece of quantum hardware as well as all other transformations it can induce. Examples of physical operations are the process of changing a single (or small subset) of parameters of the effective Hamiltonian that generates a particular gate, measurements, the turning on or off of an external field, a period of free evolution of a qubit, and a period of evolution under a static or dynamic external field.} the quantum hardware can implement.
\item \textbf{Mitigation Methods:} A (set of) mitigation method(s) designed to obtain mitigated observable estimates using the available physical operations.
\item \textbf{Compilation Scheme:} A scheme for converting target algorithms into mitigated quantum circuits.
\item \textbf{Additional Implementation-Dependent Requirements:} Depending on the hardware available and certification required (\secR{certifications}) there may be additional considerations. For example, for a scalable strategy we require that the noise level be independent of the circuit length in order to ensure the runtime scaling is independent of the circuit length (see \secR{styles}). 
\end{itemize}
The idea is that a particular hardware will have a basis set of physical operations that it can implement, each of which will have an associated noise model. Mitigation methods will generally\footnote{Identity insertion based methods (\secR{II}) can often perform mitigation without requiring any gates beyond those required for a particular algorithm.} require additional gates or physical operations to implement the mitigation, beyond those required to implement the unmitigated target circuit. The mitigation methods used within the mitigation strategy should be tailored to the hardware they are designed to mitigate and the gate set(s) they require must be implementable (and, for noise-aware (\secR{namingConvention}) strategies, characterisable) on that hardware. A mitigation method is \textbf{compatible} with a piece of hardware if the hardware can generate all the gates or physical operations required to implement the mitigation method. We introduce three strategy certification criteria:

\begin{enumerate}\listL{QCS}
\item \stepL{UniversalS}\textbf{Universal:} A universal mitigation strategy can mitigate any quantum algorithm\footnote{
If we have a `quantum accelerator', \ie a piece of hardware that is not universal but targeted at a particular task, we can relax the universality requirement and require only that the mitigation strategy can mitigate any algorithm the `quantum accelerator' is designed to implement.}. It is designed to be implemented using a particular universally-mitigatable piece of hardware, which is a piece of hardware that can generate all the gates or physical operations required to produce a mitigated circuit for any quantum algorithm. The strategy is hardware specific so the mitigation methods it contains may not be able to mitigate any possible noise model (unlike, for example, the PEC strategy of \cite{Endo2018}).  Universality requires only that the mitigation methods can mitigate the noise (or some of the noise) that occurs on the specified hardware. The advantage of this restriction is that we can target specific noise models and focus on scalability and efficacy rather than requiring our methods to be completely general.
\item \stepL{ScalableS}\textbf{Scalable:} A scalable mitigation strategy consists of only scalable mitigation methods (see \secR{styles}). This means we can mitigate arbitrarily large circuits without a diverging runtime scaling (\secR{scalability}). There is quite convincing evidence that, for stochastic noise\footnote{Pure rotational errors are generally cheaper to mitigate than stochastic noise and can be scalably mitigated even if the noise level (\secR{NL}) grows with the square root of the number of gates (see \tabR{MenagerieStatisticsRE}).}, the sampling cost of any mitigation method scales exponentially with the number of gates (\(\nGate\)) for constant (gate-wise) noise amplitude  (\(\noiseAmplitude\)), \eg see \citeRef{Takagi2022}. Thus, for fixed \(\noiseAmplitude\) no mitigation method is scalable (for stochastic noise). Therefore, we are forced to consider reductions in the noise amplitude (\eg through hardware tuning \cite{Ballance2016} or error correction \cite{Google2024}) as part of our mitigation strategy when we scale to larger circuits. It is why we consider a method scalable if the runtime scaling is finite in the limit \(\nGate\rightarrow \infty\) when the limit is taken such that the noise level (\(\noiseLevel\), \secR{NL}) is kept constant. One must ensure that the methods are still scalable when implemented using the quantum hardware specified in the strategy.

\item \textbf{Efficacious\stepL{efficaciousS}:} A mitigation strategy is efficacious if all the mitigation methods it contains are precise, efficient, unbounded, and robust (or at least quasi-hardware-robust), see \secR{styles}. This means, given a sufficiently high tolerated runtime scaling (\(\scalabilityTarget\)) it can reduce the bias (of the circuits it can mitigate) to any desired level no matter what the initial noise level, \ie we can ensure:
\begin{align}\eqL{efficacity}
\eBiasProxyM\le &\eBiasTarget,
\end{align}
for \(\scalabilityTarget\) sufficiently large, regardless of the initial value of \(\noiseLevel\). It also means that any improvements to our hardware that lead to reductions in the circuit noise level will reduce the runtime scaling required for mitigation. Lastly, quasi-hardware-robustness means that (for the hardware in our strategy) the runtime scaling and proxy bias are the same, to leading order, when actually implemented on our hardware as in the ideal case (where the noise tailoring required for mitigation does not introduce any new errors). Without quasi-hardware-robustness there is no guarantee that the other certifications will be meaningful for our strategy in practice.
\end{enumerate}
Universal strategies are required for universal quantum computation, scalable strategies are worth investing in long term, and efficacious strategies are more likely to be practical. If a strategy is universal, scalable, and efficacious then we certify it as a USEful mitigation strategy. In the next section we investigate and characterise our target noise models.

\section{Characterisation of Noise}\secL{characterisation}
Here we introduce some of our notation for error channels (\secR{notation}) as well as Pauli transfer matrices (\secR{TransferMatrix}), a useful representation for analysing noise models. Then we present characterisation data for three types of noise: stochastic noise  (SN, \secR{SN}), rotational errors (RE, \secR{RE}), and over-rotation errors (ORE, \secR{ORE}).

\subsection{Characterisation}
We consider an error channel to be characterised when we can express it as a Kraus map \cite{Preskill2018Lecture} or transfer matrix (\secR{TransferMatrix}). With access to quantum hardware one can characterise (\ie determine the structure of) error channels using gate-set tomography \cite{Greenbaum2015}, classical simulation, \etc But, for illustrative purposes, we will consider three explicit classes of noise: stochastic noise (SN, \secR{SN}), rotational errors (RE, \secR{RE}), and (stochastic) over-rotation errors (ORE, \secR{ORE}). It can be useful to model error channels in different ways for different applications.

\subsection{Circuit-Independent Noise Models}\secL{notation}
We consider noise to be associated with a gate. A noisy gate (\(\oGN{U}{}{}{}{}{}\)) is equivalent to a unitary (noise-free) gate (\(\oG{U}{}{}{}{}{}\)) followed by an error channel\footnote{Our analysis will be based on modelling error channels as following their gates but all the analysis is analogous if the channel precedes the gate.} (\(\oC{E}{}{}{}{}{}\)), which may or may not be unitary. We augment the standard circuit notation to account for noisy gates by representing noisy gates (which may not be unitaries) as gates (\eg \(\oG{U}{}{}{}{}{}\)) with a tilde atop: 
 \begin{align}\eqL{Noisygate}
\gGN{U}{}{}{}{}{} =
\gG{U}{}{}{}{}{}
\gC{E}{}{}{}{}{}.
\end{align}
We represent general channels (\eg \(\oC{C}{}{}{}{}{}\)) using a calligraphic script, while for noise-free unitaries: we use upright text for named unitaries (\eg \(\zOp\) and \(\pOpi{i}\), for the Pauli Z operator and the \(i\)th Pauli word operator, respectively) and italics for generic unitaries (\eg \(\oG{U}{}{}{}{}{}\) and \(\oG{V}{}{}{}{}{}\)). Error channels are represented by \(\oC{E}{}{}{}{}{}\) (see \secR{noiseModelling}). We reserve \(\oC{C}{}{}{}{}{}\) (see \secR{fullCircuit}) to represent the channel associated to a full circuit. We use curly brackets to denote a stochastic or quasi-stochastic ensemble, \eg the channel of \eqR{errorChannelDecomposition} of \secR{noiseModelling} can be represented by:
\begin{align}\eqL{generalNoise}
\gC{E}{}{}{}{}{}
 ~~=&~~
 \gG{}{}{}{}{}{WireVS}\Big\{
\channelCoefficienti{k}\left(\noiseAmplitude\right): \gC{N}{}{k}{}{}{}
\Big\}_{k=0}^{\nComponent}\hspace{-16 pt}\gG{}{}{}{}{}{WireVS}~~~,
\end{align}
where
\begin{align}
 \sum_{k=0}^{\nComponent}\left|\channelCoefficienti{k}\right|=&1&\!\!\!\!\!\!\implies&& \!\!\!\!\!\!\text{stochastic}&,&\sum_{k=0}^{\nComponent}\channelCoefficienti{k}=&1&\!\!\!\!\!\!\implies&&\!\!\!\!\!\! \text{quasi-stochastic}.
\end{align}

For example a (stochastic) Pauli-noise (P) channel can be defined by:
\begin{align}\eqL{pauliNoise}
\gC{E}{P}{}{}{}{}
 ~~=&~~
 \gG{}{}{}{}{}{WireVS}\Big\{
\pPaulii{k}:\gG{\pOp}{k}{}{}{}{}
\Big\}_{k=0}^{4^{\nQubit}-1}\hspace{-28 pt}\gG{}{}{}{}{}{WireVS}~~~, &\pPaulii{0}=&1-\eStoch, &\eStoch=&\sum_{k=1}^{4^{\nQubit}-1}\pPaulii{i},
\end{align}
where \(\pPaulii{k}\ge 0\) is the probability of the operator \(\pOpi{k}\) occurring  on a given run and \(\eStoch\) is the probability of any error occurring. \(\pOpi{k}\in \left\{\iOp,\xOp,\yOp, \zOp\right\}^{\otimes \nQubit}\) is the \(k\)th Pauli word (tensor product of Pauli operators) for \(\nQubit\) qubits (see \appSR{pauliWords}); \(\pOpi{0}\) is the global identity. An alternative way to represent an error channel is with a (Pauli) transfer matrix.

\subsubsection{Transfer Matrices}\secL{TransferMatrix}
The (Pauli) transfer matrix (\(\transferMatrixi{\oC{E}{}{}{}{}{}}\)) associated with a channel (\(\oC{E}{}{}{}{}{}\)) is defined by \cite{Preskill2018Lecture,Greenbaum2015}:
\begin{align}\eqL{transferMatrixDefinition}
 \left(\transferMatrixi{\oC{E}{}{}{}{}{}}\right)_{i,j}=&\frac{1}{2^{\nQubit} }\traceOp{\pOpi{i}\oC{E}{}{}{}{}{}\left(\pOpi{j}\right)},
\end{align}
where \(\nQubit\) is the number of qubits in the channel and \(\pOpi{k}\in \left\{\iOp, \xOp, \yOp,\zOp\right\}^{\otimes \nQubit}\) is the \(k\)th Pauli word operator. Transfer matrices have the nice property that the composition of two channels is just the product of their transfer matrices:
\begin{align}
\transferMatrixi{\oC{E}{}{2}{}{}{}\circ\oC{E}{}{1}{}{}{}}=\transferMatrixi{\oC{E}{}{2}{}{}{}}\transferMatrixi{\oC{E}{}{1}{}{}{}}.
\end{align}

If a transfer matrix is diagonalisable then we can write it as:
\begin{align}\eqL{DiagTransfer}
\transferMatrixi{\oC{E}{}{}{}{}{}}=\transferMatrixW\transferMatrixD\transferMatrixWd,
\end{align}
where \(\transferMatrixD\) is a diagonal matrix and \(\transferMatrixW\) is an invertible matrix. If, in addition, \(\transferMatrixW\) is independent of the noise amplitude (\(\noiseAmplitude\)) of the channel \(\oC{E}{}{}{}{}{}\); then we have a model similar to that of \secR{noiseModelling} and all the \(\noiseAmplitude\) dependence is included in the diagonal matrix (\(\transferMatrixD\)). This property is particularly useful for local mitigation strategies (see \secR{LM}) as it massively simplifies the derivation. In this case the error channel can be fully characterised by its eigenvalues (\(\eigenTransferi{m}\left(\noiseAmplitude\right)\)) and the matrix \(\transferMatrixW\). There are some useful properties of error channels that we shall discuss in the following sections.

\subsubsection{Compatibility}\secL{compatibility}
Noise channels are compatible if they commute (\ie \(\oC{N}{}{k}{}{}{}\) and \(\oC{N}{}{l}{}{}{}\) are compatible if they satisfy the relation \(\left[\oC{N}{}{k}{}{}{},\oC{N}{}{l}{}{}{}\right]_-=0\)) and an error channel decomposition is self-compatible if all its noise channels are compatible. A self-compatible error channel is one that has a diagonalisation matrix (\(\transferMatrixW\), see \secR{TransferMatrix}) that is independent of the noise amplitude, in which case it boasts a self-compatible decomposition. 

\subsubsection{Closure}\secL{closure}
An error channel decomposition (see \eqR{generalNoise} and \eqR{errorChannelDecomposition} of \secR{noiseModelling}) is closed if the span of the constituent noise channels (\ie \(\left\{\oC{N}{}{k}{}{}{}\right\}_{k=0}^{\nComponent}\)) is closed under composition. Composing two closed channels (with the same constituent noise channels) will result in a composite channel with the same constituent noise channels, but, potentially, different parameters (\eg noise amplitude).

\subsection{Stochastic Noise}\secL{SN}
(Single-parameter) stochastic noise (SN) can be expressed by: 
\begin{align}\eqL{stochasticNoise}
\gC{E}{SN}{}{}{}{}
 =
\begin{cases}
\gG{}{}{}{}{}{Wire}&\text{ with probability } 1-\eStoch,\\
\gC{N}{}{}{}{}{}&\text{ with probability } \eStoch.
\end{cases}
\end{align}
where \(\oC{N}{}{}{}{}{}\) is some noise channel (that may or may not cause an actual error to occur depending on the current state the qubits are in), which must be a physical (\ie CPTP) channel for the error channel (\(\oC{E}{}{}{}{}{}\)) to be physical. \(\eStoch\) is the probability of such a noise channel occurring and this is the noise amplitude (\ie \(\noiseAmplitude=\eStoch\)). So this channel can be characterised by (see \eqR{errorChannelDecomposition} of \secR{noiseModelling}):
\begin{align}\eqL{SNsummary}
\channelCoefficienti{0}=&1-\eStoch,&\channelCoefficienti{1}=&\eStoch,\nonumber\\
\oC{N}{}{0}{}{}{}=&\oC{I}{}{}{}{}{}, &\oC{N}{}{1}{}{}{}=&\oC{N}{}{}{}{}{},
\end{align}
where \(\oC{I}{}{}{}{}{}\) is the identity channel and does nothing. We can derive the corresponding proxy bias (upper bound on deviation from ground truth) as follows (see \eqR{ungainlyExpression} of \secR{proxyBias}):
\begin{align}
\eBiasProxyN = 2\left(1-\left(1-\eStoch\right)^{\nGate}\right).
\end{align}
So, in the limit of infinitely many applications of the channel, we have a proxy bias of two, which is indeed the maximum possible bias for a Pauli operator, so the proxy bias may well be a tight bound. The noise level (\(\eExpect\)) is the leading order component of the proxy bias (see \secR{NL}), for stochastic noise:
\begin{align}
\eExpect=\eExpectP = &\eBiasProxyN+\orderSmall\left(\eExpectP^{2}\right),
\end{align}
where:
\begin{align}\eqL{expectStoch}
\eExpectP=2\eStoch\nGate.
\end{align}
This is the same order as the noise level of \citeRef{Cai2023a} but has an additional factor of two in order that the noise level be an upper bound (to first order).

If we have uniform stochastic noise affecting \(\nGate\) gates in our circuit then the channel associated to the circuit can be represented by (see \eqR{generalUniformErrors} of \secR{fullCircuit}):
\begin{align}\eqL{noiseAmplifiedCircuitSN}
\oCN{C}{}{\nGate,\eStoch}{}{}{}=&\sum_{k=0}^{\nGate} \left(1-\eStoch\right)^{\nGate-k}\eStoch^k\chiTermi{\stochasticLabel,k},
\end{align}
where \(\chiTermi{\stochasticLabel,k}\) is independent of \(\eStoch\) (see \secR{fullCircuit}):
\begin{align}
\chiTermi{\stochasticLabel,k}=\chiTermi{\nGate-k,k},
\end{align}
\ie \(\chiTermi{\stochasticLabel,k}\) is a linear combination of all the channels that can be obtained from the original noisy circuit by replacing \(\nGate-k\) of the error channels with the identity channel (\(\oC{I}{}{}{}{}{}\)) and \(k\) with \(\oC{N}{}{}{}{}{}\). The multiplicity (see \eqR{multiplicity}) is given by:
\begin{align}
\multiplicity{\chiTermi{\stochasticLabel,k}}=\frac{\nGate!}{k!\left(\nGate-k\right)!}.
\end{align}

All stochastic-noise channels are self-compatible, since the identity channel commutes with any other channel, but they are only closed if:
\begin{align}\eqL{closureSN}
\oC{N}{}{}{}{}{}^{2}=\aParam\oC{I}{}{}{}{}{}+\left(1-\aParam\right)\oC{N}{}{}{}{}{},
\end{align}
for some \(0\le\aParam\le 1\) (\ie the product of two constituent channels is a linear combination of the existing constituent channels). The channel has eigenvalues given by:
\begin{align}\eqL{eigenFirstSN}
\eigenTransferi{\oC{E}{}{}{}{}{},m}\left(\eStoch\right)=\left(1-\eStoch\right)+\eStoch\eigenTransferi{\oC{N}{}{}{}{}{},m},
\end{align}
where \(\eigenTransferi{\oC{N}{}{}{}{}{},m}\) is the \(m\)th eigenvalue of \(\oC{N}{}{}{}{}{}\). If the error channel is closed then \(\eigenTransferi{\oC{N}{}{}{}{}{},0}=1\) and \(\eigenTransferi{\oC{N}{}{}{}{}{},1}=-\aParam\), so:
\begin{align}
\eigenTransferi{\oC{E}{}{}{}{}{},0}\left(\eStoch\right)=&1,&\eigenTransferi{\oC{E}{}{}{}{}{},1}\left(\eStoch\right)=&1-\left(1+\aParam\right)\eStoch.
\end{align}

\subsection{Rotational Errors}\secL{RE}
Rotational errors (RE) are unitary (coherent) errors generated by a Hermitian unitary operator (\(\hermitianOp\)):
\begin{align}\eqL{rotationalErrors}
\gC{E}{RE}{}{}{}{L}~
 =~
\gG{~\rotGNF{\frac{\eAngle}{2}}{\hermitianOp}}{}{}{}{}{VVL},
\end{align} 
where the noise amplitude is the rotation angle (\ie \(\noiseAmplitude =\eAngle\)) and \(\hermitianOp\) is some self-inverse operator (\(\hermitianOp^{2}=\identityGlobal\)). This channel can be characterised by (see \eqR{errorChannelDecomposition} of \secR{noiseModelling}):
\begin{align}\eqL{REsummary}
\channelCoefficienti{0}=&\cos^{2}\frac{\eAngle}{2},&\channelCoefficienti{1}=&\sin\eAngle, &\channelCoefficienti{2}=&\sin^{2}\frac{\eAngle}{2},\nonumber\\
\oC{N}{}{0}{}{}{}=&\oC{I}{}{}{}{}{}, &\oN{N}{}{1}{}{}{}\left(\rho\right)=&\mathrm{i}\frac{\hermitianOp \rho -\rho\hermitianOp}{2}, &\oC{N}{}{2}{}{}{}\left(\rho\right)=&\hermitianOp \rho \hermitianOp.
\end{align}
In this case \(\oC{N}{}{0}{}{}{}\) and \(\oC{N}{}{2}{}{}{}\) are CPTP channels, as \secR{noiseModelling} requires, but \(\oN{N}{}{1}{}{}{}\) is not. However \(\oN{N}{}{1}{}{}{}\) is normalised and can be decomposed in terms of CPTP channels:
\begin{align}
\oN{N}{}{1}{}{}{}=\frac{1}{2}\oC{N}{}{1}{}{}{}-\frac{1}{2}\oC{N}{}{3}{}{}{},
\end{align}
where \(\oC{N}{}{1}{}{}{}\) and \(\oC{N}{}{3}{}{}{}\) are the CPTP channels:
\begin{align}\eqL{REextraData}
\oC{N}{}{1}{}{}{}\left(\rho\right)=&\sOpH \rho \sOpdH,&\oC{N}{}{3}{}{}{}\left(\rho\right)=&\sOpdH \rho \sOpH, &\sOpH=&\mathrm{e}^{\mathrm{i}\frac{\uppi}{4} \hermitianOp}=\frac{\identityGlobal+\mathrm{i}\hermitianOp}{\sqrt{2}},
\end{align}
so we can perform all the analyses that we need to. The proxy bias is given by (see \eqR{ungainlyExpression} of \secR{proxyBias}):
\begin{align}
\eBiasProxyN = 1-2\cos^{2\nGate}\frac{\eAngle}{2}+\left(1+\sin\left|\eAngle\right|\right)^{\nGate}.
\end{align}
Clearly the bound is not tight in this case since the noise level tends to infinity as \(\nGate\rightarrow \infty\).  The noise level (\(\eExpect\)) is the leading order component of the proxy bias in the small noise regime (see \secR{NL}), for rotational errors:
\begin{align}
\eExpect=\eExpectR = &\eBiasProxyN+\orderSmall\left(\eExpectR^{2}\right),
\end{align}
where:
\begin{align}\eqL{expectRot}
\eExpectR=\nGate\left|\eAngle\right|.
\end{align}

If we have uniform rotational errors affecting \(\nGate\) gates in our circuit then the channel associated to the circuit can be represented by (see \eqR{generalUniformErrors} of \secR{fullCircuit}):
\begin{align}\eqL{noiseAmplifiedCircuitRE}
\oCN{C}{}{\nGate,\eAngle}{}{}{} = \sum_{k=0}^{2\nGate}\cos^{2\nGate-k}\frac{\eAngle}{2}\sin^{k}\frac{\eAngle}{2} \chiTermi{\rotationNoiseLabel,k},
\end{align}
where \(\chiTermi{\rotationNoiseLabel,k}\) is independent of \(\eAngle\), and the multiplicity (see \eqR{multiplicity}) is given by:
\begin{align}
\multiplicity{\chiTermi{\rotationNoiseLabel,k}}=\sum_{l=0}^{\left\lfloor\frac{k}{2}\right\rfloor}\frac{2^{k-2l}\left(\nGate!\right)}{l!\left(k-2l\right)!\left(\nGate+l-k\right)!}.
\end{align}
The factors of two and the summation are due to the fact there are different combinations of \(\oC{N}{}{0}{}{}{}\), \(\oN{N}{}{1}{}{}{}\), and \(\oC{N}{}{2}{}{}{}\) that lead to the same prefactor in \eqR{noiseAmplifiedCircuitRE}.

All rotational-error channels are self-compatible since \(\oC{N}{}{0}{}{}{}\), \(\oN{N}{}{1}{}{}{}\), and \(\oC{N}{}{2}{}{}{}\) commute with each other and they are also closed since:
\begin{align}\eqL{mapCompositionRE}
\oC{N}{}{0}{}{}{}\circ \oC{N}{}{k}{}{}{}=&\oC{N}{}{k}{}{}{},& \oN{N}{}{1}{}{}{}^{2}=&\frac{1}{2}\oC{N}{}{2}{}{}{}-\frac{1}{2}\oC{N}{}{0}{}{}{},&\oC{N}{}{1}{}{}{}\circ \oC{N}{}{2}{}{}{}=&-\oC{N}{}{1}{}{}{},&\oC{N}{}{2}{}{}{}^{2}=&\oC{N}{}{0}{}{}{}.
\end{align}
The channel has eigenvalues given by:
\begin{align}
\eigenTransferi{\oC{E}{}{}{}{}{},0}\left(\eAngle\right)=&1,&\eigenTransferi{\oC{E}{}{}{}{}{},1}\left(\eAngle\right)=&\mathrm{e}^{\mathrm{i}\eAngle},&\eigenTransferi{\oC{E}{}{}{}{}{},2}\left(\eAngle\right)=&\mathrm{e}^{-\mathrm{i}\eAngle}.
\end{align}

\subsection{(Stochastic) Over-Rotation Errors}\secL{ORE}
Now (stochastic) over-rotation errors (ORE) are rotational errors that occur when a gate is produced by an operation equivalent to a rotation generated by some Hermitian unitary operator (\(\hermitianOp\)) and there is an overshoot\footnote{We also have under-rotation errors where there is an undershoot and we rotate less than we aim for. For ease of notation we shall consider under-rotation as a negative over-rotation.}. So we rotate by more than the target angle. Thus over-rotation errors are generated by the same operator that generates the gate. In order to have a more realistic, and general, scenario we consider stochastic over-rotation errors. Here we have a random distribution of rotational errors, which is equivalent to a rotational-error channel followed by a stochastic-noise channel, with the same generator:
\begin{align}\eqL{SORErrors}
\gC{E}{ORE}{}{}{}{VL}
 ~~~=~
 \gG{~\rotGNF{\frac{\eAngle}{2}}{\hermitianOp}}{}{}{}{}{VVL}
\begin{cases}
\gG{}{}{}{}{}{Wire}&\text{ with probability } 1-\eStoch,\\
\gG{\hermitianOp}{}{}{}{}{}&\text{ with probability } \eStoch.
\end{cases}~~~~~~~.
\end{align}
This channel has a channel decomposition given by:
\begin{align}\eqL{SORsummary}
\channelCoefficienti{0}=&\cos^{2}\frac{\eAngle}{2}-\eStoch \cos \eAngle,&\channelCoefficienti{1}=&\left(1-2\eStoch\right)\sin\eAngle, &\channelCoefficienti{2}=&\sin^{2}\frac{\eAngle}{2}+\eStoch\cos\eAngle,\nonumber\\
\oC{N}{}{0}{}{}{}=&\oC{I}{}{}{}{}{}, &\oN{N}{}{1}{}{}{}\left(\rho\right)=&\mathrm{i}\frac{\hermitianOp \rho -\rho\hermitianOp}{2}, &\oC{N}{}{2}{}{}{}\left(\rho\right)=&\hermitianOp \rho \hermitianOp.
\end{align}
So we have the same constituent noise channels as for rotational errors (see \secR{RE}). The proxy bias is given by (see \eqR{ungainlyExpression} of \secR{proxyBias}):
\begin{align}
\eBiasProxyN = 1-2\left(\cos^{2}\frac{\eAngle}{2}-\eStoch\cos \eAngle\right)^{\nGate}+\left(1+\left|1-2\eStoch\right|\sin\left|\eAngle\right|\right)^{\nGate},
\end{align}
and we define the noise level (\(\eExpectSOR\)) as the leading order component of the proxy bias for over-rotation errors:
\begin{align}\eqL{expectSOR}
\eExpectSOR = &2\nGate\eStoch+\nGate\left|\eAngle\right|=\eExpectP+\eExpectR=\eBiasProxyN+\orderSmall\left(\eExpectR^{2},\eExpectP^{2}\right).
\end{align}

Over-rotation-error channels are self-compatible and closed since the noise channels are the same as for rotational-error channels (\cf \eqRB{REsummary}{SORsummary}). The channel has eigenvalues given by:
\begin{align}\eqL{SOREigenvalues}
\eigenTransferi{\oC{E}{}{}{}{}{},0}\left(\eStoch,\eAngle\right)=&1,&\eigenTransferi{\oC{E}{}{}{}{}{},1}\left(\eStoch,\eAngle\right)=&\left(1-2\eStoch\right)\mathrm{e}^{\mathrm{i}\eAngle},&\eigenTransferi{\oC{E}{}{}{}{}{},2}\left(\eStoch,\eAngle\right)=&\left(1-2\eStoch\right)\mathrm{e}^{-\mathrm{i}\eAngle}.
\end{align}
In the next section we will look in more detail at some mitigation methods and their components.

\section{Taxonomical Classification of Mitigation Methods}\secL{taxonomy}
Here we classify our mitigation methods based on their component parts and the variant-generation procedures they make use of. That is, the procedures used to produce the collection of noisy (variant) circuits that are used in the mitigation (\ie \(\left\{\oCN{C}{}{}{}{i}{}\right\}_{i=0}^{\nCircGlobal}\), see \eqR{linearMitigationStatesAndCircuits}). First, we discuss the naming convention we adopt for our mitigation methods (\secR{namingConvention}); then we present the components and variant-generation procedures (\secRD{VGS}{LC}{GM}{LM}) that we consider; and, finally, we present the methods that we study in this work, in the context of a taxonomical scheme (\secR{MMM}).

\subsection{Nomenclature}\secL{namingConvention}
We give our mitigation methods a four component name (see \tabR{Taxonomy} for examples):
\begin{align}
\text{[Pre-tailoring]-[Variant generation][Scope]M:[Noise Awareness]},
\end{align}
where each component uses an acronym.

The pre-tailoring strategies are used to convert the noise models of the noisy gates to different forms before the mitigation proper. In this work we introduce and use only one: local cancellation (LC), which we use to convert non-Pauli to Pauli noise (see \secR{LC}).  

In order to effect the mitigation we need to combine a number of different variants of our original circuit. Each circuit is produced by replacing (some of) our noisy gates with other gates, which we denote noisy gate variants. These have the same noise-free components as the gates they replace but different error channels. We consider four variant-generation strategies, which can also be used in combination: custom channels (C), these are tailored (\ie custom made) error channels inserted after the noisy gate in order to produce new gate variants with amplified errors (see \secR{C}); hidden inverses (HI), this is a strategy for reversing the sign (direction) of coherent error channels (see \secR{HI}); identity insertion (II), this is a noise amplification strategy that involves alternating a noisy gate and its noisy inverse (see \secR{II}); and tuned identity insertion (TII), this is a variant of identity insertion where we tune the number of identity insertions in order to reduce the runtime scaling (see \secR{TII}). 

There are three possible scopes for noise amplification: synchronous (S), where we use the same variant-generation procedure for each gate being mitigated within a particular circuit; local (L), where we have a mitigation strategy that is equivalent to mitigating each gate\footnote{The definition of a gate is used quite flexibly for local mitigation, each `gate' could be a single recognised gate, a collection of adjacent gates, or a physical operation that is part of what makes up a gate.} independently; and asynchronous (A), the most general scheme, where we can have different variant-generation procedures for different gates within the same circuit (this implicitly includes local methods).

Noise awareness denotes whether the method uses quantitative knowledge about the noise models to choose the mitigation coefficients (see \secR{LinM}), which we denote noise aware (NA); or not, which we denote knowledge free (KF). In order to specify the type of noise our method is optimised for, we can replace NA with an acronym for the noise model, \eg SN for stochastic noise, RE for rotational errors, or ORE for (stochastic) over-rotation errors. 

In the following subsections we will explore some of the components of our taxonomy in more detail.

\subsection{Variant-Generation Procedures}\secL{VGS} 
In order to effect our mitigation we make use of a number of noisy gate variants (\(\oGN{U}{}{}{}{i}{}\)):
\begin{align}\eqL{NoiseAmplifiedGates}
\gGN{U}{}{}{}{i}{} =
\gG{U}{}{}{}{}{}
\gC{E}{}{}{}{i}{},
\end{align}
which have the same noise-free component (\(\oG{U}{}{}{}{}{}\)) but different error channels (\(\oC{E}{}{}{}{i}{}\)).

\subsubsection{Type Preserving Procedures}\secL{TPP}
For global mitigation (\secR{GM}) we focus on type preserving variant-generation procedures (TPP, akin to \citeRef{Mari2021} but extended to more general noise models). These procedures result in noisy gate variants with error channels of the same type (see \secR{characterisation}) as the original channel. This means we can use the same form of model (see \secR{noiseModelling}) for the amplified channel as for the original channel, which simplifies the mitigation procedure (see \secR{GlobalMitigation}). This means we require:
\begin{align}\eqL{AmplifiedGateError}
\gC{E}{}{}{}{i}{}
=\begin{cases}
\begin{cases}
1-\eStochi{i}&:\gG{}{}{}{}{}{Wire} \\
\eStochi{i}&:
\gC{N}{}{}{}{}{}
\end{cases}&\text{ for stochastic noise,}\\
 \gG{~\rotGNF{\frac{\eAnglei{i}}{2}}{\hermitianOp}}{}{}{}{}{VVL}&\text{ for rotational errors.}
\end{cases}
~
\end{align}
with \(\eStochi{i}\ne \eStoch\) or \(\eAnglei{i}\ne \eAngle\) for stochastic noise and rotational errors, respectively. The restrictions on the channels imposed by the requirement for type preservation will vary depending on the variant-generation procedures one has access to and the error channels that need mitigating. The hidden-inverse procedure should be a TPP if it is implementable (\ie if it is possible to reverse the sign of coherent error channels, see \secR{HI}). The use of custom channels (\ie placing a single tailoring error channel after the noisy gate of the same type\footnote{If we permit custom (tailoring) channels of more general types, \ie not necessarily of the same type as the error channel they are tailoring, or we permit the use of multiple sequential custom channels, then the use of custom channels is a TPP for more general noise models.} as the original error channel) is a TPP for rotational errors (\secR{RE}); and also for stochastic noise (\secR{SN}), provided that the channel is closed (\secR{closure}). For identity insertion to be a TPP, we have the same requirements as for custom channels but we also require that the error channels of the noisy gates commute with the noise-free components of the noisy gates. There are other variant-generation procedures that we do not explicitly consider here, \eg the tuning of Hamiltonian parameters (or pulse stretching, see \citeRefs{Temme2017,Kandala2019,Kim2023}), and these will have different requirements to be TPPs. 

For local mitigation (\secR{LM}) we do not require TPPs and it is sufficient for the error channels being mitigated to be self-compatible (see \secR{compatibility}) and implementable. This means for identity insertion the noise component must commute with the noise-free component of the noisy gate. For the hidden-inverse procedure it must be possible to reverse the sign of coherent error channels.

If we use the completely general form of probabilistic error cancellation (see \citeRef{Temme2017}) then we do not explicitly require TPPs or self-compatible error channels. We only need access to a set of (noisy) gates or physical operations with our target (noise-free) gate within their span (\eg see \citeRef{Endo2018}).

\subsubsection{Hidden Inverses}\secL{HI}
Hidden inverses (HIs) are used to invert coherent error channels. The idea comes from \citeRefs{Zhang2022, Leyton2022}. Their method is designed for Hermitian (self-inverse) gates with  coherent error channels. 

Gates are implemented by a sequence of physical operations, if the order and direction of these physical operations can be reversed then we should be able to produce the inverse of our original gate, \eg if our noisy gate (\(\oGN{U}{}{}{}{}{}\)) is produced using the physical operations \(\addnoise{A}\), \(\addnoise{B}\), and \(\addnoise{D}\) \cite{Zhang2022}:
\begin{align}\eqL{physicalOperationIntro}
\oGN{U}{}{}{}{}{}= \addnoise{A}\circ \addnoise{B} \circ \addnoise{D},
\end{align}
then the inverse gate is given by:
\begin{align}
\oGDN{U}{}{}{}{}{}= \addnoisew{D^{-1}} \circ \addnoisew{B^{-1}} \circ \addnoisew{A^{-1}}.
\end{align}
If our gate (without noise) is Hermitian then the noise-free components of both implementations (\(\oGN{U}{}{}{}{}{}\) and \(\oGDN{U}{}{}{}{}{}\)) are the same (\ie \(\oG{U}{}{}{}{}{}=\oGD{U}{}{}{}{}{}\)). This means the only difference between \(\oGN{U}{}{}{}{}{}\) and \(\oGDN{U}{}{}{}{}{}\) should be the error channel. If the error is completely coherent then \(\oGDN{U}{}{}{}{}{}\) should have an error channel that is the inverse of the error channel of \(\oGN{U}{}{}{}{}{}\). So, if we have a fully coherent error channel, then the effective error channel of a hidden inverse is given by:
\begin{align}\eqL{HiddenInverseGateError}
\gC{E}{}{}{HI}{}{}
=
\Big(\!\gC{E}{}{}{}{}{}\Big)^{\!\dagger}\hspace{-15 pt}\gG{}{}{}{}{}{WireVS}
~.
\end{align}

We can expand the idea of \citeRefs{Zhang2022, Leyton2022} to encompass gates suffering from over-rotation errors (ORE, see \secR{ORE}), Hermitian or not. We assume our target rotation is \(\tAngle\) (\(0\le \tAngle\le 2\uppi\)) and that we consistently over-rotate by an angle \(\eAngle\). So the noisy gate is given by:
\begin{align}\eqL{definitionOmegaORE}
\gGN{~U\!\left(\tAngle\right)}{}{}{}{}{VL}=\gG{~\rotGNF{\frac{\tAngle+\eAngle}{2}}{\hermitianOp}}{}{}{}{}{VVL}=\gG{~U\!\left(\tAngle\right)}{}{}{}{}{VL}\gC{E}{RE}{}{}{}{VL},
\end{align}
where:
\begin{align}
\gG{~U\!\left(\tAngle\right)}{}{}{}{}{VL}~=&~ \gG{~\rotGNF{\frac{\tAngle}{2}}{\hermitianOp}}{}{}{}{}{VVL}, &\gC{E}{RE}{}{}{}{VL}=\gG{~\rotGNF{\frac{\eAngle}{2}}{\hermitianOp}}{}{}{}{}{VVL},
\end{align}
are the noise-free gate and error channel respectively. We assume that it is a true over-rotation so our target (\(\tAngle\)) and erroneous (\(\eAngle\)) rotations always have the same sign, \ie \(\tAngle\eAngle\ge 0\)  regardless of the value of \(\tAngle\) (this approach also works if we have a true under-rotation error, \ie we always have \(\tAngle\eAngle\le 0\)). In this case we define the hidden inverse (target) angle:
\begin{align}
\tAngleHI = \tAngle-2\frac{\tAngle}{\left|\tAngle\right|}\uppi, 
\end{align}
so \(\tAngleHI\tAngle<0\) and the hidden inverse is a gate with a target angle of \(\tAngleHI\) instead of \(\tAngle\):
\begin{align}
\gGN{~U\!\left(\tAngle\right)}{}{}{HI}{}{VL}=&\gGN{~~~~~~~~~U\!\left(\tAngleHI\right)}{}{}{}{}{DoubleSize}=\gGm{~\,{\rotGNF{\frac{\tAngleHI+\eAngleHI}{2}}{\hermitianOp}\vphantom{\sum\frac{A}{B}}}}{}{}{}{}{DoubleSize}=\gGm{~~~U\!(\tAngle-2\uppi\tAngle/|\tAngle|)\vphantom{\sum\frac{A}{B}}}{}{}{}{}{TripleSize}\gG{~\rotGNF{\frac{\eAngleHI}{2}}{\hermitianOp}}{}{}{}{}{VVL}\\
&=\gG{-1}{}{}{}{}{}\gG{~U\!\left(\tAngle\right)}{}{}{}{}{VL}\gC{E}{RE}{}{HI}{}{VL}=\gG{~U\!\left(\tAngle\right)}{}{}{}{}{VL}\gC{E}{RE}{}{HI}{}{VL}~~.
\end{align}
Since a global sign for a gate is irrelevant (for a quantum channel) this is equivalent to our original noisy gate but with a rotational error \(\eAngleHI\) instead of \(\eAngle\). Since we assume the sign of the rotational error is always the same as that of the target rotation (and \(\tAngleHI\) has the opposite sign to \(\tAngle\)) this means that \(\eAngleHI\) will have the opposite sign to \(\eAngle\). Thus \(\oGN{U}{}{}{HI}{}{}\) is a gate variant with the opposite rotational error to our original gate, but the same noise-free component. If the magnitude of the over-rotational error is fixed, then we have \(\eAngle =-\eAngleHI\)  and therefore \(\oC{E}{RE}{}{HI}{}{VL}=\oC{E}{RE}{}{}{}{VL}^\dagger\), as with the standard hidden-inverse procedure. However, as we will see in \secR{R}, we can tolerate some difference in the magnitudes of \(\eAngleHI\) and \(\eAngle\), as long as their signs are opposed and they have the same order of magnitude (\ie \(\eAngleHI^{2}\ll \eAngle\), \(\eAngle^{2}\ll \eAngleHI\)).

So we should be able to apply the hidden-inverse procedure to Hermitian operators/gates and operators/gates suffering from correlated over-rotation errors (\ie when the relative sign between the rotational error and the target rotation is fixed and independent of the rotation angle).

\subsubsection{Custom channels}\secL{C}
We can also produce gate variants by adding a custom (C) tailoring channel (\(\oC{E}{}{}{C}{i}{}\)) after the original noisy gate:
\begin{align}\eqL{TailoredGate}
\gGN{U}{}{}{}{i}{}~=~&\gGN{U}{}{}{}{}{}\gC{E}{}{}{C}{i}{}~=~\gG{U}{}{}{}{}{}\gC{E}{}{}{}{}{}\gC{E}{}{}{C}{i}{}~,
\end{align}
where \(\oGN{U}{}{}{}{i}{}\) is the \(i\)th tailored gate (\ie noisy gate variant), \(\oGN{U}{}{}{}{}{}\) is the original noisy gate, \(\oC{E}{}{}{}{}{}\) is the original error channel, and the effective error channel for the \(i\)th variant is given by:
\begin{align}\eqL{CustomGateError}
\gC{E}{}{}{}{i}{}~=~&\gC{E}{}{}{}{}{}\gC{E}{}{}{C}{i}{}~.
\end{align}
This idea is implemented for Pauli-noise channels in \citeRefs{Li2017,Ferracin2022,Berg2023}.

For this work we use a tailoring error channel that has the same type as the original error channel, so (\secRB{SN}{RE}):
\begin{align}\eqL{TailoringGateError}
\gC{E}{}{}{C}{i}{}
=\begin{cases}
\begin{cases}
1-\cStochi{i}&:\gG{}{}{}{}{}{Wire} \\
\cStochi{i}&:
\gC{N}{}{}{}{}{}
\end{cases}&\text{ for stochastic noise,}\\
\gG{~\rotGNF{\frac{\cAnglei{i}}{2}}{\hermitianOp}}{}{}{}{}{VVL}&\text{ for rotational errors.}
\end{cases}
\end{align}
The custom stochastic-noise channel involves doing nothing with probability \(1-\cStochi{i}\) and applying \(\oC{N}{}{}{}{}{}\) (which is the same noise channel as is present in the original error channel) with probability \(\cStochi{i}\), while the rotational channel involves deterministically applying a rotational error of \(\cAnglei{i}\) generated by the same \(\hermitianOp\) that generates the original error channel. 

If we apply the tailoring channel (\eqR{TailoringGateError}) we end up with the effective channel in \eqR{AmplifiedGateError} with:
\begin{align}\eqL{partOneEffectiveAmplitudes}
\eStochi{i}=&\eStoch+\cStochi{i}\left(\eStochMax-\eStoch\right),&\eAnglei{i}=&\eAngle+\cAnglei{i},
\end{align}
where \(\eStoch\) and \(\eAngle\) are the original unamplified errors and we have assumed that the stochastic-noise channel is closed (see \eqR{closureSN} of \secR{SN}, and \secRB{closure}{TPP}). Since \(0\le \cStochi{i}\le 1\) we have (see \eqR{closureSN}):
\begin{align}
\eStochMax = 1-\aParam\eStoch,
\end{align} 
as the largest stochastic-noise amplitude we can realise using a custom channel, assuming \(\eStoch\ll 1\).

We have found the optimal distributions for minimising the sampling cost of the mitigation are given by:
\begin{align}\eqL{optimalCustomChoice}
\cStochShifti{i} =& \frac{1-\cos \frac{i \uppi}{\nMit }}{2},&&\text{ and }&\cAnglei{0}=&0,&\cAnglei{i} =& \frac{\left(2i-1\right)\uppi}{\nMit },
\end{align}
for stochastic and rotational errors, respectively, where \(\nMit+1\) is the number of variants in the set of noisy gates (\(0\le i\le \nMit\), see \citeRef{MyLongWork} for more details). We choose \(\cStochShifti{i}\) and \(\cAnglei{i}\) independently of the original errors (\(\eStoch\) and \(\eAngle\)), since otherwise we are in the realm of feedback control rather than mitigation.

The custom channels we use are optimised to reduce the sampling cost using quantitative knowledge of the noise model. In \citeRef{Krebsbach2022} they investigate optimising (effectively) customised channels to reduce the bias, for a fixed sampling cost. They do without explicit knowledge of the noise model since they use Hamiltonian rescaling. Their work complements our own and, for stochastic noise, our custom channels could be considered as special cases of their extremal Chebyshev nodes (not their tilted Chebyshev nodes).

\paragraph{Customised channels and Hidden Inverses}
If we have access to both hidden inverses (see \secR{HI}) and custom channels (\ie CHI) then we can use:
 \begin{align}\eqL{optimalCustomChoiceHI}
\eAnglei{i}=&\eAngle+\cAnglei{i},&\cAnglei{i} =& \frac{2i\uppi}{\nCircGlobal },&\eAnglei{\nCircGlobal}=&\eAngleHI=-\eAngle, 
\end{align}
for mitigating rotational errors, where for \(i=\nCircGlobal\) we use a hidden inverse and for the other (non-zero) \(i\)s (\ie \(0< i<\nCircGlobal\)) we use custom channels.

\subsubsection{Identity Insertion}\secL{II}
Identity insertion (II, see \citeRef{Tiron2020}) is a knowledge-free noise-amplification technique. We amplify the noise by inserting resolutions of the identity, in terms of our noisy gate and its noisy inverse (which may be the same gate if the noise-free gate is Hermitian), before the original noisy gate, in the hope that this will amplify the error channel without changing its structure. II requires the noisy gate (\(\oGN{U}{}{}{}{}{}\)) and its (noisy) inverse\footnote{The noisy inverse gate (\(\oGDN{U}{}{}{}{}{}\)) is given by:
\begin{align}\eqL{NoisyInversegate}
\gGDN{U}{}{}{}{}{}=\gGD{U}{}{}{}{}{}\gC{E}{}{}{}{}{}
\end{align}
and is distinct from the inverse noisy gate (\(\left(\oGN{U}{}{}{}{}{}\right)^{-1}\)) so \(\oGN{U}{}{}{}{}{}\oGDN{U}{}{}{}{}{}\ne \oI{I}{}{}{}{}{}\), in general.} (\(\oGDN{U}{}{}{}{}{}\)) to have the same error channel (\(\oC{E}{}{}{}{}{}\)) and for this error channel to commute with both the noisy gate and its noisy inverse gate. For standard II the gate \(\oGN{U}{}{}{}{i}{}\) is produced using \(i\) identity insertions (see \eqR{Noisygate}):
\begin{align}\eqL{IdentityInsertionGate}
\scalebox{0.8}{\(\displaystyle\gGN{U}{}{}{}{i}{} =
\gG{}{}{}{}{}{WireVS}\Big(\!\gGN{U}{}{}{}{}{}\gGDN{U}{}{}{}{}{}\Big)^{\!i}\!\gGN{U}{}{}{}{}{}
 =
 \gG{}{}{}{}{}{WireVS}\Big(\!\gG{U}{}{}{}{}{}\gC{E}{}{}{}{}{}\gGD{U}{}{}{}{}{}\gC{E}{}{}{}{}{}\Big)^{\!i}\!\gG{U}{}{}{}{}{}\gC{E}{}{}{}{}{}
=
\gG{U}{}{}{}{}{}\Big(\!\gC{E}{}{}{}{}{}\Big)^{\!2i+1}\hspace{-15 pt}\gG{}{}{}{}{}{WireVS}
\)}~.
\end{align}
So the effective error channel of the \(i\)th noisy variant (\(\oGN{U}{}{}{}{i}{}\)) is:
\begin{align}\eqL{IdentityInsertionGateError}
\gC{E}{}{}{}{i}{}= \Big(\!\gC{E}{}{}{}{}{}\Big)^{\!2i+1}\hspace{-15 pt}\gG{}{}{}{}{}{WireVS}
~.
\end{align}
To first order, in the original noise amplitude (\eg \(\eStoch\) or \(\eAngle\)), the noise amplitude of the amplified gate (\(\oGN{U}{}{}{}{i}{}\)) is \(2i+1\) times the noise amplitude of the unamplified gate (\(\oGN{U}{}{}{}{}{}\)) and we have:
\begin{align}\eqL{amplifiedNoiseII}
\eStochi{i} = &\frac{1-\left(1-\eStoch\left(1+\aParam\right)\right)^{2i+1}}{1+\aParam}, &\eAnglei{i}=&\left(2i+1\right)\eAngle,
\end{align}
for stochastic noise and rotational errors, respectively, where we have assumed that the stochastic-noise channel is closed (\(0\le \aParam\le 1\), see \secRB{closure}{SN}).

\paragraph{Tuned Identity Insertion}\secL{TII}
Tuned identity insertion (TII) is a variant of the identity insertion method where, rather than using \(i\) identity insertions for the \(i\)th noisy gate variant (\(\oGN{U}{}{}{}{i}{}\)), we make the number of identity insertions, used for each variant, a free parameter (\(m_i\)):
\begin{align}\eqL{TIIExample}
\gGN{U}{}{}{}{i}{}~~ =~~
\gG{}{}{}{}{}{WireVS}\Big(\!\gGN{U}{}{}{}{}{}\gGDN{U}{}{}{}{}{}\Big)^{\!m_i}\!\gGN{U}{}{}{}{}{}~~=~~
\gG{U}{}{}{}{}{}
\Big(\!\gC{E}{}{}{}{}{}\Big)^{\!2m_i+1}\hspace{-22 pt}\gG{}{}{}{}{}{WireVS}~~~.
\end{align}
We choose \(m_i\) to minimise the runtime scaling (see \secR{scalability}). We have not yet found an analytic expression for \(m_i\) so we optimise it numerically (see \tabAR{PlotParameters} and \figRList{fullCircuitPlotSA}{fullCircuitPlotLCRB}). 
To first order the noise amplitude of the amplified gate is \(2m_i+1\) times the noise amplitude of the unamplified gate (\(\oGN{U}{}{}{}{}{}\)) as we have:
\begin{align}\eqL{amplifiedNoiseTII}
\eStochi{i} = &\frac{1-\left(1-\eStoch\left(1+\aParam\right)\right)^{2m_i+1}}{1+\aParam}, &\eAnglei{i}=&\left(2m_i+1\right)\eAngle.
\end{align}
for stochastic noise and rotational errors, respectively, where we have assumed that the stochastic-noise channel is closed (\cf \eqRB{amplifiedNoiseII}{amplifiedNoiseTII}).

\paragraph{Custom channel Identity Insertion}\secL{CII}
Custom channel identity insertion (CII) is a variant of identity insertion where we produce a single noisy variant (and its noisy inverse) of our noisy gate with amplified noise using custom error channels (we do not need to produce the original noisy gate at all). For rotational errors we have:
\begin{align}\eqL{rotatedGateDefRIILM:NA}
\gGN{U}{}{}{T}{}{}~=&~\gGN{U}{}{}{}{}{}\gC{E}{}{}{C}{}{}~=~
\gG{U}{}{}{}{}{}\gC{E}{}{}{}{}{}\gC{E}{}{}{C}{}{}~=~
\gG{U}{}{}{}{}{}\gC{E}{}{}{T}{}{},&\gC{E}{}{}{C}{}{}~= &~\gG{~\rotGNF{\frac{\cAngle}{2}}{\hermitianOp}}{}{}{}{}{VVL},
\nonumber\\
\gGDN{U}{}{}{T}{}{}~=&~\gGDN{U}{}{}{}{}{}\gC{E}{}{}{C}{}{}~=~
\gGD{U}{}{}{}{}{}\gC{E}{}{}{}{}{}\gC{E}{}{}{C}{}{}~=~
\gGD{U}{}{}{}{}{}\gC{E}{}{}{T}{}{},&\gC{E}{}{}{T}{}{}~=& ~\gG{~\rotGNF{\frac{\eAngle+\cAngle}{2}}{\hermitianOp}}{}{}{}{}{VVL}.
\end{align}
We then use identity insertion as normal but with our amplified gate (and its amplified inverse), so the noisy gate variants we use for the mitigation are:
\begin{align}
\gGN{U}{}{}{}{i}{}~ =~
\gG{}{}{}{}{}{WireVS}\Big(\!\gGN{U}{}{}{T}{}{}\gGDN{U}{}{}{T}{}{}\Big)^{\!i}\!\gGN{U}{}{}{T}{}{}~=~
\gG{U}{}{}{}{}{}
\Big(\!\gC{E}{}{}{T}{}{}\Big)^{\!2i+1}\hspace{-15 pt}\gG{}{}{}{}{}{WireVS}~.
\end{align}
Here we have assumed the noisy inverse gate is amplified by the same amount as the original noisy gate and that they have the same error channel after amplification. For rotational errors we have found that the optimal amplification is given by:
\begin{align}
\cAngle=&\frac{2\uppi}{3},&\implies&&\eAnglei{i}=&\left(2i+1\right)\left(\frac{2\uppi}{3}+\eAngle\right),
\end{align}
for unbiased local mitigation. Here \(\eAnglei{i}\) is just a variant of \eqR{amplifiedNoiseII} with the original noise level replaced by the amplified noise (\ie \(\eAngle\) replaced by \(\eAngle+\cAngle\)). If we permit different custom channels for the noisy gate and its inverse we can achieve lower runtime scalings (\secR{scalability}) at the cost of additional complexity. There is no equivalent approach for stochastic noise; as stochastic noise is not periodic.

\subsection{Pre-tailoring: Local Cancellation}\secL{LC}
Prior to performing the mitigation proper we can apply some tailoring to our gates to make them more amenable to mitigation. 

A common strategy is Pauli twirling \cite{OGorman2016,Bennett1995,Bennett1996,Knill2004} (or randomised compiling \cite{Wallman2016}), which is a knowledge-free tailoring strategy to convert non-Pauli error channels into Pauli error channels (\eqR{pauliNoise} of \secR{notation}). It involves stochastically sandwiching noisy gates between other (twirling) operators. In order for it to function correctly the twirling procedure should not introduce any non-Pauli errors, which, in general, means the twirling operators should be noise free.

A similar strategy is Clifford twirling, where one requires an enlarged twirling set, but the resulting channel is depolarising. In \citeRef{GooglePatent} they propose Clifford twirling the entire circuit, to obtain global depolarising noise, which is particularly easy to mitigate. 

We introduce local cancellation (LC) as an alternative to Pauli twirling, which achieves similar ends but has different requirements. The idea is that one finds, for each noisy gate, a stochastic combination of noisy gate variants that is equivalent to the noise-free gate followed by a stochastic-noise channel. The full procedure is given in \appSR{LC} and requires knowledge of the noise model. However, here we shall consider the special case of applying local cancellation to a gate suffering from an over-rotation error for which we can apply hidden inverses perfectly (see \secR{HI}), to invert the over-rotation-error channel. 

If we assume we have a noisy gate with an over-rotation error \(\eAngle\) then our locally-cancelled gate is given by (see \eqRC{stochasticNoise}{rotationalErrors}{HiddenInverseGateError}):
\begin{align}\eqL{exampleFirstLC}
\gGN{U}{}{}{LC}{}{}~~ =~~
\begin{cases}
\frac{1}{2}&:\gGN{U}{}{}{}{}{}\\
\frac{1}{2}&:\gGN{U}{}{}{HI}{}{}
\end{cases}~~ =~~
\gG{U}{}{}{}{}{}\begin{cases}
\frac{1}{2}&:\gC{E}{RE}{}{}{}{}\\
\frac{1}{2}&:\gC{E}{RE}{}{HI}{}{}
\end{cases}
~~=~~
\gG{U}{}{}{}{}{}
\gC{E}{SN}{}{}{}{}~,
\end{align}
where \(\oGN{U}{}{}{}{}{}\) is the original noisy gate, \(\oGN{U}{}{}{HI}{}{}\) is its hidden inverse, \(\oG{U}{}{}{}{}{}\) is its noise-free component, \(\oC{E}{RE}{}{}{}{}\) is its original rotational-error channel, \(\oC{E}{RE}{}{HI}{}{}\) is the error channel of its hidden inverse, and \(\oC{E}{SN}{}{}{}{}\) is the resultant stochastic-noise channel after local cancellation, where:
\begin{align} 
\gC{E}{RE}{}{}{}{}=&\gG{~\rotG{\eAngle}{\hermitianOp}}{}{}{}{}{VVL}&\gC{E}{RE}{}{HI}{}{}=&\gG{~\rotGD{\eAngle}{\hermitianOp}}{}{}{}{}{VVL},&\gC{E}{SN}{}{}{}{}
 =
\begin{cases}
\gG{}{}{}{}{}{Wire}&\text{ with prob. } 1-\eStochLC,\\
\gG{\hermitianOp}{}{}{}{}{}&\text{ with prob. } \eStochLC.
\end{cases}
\end{align}
So, instead of implementing our original noisy gate for every run of the circuit, we implement the original gate half the time and its hidden inverse the other half of the time. Since this is stochastic (the coefficients are true probabilities) this tailoring is free (\ie it has a runtime scaling, \eqR{scalability}, of unity), and since it is a linear `mitigation' (\secR{LinM}) it can be composed with other linear mitigation procedures as a pre-tailoring step (see \secR{composition}). The effective error probability (\(\eStochLC\)) of the effective stochastic-noise channel is given by:
\begin{align}\eqL{errorProbabilityLCCLM}
\eStochLC=\frac{1-\cos \eAngle}{2}=\frac{\eAngle^{2}}{4}+\orderZero\left(\eAngle^{4}\right),
\end{align}
(see \appSR{LC} for details). This means, if the original noise level is given by \(\eExpectR\), after local cancellation the noise level is given by (see \secR{NL}):
\begin{align}\eqL{noiseLevelReductionLC}
\eExpectP = &2\nGate \eStochLC  +\orderLarge\left(\frac{\eExpectR^{4}}{\nGate^{3}} \right)=\frac{\eExpectR^{2}}{2\nGate},&\eExpectR=&\nGate\eAngle.
\end{align}
So, for fixed noise level (\(\eExpectR\)), the effective noise level after local cancellation will tend to zero in the large-circuit limit (\secR{LargeCircuitRegime}). Therefore local cancellation will not only tailor but will also mitigate the noise in the small noise level or large-circuit regime.

\subsection{Global Mitigation}\secL{GM}
When we apply global mitigation methods we mitigate uniform noise (\ie a collection of gates with the same type of error channel, \eg all stochastic or rotational but not necessarily the same noise channel:  \(\oC{N}{}{}{}{}{}\) or Hermitian operator: \(\hermitianOp\)) with the same noise amplitude (\eg the same \(\eAngle\) or \(\eStoch\)). If we wish to mitigate multiple different types of error channels with global mitigation methods then we must compose multiple separate methods (see \secR{composition}). We will start by providing an example of the derivation of linear mitigation coefficients for the mitigation of uniform stochastic noise or rotational errors.

\subsubsection{Mitigation Coefficients}\secL{GlobalMitigation}
We consider noisy circuit variants (\eg \(\oCN{C}{}{}{}{i}{}\)) with gates suffering from stochastic noise (\secR{SN}) or rotational errors (\secR{RE}) with noise amplitude \(\eStochi{i}\) or \(\eAnglei{i}\), respectively, for circuit variant \(i\). In these cases we have a simple model of the output of the full noisy circuit (\eqRB{noiseAmplifiedCircuitSN}{noiseAmplifiedCircuitRE}):
\begin{align*}
\oCN{C}{}{}{}{i}{} = \begin{cases}
\sum_{k=0}^{\nGate} \left(1-\eStoch\right)^{\nGate-k}\eStoch^k\chiTermi{\stochasticLabel,k}&\text{ for stochastic noise,}\\
 \sum_{k=0}^{2\nGate}\cos^{2\nGate-k}\frac{\eAngle}{2}\sin^{k}\frac{\eAngle}{2} \chiTermi{\rotationNoiseLabel,k}&\text{ for rotational errors,}
 \end{cases}
\end{align*}
where \(\chiTermi{\rotationNoiseLabel,0}=\oC{C}{}{}{}{}{}=\chiTermi{\stochasticLabel,0}\) are the noise-free circuits. The mitigated circuit is a linear superposition of these noisy circuits (\eqR{linearMitigationStatesAndCircuits} of \secR{LinM}):
\begin{align*}
\oCN{C}{}{}{M}{}{} =& \sum_{i=0}^{\nCircGlobal}\coefGlobali{i}\oCN{C}{}{}{}{i}{},
\end{align*}
so the noise-free circuit can be obtained if we require:
\begin{align}
\begin{cases}
\sum_{i=0}^{\nCircGlobal}\coefGlobali{i}\left(1-\eStochi{i}\right)^{\nGate} =1,~~~\sum_{i=0}^{\nCircGlobal}\coefGlobali{i}\left(1-\eStochi{i}\right)^{\nGate-k}\eStochi{i}^k =0&\text{ for stochastic noise,}\\
\sum_{i=0}^{\nCircGlobal}\coefGlobali{i}\cos^{2\nGate}\frac{\eAnglei{i}}{2}=1,~~~ \sum_{i=0}^{\nCircGlobal}\coefGlobali{i}\cos^{2\nGate-k}\frac{\eAnglei{i}}{2}\sin^{k}\frac{\eAnglei{i}}{2} =0,&\text{ for rotational errors,}
 \end{cases}
\end{align}
for \(0<k\le  \nMit\) where:
\begin{align}
\nMit=\begin{cases}
\nGate,&\text{ for stochastic noise,}\\
 2\nGate,&\text{ for rotational errors,}
 \end{cases}
\end{align}
which will result in unbiased mitigation. 

If we can accept a biased mitigation we could use:
\begin{align}\eqL{biasedMitSN}
\sum_{i=0}^{\nMit }\coefGlobali{i}  =&1,&\sum_{i=0}^{\nMit }\coefGlobali{i} \eStochi{i}^{k} =&0,
\end{align}
for \(0<k\le \nMit\le \nGate\) for stochastic noise and:
\begin{align}\eqL{biasedMitRE}
\sum_{i=0}^{\nMit }\coefGlobali{i}  =& 1,&
\sum_{i=0}^{\nMit }\coefGlobali{i} \sin \eAnglei{i} \sin^{2\left(m-1\right)} \frac{\eAnglei{i}}{2}=& 0,&\sum_{i=0}^{\nMit }\coefGlobali{i} \sin^{2m} \frac{\eAnglei{i}}{2}= &0,
\end{align}
for rotational errors. In the latter case we use the first \(\nMit\le 2\nGate\) equations (iterating from left to right of \eqR{biasedMitRE} for sequential values of \(m\) until we reach \(\nMit+1\) equations). This should neglect the mitigation of higher order terms\footnote{In order for us to achieve exponential suppression of errors we require \(\eStochi{i}^2\ll \eStoch\) and/or \(\eAnglei{i}^{2}\ll \eAngle\) for all \(i\).} (\eg  terms of the same order as \(\orderZero\left(\eStoch^{\nMit+1}\right)\) or \(\orderZero\left(\eAngle^{\nMit+1}\right)\)), which will have a smaller contribution to the overall error.

The mitigation coefficients that satisfy \eqRB{biasedMitSN}{biasedMitRE} are given by:
\begin{align}\eqL{coefficientFormRichardZNE}
\coefGlobali{i}=\begin{cases}
\prod_{m=0:m\ne i}^{\nMit }\frac{\eStochi{m}}{\eStochi{m}-\eStochi{i}},&\text{ for stochastic noise,}\\
 \prod_{j=0,j\ne i}^{\nMit} \frac{\sin \frac{\eAnglei{j}}{2}}{\sin\frac{ \eAnglei{j}- \eAnglei{i}}{2}}&\text{ for rotational errors,}
\end{cases}
\end{align}
(where the rotational expression only holds for even \(\nMit\)). The strategy outlined above is for synchronous mitigation strategies, where all the noisy gates are amplified by the same amount. So all the noisy circuits (\(\oCN{C}{}{}{}{i}{}\)) suffer from uniform noise. For asynchronous strategies we amplify each noisy gate by a different amount; so even if the original circuit suffered from uniform noise the variant circuits (\(\oCN{C}{}{}{}{i}{}\)) may not. All concatenated local methods (see \secR{concatenation}) are examples of asynchronous methods but the only global non-synchronous method we consider is identity-insertion-assisted asynchronous mitigation (IIAM) otherwise known as RIIM \cite{He2020,Pascuzzi2022}. In \appSR{IIAM} we provide mitigation coefficients for IIAM that can be used for mitigating both stochastic noise and rotational errors (and perhaps other error types), generalising \citeRef{He2020} (which targets stochastic noise).

\subsubsection{Mitigation Composition}\secL{composition}
Mitigation composition involves (effectively) applying mitigation methods sequentially to the same circuit to strip away one uniform noise layer at a time. The full implementation is discussed in \citeRef{MyLongWork} but we will present a brief example. We assume that we have a circuit (\(\oCN{C}{SN,RE}{}{}{}{}\)) suffering from independent stochastic noise and rotational errors. First, we derive two independent mitigation methods. The first one mitigates stochastic noise (see \eqR{linearMitigationStatesAndCircuits}):
\begin{align}
\oCN{C}{SN}{}{M}{}{}= \sum_{i=0}^{\nMitSN}\coefSNi{i}\oCN{C}{SN}{}{}{i}{} =\oC{C}{}{}{}{}{} +\orderSmall\left(\eExpectP^{\nMitSN+1}\right),
\end{align}
where \(\oCN{C}{SN}{}{M}{}{}\) is the mitigated version of a circuit suffering from stochastic noise, \(\nMitSN\) is the order of the stochastic mitigation, \(\oCN{C}{SN}{}{}{i}{}\) is the \(i\)th noisy circuit variant  (suffering from stochastic noise) that we use in our mitigation, and \(\oC{C}{}{}{}{}{}\) is the noise-free (target) circuit. The second method mitigates rotational errors (see \eqR{linearMitigationStatesAndCircuits}):
\begin{align}
\oCN{C}{RE}{}{M}{}{} = \sum_{j=0}^{\nMitRE}\coefREi{j}\oCN{C}{RE}{}{}{j}{} =\oC{C}{}{}{}{}{}+\orderSmall\left(\eExpectR^{\nMitRE+1}\right),
\end{align}
where \(\oCN{C}{RE}{}{M}{}{}\) is the mitigated version of a circuit suffering from rotational errors, \(\nMitRE\) is the order of the rotational mitigation, and \(\oCN{C}{RE}{}{}{j}{}\) is the \(j\)th noisy circuit variant (suffering from rotational errors) that we use in our mitigation. Now, we define variant-generation procedures \(\oM{G}{SN}{}{}{i}{}\) and \(\oM{G}{RE}{}{}{j}{}\) where:
 \begin{align}\eqL{firstVariantGenerator}
\oCN{C}{SN}{}{}{i}{}=& \oM{G}{SN}{}{}{i}{}\circ \oCN{C}{SN}{}{}{}{},&\oCN{C}{RE}{}{}{j}{}=& \oM{G}{RE}{}{}{j}{}\circ \oCN{C}{RE}{}{}{}{},
 \end{align}
 where \(\oCN{C}{SN}{}{}{}{}\) and \(\oCN{C}{RE}{}{}{}{}\) are the original noisy circuits suffering from stochastic noise or rotational errors, respectively, and \(\oCN{C}{SN}{}{}{j}{}\) and \(\oCN{C}{RE}{}{}{j}{}\) are the \(j\)th amplified (or otherwise modified)  circuits used for the mitigations (see \secRB{LinM}{VGS}). So, for example, \(\oM{G}{RE}{}{}{j}{}\) could be a procedure that amplifies all the rotational-error amplitudes from \(\eAngle\) to \(\eAnglei{j}\) (see \secR{C}), replaces all the gates with rotational errors with their hidden inverses (see \secR{HI}), or applies identity insertion \(j\) times before each gate with a rotational error (see \secR{II}). 

Now, if we have a circuit that suffers from both rotational errors and stochastic noise (\(\oCN{C}{SN,RE}{}{}{}{}\)) then we simply apply the mitigation method for rotational errors followed by\footnote{We assume in \eqRB{composedMitigation}{composedMitigationInformation} that the two constituent mitigation methods commute, \ie \(\oM{G}{SN}{}{}{i_2}{}\) and \(\oM{G}{RE}{}{}{i_1}{}\) commute for all \(i_1\), \(i_2\). This may not be the case if we use identity insertion to amplify the errors. If the mitigation of one type of error changes the amplitude or type of another we must take this into account when applying our subsequent mitigation strategies.} the one for stochastic noise. The composed mitigation strategy is given by:
\begin{align}\eqL{composedMitigation}
\oCN{C}{SN,RE}{}{M}{}{} = \sum_{i=0}^{\nCircGlobal}\coefGlobali{i}\oCN{C}{SN,RE}{}{}{i}{} =\oC{C}{}{}{}{}{}+\orderSmall\left(\eExpectP^{\nMitSN+1},\eExpectR^{\nMitRE+1}\right),
\end{align}
where:
\begin{align}\eqL{composedMitigationInformation}
\oCN{C}{SN,RE}{}{}{i}{}=&\oM{G}{SN}{}{}{i_2}{}\circ\oM{G}{RE}{}{}{i_1}{}\circ\oCN{C}{SN,RE}{}{}{}{},&\coefGlobali{i} = &\coefSNi{i_2}\coefREi{i_1},\\
\nCircGlobal=&\left(\nMitSN+1\right)\left(\nMitRE+1\right)-1&
i_1=&i \bmod\nMitRE,&i_2=&\left\lfloor\frac{i}{\nMitRE}\right\rfloor,
\end{align}
where \(\left\lfloor\frac{i}{\nMitRE}\right\rfloor \) is the integer part of the division of \(i\) by \(\nMitRE\) and \(i \bmod \nMitRE\) is the remainder. 

In general, our sampling cost is the product of the sampling costs of the constituent mitigation methods. So, for our example, we have (\eqR{costScale}):
\begin{align}\eqL{composedCost}
\costGlobal=&\sum_{i=0}^{\nCircGlobal}\left|\coefGlobali{i}\right|=\costSN\costRE,&
\costSN=&\sum_{j=0}^{\nMitSN}\left|\coefSNi{j}\right|,&\costRE=&\sum_{j=0}^{\nMitRE}\left|\coefREi{j}\right|,
\end{align}
and the length scale factor is given by the sum of the length scale factors of the constituent mitigation methods minus one less than the number of constituent mitigation methods, so for our example we have (\eqR{costScale}):
\begin{align}
\oV{F}{Gl,L}{}{}{}{}=&\frac{\sum_{i=0}^{\nCircGlobal}\left|\coefGlobali{i}\right|}{\costGlobal}\frac{\timeExeci{i}}{\timeExec}=\oV{F}{SN,L}{}{}{}{}+\oV{F}{RE,L}{}{}{}{}-1,\\
\oV{F}{SN,L}{}{}{}{}=&\frac{\sum_{j=0}^{\nMitSN}\left|\coefSNi{j}\right|}{\costSN}\frac{\timeExecSNi{j}}{\timeExec},&
\oV{F}{RE,L}{}{}{}{}=&\frac{\sum_{j=0}^{\nMitRE}\left|\coefREi{j}\right|}{\costRE}\frac{\timeExecREi{j}}{\timeExec},
\end{align}
where \(\timeExec\) is the runtime of the unmodified circuit, \(\timeExeci{i}\) is the runtime of the \(i\)th amplified circuit (\(\oCN{C}{SN,RE}{}{}{i}{}\)), \(\timeExecSNi{j}\) is the runtime of the \(j\)th stochastically amplified circuit (\(\oCN{C}{SN}{}{}{j}{}\)), \(\timeExecREi{j}\) is the runtime of the \(j\)th rotationally amplified circuit (\(\oCN{C}{RE}{}{}{j}{}\)).

To the most significant order, the mitigated noise level is the sum of the noise levels of the constituent mitigation strategies, so, for our example:
\begin{align}
\eBiasProxyM = \orderSmall\left(\eExpectP^{\nMitSN+1},\eExpectR^{\nMitRE+1}\right),
\end{align}
assuming \(\eExpectP\) and \(\eExpectR\) are of the same order.

\subsection{Local Mitigation}\secL{LM}
When we apply local mitigation methods we (effectively) mitigate each gate (which could be a composite gate) independently, defining:
\begin{align}\eqL{LocalMitigationDiagram}
\gGN{U}{}{}{M}{}{}~ =~
\gG{}{}{}{}{}{WireVS}\Big\{
\coefLocali{i}:\gGN{U}{}{}{}{i}{}
\Big\}_{i=0}^{\nCircLocal}\hspace{-18 pt}\gG{}{}{}{}{}{WireVS}\gG{}{}{}{}{}{WireVS}~=~
\gG{}{}{}{}{}{WireVS}\gG{U}{}{}{}{}{}\Big\{
\coefLocali{i}:\gC{E}{}{}{}{i}{}
\Big\}_{i=0}^{\nCircLocal}\hspace{-18 pt}\gG{}{}{}{}{}{WireVS}\gG{}{}{}{}{}{WireVS}
~\approx~
\gG{U}{}{}{}{}{},
\end{align}
where the definition of the noisy gate variant \(\oGN{U}{}{}{}{i}{}\) and the values of \(\coefLocali{i}\) will depend on the mitigation method we use (see \tabR{Taxonomy}). \(\oG{U}{}{}{}{}{}\) is the noise-free component of the noisy gate and \(\oC{E}{}{}{}{i}{}\) is the error channel associated to the \(i\)th noisy gate variant. We use \(\nCircLocal\) extra noisy gate variants (in addition to the original noisy gate). The curly brackets denote a quasi-stochastic combination, so the channel \(\oCN{U}{}{}{M}{}{}\) (equivalent to the effective gate we denote \(\oGN{U}{}{}{M}{}{}\)) is given by (\eqR{NoiseAmplifiedGates}):
\begin{align}\eqL{mitigatedGateLocal}
\oCN{U}{}{}{M}{}{} =& \sum_{i=0}^{\nCircLocal}\coefLocali{i} \oCN{U}{}{}{}{i}{}, &\text{ where }&&\oCN{U}{}{}{}{i}{}\left(\rho\right)=&\oC{E}{}{}{}{i}{}\left( \oG{U}{}{}{}{}{}\rho\oGD{U}{}{}{}{}{}\right).
\end{align}
All unbiased local mitigation methods are examples of probabilistic error cancellation (PEC, see \citeRef{Piveteau2022, Temme2017,Endo2018}) but rather than using the universal mitigation gate set of \citeRef{Endo2018} or allowing complete generality (see \citeRef{Takagi2022}), we focus on finding optimal gate sets for particular noise models, building on the work of \citeRef{Mari2021}.

\subsubsection{Eigenvalue Mitigation}\secL{eigenvalueMitigation}
If the transfer matrices (\eg \(\transferMatrixi{\oC{E}{}{}{}{i}{}}\), see \secR{TransferMatrix}) of all our noisy gate variant error channels (\eg \(\oC{E}{}{}{}{i}{}\)) can be diagonalised in the same basis, \ie we have (\eqR{DiagTransfer}):
\begin{align*}
\transferMatrixi{\oC{E}{}{}{}{i}{}}=\transferMatrixW\transferMatrixDi{i}\transferMatrixWd,
\end{align*}
where \(\transferMatrixW\) is independent of \(i\), then we can perform eigenvalue mitigation. \eqR{LocalMitigationDiagram} (with equality instead of approximate equality) is equivalent to the requirement that the quasi-stochastic sum of our error channels is equivalent to the identity channel, which means:
\begin{align}\eqL{LocalMitigation}
\transferMatrixi{\oC{I}{}{}{}{}{}} = &\sum_{i=0}^{\nCircLocal}\coefLocali{i}\transferMatrixi{\oC{E}{}{}{}{i}{}}=\transferMatrixW\left(\sum_{i=0}^{\nCircLocal}\coefLocali{i}\transferMatrixDi{i}\right)\transferMatrixWd&\implies \sum_{i=0}^{\nCircLocal}\coefLocali{i}\transferMatrixDi{i}=&\transferMatrixi{\oC{I}{}{}{}{}{}}&\implies \sum_{i=0}^{\nCircLocal}\coefLocali{i}\eigenTransferi{i,k}=1,
\end{align}
where \(\transferMatrixi{\oC{I}{}{}{}{}{}}\) is the identity-channel transfer matrix, \(\coefLocali{i}\) is the \(i\)th mitigation coefficient, and \(\eigenTransferi{i,k}\) is the \(k\)th eigenvalue of the \(i\)th noisy gate variant used in our mitigation. So we require (at most) one (distinct) noisy gate variant for each eigenvalue of our noisy transfer matrix, for unbiased mitigation.

\subsubsection{Mitigation Concatenation}\secL{concatenation}
To mitigate an entire circuit we must concatenate the mitigation strategies for the individual gates. This can be considered as a special case of mitigation composition (see \secR{composition}) where each of the constituent strategies applies only to a single gate. Concatenated local mitigation (see \cite{Cai2021}) involves determining mitigation coefficients (\(\coefLocali{j}\)) and producing noisy gate variants (\( \oGN{U}{}{}{}{j}{}\)) for each gate independently and then concatenating these (locally) mitigated gates in order to produce a complete mitigated circuit. As an example we consider the case of a noisy circuit (\(\oCN{C}{}{}{}{}{}\)) consisting of two noisy gates:
\begin{align}\eqL{VunitaryFirst}
\gCN{C}{}{}{}{}{}~=&~ \gGN{V}{}{1}{}{}{}\gGN{V}{}{2}{}{}{},
&\gGN{V}{}{1}{}{}{}~=&~\gG{V}{}{1}{}{}{}\gC{E}{}{1}{}{}{},
&\gGN{V}{}{2}{}{}{}~=&~\gG{V}{}{2}{}{}{}\gC{E}{}{2}{}{}{}.
\end{align}
We replace each noisy gate with a mitigated version:
 \begin{align}
 \gGN{V}{}{1}{M}{}{}~=&~
 \gG{}{}{}{}{}{WireVS}
\Big\{
\coefLocali{1,j}: \gGN{V}{}{1}{}{j}{}
\Big\}_{j=0}^{\nCircLocali{1}}
\hspace{-25 pt}\gG{}{}{}{}{}{WireVS}~~~~,
&\gGN{V}{}{2}{M}{}{}=&
 \gG{}{}{}{}{}{WireVS}
\Big\{
\coefLocali{2,j}: \gGN{V}{}{2}{}{j}{}
\Big\}_{j=0}^{\nCircLocali{2}}
\hspace{-25 pt}\gG{}{}{}{}{}{WireVS}~~~~,
 \end{align}
 where \(\nCircLocali{m}\) is the number of extra gate variants required to mitigate the \(m\)th gate and \(\oGN{V}{}{m}{}{j}{}\) is the \(j\)th noisy variant of the \(m\)th gate in our circuit. We then construct the mitigated circuit as the concatenation of the two mitigated gates:
 \begin{align}
\gCN{C}{}{}{M}{}{} ~=&~\gGN{V}{}{1}{M}{}{}\gGN{V}{}{2}{M}{}{}=
\gG{}{}{}{}{}{WireVS}\Big\{
\coefLocali{1,j}: \gGN{V}{}{1}{}{j}{}
\Big\}_{j=0}^{\nCircLocali{1}}
\hspace{-25 pt}\gG{}{}{}{}{}{Wire}
\Big\{
\coefLocali{2,j}: \gGN{V}{}{2}{j}{}{}
\Big\}_{j=0}^{\nCircLocali{2}}
\hspace{-25 pt}\gG{}{}{}{}{}{WireVS}\nonumber\\
=&\Big\{
\coefGlobali{i}: \gCN{C}{}{}{}{i}{} 
\Big\}_{i=0}^{\nCircGlobal},
\end{align}
 where:
\begin{align}
\nCircGlobal=&\left(\nCircLocali{1}+1\right)\left(\nCircLocali{2}+1\right)-1,& \coefGlobali{i} = &\coefLocali{1,i_1}\coefLocali{2,i_2}, &\gCN{C}{}{}{}{i}{}  ~=&~ \gGN{V}{}{1}{}{i_1}{}\gGN{V}{}{2}{}{i_2}{},\nonumber\\
i_1=&i \bmod\nCircLocali{1},&i_2=&\left\lfloor\frac{i}{\nCircLocali{1}}\right\rfloor,
\end{align}
where \(\left\lfloor\frac{i}{\nCircLocali{1}}\right\rfloor \) is the integer part of the division of \(i\) by \(\nCircLocali{1}\) and \(i \bmod \nCircLocali{1}\) is the remainder. \(\nCircGlobal\) is the total number of extra circuits required for the global mitigation.

The sampling cost (see \secR{scalability}) for a concatenated local mitigation with \(\nGate\) gates is given by:
\begin{align}\eqL{concatenatedCost}
\costGlobal=&\prod_{m=1}^{\nGate}\costLocali{m},&
\costLocali{m}=&\sum_{j=0}^{\nCircLocali{m}}\left|\coefLocali{m,j}\right|,
\end{align}
where \(\costLocali{m}\) is the sampling cost for the \(m\)th gate. The length scale factor (see \secR{scalability}) is given by:
\begin{align}\eqL{concatenatedLengthScale}
\oV{F}{Gl,L}{}{}{}{}=&\sum_{m=1}^{\nGate}\oV{F}{L,L}{m}{}{}{}\frac{\timeExecLocali{m}}{\timeExec}, &\oV{F}{L,L}{m}{}{}{}=&\frac{\sum_{j=0}^{\nCircLocali{m}}\left|\coefLocali{m,j}\right|\timeExecLocali{m,j}}{\costLocali{m}\timeExecLocali{m}},&\timeExec=\sum_{m=1}^{\nGate}\timeExecLocali{m},
\end{align}
where \(\oV{F}{L,L}{m}{}{}{}\) is the length scale factor of the \(m\)th gate, \(\timeExecLocali{m}\) is the unmitigated execution time of the \(m\)th gate, \(\timeExecLocali{m,j}\) is the execution time of the \(j\)th noisy variant of the \(m\)th gate, and \(\timeExec\) is the unmitigated execution time of the entire circuit. If we have a uniform noise model and local execution time (\(\timeExec=\timeExecLocal\nGate\) and \(\timeExecLocali{m}=\timeExecLocali{}\) for all \(m\)); such that the sampling costs and length scale factors are identical for every gate (\(\oV{F}{L,L}{m}{}{}{}=\oV{F}{L,L}{}{}{}{}\) and \(\costLocali{m}=\costLocal\) (we could use the geometric mean for rigour)), we find:
\begin{align}
\oV{F}{Gl,L}{}{}{}{}=&\oV{F}{L,L}{}{}{}{}, &\costGlobal=&\costLocal^{\nGate}.
\end{align}
So, if the local cost is linear in the noise amplitude (\(\noiseAmplitude\) is \(\eStoch\) or \(\eAngle\) for stochastic noise or rotational errors, respectively), we find:
\begin{align}\eqL{bFirst}
&\costLocal=1+\bValue\noiseAmplitude+\orderZero\left(\noiseAmplitude^{2}\right)=\mathrm{e}^{\bValue\noiseAmplitude}\left(1+\orderZero\left(\noiseAmplitude^{2}\right)\right)
&\implies \costGlobal=&\mathrm{e}^{\bValue\nGate \noiseAmplitude}\left(1+\orderZero\left(\nGate\noiseAmplitude^{2}\right)\right),
\end{align}
that the global cost grows exponentially with the noise level. So a linear local cost is sufficient to ensure a finite cost in the large gate limit (\(\nGate\rightarrow \infty\)) provided we have a constant noise level. At least for stochastic noise and rotational errors where the noise level is proportional to \(\nGate \noiseAmplitude\) (see \secRC{NL}{SN}{RE}).

In order to determine the most significant contributions to the proxy bias for concatenated mitigation methods we separate our circuit into collections of gates with uniform noise models, determine the most significant term in the proxy bias of each and then add them up to obtain the total noise level (see \secR{composition}). This requires modelling the global effect of our mitigated channels on the circuit (see \secRC{noiseModelling}{fullCircuit}{NL}), which will, in general, be qualitatively different to the original unmitigated channels. If, however, the local mitigation strategy preserves the structure of the channel and only modifies the noise amplitude (\(\noiseAmplitude\) is reduced to \(\noiseAmplitudeM\)) then the leading contribution to the proxy bias is simply the noise level (\eqR{ungainlyExpression}) with \(\noiseAmplitude\) replaced by \(\noiseAmplitudeM\). So we have (see \secRB{SN}{RE}):
\begin{align}\eqL{expectAllLocal}
\eBiasProxyM = \begin{cases}
2\eStochM\nGate+\orderLargeSmall\left(\left(\eStochM\nGate\right)^{2}\right)&\text{ for stochastic noise,} \\
\nGate\left|\eAngleM\right|+\orderLargeSmall\left(\left(\nGate\eAngleM\right)^{2}\right)&\text{ for rotational errors,}
\end{cases}
\end{align}
where \(\eStochM\) is the stochastic-noise amplitude after mitigation and \(\eAngleM\) is the rotational-error amplitude after mitigation.

\subsection{Mitigation Method Catalogue}\secL{MMM}
We now summarise a collection of mitigation methods derived from combinations of the variant-generation procedures discussed above, which form the foundation of our taxonomy. \tabR{Taxonomy} summarises the linear mitigation methods examined in this work, detailing their characteristics and their locations in the appendix. Each method can be combined with local cancellation (see \secR{LC}) when addressing non-stochastic noise. \figR{Taxonomy} contextualises these methods within the overall taxonomy, illustrating that synchronised and concatenated local methods are subset of asynchronous methods, as the latter encompasses all possible approaches for generating circuit variants. Local mitigation methods can also (sometimes) be considered as synchronised mitigation applied to a single gate, this equivalence holds if we use a type preserving procedure (\secR{TPP}). In the following section we will certify our mitigation methods with respect to the criteria in \secR{styles}.

\begin{table}[htbp]
    \centering
    \caption{\textbf{Table of mitigation methods.} The naming convention is explained in \secR{namingConvention}. Key: C-Custom channel (\secR{C}), HI-Hidden Inverses (\secR{HI}), T-Tuned (\secR{TII}), II-Identity Insertion (\secR{II}); AM-Asynchronous Mitigation (\secR{namingConvention}), SM-Synchronous Mitigation (\secR{namingConvention}), LM-Local Mitigation (\secR{LM}); KF-Knowledge Free (\secR{namingConvention}), NA-Noise Aware (\secR{namingConvention}). Where neither NA nor KF are specified the method is noise aware; except for IIAM, which is knowledge free. LC- can be prefixed to the other methods, which indicates we perform local cancellation (\secR{LC}) before we carry out the other method (LC is knowledge free if we have rotational errors and perfect hidden inverses). An unbiased method has no (intrinsic) bias (\secR{proxyBias}) after mitigation so the only bias will be due to uncertainty in the characterisation or implementation of the mitigation.}
    \tabL{Taxonomy}
    \scalebox{0.75}{
    \begin{tabular}{lp{53.5pt}|>{\raggedright}p{40pt}p{34pt}>{\raggedright}p{42.5pt}|cc|>{\raggedright}p{29.5pt}|c|l}
        \hline
Name			& Alternative Acronym 		& Custom Channel	& Hidden Inverse & Identity Insertion	&Local			&Synchronous		& Noise Aware 		& Biased		& Coefficients \\
        \hline
CHILM			& 						&\yesmark			&\yesmark			&\nomark			&\yesmark			&\nomark			&\yesmark			&\nomark		&\appSR{CHILM}\\	
CHISM			& 						&\yesmark			&\yesmark			&\nomark			&\nomark			&\yesmark			&\yesmark			&\nomark		&\appSR{CHISM}\\	
CIILM			& 						&\yesmark			&\nomark			&\yesmark			&\yesmark			&\nomark			&\yesmark			&\nomark		&\appSR{CIILM}\\
CLM				& 						&\yesmark			&\nomark			&\nomark			&\yesmark			&\nomark			&\yesmark			&\nomark		&\appSR{CLM}\\
CSM				& 						&\yesmark			&\nomark			&\nomark			&\nomark			&\yesmark			&\yesmark			&\nomark		&\appSR{CSM}\\
IIAM				& RIIM \cite{He2020}			&\nomark			&\nomark			&\yesmark			&\nomark			&\nomark			&\nomark			&\yesmark		&\appSR{IIAM}\\
IILM:KF			& 						&\nomark			&\nomark			&\yesmark			&\yesmark			&\nomark			&\nomark			&\yesmark		&\appSR{IILM}\\
IILM:NA			& 						&\nomark			&\nomark			&\yesmark			&\yesmark			&\nomark			&\yesmark			&\nomark		&\appSR{IILM}\\
IISM:KF			& FIIM \cite{He2020}			&\nomark			&\nomark			&\yesmark			&\nomark			&\yesmark			&\nomark			&\yesmark		&\appSR{IISM}\\
IISM:NA			& 						&\nomark			&\nomark			&\yesmark			&\nomark			&\yesmark			&\yesmark			&\yesmark		&\appSR{IISM}\\
LC-				& 						&\nomark			&\yesmark			&\nomark			&\yesmark			&\nomark			&\yesmark			&\yesmark		&\appSR{LC}\\
TIILM			& 						&\nomark			&\nomark			&\yesmark			&\yesmark			&\nomark			&\yesmark			&\nomark		&\appSR{TIILM}\\
        \hline		
    \end{tabular}
    }
    
\end{table}

\begin{landscape}
\begin{figure}[htbp]
    \centering
    \scalebox{0.76}{
\begin{tikzpicture}[
roundnode/.style={ellipse, draw=black, fill=white, very thick, minimum size=7mm, align=center, text width=3cm},
squarednode/.style={rectangle, draw=black, fill=white, very thick, minimum size=7mm, align=center, text width=3.7cm},
curvednode/.style={rounded rectangle, draw=black, fill=white, very thick, minimum size=7mm, align=center, text width=3.7cm},
]
\node[curvednode]	(NT)								{Noise Tailoring};

\node[curvednode]	(LT)		[below= of NT]				{Local Tailoring};
\node[curvednode]	(CT)		[right=6 cm of LT]				{\textbf{Global Clifford Twirling \cite{GooglePatent}}};
\node[squarednode]	(GM)	[right=16 cm of LT]			{Global Mitigation\\ (\secR{GM})};

\node[squarednode]	(AM)		[below=of GM]				{Asynchronous Mitigation (AM)\\ (\appSR{scope})};
\node[roundnode]	(RC)		[below=2.5 cm of LT]				{\textbf{Randomised Compiling \cite{Wallman2016}}};
\node[curvednode]	(LC)		[right= of RC]				{\color{blue}{Local \\Cancellation (LC) (\appSR{LC})}};

\node[squarednode]	(SM)		[below=of AM]				{Synchronised Mitigation (SM)\\ (\appSR{scope})};
\node[squarednode]	(CatLM)	[left=6 cm of SM]				{Concatenated Local Mitigation \\ (\cite{Mari2021}; \secR{concatenation})};
\node[roundnode]	(PT)		[below=of RC]				{\textbf{Pauli Twirling \cite{OGorman2016,Bennett1995,Bennett1996,Knill2004}}};
\node[squarednode]	(IIAMcKF)	[right=of SM]				{\textbf{\color{violet}{IIAM (RIIM)}\\ (\cite{He2020,Pascuzzi2022}; \appSR{IIAM})}};

\node[squarednode]	(IISMcKF)	[below=of CatLM]			{\textbf{IISM:KF (FIIM)\\ (\cite{He2020}; \appSR{IISM})}};
\node[squarednode]	(LM)		[left=of IISMcKF]				{Local Mitigation (LM) (\secR{LM})};

\node[squarednode]	(IILMcKF)	[below=of LM]				{\textbf{\color{violet}{IILM:KF} \\ (\appSR{IILM})}};
\node[squarednode]	(PEC)		[left=of IILMcKF]				{Probabilistic Error Cancellation (PEC)\\ \cite{Temme2017,Endo2018}};
\node[squarednode]	(LC-CSM)	[right=of IILMcKF]			{\color{blue}{LC-CSM}};
\node[squarednode]	(CSM)	[right=of LC-CSM]			{\color{violet}{CSM (\cite{Li2017, Tiron2020}; \appSR{CSM})}};
\node[squarednode]	(CHISM)	[right=of CSM]				{\color{violet}{CHISM\\ (\appSR{CHISM})}};
\node[squarednode]	(IISMcNA)	[right=of CHISM]				{\color{violet}{IISM:NA\\ (\appSR{IISM})}};

\node[squarednode]	(CLM)		[below=of CSM]				{\color{violet}{CLM} (\cite{Mari2021, Cai2021, Takagi2022}; \appSR{CLM})};
\node[squarednode]	(LC-CLM)	[below=of LC-CSM]			{\color{blue}{LC-CLM}};
\node[squarednode]	(CHILM)	[below=of CHISM]			{\color{violet}{CHILM\\ (\appSR{CHILM})}};
\node[squarednode]	(CIILMcNA)	[left=of LC-CLM]				{\color{blue}{CIILM \\(\appSR{CIILM})}};
\node[squarednode]	(TIILMcNA)	[left=of CIILMcNA]			{\color{blue}{TIILM}\\(\appSR{TIILM})};

\node[squarednode]	(IILMcNA)	[below= 2.7 cm of IISMcNA]		{\color{violet}{IILM:NA \\ (\cite{Mari2021}; \appSR{IILM})}};

\draw[-] (NT.south) -- (LT.north);
\draw[-] (NT.south) -- (CT.north);
\draw[-] (NT.south) -- (GM.north);

\draw[-] (LT.south)	-- 	(LM.north);
\draw[-] (LT.south)	-- 	(RC.north);
\draw[-] (LT.south)	-- 	(LC.north);

\draw[-] (CatLM.south)	-- 	(LM.north);
\draw[-] (RC.south)	-- 	(PT.north);
\draw[-] (LM.south)	-- 	(PEC.north);
\draw[-] (LM.south)	-- 	(IILMcKF.north);
\draw[dotted] (IISMcKF.south)	-- 	(IILMcKF.north);
\draw[-] (AM.south)	-- 	(SM.north);
\draw[-] (AM.south)	-- 	(CatLM.north);
\draw[-] (GM.south)	-- 	(AM.north);

\draw[-] (AM.south)	-- 	(IIAMcKF.north);
\draw[-] (PEC.south)	-- 	(CLM.north);
\draw[-] (PEC.south)	-- 	(CHILM.north);
\draw[-] (PEC.south)	-- 	(TIILMcNA.north);
\draw[-] (SM.south)	-- 	(CSM.north);
\draw[-] (SM.south)	-- 	(LC-CSM.north);
\draw[-] (SM.south)	-- 	(IISMcKF.north);
\draw[-] (SM.south)	-- 	(IISMcNA.north);

\draw[-] (TIILMcNA.south)	-- 	(IILMcNA.north);
\draw[-] (SM.south)	-- 	(CHISM.north);
\draw[dotted] (CSM.south)	-- 	(CLM.north);
\draw[-] (PEC.south)	-- 	(CIILMcNA.north);
\draw[-] (PEC.south)	-- 	(LC-CLM.north);
\draw[dotted] (LC-CSM.south)	-- 	(LC-CLM.north);

\draw[dotted] (IISMcNA.south)	-- 	(IILMcNA.north);
\draw[dotted] (CHISM.south)	-- 	(CHILM.north);
\end{tikzpicture}}
\caption{Taxonomy of the mitigation methods in \tabR{Taxonomy} (this is far from a complete taxonomy, \eg see \cite{Endo2021,Cai2023a}): Rectangles are for mitigation methods, ellipses for tailoring methods designed to produce Pauli error channels from non-Pauli error channels, and rounded-vertex rectangles are for more general tailoring methods. Text in blue indicates methods developed for this work and text in violet indicates methods adapted or optimised within this work. Methods in bold font are knowledge free. A method that is below and linked to another method with a solid line is a special case of that method, if it is linked with a dotted line it is a special case only when the noise type is preserved by the variant-generation procedure (see \secR{TPP}).}
\figL{Taxonomy}
\end{figure}
\end{landscape}

\section{Case Study: Method Certification}\secL{evaluation}
Here we certify the mitigation methods presented in \tabR{Taxonomy} 
 for stochastic noise (\secR{StochasticCertification}) and rotational errors (\secR{RotationalCertification}). We first present the quantitative metrics (\tabRB{MenagerieStatisticsSN}{MenagerieStatisticsRE}) from \secR{metrics} in appropriate regimes (\secR{regimes}) then use these metrics to certify the methods (\tabRB{QualitativeSN}{QualitativeRE}) with respect to the criteria from \secR{styles}.

We conclude by providing a numerical analysis (\secR{numericalAnalysis}) of the methods for a simple simulated circuit (\secR{benchCircuit}) suffering from stochastic over-rotation errors, to elucidate our analytical discussion.

\subsection{Certification for Stochastic Noise}\secL{StochasticCertification}
We consider uniform closed stochastic noise  (see \secRB{SN}{closure}). We gauge the strength of the noise by the size of \(\eExpectP\) (\eqR{expectStoch}):
\begin{align*}
\eExpectP = &2\eStoch\nGate,
\end{align*}
the noise level (\secR{NL}) of the circuit being mitigated (in the small-noise-level regime, \secR{LowNoiseRegime}). Here \(\eStoch\) is the (gate-wise) noise amplitude, \ie the error probability for each gate, and \(\nGate\) is the number of noisy gates.

In \tabR{MenagerieStatisticsSN} we provide the leading order expressions for the proxy bias (\ie the (mitigated) noise level, see \secR{proxyBiasMitigated}), runtime scaling (\secR{scalability}), and noise boundary (\secR{Thresholds}) in the appropriate regimes (see \secR{regimes}). We then use these quantitative metrics in order to certify our mitigation methods in \tabR{QualitativeSN}.
\begin{table}[bhtbp]
    \centering
     \caption{\textbf{Quantitative Mitigation Metrics for Stochastic Noise.} The noise level (\ie first order term of the proxy bias, see \secR{proxyBias}),  runtime scaling (\secR{scalability}), and noise boundary (\secR{Thresholds}) are recorded for a number of mitigation methods. We consider both the small noise (see \secR{LowNoiseRegime}) and large-circuit (see \secR{LargeCircuitRegime}) regimes. For the runtime (\secR{scalability}) calculations we assume that all gates and custom error channels (\secR{C}) take the same time to implement. \(\nMit\), where given, is the order of the mitigation (see \secR{Thresholds}). All results are for the mitigation of uniform closed (see \secR{closure}) stochastic noise (see \secR{SN}). \(^{*}\) Unmitigated results relate to the original noisy circuit. \(^{**}\) IIAM results are given for the combined large-circuit and small-noise regime, for more information see \appSR{IIAM}. \(^{***}\) IISM results apply to both the knowledge-free (KF) and noise-aware (NA) variants and are given in the small-noise regime. The runtime scaling is given to leading order for \(\nMit\gg 1\), for more information see \citeRef{MyLongWork}. \(^{****}\) \(0\le \aParam\le 1\) is a property of the original noise model (see \eqR{closureSN} of \secR{SN}). Further details of each method can be found in \tabR{Taxonomy} and \figR{Taxonomy} of \secR{taxonomy}. The naming convention is explained in \secR{namingConvention}.}
    \tabL{MenagerieStatisticsSN}
    \scalebox{0.75}{
    \begin{tabular}{l|ll|ll}
        \hline
Metric			&Mitigated Noise Level													&Excess Runtime Scaling							&Runtime Scaling 							& Noise Boundary		\tabularnewline		
			&(\(\noiseLevelM\))													&	(\(\oV{S}{Gl}{}{}{}{}-1\))							&		(\(\oV{S}{Gl}{}{}{}{}\))						&  (\(\oV{e}{B}{}{}{}{}\))			\tabularnewline	
Regime				&Small Noise												&Small Noise									&Large Circuit									&Large Circuit		\tabularnewline
				&	(\(\eExpectP\ll 1\))												&(\(\eExpectP\ll 1\))									& (\(\nGate\gg 1\))										& (\(\nGate\gg 1\))			\tabularnewline
        \hline
Unmitigated\(^{*}\)		&\(\eExpectP\)													&\(0\)												&\(1\)											&\(\eBiasTarget\) 			\\	
CLM			&0															&\(\left(2+\frac{1}{2\nGate}\right)\eExpectP\)					&\(\mathrm{e}^{2\eExpectP}\)							&\(\frac{\ln\left(\scalabilityTarget\right)}{2}\) 			\\
CSM			&0															&\(\left(\frac{8}{3}-\frac{1}{6\nGate^{2}}\right)\nGate\eExpectP\)									&\( 2\cosh^{2}\sqrt{2\nGate\eExpectP}\)				& \(\frac{\ln^2\left(2\scalabilityTarget\right)}{8\nGate}\)			\\
IIAM	\(^{**}\)	&\(\frac{\eExpectP^{\nMit+1}}{\left(\nMit +1\right)!}\)							&\( \frac{\nGate^{2\nMit}}{\left(\nMit!\right)^2}-1\)					& \(\frac{\nGate^{2\nMit}}{\left(\nMit!\right)^2}\)				& \(\eBiasTarget\)			\\
IILM:KF	&\(\frac{\eExpectP^{\nMit+1}}{\nGate^{\nMit}}\left(\frac{1+\aParam}{2}\right)^{\nMit} \frac{\prod_{i=0}^{\nMit}\left(2i+1\right)}{\left(\nMit+1\right)!}\)&\(\ge \frac{3\times4^{\nGate}}{2}-1\)&\(\ge\frac{3\times4^{\nGate}}{2}\)&\(\eBiasTarget\)\\	
IILM:NA			&\(0\)															&\(\frac{3\times4^{\nGate}}{2}-1\)&\(\frac{3\times4^{\nGate}}{2}\mathrm{e}^{\frac{3}{4}\left(1+\aParam\right)\eExpectP}\)	& \(\eBiasTarget\)			\\
IISM\(^{***}\)		&\(\frac{\eExpectP^{\nMit+1}\nGate!\prod_{n=0}^{\nMit }\left(2i+1\right)}{\nGate^{\nMit+1}\left(\nGate-\nMit-1\right)! \left(\nMit+1\right)!} \)& \(\approx\frac{4^{\nMit +1}}{\uppi }-1\)						&  \(\approx\frac{4^{\nMit +1}}{\uppi }\)					&  \(\approx\frac{4^{\frac{\ln \left(2\eBiasTarget\right)}{\ln \left(\uppi\scalabilityTarget\right)}}}{2}\)\\
TIILM\(^{****}\)		&\(0\)	& \(1\)						&  \(\mathrm{e}^{4\left(1+\aParam\right)\eExpectP}\left(1+2\ln2\right)\)					&  \(\frac{\ln \frac{\scalabilityTarget}{1+2\ln 2}}{4\left(1+\aParam\right)}\)\\
        \hline		
    \end{tabular}
    }   
\end{table}

\begin{table}[htbp]
    \centering
    \caption{\textbf{Qualitative Certification Criteria for Stochastic Noise.} We assess each mitigation method against the criteria in \secR{styles}. \yesmark\textsubscript{q} indicates a method is quasi-robust. \yesmark\textsubscript{S} and \yesmark\textsubscript{q}\textsubscript{S} indicate a method is robust or quasi-robust, respectively, in the small-noise regime (see \secR{LowNoiseRegime}). Further details of each method can be found in \tabR{Taxonomy} and \figR{Taxonomy} of \secR{taxonomy}. Notice we have one extra method compared to \tabR{MenagerieStatisticsSN} because IISM:KF and IISM:NA have qualitatively different robustness but identical quantitative metrics. The naming convention is explained in \secR{namingConvention}.}
    \tabL{QualitativeSN}
    \scalebox{1}{
    \begin{tabular}{l|ccccc}
        \hline
Criteria			&Scalable		&Unbounded		& Precise			& Efficient			& Robust\\
        \hline
Unmitigated			&\yesmark		&\nomark		&\nomark			&\yesmark			&\yesmark\\
CLM				&\yesmark		&\yesmark		& \yesmark			& \yesmark			&\yesmark\\
CSM				&\nomark		&\nomark		& \yesmark			& \yesmark			&\yesmark\textsubscript{S}\\
IIAM			&\nomark		&\nomark		& \yesmark			& \nomark			&\yesmark\\
IILM:KF			&\nomark		&\nomark		& \yesmark			& \nomark			&\yesmark\\
IILM:NA			&\nomark		&\nomark		& \yesmark			& \nomark			&\yesmark\textsubscript{q}\\	
IISM:KF			&\yesmark		&\nomark		& \yesmark			& \nomark			&\yesmark\\
IISM:NA			&\yesmark		&\nomark		& \yesmark			& \nomark			&\yesmark\textsubscript{qS}\\
TIILM			&\yesmark		&\yesmark		& \yesmark			& \nomark			&\yesmark\textsubscript{q}\\
        \hline		
    \end{tabular}
    }
\end{table}

As we see in \tabR{QualitativeSN} the only method that meets all the criteria, and therefore can be certified SUPER (see \secR{styles}), is CLM (local mitigation with custom channels, see \appSR{CLM}). So this is the optimal method to use, if it is available. CLM and TIILM are the only methods that could be expected to mitigate large noise levels (\ie of order unity or above, see noise boundary column of \tabR{MenagerieStatisticsSN}). If we lack knowledge about the noise model then we should use IISM:KF but this will only be effective if the noise level is reasonably small (\ie \(\eExpectP<\frac{1}{2}\)) and IISM:KF is much less efficient than CLM (see runtime scaling, small noise column of \tabR{MenagerieStatisticsSN}).

\subsection{Certification for Rotational Errors}\secL{RotationalCertification}
For rotational errors (see \secR{RE}), we gauge the strength of the noise by the size of \(\eExpectR\) (\eqR{expectRot}):
\begin{align*}
\eExpectR = &\nGate\left|\eAngle\right|.
\end{align*}
The noise level (\secR{NL}) of the circuit being mitigated (in the small-noise regime, \secR{LowNoiseRegime}). Here \(\eAngle\) is the (gate-wise) noise amplitude, \ie the rotational error for each gate, and \(\nGate\) is the number of noisy gates.

In \tabR{MenagerieStatisticsRE} we provide the leading order expressions for the  proxy bias (\ie the (mitigated) noise level, see \secR{proxyBias}), runtime scaling (\secR{scalability}), and noise boundary (\secR{Thresholds}) in the appropriate regimes (see \secR{regimes}). We then use these quantitative metrics in order to certify our mitigation methods in \tabR{QualitativeRE}.
\begin{table}[htbp]
    \centering
     \caption{Quantitative Mitigation Metrics: Rotational Errors. The noise level (\ie first order term of the proxy bias, see \secR{proxyBias}),  runtime scaling (\secR{scalability}), and noise boundary (\secR{Thresholds}) are recorded for a number of mitigation methods. We consider both the small noise (see \secR{LowNoiseRegime}) and large-circuit (see \secR{LargeCircuitRegime}) regimes. For the runtime (\secR{scalability}) calculations we assume that all gates and custom error channels (\secR{C}) take the same time to implement. \(\nMit\), where given, is the order of the mitigation  (see \secR{Thresholds}). All results are for the mitigation of uniform rotational errors (see \secR{RE}).  \(^{*}\) Unmitigated results relate to the original noisy circuit. \(^{**}\) IIAM results are given for the combined large-circuit and small-noise regime, for more information see \appSR{IIAM}. \(^{***}\) IISM results apply to both the knowledge-free (KF) and noise-aware (NA) variants and are given in the small-noise regime. The runtime scaling is given to leading order for \(\nMit\gg 1\), for more information see \citeRef{MyLongWork}. \(^{****}\) CHISM and CSM results are shown for \(\eExpectR\le \frac{\uppi}{2}\). We do not have analytic results for larger noise levels for these methods but the best strategy would be to concatenate (\secRB{composition}{concatenation}) multiple instances together, splitting the noisy gates into sets with lower noise levels; so that, for each set, \(\eExpectR\le \frac{\uppi}{2}\). For CSM to reach the boundary (\(\oV{e}{B}{}{}{}{}=\frac{\uppi}{2}\)) we require \(\scalabilityTarget\ge \frac{3+\sqrt{10}}{2}\) (see \citeRef{MyLongWork}). \(^{*****}\) The LC-CSM noise boundary is only valid for large noise levels (\(\eExpectR\gg 1\), see \citeRef{MyLongWork}). Further details of each method can be found in \tabR{Taxonomy} and \figR{Taxonomy} of \secR{taxonomy}.  The naming convention is explained in \secR{namingConvention}.}
    \tabL{MenagerieStatisticsRE}
    \scalebox{0.65}{
    \begin{tabular}{l|ll|ll}
        \hline
Metric			&Mitigated Noise Level 													&Excess Runtime Scaling							&Runtime Scaling 							& Noise Boundary		\tabularnewline		
			&(\(\noiseLevelM\))													&	(\(\oV{S}{Gl}{}{}{}{}-1\))							&		(\(\oV{S}{Gl}{}{}{}{}\))						&  (\(\oV{e}{B}{}{}{}{}\))			\tabularnewline	
Regime				&Small Noise												&Small Noise									&Large Circuit									&Large Circuit		\tabularnewline
				&	(\(\eExpectP\ll 1\))												&(\(\eExpectP\ll 1\))									& (\(\nGate\gg 1\))										& (\(\nGate\gg 1\))			\tabularnewline
        \hline
Unmitigated\(^{*}\)	&\(\eExpectR\)																	&\(0\)																&\(1\)												&\(\eBiasTarget\) 								\tabularnewline	
CHILM			&0																		&\(\left(1+\frac{1}{4\nGate}\right)\frac{\eExpectR^{2}}{\nGate}\)								&\(1+\frac{\oV{e}{RE}{}{}{}{}^{2}}{\nGate}\)												&\(\sqrt{\nGate\left(\scalabilityTarget-1\right)}\) 				\tabularnewline	
CHISM\(^{****}\)	&0																		&\(\left(\frac{4}{3}-\frac{1}{12\nGate^{2}}\right)\eExpectR^{2}\)								&\(\frac{3}{2\cos^{2}\eExpectR}-\frac{\tan\eExpectR}{2\eExpectR}\)				&\(\frac{\uppi}{2}-\sqrt{\frac{3}{2\scalabilityTarget}}\) 				\tabularnewline	
CIILM			&0																		&\(5\)																&\(6\mathrm{e}^{4\sqrt{3}\eExpectR}\)							& \(\frac{\ln\left(\frac{\scalabilityTarget}{6}\right)}{4\sqrt{3}}\) 			\tabularnewline
CLM			&0																		&\(\left(2+\frac{1}{\nGate}\right)\eExpectR\)											&\(\mathrm{e}^{2\eExpectR}\)								&\(\frac{\ln\scalabilityTarget}{2}\)	 				\tabularnewline
CSM\(^{****}\)	&0																		&\(3\eExpectR\)														&\(1+\sin\eExpectR\left(3\cos \eExpectR+\sin \eExpectR\right)\)				&\(\frac{\uppi}{2}\) 								\tabularnewline
IIAM\(^{**}\)		&\(\frac{\eExpectR^{\nMit+1}}{\left(\nMit +1\right)!}\)											&\( \frac{\nGate^{2\nMit}}{\left(\nMit!\right)^2}-1\)										& \(\frac{\nGate^{2\nMit}}{\left(\nMit!\right)^2}\)						& \(\eBiasTarget\)										\tabularnewline
IILM:KF			&\(\frac{\eExpectR^{\nMit+1}}{\nGate^{\nMit}}\frac{\prod_{i=0}^{\nMit}\left(2i+1\right)}{\left(\nMit+1\right)!}\)				&\(\ge \frac{3\times4^{\nGate}}{2}-1\)											&\(\ge \frac{3\times4^{\nGate}}{2}\)								&\(\eBiasTarget\)										\tabularnewline	
IILM:NA		&0																		&\(\frac{15}{7}\times\left(\frac{7}{2}\right)^{2\nGate}-1\)									&\(\frac{15}{7}\times\left(\frac{7}{2}\right)^{2\nGate}\)					&\(\eBiasTarget\)										\tabularnewline
IISM\(^{***}\)		&\(\frac{\eExpectR^{\nMit+1}\nGate!\prod_{n=0}^{\nMit }\left(2i+1\right)}{\nGate^{\nMit+1}\left(\nGate-\nMit-1\right)! \left(\nMit+1\right)!} \)							& \(\approx\frac{4^{\nMit +1}}{\uppi }-1\)											& \(\approx\frac{4^{\nMit +1}}{\uppi }\)							&  \(\approx\frac{4^{\frac{\ln \left(\eBiasTarget\right)}{\ln \left(\uppi\scalabilityTarget\right)}}}{2}\)\tabularnewline
LC-			&\(\frac{\eExpectR^{2}}{2\nGate}\)														&\(0\)																& \(1\)												&\(\sqrt{2\nGate\eBiasTarget}\)							\tabularnewline
LC-CLM			&0																		&\(\left(1+\frac{1}{4\nGate}\right)\frac{\eExpectR^{2}}{\nGate}\)								&\(1+\frac{\eExpectR^{2}}{\nGate}\)												&\(\sqrt{\nGate\left(\scalabilityTarget-1\right)}\)					\tabularnewline
LC-CSM\(^{*****}\)		&0																		&\(\left(\frac{4}{3}-\frac{1}{12\nGate^{2}}\right)\eExpectR^{2}\)								&\( \left(2-\frac{\tanh\eExpectR}{\eExpectR}\right)\cosh^{2} \eExpectR\) 			& \(\frac{\ln\left(2\scalabilityTarget\right)}{2}\)			\tabularnewline
LC-IIAM		&\(\frac{\eExpectR^{2\left(\nMit+1\right)}}{\left(2\nGate\right)^{\nMit+1}\left(\nMit +1\right)!}\)						&\( \frac{\nGate^{2\nMit}}{\left(\nMit!\right)^2}-1\)										& \(\frac{\nGate^{2\nMit}}{\left(\nMit!\right)^2}\)						& \(\eBiasTarget\)										\tabularnewline
LC-IISM\(^{***}\)	&\(\frac{\eExpectR^{2\left(\nMit+1\right)}\nGate!\prod_{n=0}^{\nMit }\left(2i+1\right)}{2^{\nMit+1}\nGate^{2\left(\nMit+1\right)}\left(\nGate-\nMit-1\right)! \left(\nMit+1\right)!} \)	& \(\approx\frac{4^{\nMit +1}}{\uppi }-1\)											& \(\approx\frac{4^{\nMit +1}}{\uppi }\)							&  \(\approx\sqrt{\nGate 4^{\frac{\ln \left(\eBiasTarget\right)}{\ln \left(\uppi\scalabilityTarget\right)}}}\) \tabularnewline
LC-TIILM	&\(0\)	& \(1\)											& \(2\left(1+\sqrt{\frac{2}{\nGate}}\eExpectR\right)\)							&  \(\sqrt{\frac{\nGate}{2}}\left(\frac{\scalabilityTarget}{2}-1\right)\) \tabularnewline
TIILM			&\(0\)																		& \(1\)																& \(\mathrm{e}^{2\eExpectR}\left(1+\uppi\right)\)						&  \(\frac{\ln\frac{\scalabilityTarget}{1+\uppi}}{2}\) \tabularnewline
        \hline		
    \end{tabular}
    }   
\end{table}

\begin{table}[htbp]
    \centering
    \caption{Qualitative Certification Criteria: Rotational Errors. We assess each mitigation method against the criteria in \secR{styles}. \yesmark\textsubscript{q} indicates a method is quasi-robust. \yesmark\textsubscript{S} and \yesmark\textsubscript{q}\textsubscript{S} indicate a method is robust or quasi-robust, respectively, in the small-noise regime (see \secR{LowNoiseRegime}). \(^{*}\) Unmitigated results relate to the original noisy circuit. \(^{****}\) CSM and CHISM are unbounded provided we are willing to concatenate (\secRB{composition}{concatenation}) multiple instances together to deal with large noise levels. \(^{*****}\) results for CIILM are shown using the same custom channel for the noisy gate and its inverse, if we use separate channels for each we obtain lower sampling costs and an efficient mitigation, see 
 \citeRef{MyLongWork}. Further details of each method can be found in \tabR{Taxonomy} and \figR{Taxonomy} of \secR{taxonomy}.  Notice we have one extra method compared to \tabR{MenagerieStatisticsRE} because IISM:KF and IISM:NA have qualitatively different robustness but identical quantitative metrics. The naming convention is explained in \secR{namingConvention}.}
    \tabL{QualitativeRE}
    \scalebox{0.95}{
    \begin{tabular}{l|ccccc}
        \hline
Criteria			&Scalable		&Unbounded		& Precise		& Efficient					& Robust \\
        \hline
Unmitigated\(^{*}\)	&\yesmark		&\nomark		&\nomark		&\yesmark					&\yesmark\\
CHILM			&\yesmark		&\yesmark		& \yesmark		& \yesmark					&\yesmark\textsubscript{q}\\	
CHISM\(^{****}\)	&\yesmark		&\yesmark		& \yesmark		& \yesmark					&\yesmark\textsubscript{q}\\	
CIILM\(^{*****}\)	&\yesmark		&\yesmark		& \yesmark		& \nomark					&\yesmark\textsubscript{q}\\
CLM			&\yesmark		&\yesmark		& \yesmark		& \yesmark					&\yesmark\\
CSM\(^{****}\)	&\yesmark		&\yesmark		& \yesmark		& \yesmark					&\yesmark\textsubscript{S}\\
IIAM			&\nomark		&\nomark		& \yesmark		& \nomark					&\yesmark\\
IILM:KF			&\nomark		&\nomark		& \yesmark		& \nomark					&\yesmark\\
IILM:NA		&\nomark		&\nomark		& \yesmark		& \nomark					&\yesmark\textsubscript{q}\\
IISM:KF		&\yesmark		&\nomark		& \yesmark		& \nomark					&\yesmark\\
IISM:NA		&\yesmark		&\nomark		& \yesmark		& \nomark					&\yesmark\textsubscript{q}\textsubscript{S}\\
LC-			&\yesmark		&\yesmark		& \nomark		& \yesmark					&\yesmark\textsubscript{q}\\
LC-CLM			&\yesmark		&\yesmark		& \yesmark		& \yesmark					&\yesmark\textsubscript{q}\\
LC-CSM			&\yesmark		&\yesmark		& \yesmark		& \yesmark					&\yesmark\textsubscript{q}\textsubscript{S}\\
LC-IIAM		&\nomark		&\nomark		& \yesmark		& \nomark					&\yesmark\textsubscript{q}\\
LC-IISM			&\yesmark		&\yesmark		& \yesmark		& \nomark					&\yesmark\textsubscript{q}\textsubscript{S}\\
LC-TIILM		&\yesmark		&\yesmark		& \yesmark		& \nomark					&\yesmark\textsubscript{q}\\
TIILM			&\yesmark		&\yesmark		& \yesmark		& \nomark					&\yesmark\textsubscript{q}\\
        \hline		
    \end{tabular}
    }
\end{table}

As we see in \tabR{QualitativeRE} there are a number of methods that meet all criteria, and therefore can be certified SUPER (see \secR{styles}). As for stochastic noise, CLM is featured but CHILM and LC-CLM are much more efficient and have noise boundaries with much better scaling (see runtime scaling and noise boundary column of \tabR{MenagerieStatisticsRE}). In fact, for a fixed noise level the sampling costs of both of these methods tends to unity in the large-circuit limit (\(\nGate\rightarrow \infty\)), so they would be effectively free in this limit. If we lack knowledge about the noise model then we could use IISM:KF but this will only be effective if the noise level is reasonably small (\ie \(\eExpectR<\frac{1}{2}\)). A better choice would be LC-IISM:KF (if it is available) as this has a noise boundary that scales with the square root of the number of gates  (see noise boundary column of \tabR{MenagerieStatisticsRE}). So, though not efficient, it satisfies all the other criteria. It is worth noting that LC used on its own (see LC- row of \tabR{MenagerieStatisticsRE}) gives unbiased results in the large-circuit limit (\(\nGate\rightarrow \infty\), for fixed \(\eExpectR\)), with no excess sampling cost.

\subsection{Numerical Analysis}\secL{numericalAnalysis}
For our numerical analysis we simulate the mitigation of the circuit in \secR{benchCircuit}, suffering from (stochastic) over-rotation errors (see \secR{ORE}), for a range of mitigation methods. Details about the simulation methods used can be found in \appSR{numerical}. Our numerical results support our analytical analysis. 

\subsubsection{Benchmarking Circuit}\secL{benchCircuit}
The circuit we use for benchmarking is:
\begin{align}\eqL{depolarisingFullCircuit}
\scalebox{0.665}{\(\displaystyle 
\begin{aligned}
&\gS{0}{}{}{}{}{Blank}\\[-0.76001 em]
&\gS{0}{}{}{}{}{Blank}\\[-0.76001 em]
&\gS{0}{}{}{}{}{Blank}\\[-0.76001 em]
&\gS{0}{}{}{}{}{Blank}
\end{aligned}
\hspace{-1.1 em}
\left(\begin{aligned}
&\gGN{\yOp}{}{}{}{}{}	\gGN{\tOp}{}{1}{}{}{}	\gGN{}{}{}{}{}{2T}		\gG{}{}{}{}{}{Wire}		\gG{}{}{}{}{}{Wire}		\gG{}{}{}{}{}{Wire}	\gG{}{}{}{}{}{Wire}	\gG{}{}{}{}{}{Wire}	\gG{}{}{}{}{}{Wire}	\gG{}{}{}{}{}{Wire}	\gG{}{}{}{}{}{Wire}	\gG{}{}{}{}{}{Wire}	\gG{}{}{}{}{}{Wire}	\gG{}{}{}{}{}{Wire}	\gG{}{}{}{}{}{Wire}		\gGN{}{}{1}{}{}{2T}	\gGDN{\tOp}{}{1}{}{}{}	\gGN{\yOp}{}{}{}{}{}			\\[-0.76001 em]
&\gG{}{}{}{}{}{Wire}	\gG{}{}{}{}{}{Wire}		\gGN{\sOp}{}{15}{}{}{2B}	\gGN{\tOp}{}{1}{}{}{}	\gGN{}{}{}{}{}{2T}		\gG{}{}{}{}{}{Wire}	\gG{}{}{}{}{}{Wire}	\gG{}{}{}{}{}{Wire}	\gG{}{}{}{}{}{Wire}	\gG{}{}{}{}{}{Wire}	\gG{}{}{}{}{}{Wire}	\gG{}{}{}{}{}{Wire}	\gG{}{}{}{}{}{Wire}	\gGN{}{}{}{}{}{2T}	\gGDN{\tOp}{}{1}{}{}{}	\gGDN{\sOp}{}{15}{}{}{2B}		\gG{}{}{}{}{}{Wire}		\gG{}{}{}{}{}{Wire}
	\\[-0.76001 em]
&\gG{}{}{}{}{}{Wire}	\gG{}{}{}{}{}{Wire}		\gG{}{}{}{}{}{Wire}		\gG{}{}{}{}{}{Wire}		\gGN{\sOp}{}{15}{}{}{2B}	\gGN{\tOp}{}{1}{}{}{}		\gGN{}{}{}{}{}{2T}			\gG{}{}{}{}{}{Wire}	\gG{}{}{}{}{}{Wire}	\gG{}{}{}{}{}{Wire}	\gG{}{}{}{}{}{Wire}	\gGN{}{}{}{}{}{2T}			\gGDN{\tOp}{}{1}{}{}{}		\gGDN{\sOp}{}{15}{}{}{2B}		\gG{}{}{}{}{}{Wire}	\gG{}{}{}{}{}{Wire} 	\gG{}{}{}{}{}{Wire}	\gG{}{}{}{}{}{Wire}	\\[-0.76001 em]
&\gG{}{}{}{}{}{Wire}\gG{}{}{}{}{}{Wire}	\gG{}{}{}{}{}{Wire}	\gG{}{}{}{}{}{Wire}	\gG{}{}{}{}{}{Wire}	\gG{}{}{}{}{}{Wire}	\gGN{\sOp}{}{15}{}{}{2B}		\gGN{\tOp}{}{1}{}{}{}		\gGN{\yOp}{}{}{}{}{}			\gGN{\yOp}{}{}{}{}{}			\gGDN{\tOp}{}{1}{}{}{}		\gGDN{\sOp}{}{15}{}{}{2B}	 	\gG{}{}{}{}{}{Wire}	\gG{}{}{}{}{}{Wire}	\gG{}{}{}{}{}{Wire}	\gG{}{}{}{}{}{Wire} 	\gG{}{}{}{}{}{Wire}	\gG{}{}{}{}{}{Wire}
\end{aligned}\right)^{\nRepeat}
\hspace{-0.9 em}
\begin{aligned}
&\gG{\obsOp}{}{}{}{}{Meas}\\[-0.76001 em]
&\gG{}{}{}{}{}{Blank}\\[-0.76001 em]
&\gG{}{}{}{}{}{Blank}\\[-0.76001 em]
&\gG{}{}{}{}{}{Blank}
\end{aligned}\)}
\end{align}
where \(\xVac\) is the \(+1\) eigenstate of the Pauli Z operator; \(\nRepeat\) is the number of sets of gates we use in our circuit; X, Y, and Z are the standard Pauli operators; and:
\begin{align}\eqL{benchGateSet}
\tOpi{1}= &\mathrm{e}^{\mathrm{i}\frac{\uppi}{8}\xOp},&\sOpi{15}=&\mathrm{e}^{\mathrm{i}\frac{\uppi}{4}\zOp\otimes \zOp},
\end{align}
where the \(\sOpi{15}\) acts on two qubits, in this case always nearest neighbours. We define our benchmarking bias as:
\begin{align}\eqL{biasScaleBench}
\eBias = &\sqrt{\frac{\eBiasi{\xOp}^{2}+\eBiasi{\yOp}^{2}+\eBiasi{\zOp}^{2}}{3}},
\end{align}
where (see \eqRB{bias}{observablesFullSet}):
\begin{align}\eqL{operatorBiasCalc}
\eBiasi{\obsOp}=\traceOp{\obsOp\left(\oC{C}{}{}{}{}{}\left(\initialState\right)-\oCN{C}{}{}{M}{}{}\left(\initialState\right)\right)},
\end{align}
where  \(\initialState=\bigotimes_{n=1}^{4}\xVac\) is the initial state, \(\oC{C}{}{}{}{}{}\) is the channel corresponding to the noise-free circuit, \(\oCN{C}{}{}{M}{}{}\) is the (effective) channel corresponding to the mitigated circuit (see \secR{LinM}), and \(\obsOp\) corresponds to the Pauli X, Y, or Z operator measured on the first qubit. We use \eqR{biasScaleBench} rather than the fidelity (as suggested in \citeRef{Cai2023b}) since our noise-free state (\(\oC{C}{}{}{}{}{}\left(\initialState\right)=\initialState\)) is a product state of \(+1\) Z eigenstates, so the fidelity would not adequately account for errors in other degrees of freedom. The runtime scaling is determined in the negligible initialisation and readout time limit (see \eqR{scalability}):
\begin{align}\eqL{scalabilityBench}
\scalabilityGlobal =\costGlobal^{2}\oV{F}{Gl,L}{}{}{}{}.
\end{align}
For our numerical analysis we consider two different length circuits, with \(\nRepeat=1\) (18 gates) and \(\nRepeat=10\) (180 gates), to demonstrate the scalability (or otherwise) of our methods. Scalable methods are those where the runtime scaling is independent of the number of gates (\(\nGate\)) for a fixed noise level (\(\eExpectP\) or \(\eExpectR\)). 

We choose these circuit and observable definitions because it ensures that all the error channels contribute to the final bias and thus all must be mitigated to mitigate the circuit completely.

\subsubsection{Noise Model}\secL{noiseModel}
For our numerical analysis we use a uniform (stochastic-)over-rotation-error (see \secR{ORE}) noise model. So every gate is followed by a stochastic-noise channel (\secR{SN}), with noise amplitude \(\eStoch\), and a rotational-error channel (\secR{RE}), with noise amplitude \(\eAngle\). These error channels are both generated by the same operator as the gate itself; \ie an error channel generated by \(\xOp\) follows each \(\tOpi{1}\) (acting on the same qubit), an error channel generated by \(\yOp\) follows each \(\yOp\) (acting on the same qubit), and an error channel generated by \(\pOpi{15}=\zOp\otimes\zOp\) follows \(\sOpi{15}\) (acting on the same qubits). 

To further simplify the analysis we consider the two extremes: pure stochastic noise, \ie \(\eAngle=0\) (see \secR{BenchSN}), and pure rotational errors, \ie \(\eStoch=0\) (see \secR{BenchRE}). All rotational errors act in the same direction\footnote{If the inverse gates had rotational-error channels that were the inverse of those of the forward acting gates then these error channels would cancel and we would need only to mitigate the \(\addnoise{\yOp}\) gates (which are self-inverse) to mitigate the entire circuit. This is a known disadvantage of using a circuit that is equivalent to the identity in the absence of noise.} (\ie have the same sign). 

\subsubsection{Marking the Bench: Stochastic Noise}\secL{BenchSN}
In \figRB{fullCircuitPlotSA}{fullCircuitPlotSB} we plot the runtime scaling (\eqR{scalabilityBench}) against the bias (\eqR{biasScaleBench}) for various mitigation methods, demonstrating the superiority of CLM for almost all regimes.

The vertical black lines in \figRB{fullCircuitPlotSA}{fullCircuitPlotSB} show the bias before mitigation, horizontal lines are the runtime scalings of unbiased mitigation methods, and each point represents a different order of mitigation (\(\nMit\)) for the biased methods (some biased methods become unbiased for sufficiently large \(\nMit\)). Increasing \(\nMit\), generally, results in lower bias but higher sampling cost. Some methods have both unbiased and biased variants. Methods not featured have runtime scalings outside the range of the plot. For example, IILM has a very large runtime scaling (see \tabR{MenagerieStatisticsSN}). For scalable methods (see \secR{styles}) the runtime scaling for a mitigation with a given bias should not increase with the number of gates, for a constant noise level, so the top and bottom plots of \figRB{fullCircuitPlotSA}{fullCircuitPlotSB} (with the same noise level: \(\eExpectP\)) should look the same for scalable methods.

This numerical analysis also enables us to study mitigation methods that we have not yet derived analytic formula for. We consider a biased version of CSM, where we use \(\nMit<\nGate\) (see \secRB{C}{GlobalMitigation} and \eqRB{biasedMitSN}{biasedMitRE}), to compliment the unbiased version that features in \tabRB{MenagerieStatisticsSN}{QualitativeSN}. At present there is not an analytic expression for the proxy bias of this method but the numerical results look promising. Though the unbiased method is not scalable for stochastic noise (see \tabRB{MenagerieStatisticsSN}{QualitativeSN}), the biased version may well be, since the runtime scaling required for a given residual bias (bias after mitigation) seems to depend only on the noise level (\(\eExpect=\eExpectP\)) and not the number of gates (\(\nGate=18\nRepeat\)), see \figRB{fullCircuitPlotSA}{fullCircuitPlotSB}.

\begin{figure}[htbp]
	\begin{subfigure}[b]{\linewidth*\real{0.9}}
		\centering
		\includegraphics[width=\linewidth]{
		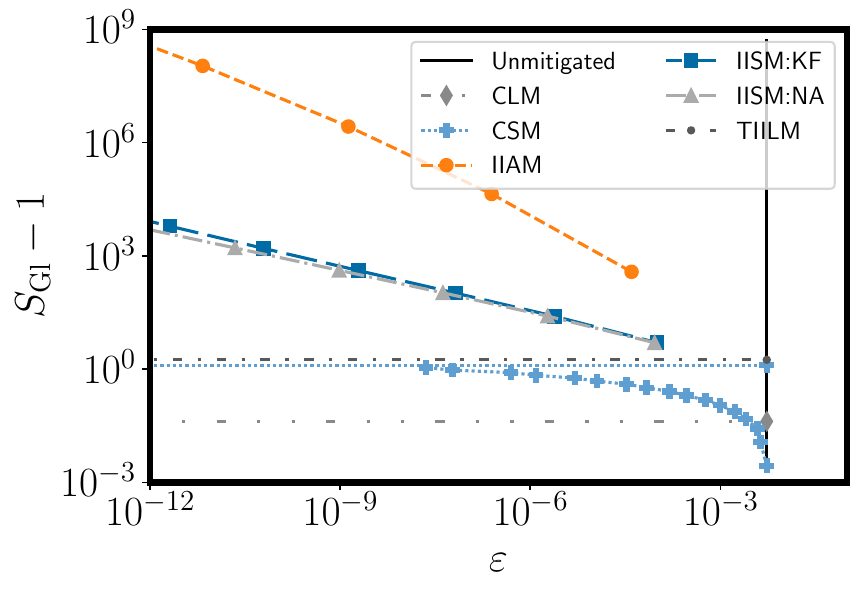
		}
		\caption{ \(\eExpectP=0.02\), \(\eExpectR=0\), \(\nRepeat=1\) (for TIILM \(m_0=0\), \(m_1=246\)).}
		\figL{FCD001r1}
	\end{subfigure}\\
	\begin{subfigure}[b]{\linewidth*\real{0.9}}
		\centering
		\includegraphics[width=\linewidth]{
		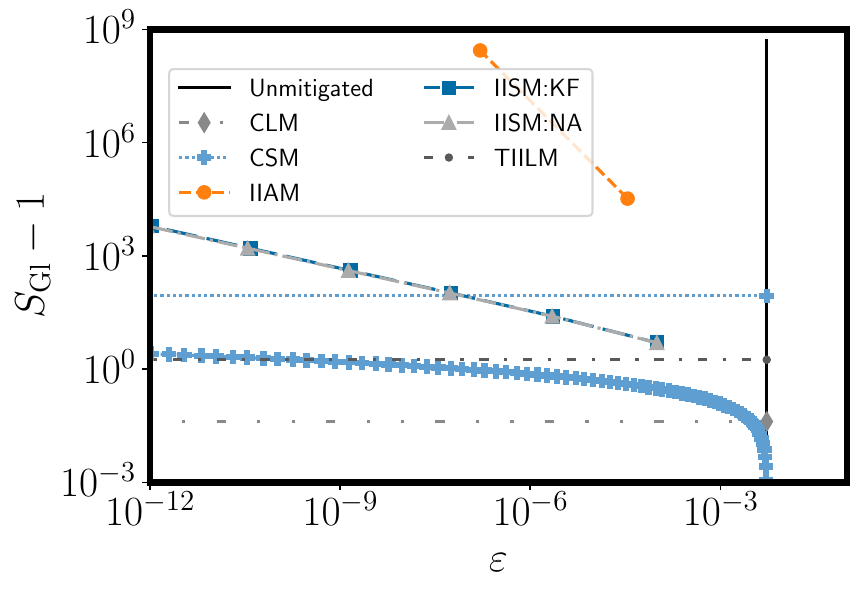
		}
		\caption{ \(\eExpectP=0.02\), \(\eExpectR=0\), \(\nRepeat=10\) (for TIILM \(m_0=0\), \(m_1=2478\)).}
		\figL{FCD001r10}
	\end{subfigure}
	\caption{Runtime scaling (\eqR{scalabilityBench}) versus bias (\eqR{biasScaleBench}) for a circuit (\secR{benchCircuit}) with stochastic noise (see \secRB{SN}{noiseModel}) with noise amplitude \(\eStoch=\frac{\eExpectP}{2\nGate}\), where \(\nGate=18\nRepeat\) is the number of gates. The vertical black lines show the bias before mitigation, horizontal lines are the sampling costs of unbiased methods, and each point represents a different order of mitigation (\(\nMit\)). The methods from \tabR{MenagerieStatisticsSN} that are not featured have runtime scalings higher than \(10^9\), so they do not fit in the plot. \(m_0\) and \(m_1\) are the parameters used for TIILM (see \secR{TII}).}
	\figL{fullCircuitPlotSA}
\end{figure}
\begin{figure}[htbp]
	\begin{subfigure}[b]{\linewidth*\real{0.9}}
		\centering
		\includegraphics[width=\linewidth]{
		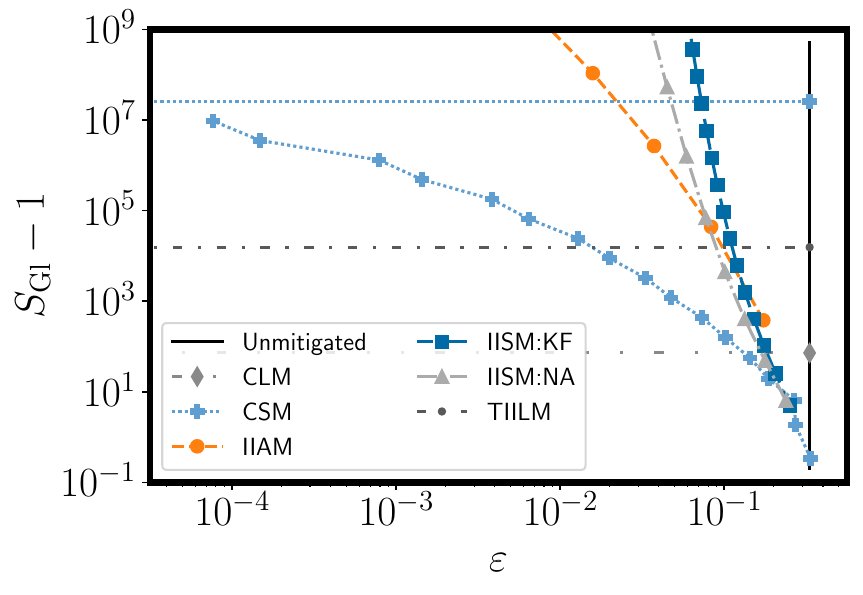
		}
		\caption{ \(\eExpectP=2\), \(\eExpectR=0\), \(\nRepeat=1\) (for TIILM \(m_0=0\), \(m_1=16\)).}
		\figL{FCD1r1}
	\end{subfigure}\\
	\begin{subfigure}[b]{\linewidth*\real{0.9}}
		\centering
		\includegraphics[width=\linewidth]{
		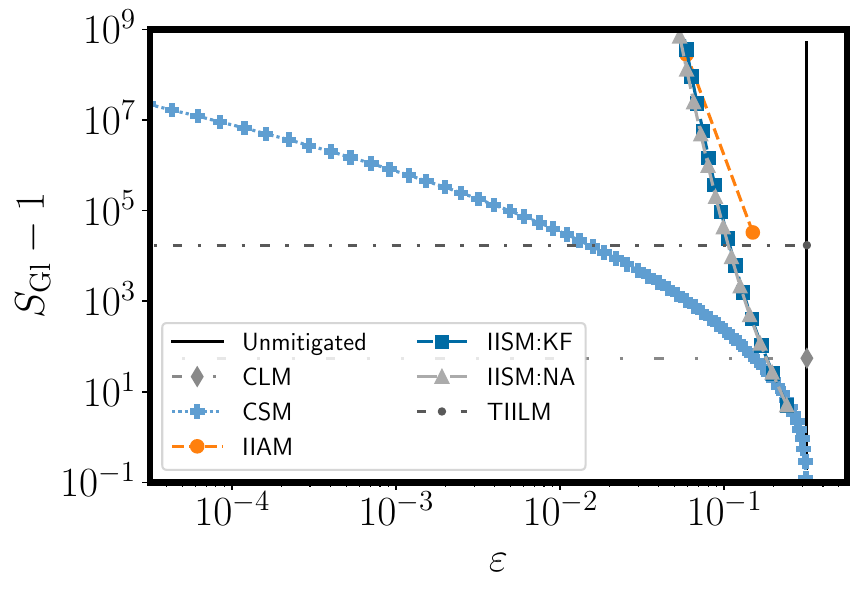
		}
		\caption{\(\eExpectP=2\), \(\eExpectR=0\), \(\nRepeat=10\) (for TIILM \(m_0=0\), \(m_1=169\)).}
		\figL{FCD1r10}
	\end{subfigure}
	\caption{Runtime scaling (\eqR{scalabilityBench}) versus bias (\eqR{biasScaleBench}) for a circuit (\secR{benchCircuit}) with stochastic noise (see \secRB{SN}{noiseModel}) with noise amplitude \(\eStoch=\frac{\eExpectP}{2\nGate}\), where \(\nGate=18\nRepeat\) is the number of gates. The vertical black lines show the bias before mitigation, horizontal lines are the sampling costs of unbiased methods, and each point represents a different order of mitigation (\(\nMit\)). The methods from \tabR{MenagerieStatisticsSN} that are not featured have runtime scalings higher than \(10^9\), so they do not fit in the plot. \(m_0\) and \(m_1\) are the parameters used for TIILM (see \secR{TII}).}
	\figL{fullCircuitPlotSB}
\end{figure}

\subsubsection{Marking the Bench: Rotational Errors}\secL{BenchRE}
In \figRB{fullCircuitPlotRA}{fullCircuitPlotRB} we plot the runtime scaling (\eqR{scalabilityBench}) against the bias (\eqR{biasScaleBench}) for various mitigation methods, demonstrating the superiority of CHILM for almost all regimes.

\begin{figure}[htbp]
	\begin{subfigure}[b]{\linewidth*\real{0.9}}
		\centering
		\includegraphics[width=\linewidth]{
		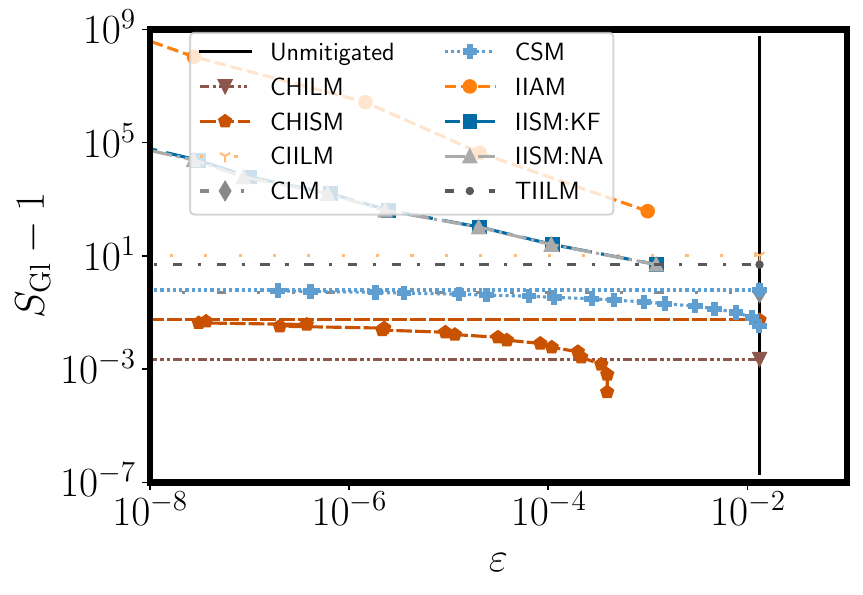
		}
		\caption{ \(\eExpectP=0\), \(\eExpectR=0.2\), \(\nRepeat=1\) (for TIILM \(m_0=0\), \(m_1=58\), \(m_2=200\)).}
		\figL{FCO01r1}
	\end{subfigure}\\
	\begin{subfigure}[b]{\linewidth*\real{0.9}}
		\centering
		\includegraphics[width=\linewidth]{
		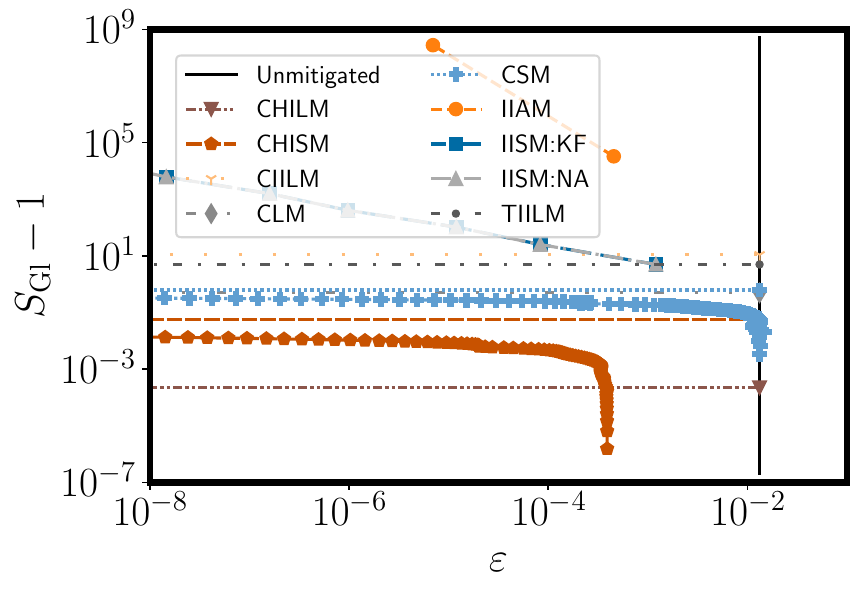
		}
		\caption{ \(\eExpectP=0\), \(\eExpectR=0.2\), \(\nRepeat=10\) (for TIILM \(m_0=0\), \(m_1=589\), \(m_2=2002\)).}
		\figL{FCO01r10}
	\end{subfigure}
	\caption{Runtime scaling (\eqR{scalabilityBench}) versus bias (\eqR{biasScaleBench}) for a circuit (\secR{benchCircuit}) with rotational errors (see \secRB{RE}{noiseModel}) with noise amplitude \(\eAngle=\frac{\eExpectR}{\nGate}\), where \(\nGate=18\nRepeat\) is the number of gates. The vertical black lines show the bias before mitigation, horizontal lines are the sampling costs of unbiased methods, and each point represents a different order of mitigation (\(\nMit\)). The methods from \tabR{MenagerieStatisticsRE} that are not featured have runtime scalings higher than \(10^9\), so they do not fit in the plot; or use LC pre-tailoring and are featured in \figR{fullCircuitPlotLCRA}. \(m_0\), \(m_1\), and \(m_2\) are the parameters used for TIILM (see \secR{TII}).}
	\figL{fullCircuitPlotRA}
\end{figure}
\begin{figure}[htbp]
	\begin{subfigure}[b]{\linewidth*\real{0.9}}
		\centering
		\includegraphics[width=\linewidth]{
		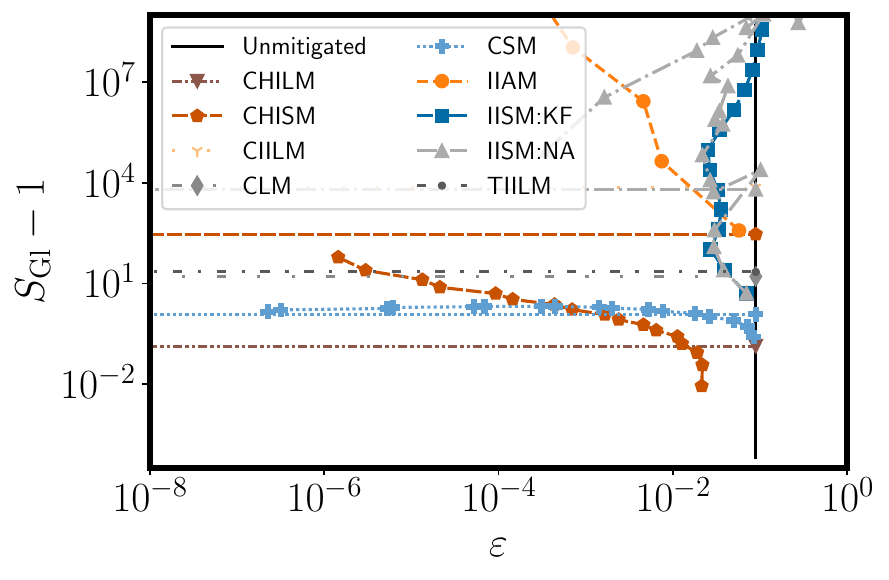
		}
		\caption{ \(\eExpectP=0\), \(\eExpectR=1.5\), \(\nRepeat=1\) (for TIILM \(m_0=0\), \(m_1=15\), \(m_2=34\)).}
		\figL{FCO075r1}
	\end{subfigure}\\
	\begin{subfigure}[b]{\linewidth*\real{0.9}}
		\centering
		\includegraphics[width=\linewidth]{
		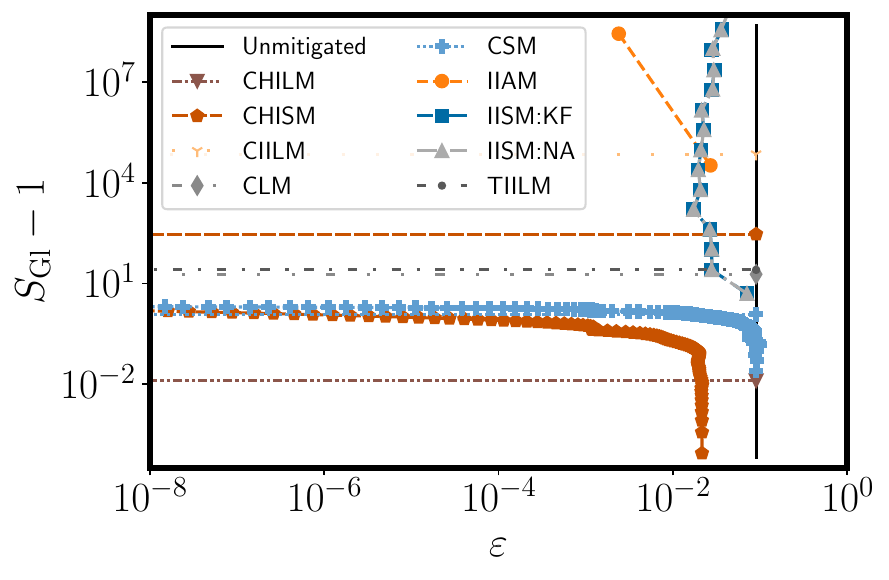
		}
		\caption{\(\eExpectP=0\), \(\eExpectR=1.5\), \(\nRepeat=10\) (for TIILM  \(m_0=0\), \(m_1=152\), \(m_2=341\)).}
		\figL{FCO075r10}
	\end{subfigure}
	\caption{Runtime scaling (\eqR{scalabilityBench}) versus bias (\eqR{biasScaleBench}) for a circuit (\secR{benchCircuit}) with rotational errors (see \secRB{RE}{noiseModel}) with noise amplitude \(\eAngle=\frac{\eExpectR}{\nGate}\), where \(\nGate=18\nRepeat\) is the number of gates. The vertical black lines show the bias before mitigation, horizontal lines are the sampling costs of unbiased methods, and each point represents a different order of mitigation (\(\nMit\)). The methods from \tabR{MenagerieStatisticsRE} that are not featured have runtime scalings higher than \(10^9\), so they do not fit in the plot; or use LC pre-tailoring and are featured in \figR{fullCircuitPlotLCRB}. \(m_0\), \(m_1\), and \(m_2\) are the parameters used for TIILM (see \secR{TII}).}
	\figL{fullCircuitPlotRB}
\end{figure}

\begin{figure}[htbp]
	\begin{subfigure}[b]{\linewidth*\real{0.9}}
		\centering
		\includegraphics[width=\linewidth]{
		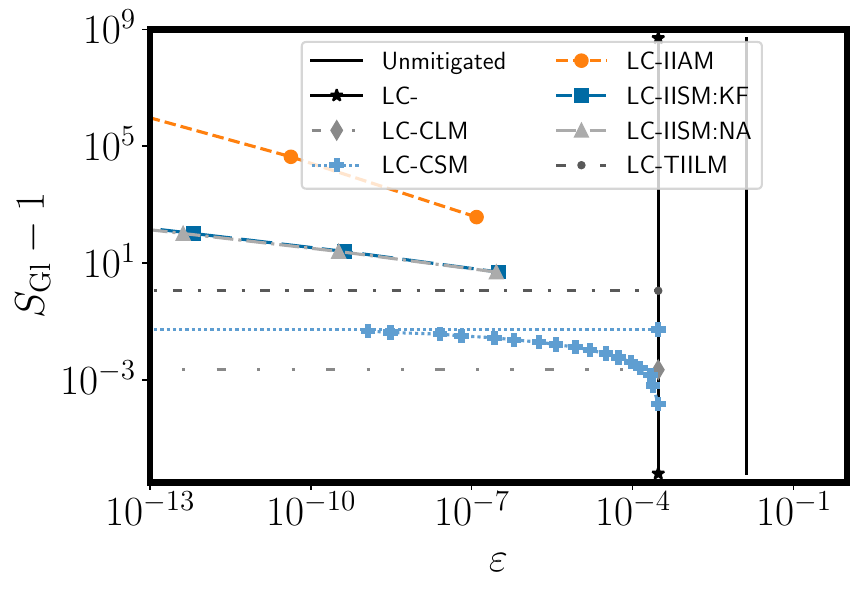
		}
		\caption{ \(\eExpectP=0\), \(\eExpectR=0.2\), \(\nRepeat=1\) (for TIILM \(m_0=0\), \(m_1=1066\)).}
		\figL{LCFCO01r1}
	\end{subfigure}\\
	\begin{subfigure}[b]{\linewidth*\real{0.9}}
		\centering
		\includegraphics[width=\linewidth]{
		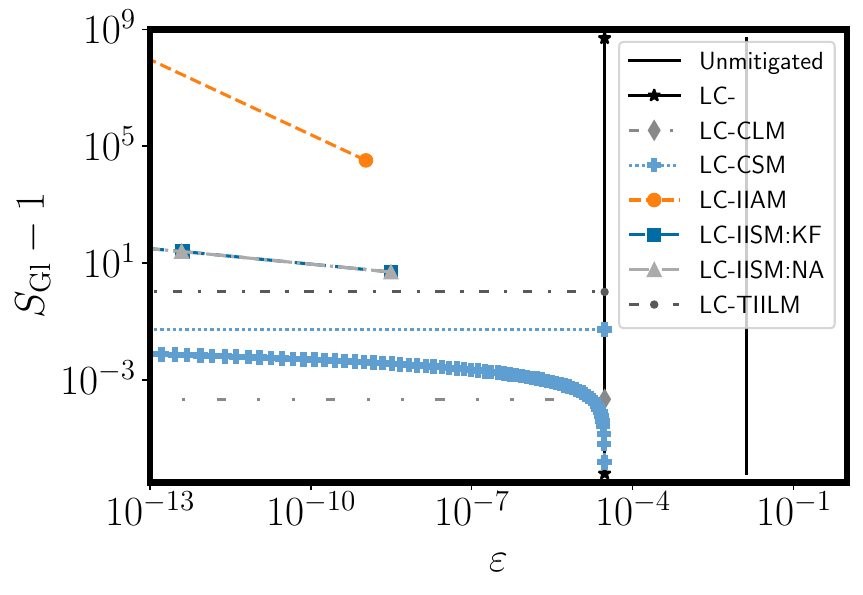
		}
		\caption{ \(\eExpectP=0\), \(\eExpectR=0.2\), \(\nRepeat=10\) (for TIILM \(m_0=0\), \(m_1=34068\)).}
		\figL{LCFCO01r10}
	\end{subfigure}
	\caption{Runtime scaling (\eqR{scalabilityBench}) versus bias (\eqR{biasScaleBench}) after local cancellation (LC) for a circuit (\secR{benchCircuit}) with rotational errors (see \secRB{RE}{noiseModel}) with noise amplitude \(\eAngle=\frac{\eExpectR}{\nGate}\), where \(\nGate=18\nRepeat\) is the number of gates. The vertical black lines show the bias before mitigation (the one with stars is after we apply LC but before subsequent mitigation methods), horizontal lines are the sampling costs of unbiased methods, and each point represents a different order of mitigation (\(\nMit\)). The methods from \tabR{MenagerieStatisticsRE} that are not featured have runtime scalings higher than \(10^9\), so they do not fit in the plot; or do not use LC pre-tailoring and are featured in \figR{fullCircuitPlotRA}. \(m_0\) and \(m_1\) are the parameters used for TIILM (see \secR{TII}).}
	\figL{fullCircuitPlotLCRA}
\end{figure}
\begin{figure}[htbp]
	\begin{subfigure}[b]{\linewidth*\real{0.9}}
		\centering
		\includegraphics[width=\linewidth]{
		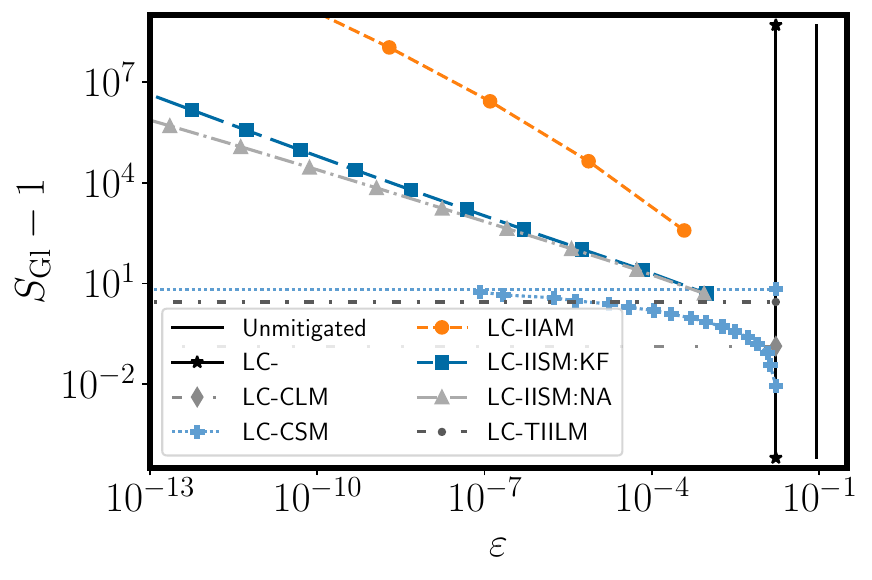
		}
		\caption{ \(\eExpectP=0\), \(\eExpectR=1.5\), \(\nRepeat=1\), (for TIILM \(m_0=0\), \(m_1=135\)).}
		\figL{LCFCO075r1}
	\end{subfigure}\\
	\begin{subfigure}[b]{\linewidth*\real{0.9}}
		\centering
		\includegraphics[width=\linewidth]{
		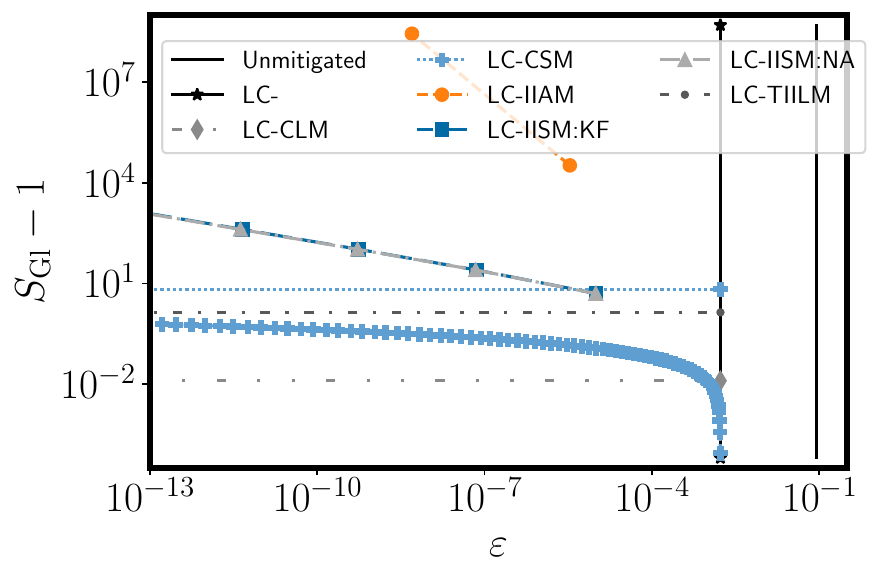
		}
		\caption{\(\eExpectP=0\), \(\eExpectR=1.5\), \(\nRepeat=10\) (for TIILM \(m_0=0\), \(m_1=4487\)).}
		\figL{LCFCO075r10}
	\end{subfigure}
	\caption{Runtime scaling (\eqR{scalabilityBench}) versus bias (\eqR{biasScaleBench}) after local cancellation (LC) for a circuit (\secR{benchCircuit}) with rotational errors (see \secRB{RE}{noiseModel}) with noise amplitude \(\eAngle=\frac{\eExpectR}{\nGate}\), where \(\nGate=18\nRepeat\) is the number of gates. The vertical black lines show the bias before mitigation (the one with stars is after we apply LC), horizontal lines are the sampling costs of unbiased methods, and each point represents a different order of mitigation (\(\nMit\)). The methods from \tabR{MenagerieStatisticsRE} that are not featured have runtime scalings higher than \(10^9\), so they do not fit in the plot; or do not use LC pre-tailoring and are featured in \figR{fullCircuitPlotRB}. \(m_0\) and \(m_1\) are the parameters used for TIILM (see \secR{TII}).}
	\figL{fullCircuitPlotLCRB}
\end{figure}

What is more, for CHILM, the larger the circuit we wish to mitigate the smaller the runtime scaling (for fixed noise level: \(\eExpectR\)). This means CHILM is not just scalable it improves as we scale up the circuit size (for constant noise level). So we can classify CHILM as super-scalable. The vertical black lines in \figRB{fullCircuitPlotRA}{fullCircuitPlotRB} show the bias before mitigation, horizontal lines are the runtime scalings of unbiased mitigation methods, and each point represents a different order of mitigation (\(\nMit\)) for the biased methods (some biased methods become unbiased for sufficiently large \(\nMit\)). Increasing \(\nMit\), generally, results in lower bias but higher sampling cost.

Methods not featured have runtime scalings outside the range of the plot. For example, IILM has a very large runtime scaling (see \tabR{MenagerieStatisticsRE}). It is worth noting that, for large enough noise levels, increasing the order of mitigation can result in a lower quality mitigation, as is the case for IISM (see \figR{fullCircuitPlotRB}). For scalable methods (see \secR{styles}) the runtime scaling for a mitigation with a given bias should not increase with the number of gates, for a constant noise level, so the top and bottom plots of \figRB{fullCircuitPlotRA}{fullCircuitPlotRB} (with the same noise level: \(\eExpectR\)) should look the same; or, if we have a super-scalable method like CHILM, the sampling cost will be lower in the bottom plots.

In \figRB{fullCircuitPlotLCRA}{fullCircuitPlotLCRB} we provide a similar analysis but make use of local cancellation (LC) to tailor the noise model before applying the mitigation methods (for stochastic noise). We find LC-CLM performs identically to CHILM, which is not surprising as they make use of the same noisy gate variants, and all other methods perform much better with local cancellation. The black line with a star on either end is the bias after applying local cancellation, without additional mitigation.

In the next section we will use these results to develop a mitigation strategy for a piece of hardware suffering from (stochastic) over-rotation errors (\secR{ORE}) and certify that it is USEful (see \secR{certificationCriteria}).

\section{Case Study: Strategy Certification}\secL{USEM:ORE}
Here we present (\secR{strategy}) and certify (\secRB{MC}{certificationCriteria}) USEM:ORE
(Universal, Scalable, and Efficacious Mitigation for Over-Rotation Errors), a mitigation strategy for a piece of hardware suffering from (stochastic) over-rotation errors. We certify it as USEful (universal, scalable, and efficacious) using the certification criteria of \secR{certificationStrategies}.

\subsection{The USEM:ORE strategy}\secL{strategy}
In order to construct a mitigation strategy (\secR{certificationStrategies}) we need to carry out the following steps:
\begin{enumerate}\listL{Strategy}
\item \textbf{Choose Application:} specify our target application and select appropriate metrics (\eg \secR{metrics}) and certifications (\eg \secR{certifications}) for our strategy (\secR{specify}).
\item \stepL{Characterise Hardware}\textbf{Characterise Hardware:} determine the elementary physical operations our hardware can implement and their associated error channels (\secR{hardwareCharacterisation}). 
\item \textbf{Compare Mitigation Methods:} select the mitigation methods for our mitigation strategy; using information from the hardware characterisation, metrics, and certification (\secR{CMM}). 
\item \textbf{Collate Gate Set:} designate a set of gates (or physical operations) that spans the space of gates needed for our calculations and mitigations (\secR{GS}).
\item \textbf{Compilation Scheme:} design a compilation scheme that permits the production of mitigated versions of the algorithms we wish to implement (\secR{CS}).
\end{enumerate}

\subsubsection{Application Specification}\secL{specify}
Our application is implementing mitigated versions of any quantum algorithm, taking maximal advantage of improvements in hardware. Therefore we want USEM:ORE to be universal, scalable, and efficacious (see \secR{certificationStrategies}) for (stochastic) over-rotation errors. This means we will be interested in finding SUPER (see \secR{styles}) mitigation methods. This requires the analysis of the proxy bias and runtime scaling of mitigation methods (see \secR{metrics}) in the large-circuit and small-noise regimes (see \secR{regimes}).

\subsubsection{Hardware Characterisation}\secL{hardwareCharacterisation}
For illustrative purposes we consider a piece of quantum hardware for which the native physical operations are rotations generated by any Pauli word operator (\(\pOpi{i}\)) of any angle (\(\tAngle\)), \eg we can produce:
\begin{align}\eqL{pauliDefinition}
\mathrm{e}^{\mathrm{i}\frac{\tAngle}{2} \pOpi{i}},& &\text{ for any }&&\pOpi{i} =& \bigotimes_{n=1}^{\nQubit}\vec{\sigma}_{\acton{n},\left\lfloor\frac{i}{4^{n-1}}\right\rfloor \bmod{4}},
\end{align}
where \(\tAngle\) is the target rotation angle and \(\nQubit\) is the number of qubits the physical operation acts on. \(\vec{\sigma}_{\acton{n},j}\) is the \(j\)th single qubit Pauli operator for qubit \(n\) (\(\vec{\sigma}_{\acton{n},0}=\iOp\), \(\vec{\sigma}_{\acton{n},1}=\xOp\), \(\vec{\sigma}_{\acton{n},2}=\yOp\), and \(\vec{\sigma}_{\acton{n},3}=\zOp\)), \(\left\lfloor\frac{i}{4^{n-1}}\right\rfloor \) is the integer part of the division of \(i\) by \(4^{n-1}\), and \(k \bmod 4\) is the remainder after dividing \(k\) by \(4\). This means we have\footnote{Operators (\(\vec{\sigma}_{\acton{n}}\)) acting on higher index (larger \(n\)) qubits are denoted to the left of those acting on lower index qubits, in our notation for Pauli words, so that the \(i\) index is independent of the number of qubits (\(\nQubit\)) for sufficiently large \(\nQubit\) (\ie \(\nQubit > \log_4 i\), otherwise \(\pOpi{i}\) is not well defined). We neglect to denote leading order identities for notational convenience.} \(\pOpi{0}=\mathbb{I}\), \(\pOpi{1}=\xOp\), \(\pOpi{2}=\yOp\), \(\pOpi{3}=\zOp\), \(\pOpi{14}=\zOp\otimes\yOp\), and \(\pOpi{15}=\zOp\otimes\zOp\) (for \(\nQubit\ge 2\), where we suppress leading identities in the tensor product, see \eqR{pauliDefinition}). To simplify our notation we define:
\begin{align}\eqL{physicalOperators}
\pOpi{i}=&\mathrm{e}^{\mathrm{i}\frac{\uppi}{2} \pOpi{i}},&\sOpi{i}=&\mathrm{e}^{\mathrm{i}\frac{\uppi}{4} \pOpi{i}}=\sqrt{ \pOpi{i}}, & \tOpi{i}=&\mathrm{e}^{\mathrm{i}\frac{\uppi}{8} \pOpi{i}}=\left( \pOpi{i}\right)^{\frac{1}{4}}.
\end{align}

For this piece of hardware we assume each physical operation is noisy and suffers from (stochastic) over-rotation errors. This means each time we try to implement one of the native physical operations we rotate slightly too far or not far enough and the amount of over-rotation for each implementation is random (see \secR{ORE}). We assume that the over-rotations follow a Gaussian distribution (with mean \(\gausMean\) and standard deviation \(\gausSD\)) and therefore a noisy gate can be modelled by:
\begin{align}\eqL{SORErrorsCharac}
\gGN{~\rotG{\tAngle}{\pOpi{i}}}{}{}{}{}{VVL}
=
\gG{~\rotG{\tAngle}{\pOpi{i}}}{}{}{}{}{VVL}
 \gG{~\rotG{\eAngle}{\pOpi{i}}}{}{}{}{}{VVL}
\begin{cases}
\gG{}{}{}{}{}{Wire}&\text{ with probability } 1-\eStoch\\
\gG{\pOpi{i}}{}{}{}{}{}&\text{ with probability } \eStoch,
\end{cases}
\end{align}
where \(\eAngle\) is the noise amplitude of the rotational (component of the) error channel and \(\eStoch\) is the noise amplitude of the stochastic (component of the) error channel:
\begin{align}\eqL{gaussianSOR}
\eAngle =& \gausMean, & \eStoch =& \frac{1-\mathrm{e}^{-\frac{\gausSD^{2}}{2}}}{2}=\frac{\gausSD^{2}}{4}+\orderZero\left(\gausSD^{4}\right).
\end{align}

The error channels have eigenvalues of the form (\eqR{SOREigenvalues} of \secR{ORE}):
\begin{align*}
\eigenTransferi{\oC{E}{}{}{}{}{},0}\left(\eStoch,\eAngle\right)=&1,&\eigenTransferi{\oC{E}{}{}{}{}{},1}\left(\eStoch,\eAngle\right)=&\left(1-2\eStoch\right)\mathrm{e}^{\mathrm{i}\eAngle},&\eigenTransferi{\oC{E}{}{}{}{}{},2}\left(\eStoch,\eAngle\right)=&\left(1-2\eStoch\right)\mathrm{e}^{-\mathrm{i}\eAngle}.
\end{align*}
We will assume there can be different noise amplitudes (\(\eAngle\) and \(\eStoch\)) for each physical operation that we mitigate so we benefit from using a local mitigation method. 

\subsubsection{Mitigation Methods}\secL{CMM}
There are a variety of local mitigation methods that we could choose from (see \tabR{Taxonomy} of \secR{MMM}). Since our noise has both stochastic and rotational components (see \eqR{SORErrorsCharac}) it seems reasonable to choose a method that performs well for both types of noise (see \tabRB{QualitativeSN}{QualitativeRE} of \secRB{StochasticCertification}{RotationalCertification}, respectively). The obvious choice is CLM but CHILM is much more efficient for rotational errors. So we propose C(HI)LM, that is, if we have access to hidden inverse (for a particular physical operation) we will use CHILM but if not we will settle for CLM. We will use the version designed to mitigate rotational errors (but modify the mitigation coefficients to account for the stochastic contribution). Since the noise channels for stochastic over-rotation errors and rotational errors are the same (\cf \secRB{RE}{ORE}).

This means we find a linear combination of noisy gate variants that is equivalent to the noise-free target gate (\eqR{LocalMitigationDiagram} of \secR{LM}):
\begin{align}\eqL{mitigatedGateVersions}
\gGN{U}{}{}{M}{}{}~~ =~~
\gG{}{}{}{}{}{WireVS}\Big\{
\coefLocali{j}:\gGN{U}{}{}{}{j}{}
\Big\}_{j=0}^{2}\hspace{-18 pt}\gG{}{}{}{}{}{WireVS}\gG{}{}{}{}{}{WireVS}
~~=~~
\gG{U}{}{}{}{}{},
\end{align}
where we use \(3\) gate variants in our mitigation (\ie \(\nCircLocal=2\)), with indices \(0\), \(1\), and \(2\). The noisy gates to be mitigated have the form (\eqR{SORErrorsCharac}):
\begin{align}\eqL{noisyGateDefinitionStrategy}
\gGN{U}{}{}{}{}{}~=&~\gG{U}{}{}{}{}{}\gC{E}{}{}{}{}{},&\text{ where }&&\gG{U}{}{}{}{}{}~=&~\gG{~\rotG{\tAngle}{\pOpi{i}}}{}{}{}{}{VVL},
\end{align}
is the noise-free component of the gate and \(\oC{E}{}{}{}{}{}\) is the error channel:
\begin{align}\eqL{noisyGateDefinitionStrategyChannel}
\gC{E}{}{}{}{}{}~=&~\gG{~\rotG{\eAngle}{\pOpi{i}}}{}{}{}{}{VVL}\begin{cases}
\gG{}{}{}{}{}{Wire}&\text{ with probability } 1-\eStoch,\\
\gG{\pOpi{i}}{}{}{}{}{}&\text{ with probability } \eStoch.
\end{cases}
\end{align}
The noisy variants (\(\oGN{U}{}{}{}{j}{}\)) that we use for the mitigation are generated using custom channels (for CLM) or custom channels and a hidden inverse (for CHILM) and are given by\footnote{Here, for simplicity, we assume our custom amplifying channels can be implemented perfectly without introducing additional errors, for a more robust analysis see \secR{R}.}:
\begin{align}\eqL{amplifiedChannels}
\begin{cases}
\gGN{U}{}{}{}{j}{}~=~\gGN{U}{}{}{}{}{}\gG{~\rotG{\cAnglei{j}}{\pOpi{i}}}{}{}{}{}{VVL}&\text{ for CLM (where \(0\le j\le 2\))},\\
\gGN{U}{}{}{}{j}{}~=~\gGN{U}{}{}{}{}{}\gG{~\rotG{\cAnglei{j}}{\pOpi{i}}}{}{}{}{}{VVL},~~~\gGN{U}{}{}{}{2}{}~=~\gGN{U}{}{}{HI}{}{}&\text{ for  CHILM (where \(0\le j<2\))},
\end{cases}
\end{align}
where the controlled rotations (\(\cAnglei{j}\)) we apply after our noisy gate are (\eqRB{optimalCustomChoice}{optimalCustomChoiceHI}):
\begin{align}\eqL{rotationAngles}
\begin{cases}
\cAnglei{0}=0,~~~\cAnglei{1}=\frac{\uppi}{2},~~~\cAnglei{2}=\frac{3\uppi}{2}&\text{ for CLM},\\
\cAnglei{0}=0,~~~\cAnglei{1}=\uppi,~~~\cAnglei{2}=0&\text{ for CHILM},
\end{cases}
\end{align}
and for CHILM we use a hidden inverse for the last (unamplified) gate variant, so the total rotational errors\footnote{The total rotational errors of the tailored noisy gates are given by \(\eAnglei{j}=\eAngle+\cAnglei{j}\) for gate variants without hidden inverses and \(\eAnglei{j}=-\eAngle+\cAnglei{j}\) when we use the hidden inverse. \(\eAngle\) is the rotational error of the original noisy gate and \(\cAnglei{j}\) is the rotational angle of the custom channel.} of the gate variants are given by:
\begin{align}\eqL{noiseAngles}
\begin{cases}
\eAnglei{0}=\eAngle,~~~\eAnglei{1}=\eAngle+\frac{\uppi}{2},~~~\eAnglei{2}=\eAngle+\frac{3\uppi}{2}&\text{ for CLM},\\
\eAnglei{0}=\eAngle,~~~\eAnglei{1}=\eAngle+\uppi,~~~\eAnglei{2}=-\eAngle&\text{ for CHILM},
\end{cases}
\end{align}
where:
\begin{align}\eqL{variantGenerationStrat}
\gGN{U}{}{}{}{j}{}~=&~\gG{U}{}{}{}{}{}\gC{E}{}{}{}{j}{},&\gC{E}{}{}{}{j}{}~=&~\gG{~\rotG{\eAnglei{j}}{\pOpi{i}}}{}{}{}{}{VVL}\begin{cases}
\gG{}{}{}{}{}{Wire}&\text{ with probability } 1-\eStoch,\\
\gGN{\pOpi{i}}{}{}{}{}{}&\text{ with probability } \eStoch.
\end{cases}
\end{align}
We assume that the stochastic noise level remain unchanged, \ie \(\eStochi{j}=\eStoch\) for both methods. 

Now the eigenvalues associated with the transfer matrices (\(\transferMatrixi{\oC{E}{}{}{}{j}{}}\)) that characterise the effect of the (amplified) error channels (\(\oC{E}{}{}{}{j}{}\)) are given by (see \eqR{SOREigenvalues}):
\begin{align}\eqL{stratEigenValuesTwo}
\eigenTransferi{j,0}=&\eigenTransferi{0}\left(\eStochi{j},\eAnglei{j}\right)=1, &\eigenTransferi{j,1}=&\eigenTransferi{1}\left(\eStochi{j},\eAnglei{j}\right)=\left(1-2\eStochi{j}\right)\mathrm{e}^{\mathrm{i} \eAnglei{j}}, \\
&&\eigenTransferi{j,2}=&\eigenTransferi{2}\left(\eStochi{j},\eAnglei{j}\right)=\left(1-2\eStochi{j}\right)\mathrm{e}^{-\mathrm{i} \eAnglei{j}}.
\end{align}
In order to perform an unbiased mitigation we must determine mitigation coefficients that satisfy (see \eqR{LocalMitigation} of \secR{eigenvalueMitigation}):
\begin{align*}
\sum_{j=0}^{\nCircLocal}\coefLocali{j}\eigenTransferi{j,k}=1,
\end{align*}
for (integer) \(0\le k\le 2\), \ie we require:
\begin{align}\eqL{coefficients}
\begin{cases}
\coefLocali{0}=\frac{\cos\eAngle}{1-2\eStoch},~~~\coefLocali{1}=\frac{1-\cos\eAngle-2\eStoch-\sin\eAngle}{2\left(1-2\eStoch\right)},~~~\coefLocali{2}=\frac{1-\cos\eAngle-2\eStoch+\sin\eAngle}{2\left(1-2\eStoch\right)}&\text{ for CLM},\\
\coefLocali{0}=\frac{1}{2},~~~~~~~\coefLocali{1}=\frac{1-\left(\cos\eAngle\right)^{-1}-2\eStoch}{2\left(1-2\eStoch\right)},~~~~~\coefLocali{2}=\frac{1}{2\left(1-2\eStoch\right)\cos\eAngle}&\text{ for CHILM},
\end{cases}
\end{align}
as the coefficients of our mitigation methods.

\subsubsection{Gate Set}\secL{GS}
We desire a universal (see \secR{certificationStrategies}) mitigation strategy so we require a gate set that includes: a collection of gates (or physical operations) that is sufficient to implement any algorithm we wish to apply and, also, all the gates (or physical operations) needed to mitigate that collection of gates. There are multiple different gate sets we could choose that would satisfy our criteria. We choose the following (\eqR{pauliDefinition}):
\begin{align}
\tOp=\tOpi{3}= &\mathrm{e}^{\mathrm{i}\frac{\uppi}{8}\zOp},&\sOp=\sOpi{3}= &\mathrm{e}^{\mathrm{i}\frac{\uppi}{4}\zOp},&\zOp=\pOpi{3} =&\mathrm{e}^{\mathrm{i}\frac{\uppi}{2}\zOp},\eqL{completeGateSetA}\\
&&\sOpi{1}= &\mathrm{e}^{\mathrm{i}\frac{\uppi}{4}\xOp},&\xOp= \pOpi{1}=&\mathrm{e}^{\mathrm{i}\frac{\uppi}{2}\xOp},\eqL{completeGateSetB}\\
&&\sOpi{15}=&\mathrm{e}^{\mathrm{i}\frac{\uppi}{4}\zOp\otimes \zOp}\eqL{completeGateSetC},
\end{align}
\ie we choose the standard T, S, and Pauli Z gate (\eqR{completeGateSetA}); the Pauli X gate and its square root (\(\sOpi{1}=\sqrt{\xOp}\), \eqR{completeGateSetB}); and one two-qubit gate: \(\sOpi{15}\). We require a copy of each of the single qubit gates for each of the qubits that we wish to act on and a copy of the two qubit gate for each pair of qubits we wish to act on. We could do without \(\xOp\), \(\zOp\), and/or \(\sOp\) if we wished to, since \(\tOp^2=\sOp\), \(\sOp^2=\zOp\), and \(\sOpi{1}^2=\xOp\).

This gate set is also sufficient for our mitigation since the amplification channels (see \eqRB{amplifiedChannels}{rotationAngles}) we need for C(HI)LM can be decomposed as:
\begin{align}
\mathrm{e}^{\mathrm{i}\frac{\uppi}{4}\pOpi{3}}=&\sOpi{3}=\sOp,&\mathrm{e}^{\mathrm{i}\frac{\uppi}{2}\pOpi{3}}=&\pOpi{3}=\zOp,&\mathrm{e}^{\mathrm{i}\frac{3\uppi}{4}\pOpi{3}}=&\sOp\zOp,\eqL{completeGateSetAMit}\\
\mathrm{e}^{\mathrm{i}\frac{\uppi}{4}\pOpi{1}}=&\sOpi{1},&\mathrm{e}^{\mathrm{i}\frac{\uppi}{2}\pOpi{1}}=& \pOpi{1}=\xOp,&\mathrm{e}^{\mathrm{i}\frac{3\uppi}{4}\pOpi{1}}=&\sOpi{1}\xOp,\eqL{completeGateSetBMit}\\
\mathrm{e}^{\mathrm{i}\frac{\uppi}{4}\pOpi{15}}=&\sOpi{15},&\mathrm{e}^{\mathrm{i}\frac{\uppi}{2}\pOpi{15}}=&\pOpi{15}=\zOp\otimes \zOp,&\mathrm{e}^{\mathrm{i}\frac{3\uppi}{4}\pOpi{15}}=&\sOpi{15}\left(\zOp\otimes \zOp\right)\eqL{completeGateSetCMit},
\end{align}
where each row of operators can be used to mitigate (all the gates in) the corresponding row of our gate set (\eqRC{completeGateSetA}{completeGateSetB}{completeGateSetC}, see \secR{CMM}).

If we wish to mitigate any gate using CHILM we do not require the \(\sOpi{i}\) gates for the mitigation but we do require the hidden inverse implementation of the gate we wish to mitigate, which is the same as the original gate except that the rotational error of the error channel has its sign reversed (see \secRB{HI}{hardwareCharacterisation}), \eg if we have the noisy gate (\eqR{noisyGateDefinitionStrategy}):
\begin{align}\eqL{hiddenInverseStrat}
\gGN{U}{}{}{}{}{}=&~\gGN{~\rotG{\tAngle}{\pOpi{i}}}{}{}{}{}{VVL}~=~
\gG{~\rotG{\tAngle}{\pOpi{i}}}{}{}{}{}{VVL}
\gG{~\rotG{\eAngle}{\pOpi{i}}}{}{}{}{}{VVL}
\begin{cases}
\gG{}{}{}{}{}{Wire}&\text{ with probability } 1-\eStoch,\\
\gG{\pOpi{i}}{}{}{}{}{}&\text{ with probability } \eStoch,
\end{cases}
\end{align}
then the hidden inverse of this noisy gate is given by:
\begin{align}\eqL{hiddenInverseStratTwo}
\gGN{U}{}{}{HI}{}{}=&~\gGN{~\rotG{\tAngle}{\pOpi{i}}}{}{}{HI}{}{VVL}~=~
\gG{~\rotG{\tAngle}{\pOpi{i}}}{}{}{}{}{VVL}
\gG{~\rotGD{\eAngle}{\pOpi{i}}}{}{}{}{}{VVL}
\begin{cases}
\gG{}{}{}{}{}{Wire}&\text{ with probability } 1-\eStoch,\\
\gG{\pOpi{i}}{}{}{}{}{}&\text{ with probability } \eStoch.
\end{cases}
\end{align}
So we need to add the hidden-inverse implementation, of any gate we wish to mitigate using CHILM, to our gate set (\eqRC{completeGateSetA}{completeGateSetB}{completeGateSetC}), which could double the number of gates in our gate set.

\subsubsection{Compilation Scheme}\secL{CS}
The compilation scheme is simple. We produce a quantum circuit that is equivalent to the algorithm we wish to implement (how we go about doing this is beyond the scope of this work). Then we decompose every gate in this circuit in terms of the gates within our gate set (\eqRC{completeGateSetA}{completeGateSetB}{completeGateSetC}). If we wish to be more efficient we can combine the previous two steps and optimise the compilation scheme in order to minimise the circuit length and/or runtime scaling. Once we have a circuit that implements our algorithm, written in terms of gates from our gate set, we simply replace every gate with its mitigated version (\eqR{mitigatedGateVersions}). This procedure means we end up with a mitigated circuit that only requires us to make use of gates that we can implement on our hardware (see \secRB{CMM}{GS}), and this mitigated circuit is the circuit we will run on our quantum hardware to implement our algorithm.

\subsubsection{Additional Requirements}
Since we desire a scalable mitigation strategy we require the noise level to be independent of circuit length (see \secR{styles}). This means some level of hardware tuning, \ie improving gate implementation to reduce errors; or error correction, \ie actively correcting the errors that occur during the application of logical gates, might be required in order to permit the mitigation of larger circuits.

\subsection{Method Certification}\secL{MC}
We can certify our mitigation methods (CLM and CHILM, see \secR{CMM}) with regards to the criteria introduced in \secR{styles}, but first we need to calculate the quantitative metrics (\secR{metrics}).

\subsubsection{Quantitative Mitigation Metrics}
The proxy bias (\secR{proxyBias}) is zero for both our methods (\(\eBiasProxyM=0\)) since they are (intrinsically) unbiased and if our characterisation is perfect the mitigated gates will be error free.

To calculate the runtime scaling (\secR{scalability}) we must first calculate the (local) sampling cost\footnote{For simplicity we assume every gate suffers from the same noise amplitude (\ie we have a uniform noise model). This simplification can be easily relaxed by replacing the local sampling cost with the geometric mean of the sampling costs for each gate, but, as long as the noise amplitudes are all of similar orders of magnitude, our derivation is relevant.} (see \eqRC{costScale}{concatenatedCost}{coefficients}):
\begin{align}\eqL{costLocal}
\costLocal=&\sum_{j=0}^{2}\left|\coefLocali{j}\right|=\begin{cases}
\begin{cases}
\frac{\cos \eAngle+\left|\sin\eAngle\right|}{1-2\eStoch}&\text{ for } \left|\eAngle\right|>2\eStoch,\\
\frac{2\cos\eAngle-1+2\eStoch}{1-2\eStoch}&\text{ for } \left|\eAngle\right|<2\eStoch,\\
\end{cases}
&\text{ for CLM},\\
\frac{1}{\left(1-2\eStoch\right)\cos \eAngle}&\text{ for CHILM},
\end{cases}
\end{align}
and the (local) length scale factor (\eqRC{costScale}{concatenatedLengthScale}{coefficients}):
\begin{align}\eqL{lengthLocal}
\oV{F}{L,L}{}{}{}{}=&\frac{\sum_{j=0}^{2}\left|\coefLocali{j}\right|\timeExecLocali{j}}{\costLocal\timeExecLocal}=\begin{cases}
\begin{cases}
\frac{3\cos \eAngle-1+2\eStoch+5\left|\sin\eAngle\right|}{2\left(\cos \eAngle+\left|\sin \eAngle\right|\right)}&\text{ for } \eAngle<-2\eStoch,\\
\frac{7\cos \eAngle-5+10\eStoch-\sin\eAngle}{2\left(2\cos \eAngle-2+4\eStoch\right)}&\text{ for } \left|\eAngle\right|<2\eStoch,\\
\frac{\cos \eAngle+1-2\eStoch+5\left|\sin\eAngle\right|}{2\left(\cos \eAngle+\left|\sin \eAngle\right|\right)}&\text{ for } \eAngle>2\eStoch,\\
\end{cases}
&\text{ for CLM},\\
\frac{3-\left(1-2\eStoch\right)\cos \eAngle}{2}&\text{ for CHILM},
\end{cases}
\end{align}
where we assume every gate takes the same amount of time\footnote{In this section we assume every gate takes the same amount of time to implement, this is a slightly different approach to that used in deriving \tabR{MenagerieStatisticsRE} where we assumed that every custom channel takes the same amount of time to implement as the gate it amplifies. So for CLM we used \(\frac{\timeExecLocali{2}}{\timeExecLocal}=2\) in \tabR{MenagerieStatisticsRE} rather than \(\frac{\timeExecLocali{2}}{\timeExecLocal}=3\) as used in \eqRB{lengthLocal}{timeLocal}, see \secR{scalability} for more information on the derivation.}  to implement so we use (see \eqRC{completeGateSetAMit}{completeGateSetBMit}{completeGateSetCMit}):
\begin{align}\eqL{timeLocal}
\begin{cases}
\frac{\timeExecLocali{0}}{\timeExecLocal}=1,~~~\frac{\timeExecLocali{1}}{\timeExecLocal}=2,~~~\frac{\timeExecLocali{2}}{\timeExecLocal}=3
&\text{ for CLM},\\
\frac{\timeExecLocali{0}}{\timeExecLocal}=1,~~~\frac{\timeExecLocali{1}}{\timeExecLocal}=2,~~~\frac{\timeExecLocali{2}}{\timeExecLocal}=1&\text{ for CHILM}.
\end{cases}
\end{align}
The full runtime scaling (for negligible initialisation time, see \eqRC{scalability}{concatenatedCost}{concatenatedLengthScale}): \(\scalabilityGlobal=\costLocal^{2\nGate}\oV{F}{L,L}{}{}{}{}\), is not particularly informative so we will instead expand it in the small-noise regime (\(\eExpectP,\eExpectR\ll 1\)):
\begin{align}\eqL{runtimeSmalll}
\scalabilityGlobal=&\begin{cases}
\begin{cases}
1+2\left(1+\frac{1}{4\nGate}\right)\eExpectP+2\left(1+\frac{3}{4\nGate}\right)\eExpectR&\text{ for } \eAngle<-2\eStoch,\\
1+4\left(1+\frac{3}{8\nGate}\right)\eExpectP+\frac{\eExpectR}{2\nGate}&\text{ for } -2\eStoch<\eAngle<0,\\
1+4\left(1+\frac{3}{8\nGate}\right)\eExpectP-\frac{\eExpectR}{2\nGate}&\text{ for } 0<\eAngle<2\eStoch,\\
1+2\left(1-\frac{1}{4\nGate}\right)\eExpectP+2\left(1+\frac{3}{4\nGate}\right)\eExpectR&\text{ for } \eAngle>2\eStoch,\\
\end{cases}
&\text{ for CLM},\\
1+2\left(1+\frac{1}{4\nGate}\right)\eExpectP+\left(1+\frac{1}{4\nGate}\right)\frac{\eExpectR^{2}}{\nGate}&\text{ for CHILM},
\end{cases}
\end{align}
(where we have neglected terms of order \(\orderSmall\left(\eExpectP^{2},\eExpectP\eExpectR,\eExpectR^{2}\right)\) for CLM and those of order \(\orderSmall\left(\eExpectP^{2},\eExpectP\eExpectR^{2},\eExpectR^{4}\right)\) for CHILM). In the large-circuit regime (\(\nGate\gg 1\)) the runtime scaling is given by:
\begin{align}\eqL{runtimeLarge}
\scalabilityGlobal=&\begin{cases}
\begin{cases}
\mathrm{e}^{2\left(\eExpectP+\eExpectR\right)}\left(1+\orderLarge\left(\frac{\eExpectP}{\nGate},\frac{\eExpectP^{2}}{\nGate},\frac{\eExpectR}{\nGate},\frac{\eExpectR^{2}}{\nGate}\right)\right)&\text{ for } \eExpectR>\eExpectP,\\
\mathrm{e}^{4\eExpectP}\left(1+\orderLarge\left(\frac{\eExpectP}{\nGate},\frac{\eExpectR}{\nGate},\frac{\eExpectR^{2}}{\nGate}\right)\right)&\text{ for } \eExpectR<\eExpectP,
\end{cases}
&\text{ for CLM},\\
\mathrm{e}^{2\eExpectP}\left(1+\orderLarge\left(\frac{\eExpectP}{\nGate},\frac{\eExpectP^{2}}{\nGate},\frac{\eExpectR^{2}}{\nGate}\right)\right)&\text{ for CHILM}.
\end{cases}
\end{align}
Here \(\nGate\) is the number of mitigated gates and we have used (\eqRB{expectStoch}{expectRot}):
\begin{align}
\eExpectP=&2\eStoch\nGate,&\eExpectR = &\nGate\left|\eAngle\right|.
\end{align}
Interestingly, \eqR{runtimeLarge} means, if we have access to hidden inverses, the rotational noise level is largely irrelevant in the large-circuit regime (provided the rotational noise level is independent of \(\nGate\)).

Determining the noise boundary (see \secR{Thresholds}) is now just a case of inverting the large-circuit-regime runtime equation (\eqR{runtimeLarge}) in the large-circuit limit (\(\nGate\rightarrow \infty\)). We find:
\begin{align}\eqL{largeBoundary}
\begin{cases}
\begin{cases}
\eThreshold= \frac{\ln \scalabilityTarget}{2} &\text{ for } \eExpectR>\eExpectP,\\
\eThresholdP= \frac{\ln \scalabilityTarget}{4}&\text{ for } \eExpectR<\eExpectP,
\end{cases}
&\text{ for CLM},\\
\eThresholdP= \frac{\ln \scalabilityTarget}{2}&\text{ for CHILM},
\end{cases}
\end{align}
where \(\eThreshold\) is the total noise boundary and \(\eThresholdP\) is the noise boundary for stochastic noise. So, if we have a tolerated runtime scaling of \(\scalabilityTarget\) we can mitigate circuits with any (finite) noise level (\(\eExpectP+\eExpectR\)) such that:
\begin{align}
\begin{cases}
\eExpectR<\eExpectP<\eThresholdP~\text{ or }~\left(\eExpectR+\eExpectP<\eThreshold ~\text{ and } \eExpectR>\eExpectP\right)&\text{ with CLM},\\
\eExpectP<\eThresholdP \text{ and } \eExpectR\text{ is finite}&\text{ with CHILM},
\end{cases}
\end{align}
in the large-circuit regime. So CHILM offers a distinct advantage, particularly if \(\eExpectR\) is the most significant contribution to the noise.

\subsubsection{Hardware Robustness}\secL{R}
The test for hardware robustness is relatively simple (see \secR{styles}). We assume that every gate has independent (\ie non uniform) noise, which we assume has been well-characterised. Then we calculate the proxy bias (which is still zero) and runtime scaling (\secR{scalability}) of mitigation using these noisy gates for variant generation rather than the noise-free amplification gates (\eqRB{noiseAngles}{variantGenerationStrat}) and hidden-inverse procedures (\eqR{hiddenInverseStrat}) that we considered in \secRC{MC}{GS}{CMM}, which did not add any uncontrolled errors into our circuit beyond those present in the original noisy gate. If the runtime scaling does not change (to leading order) then the method is robust for our hardware.

The two mitigation protocols we implement are given by (see \eqRD{physicalOperators}{mitigatedGateVersions}{amplifiedChannels}{noiseAngles}): 
\begin{align}\eqL{USEMGateMitigation}
\gGN{U}{}{}{M}{}{}=
\gGN{U}{}{}{}{}{}\begin{rcases}\begin{dcases}
\coefLocalCLMni{0}:~
\gG{}{}{}{}{}{Wire}
\\
\coefLocalCLMni{1}:~\gGN{\sOpi{i}}{}{}{}{}{}
\\
\coefLocalCLMni{2}:~\gGN{\pOpi{i}}{}{}{}{}{}\gGN{\sOpi{i}}{}{}{}{}{}
\end{dcases}\end{rcases}
=\begin{rcases}\begin{dcases}
\coefLocalCHIni{0}:~\gGN{U}{}{}{}{}{}
\\
\coefLocalCHIni{1}:~\gGN{U}{}{}{}{}{}\gGN{\pOpi{i}}{}{}{}{}{}
\\
\coefLocalCHIni{2}:~\gGN{U}{}{}{HI}{}{}
\end{dcases}\end{rcases}=\gG{U}{}{}{}{}{},
\end{align}
where the first decomposition is CLM and the second CHILM. Now, we assume our noisy gate \(\oGN{U}{}{}{}{}{}\) has rotational-error and stochastic-noise amplitudes \(\eAngle\) and \(\eStoch\), respectively (see \eqR{SORErrorsCharac}), as before. However, we also assume that our mitigation gates and the hidden-inverse procedure are noisy, so we have (see \eqR{SORErrors}):
\begin{align}\eqL{MitigatorErrors}
\gGN{U}{}{}{}{}{}~=~&\gG{U}{}{}{}{}{}
\gGm{~~\channelOperator{E}_{\mathrm{ORE}:\,\eStoch,\eAngle}\vphantom{\sum\frac{a}{B}}}{}{}{}{}{DoubleSize},
&\gGN{U}{}{}{HI}{}{}~=~&\gG{U}{}{}{}{}{}
\gGm{~~~\channelOperator{E}_{\mathrm{ORE}:\,\noiseFactorSHI\eStoch,-\noiseFactorRHI\eAngle}\vphantom{\sum\frac{a}{B}}}{}{}{}{}{TripleSize},\\
\gGN{\pOpi{i}}{}{}{}{}{}~=~&\gG{\pOpi{i}}{}{}{}{}{}
\gGm{~~~\channelOperator{E}_{\mathrm{ORE}:\,\noiseFactorRP\eStoch,\noiseFactorRP\eAngle}\vphantom{\sum\frac{a}{B}}}{}{}{}{}{TripleSize},
&\gGN{\sOpi{i}}{}{}{}{}{}~=~&\gG{\sOpi{i}}{}{}{}{}{}
\gGm{~~~\channelOperator{E}_{\mathrm{ORE}:\,\noiseFactorRS\eStoch,\noiseFactorRS\eAngle}\vphantom{\sum\frac{a}{B}}}{}{}{}{}{TripleSize}.
\end{align}
\(\oON{\pOpi{i}}{}{}{}{}{}\); \(\oON{\sOpi{i}}{}{}{}{}{}\); and \(\oGN{U}{}{}{HI}{}{}\) are noisy, with noise amplitudes \(\noiseFactorRP\eAngle\) and \(\noiseFactorSP\eStoch\); \(\noiseFactorRS\eAngle\) and \(\noiseFactorSS\eStoch\); and \(-\noiseFactorRHI\eAngle\) (where \(\noiseFactorRHI>0\), otherwise the hidden inverse failed) and \(\noiseFactorSHI\eStoch\), respectively. This means the rotational-error amplitudes of our noisy variants are given by:
\begin{align}\eqL{noiseAnglesN}
\begin{cases}
\eAnglei{0}=\eAngle,~~~\eAnglei{1}=\left(1+\noiseFactorRS\right)\eAngle+\frac{\uppi}{2},~~~\eAnglei{2}=\left(1+\noiseFactorRS+\noiseFactorRP\right)\eAngle+\frac{3\uppi}{2}&\text{ for CLM},\\
\eAnglei{0}=\eAngle,~~~\eAnglei{1}=\left(1+\noiseFactorRP\right)\eAngle+\uppi,~~~\eAnglei{2}=-\noiseFactorRHI\eAngle&\text{ for CHILM},
\end{cases}
\end{align}
(where the gates are numbered from top to bottom in \eqR{USEMGateMitigation}) and their stochastic-noise amplitudes are:
\begin{align}\eqL{noiseStochN}
\begin{cases}
\eStochi{0}=\eStoch,~~~\eStochi{1}=\left(1+\noiseFactorSS\right)\eStoch,~~~\eStochi{2}=\left(1+\noiseFactorSS+\noiseFactorSP\right)\eStoch&\text{ for CLM},\\
\eStochi{0}=\eStoch,~~~\eStochi{1}=\left(1+\noiseFactorSP\right)\eStoch,~~~\eStochi{2}=\noiseFactorSHI\eStoch&\text{ for CHILM}.
\end{cases}
\end{align}
So:
\begin{align}
\gGN{U}{}{}{}{j}{}=\gG{U}{}{}{}{}{}\gC{E}{}{}{}{j}{}=\gG{U}{}{}{}{}{}\gG{~\rotG{\eAnglei{j}}{\pOpi{i}}}{}{}{}{}{VVL}\begin{cases}
\gG{}{}{}{}{}{Wire}&\text{ with probability } 1-\eStochi{j},\\
\gG{\pOpi{i}}{}{}{}{}{}&\text{ with probability } \eStochi{j},
\end{cases}
\end{align}
is the \(j\)th gate variant used to mitigate the noisy gate \(\oGN{U}{}{}{}{}{}\).

Now the eigenvalues associated with the transfer matrices (\(\transferMatrixi{\oC{E}{}{}{}{j}{}}\)) that characterise the effect of the (amplified) error channels (\(\oC{E}{}{}{}{j}{}\)) are given by \eqR{stratEigenValuesTwo} as before and we must solve \eqR{LocalMitigation} to find our coefficients (where \(\coefLocali{j}=\coefLocalCLMni{j}\) for CLM and \(\coefLocali{j}=\coefLocalCHIni{j}\) for CHILM). The resulting coefficients are too cumbersome to write in full here but we can examine their first order expansions (in the noise amplitudes), \eg we  neglect terms of order \(\orderZero\left(\eStoch^{2},\eStoch\eAngle,\eAngle^{2}\right)\) for CLM. We also make the assumption that all the noise amplitudes have the same order of magnitude, \ie \(\left(1+\noiseFactorRS+\noiseFactorRP+\noiseFactorRHI\right)\eAngle,\left(1+\noiseFactorSS+\noiseFactorSP+\noiseFactorSHI\right)\eStoch\ll 1\).

The coefficients for CLM (to leading order) are:
\begin{align}\eqL{coefficientsNoisyCLM}
\coefLocalCLMni{0}=&1+2\eStoch,&\coefLocalCLMni{1}=&-\eStoch-\frac{\eAngle}{2},&\coefLocalCLMni{2}=&-\eStoch+\frac{\eAngle}{2}.
\end{align}
These are independent of \(\noiseFactorSS,\noiseFactorRS,\noiseFactorSP,\noiseFactorRP\) so CLM is robust to using noisy amplification gates; since the runtime scaling is a function of the coefficients in \eqR{coefficientsNoisyCLM} (to leading order).

The coefficients for CHILM (to leading order) are:
\begin{align}
\coefLocalCHIni{0}=&\frac{\noiseFactorRHI}{1+\noiseFactorRHI}+\frac{\left(1+\noiseFactorRHI\right)\left(\noiseFactorRHI-\noiseFactorSHI\right)-\left(\noiseFactorRHI+\noiseFactorSHI\right)\noiseFactorRP}{\left(1+\noiseFactorRHI\right)^{2}}\eStoch+\frac{\noiseFactorRHI\left(\noiseFactorRHI-3\noiseFactorRP-1\right)}{12\left(1+\noiseFactorRHI\right)}\eAngle^{2},\nonumber\\
\coefLocalCHIni{1}=&-\frac{\noiseFactorRHI+\noiseFactorSHI}{1+\noiseFactorRHI}\eStoch-\frac{\noiseFactorRHI}{4}\eAngle^{2},\nonumber\\
\coefLocalCHIni{2}=&\frac{1}{1+\noiseFactorRHI}+\frac{2\noiseFactorSHI\left(1+\noiseFactorRHI\right)+\left(\noiseFactorRHI+\noiseFactorSHI\right)\noiseFactorRP}{\left(1+\noiseFactorRHI\right)^{2}}\eStoch+\frac{\noiseFactorRHI\left(2\noiseFactorRHI+3\noiseFactorRP+4\right)}{12\left(1+\noiseFactorRHI\right)}\eAngle^{2}.
\end{align}
These do depend on \(\noiseFactorSHI,\noiseFactorRHI\) and \(\noiseFactorRP\). Thus we must calculate the runtime scaling explicitly to determine if CHILM is robust. CHILM's runtime scaling is given by: 
\begin{align}\eqL{runtimeNoisy}
\scalabilityGlobal=&1+2\left(1+\frac{1}{4\nGate}\right)\frac{\noiseFactorSHI+\noiseFactorRHI}{1+\noiseFactorRHI}\eExpectP+\left(1+\frac{1}{4\nGate}\right)\noiseFactorRHI\frac{\eExpectR^{2}}{\nGate}+\orderSmall\left(\eExpectP^{2},\eExpectP\eExpectR^{2},\eExpectR^{4}\right)\nonumber\\
=&
\mathrm{e}^{2\frac{\noiseFactorSHI+\noiseFactorRHI}{1+\noiseFactorRHI}\eExpectP}\left(1+\orderLarge\left(\frac{\eExpectP}{\nGate},\frac{\eExpectP^{2}}{\nGate},\frac{\eExpectR^{2}}{\nGate}\right)\right),
\end{align}
and the noise boundary is given by:
\begin{align}\eqL{largeBoundaryNoisy}
\eThresholdP= \frac{1+\noiseFactorRHI}{\noiseFactorSHI+\noiseFactorRHI}\frac{\ln \scalabilityTarget}{2}.
\end{align}
So, though CHILM is not robust, it is quasi-robust (see \secR{styles}) since the orders of the most significant order terms for its noise boundary and runtime scaling are independent of \(\noiseFactorSHI\), \(\noiseFactorRHI\), and \(\noiseFactorRP\); even though the values of the coefficients are not. The runtime scaling and noise boundary are robust to noise in the custom channel (\ie the values of \(\noiseFactorRP\) and \(\noiseFactorSP\)) and, if \(\noiseFactorSHI<1\), using imperfect hidden inverses could actually perform better than using noise-free hidden inverses.

\subsubsection{Method Certification}\secL{QCMeasure}
We can use our metrics to certify our mitigation methods (see \secR{styles}).
\begin{enumerate}\listL{QCOut}
\item \stepL{ScalableOut}\textbf{Scalable}: Both methods are scalable since their runtime scaling in the large-circuit regime depends only on the noise level (\(\eExpectP\) and/or \(\eExpectR\)) and not the number of gates (\(\nGate\), see \eqR{runtimeLarge}). However, CHILM scales better than CLM as CHILM's runtime is independent of the rotational noise level (\(\eExpectR\)) in the large-circuit limit.
\item \textbf{Unbounded} (in principle): The noise boundary (see \eqR{largeBoundary}) grows as the logarithm of the tolerated runtime scaling (\(\scalabilityTarget\)) for both methods. So we can mitigate arbitrary noise levels with either methods, if we are willing to wait long enough. Thus both methods are unbounded but CHILM is independent of the rotational noise level, so, again, it has a slight edge. 
\item \textbf{Precise}: Since both methods are unbiased they are both precise and can be used to reduce the bias to arbitrarily small levels, if we are willing to wait long enough.
\item \textbf{Efficient}: Both methods are also efficient since the runtime scaling in the small-noise limit (\(\eExpectP,\eExpectR\rightarrow 0\)) is unity (see \eqR{runtimeSmalll}). So they can make good use of any reduction in noise level. However, CHILM is more efficient than CLM since it is independent of \(\eExpectR\) to first order.
\item \textbf{Robust} (see \secR{R}): CLM is robust for our hardware since noise in the mitigation gates does not change the runtime cost, to leading order in the large-circuit or small-noise regime; provided the noise is characterised and the same order of magnitude as that of the gate being mitigated. CHILM is quasi-robust (see \secR{styles}) since noise in the hidden-inverse procedure can change the effective size (but not order) of the stochastic noise level, but it is robust to noise in the amplification channel. 
\end{enumerate}
So CLM can be certified SUPER (scalable, unbounded, precise, efficient, and robust) and CHILM is a SUPEqR (scalable, unbounded, precise, efficient, and quasi-robust) mitigation method. Though CHILM is not robust it does scale better than CLM in general. In the next section we shall certify the mitigation strategy as a whole.

\subsection{Strategy Certification}\secL{certificationCriteria}
We can now certify our full mitigation strategy according to the criteria in  \secR{certificationStrategies}.

\begin{enumerate}\listL{QCSOut}
\item \textbf{Universal}: We have demonstrated (see \secR{CMM}) that we can mitigate any gate within our gate set (\eqRC{completeGateSetA}{completeGateSetB}{completeGateSetC}) using only gates that are already within the gate set to perform the mitigation (see \eqRC{completeGateSetAMit}{completeGateSetBMit}{completeGateSetCMit} of \secR{GS}, the exception being for CHILM where we also require hidden inverse implementations of the gate to be mitigated), if all the gates suffer from stochastic over-rotation errors (see \secR{R}). Now, all that remains is to show that our gate set is universal. To do this we will provide decompositions of the gates within a universally-recognised universal gate set (T-Z45, H-Hadamard, CX-Controlled Not; \eg see \citeRefs{Sousa2007,Steffen2013Thesis}) in terms of our (universally mitigatable) gate set (\eqRC{completeGateSetA}{completeGateSetB}{completeGateSetC}):
\begin{align}
\tOp=&\tOpi{3},&\mathrm{CZ} =& \mathrm{e}^{-3\mathrm{i}\frac{\uppi}{4}}\left(\sOpi{3} \zOp\otimes \sOpi{3} \zOp\right) \sOpi{15},&
\hOp=&\mathrm{e}^{\mathrm{i}\frac{\uppi}{2}}\sOpi{1}\sOpi{3}\sOpi{1},&\mathrm{CX}=&\left(\iOp\otimes\hOp\right)\mathrm{CZ} \left(\iOp\otimes\hOp\right).\eqL{compositionsSORUniversalB}
\end{align}
\item \textbf{Scalable}: Since C(HI)LM is scalable (see \secR{QCMeasure}) even when using imperfect custom channels and hidden inverses to effect the mitigation (see \secR{R}) the full mitigation strategy is scalable.
\item \textbf{Efficacious}: C(HI)LM is SUPE(q)R (see \secR{QCMeasure}), so it is unbounded, precise, efficient and (quasi-)robust for hardware that suffers from over-rotation errors (see \secR{R}), therefore it is efficacious (the quasi-robustness ensures that the method certifications are still relevant for the noisy hardware we consider).
\end{enumerate}
Thus our mitigation strategy (USEM:ORE) can be certified USEful for hardware suffering from stochastic over-rotation errors and we can successfully conclude our certification.

\section{Concluding Remarks}\secL{conclude}
In this work, we introduced a unified framework for the classification and certification of quantum error mitigation (QEM) methods. Our approach combines quantitative metrics\ \textendash\ such as proxy bias, runtime scaling, and noise boundaries\ \textendash\ with qualitative criteria including scalability, robustness, and efficiency. These tools enable a systematic evaluation of linear mitigation methods and their applicability across different noise models.

Through our taxonomy and case studies, we demonstrated how the proposed certification scheme can be applied to compare and design mitigation strategies tailored to specific quantum hardware and noise profiles.

\subsection{Further Work}
Future efforts should aim to:
\begin{enumerate}[label = (\roman*)]
\item Extend the certification framework to non-linear and hybrid error mitigation methods as well as additional noise models (\eg amplitude damping \cite{Preskill2018Lecture}). In \citeRef{MyLongWork} we consider virtual distillation\ \textendash\ a non-linear mitigation method \cite{Koczor2021a,Huggins2021, Czarnik2021V}\ \textendash\  applied to circuits suffering from global depolarising noise and calculate the relevant quantitative metrics (runtime scaling, proxy bias, and noise boundary).
\item Incorporate realistic experimental imperfections into the certification process and make use of real noise characterisation data obtained by tomography.
\item Explore integration with other approaches to noise suppression, such as error correction and/or hardware tuning (\eg incorporate metrics such as energy consumption \cite{Fellous2023}).
\end{enumerate}
A promising direction is the development of automated standardised certification pipelines combining theoretical metrics and experimental data to enable adaptive mitigation strategies in near-term quantum devices.



\section*{Acknowledgments}
Plots and simulations were produced using Python (using matplotlib, mpmath, numpy, scipy, and sympy) and the analytic analysis made use of Mathematica. This publication was made possible by the use of the FactoryIA supercomputer, financially supported by the Ile-de-France Regional Council. Thanks to David Herrera-Mart\'{i} for securing the funding that made this work possible and for his contributions throughout. Many thanks to Vasily Sazonov for his part in developing the ideas that motivated and formed the basis of this work, his advice, and his comments on the manuscript.
\addcontentsline{toc}{section}{References}

\renewcommand*{\bibfont}{\raggedright}
\bibliographystyle{quantum}
\bibliography{bibliography/biblio}
\newpage


\appendix

\setlength{\headheight}{13.59999pt}
\fancyhead[C]{\textbf{Appendix}}
\renewcommand{\theequation}{\thesection.\arabic{equation}}
\phantomsection
\addcontentsline{toc}{section}{Appendices}
\stopcontents[sections]
\startcontents[appendices]
\printcontents[appendices]{l}{1}{\section*{Appendices}\setcounter{tocdepth}{2}}

%
\newpage
\section{Glossary of Terms}\appL{glossary}
Here we present a glossary of terms that we use in this work in alphabetical order, with references to where they are used in the main text.

\scalebox{0.75}{
\begin{tabular}{l>{\raggedright}p{280pt}>{\raggedright}p{55pt}}
        \hline
Term					& Description 						&  Reference \tabularnewline
        \hline
Area Scale Factor (\(\oV{F}{Gl,A}{}{}{}{}\))		& The product of the length and width scale factors, representing the resource overhead of mitigation per run of the circuit. & \secR{scalability}\tabularnewline	
Asynchronous Mitigation (AM)	& Mitigation where different gates can be amplified by different amounts in each circuit variant.  & \secR{namingConvention}\tabularnewline	
Bias (\(\varepsilon\))		& Deviation of the measured observable expectation value from the true noiseless value. &  \secR{LinM}\tabularnewline	
Certification				& Process of assessing a mitigation method using predefined quantitative and qualitative metrics.&\secR{certifications}\tabularnewline	
Circuit Variant				& One of a selection of noisy implementations of a noise-free target circuit.   &\secR{VGS} \tabularnewline	
CPTP					&Completely Positive and Trace Preserving&\citeRef{BreuerBook2002} \tabularnewline	
Custom Channel (C)		& A tailored error channel inserted after a gate to amplify or modify its noise characteristics. &\secR{C}\tabularnewline	
Efficacious				& Mitigation strategy certification criteria for when all mitigation methods within the strategy are SUPER or SUPEqR. & \secR{certificationStrategies}\tabularnewline	
Efficient					& Property of a mitigation method with unit runtime scaling in the limit of vanishing noise. & \secR{styles}\tabularnewline	
Error Channel (\(\mathcal{E}\))	& Mathematical model representing the effect of errors in a quantum gate. & \secR{noiseModelling}\tabularnewline
Error Mitigation			& Technique to reduce the impact of noise on quantum computations without reducing the noise level of any individual run. &\secR{intro} \tabularnewline	
Gate Variant				&  One of a selection of noisy implementations of a noise-free target gate.   &\secR{VGS} \tabularnewline	
Hidden Inverse (HI)		& A gate variant constructed to have the opposite coherent error to the original gate, used for local cancellation.  & \secR{HI}\tabularnewline	
Identity Insertion (II)		& A technique that amplifies noise by inserting sequences of gates and their inverses. & \secR{II} \tabularnewline	
Knowledge Free (KF)		&  Mitigation methods that do not require quantitive knowledge of the noise model. & \secR{namingConvention}\tabularnewline	
Length Scale Factor (\(\oV{F}{Gl,L}{}{}{}{}\))		& The average increase in circuit runtime due to mitigation. & \secR{scalability} \tabularnewline	
Linear Mitigation			& Mitigation methods that use linear combinations of noisy circuit outputs to estimate noise-free results.& \secR{LinM}\tabularnewline	
Local Cancellation (LC) 			& A tailoring method that converts coherent noise to incoherent/stochastic noise. &\secR{LC}\tabularnewline	
Local Mitigation (LM)		& Mitigation applied independently to each gate or small group of nearby gates.  &\secR{namingConvention}\tabularnewline	
Mitigation Method 	& A procedure for mitigating the noise of a quantum circuit.							& \secR{taxonomy}\tabularnewline	
Mitigation Strategy & A procedure for producing mitigated circuits for quantum algorithms implemented on a selected and characterised piece of quantum hardware. May consist of multiple mitigation methods.	& \secR{certificationStrategies}\tabularnewline	
NISQ 					& Noisy Intermediate-Scale Quantum: The current era of quantum devices with limited qubit counts and significant noise. & \citeRef{Preskill2018}\tabularnewline	
Noise Amplitude (\({x}\), \(p\) or \(\phi\)) & Parameter quantifying the strength of noise in a quantum channel. & \secR{noiseModelling}\tabularnewline	
Noise Aware (NA)			& Mitigation methods that use quantitative knowledge of the noise model to optimise performance. &\secR{namingConvention} \tabularnewline	
Noise Boundary (\(\eThreshold\))	& The maximum noise level that can be mitigated for a given tolerated runtime scaling and target bias. & \secR{Thresholds}\tabularnewline	
Noise Channel (\(\mathcal{N}\))	& A CPTP map that is part of a linear decomposition of an error channel.	&\secR{noiseModelling}\tabularnewline	
Noise Level	(\(\noiseLevel\))	& Parameter quantifying the strength of noise in a quantum circuit to leading order. & \secR{NL}\tabularnewline	
ORE					& Over-Rotation Errors: An error channel equivalent to that produced by a rotation in Hilbert space rotating too far. & \secR{ORE}\tabularnewline	
\end{tabular}
}

\scalebox{0.75}{
\begin{tabular}{l>{\raggedright}p{280pt}>{\raggedright}p{55pt}}
        \hline
Term					& Description 						&  Reference \tabularnewline
        \hline
PEC					& Probabilistic Error Cancellation: A linear mitigation method using quasi-probability decompositions. &\citeRef{Temme2017} \tabularnewline	
Precise			& Property of a mitigation method that can reduce the effective noise level to an arbitrarily small amount. & \secR{styles}\tabularnewline	
Proxy Bias (\(\eBiasProxyN\))	& An upper bound to the true bias, calculable without explicit circuit simulation.& \secR{proxyBias}\tabularnewline	
QEM					& Quantum Error Mitigation: See Error Mitigation& \tabularnewline
Robust					& Property of a mitigation method with runtime scaling independent of small characterised imperfections in method implementation. & \secR{styles}\tabularnewline	
Rotational Errors (RE)		& A coherent error channel equivalent to a (unitary) rotation generated by a Hermitian unitary (self-inverse) operator. & \secR{RE}\tabularnewline		
Runtime Scaling (\(S\))		& The ratio of the total runtime required for a mitigated circuit to that of the unmitigated circuit to achieve the same precision. & \secR{scalability}\tabularnewline
Sampling Cost (\({C}\))		& The increase in the number of samples (shots) required for a mitigated circuit to preserve the statistical precision scaling of the original circuit. & \secR{scalability}\tabularnewline	
Scalable			& Property of a mitigation method that allows it to maintain performance as circuit size increases for fixed noise level. & \secRB{styles}{certificationStrategies}\tabularnewline	
SPAM Errors					& State Preparation and Measurement Errors: Errors associated with initialising and measuring quantum states. &\citeRef{Merkel2013}\tabularnewline
Stochastic Noise (SN)		&Random noise model where errors occur probabilistically. &\secR{SN} \tabularnewline	
SUPEqR				&  Scalable, Unbounded, Precise, Efficient, quasi-Robust: Certification for mitigation methods that are robust to leading order. &\secR{certifications}\tabularnewline
SUPER				&  Scalable, Unbounded, Precise, Efficient, Robust: Certification for the highest standard of mitigation methods.	&\secR{certifications}\tabularnewline
Synchronous Mitigation (SM) & A mitigation method where all gates, within a particular circuit variant, are amplified by the same amount.  &\secR{namingConvention}\tabularnewline	
Taxonomy (of mitigation methods) 	&Structured classification of error mitigation techniques based on their characteristics and requirements.&\secR{taxonomy}\tabularnewline
Tuned Identity Insertion (TII)		&An optimised version of identity insertion where the number of identity insertions is chosen to minimise the runtime scaling. &\secR{TII}\tabularnewline
Unbounded			& Property of a mitigation method that can mitigate arbitrarily high noise levels. & \secR{styles}\tabularnewline	
Uniform Noise Model	& Noise model where all gates have the same type of error channel and the same noise amplitude. & \secR{fullCircuit}\tabularnewline	
Universal				& Property of a mitigation strategy that allows it to provide a mitigated circuit for any quantum algorithm. & \secR{certificationStrategies}\tabularnewline	
USEful				&  Certification that a mitigation strategy is Universal, Scalable, and Efficacious	&\secR{certificationStrategies}\tabularnewline	
USEM:ORE			& Universal, Scalable, Efficacious, Mitigation for Over-Rotation Errors: A mitigation strategy certified USEful for hardware with over-rotation errors. &\secR{strategy}\tabularnewline
Width Scale Factor	(\(\oV{F}{Gl,W}{}{}{}{}\))		& The average increase in qubits required per circuit due to mitigation. & \secR{scalability} \tabularnewline	
\end{tabular}
}
	
\newpage
\section{Worked Example: IISM:NA}\appSL{workedExample}
Here we provide calculations of proxy bias, before and after mitigation, and the sampling cost; for hardware suffering from uniform dephasing noise, mitigated using IISM:NA (noise-aware identity-insertion-assisted synchronous mitigation).

\subsection{Notation}
We use Pauli words (\appSR{pauliWords}) and transfer matrices (\appSR{transferMatrices}).

\subsubsection{Pauli Words and Bloch Decomposition}\appSL{pauliWords}
\(\vec{\pOp}\) is the Pauli word vector and:
\begin{align}
\pOpi{i}=&\left(\vec{\pOp}\right)_{i},&\pOpi{i}\in &\left\{\iOp,\xOp,\yOp, \zOp\right\}^{\otimes \nQubit},
\end{align}
where \(\pOpi{i}\) is the \(i\)th Pauli word and the \(i\)th element of the Pauli word vector. For concreteness we define:
\begin{align}\eqAL{pauliWordVector}
\left(\vec{\pOp}\right)_{i}=\pOpi{i}=\bigotimes_{n=1}^{\nQubit}\left(\vec{\sigma}_{\acton{n}}\right)_{\left\lfloor\frac{i}{4^{n-1}}\right\rfloor \bmod{4}}, 
\end{align}
where a square-bracketed subscript \(\acton{n}\) denotes that an operator acts on the \(n\)th qubit and  \(\vec{\sigma}_{\acton{n}}\) is the Pauli vector for the \(n\)th qubit: \(\left(\vec{\sigma}_{\acton{n}}\right)_0 =\iOp\), the two by two identity matrix; \(\left(\vec{\sigma}_{\acton{n}}\right)_1 =\xOp\), the two by two Pauli X matrix; \(\left(\vec{\sigma}_{\acton{n}}\right)_2 = \yOp\), the two by two Pauli Y matrix; and \(\left(\vec{\sigma}_{\acton{n}}\right)_3 = \zOp\), the two by two Pauli Z matrix. If we wish to denote the \(n\)th qubit operator of the \(i\)th component of \(\vec{\pOp}\) we use the notation:
\begin{align}
\pOpi{i,n}=\left(\vec{\sigma}_{\acton{n}}\right)_{\left\lfloor\frac{i}{4^{n-1}}\right\rfloor \bmod{4}}. 
\end{align}
Some examples of Pauli words are:
\begin{align}
\pOpi{0}=&\identityGlobal=\iOp\otimes\dots \iOp,&\pOpi{1}=&\iOp\otimes\dots \xOp,&\pOpi{2}=&\iOp\otimes\dots \yOp,&\pOpi{3}=&\iOp\otimes\dots \zOp,\\
&&\pOpi{15}=&\iOp\otimes\dots \zOp\otimes \zOp,&\pOpi{27}=&\iOp\otimes\dots \xOp\otimes\yOp\otimes \zOp,
\end{align}
though generally we neglect leading identities.

The (generalised) Bloch decomposition is a vector representation of a density matrix in the basis of Pauli words:
\begin{align} \eqAL{generalisedBlochForm}
\rho = \frac{{\blochOperator} \cdot \vec{\pOp}}{2^{\nQubit}},
\end{align}
where \(\nQubit\) is the number of qubits and 
\({\blochOperator}\), the generalised Bloch vector, is a vector of expectations for each of the Pauli words, so:
\begin{align}\eqAL{iThElement}
\left({\blochOperator}\right)_i=\langle\pOpi{i}\rangle = \traceOp{ \rho\pOpi{i}},
\end{align}
and therefore \(-1\le \left({\blochOperator}\right)_i\le 1\).

\subsubsection{Transfer Matrices}\appSL{transferMatrices}
The output of a CPTP (completely-positive and trace-preserving) error channel (\(\oC{E}{}{}{}{}{}\) with input state \(\rho\)) is a density matrix:
\begin{align}
\rhoN =\oC{E}{}{}{}{}{}\left(\rho\right).
\end{align}
If \(\blochOperator\) is the Bloch vector of \(\rho\) and \(\blochOperatorN\) is the Bloch vector of \(\rhoN\) we have:
\begin{align}
\frac{1}{2^{\nQubit }}\sum_{j=0}^{4^{\nQubit }-1}\left({\blochOperatorN}\right)_j \pOpi{j}=\frac{{\blochOperatorN} \cdot \vec{\pOp}}{2^{\nQubit }}=\rhoN =\oC{E}{}{}{}{}{}\left(\rho\right)= \frac{\oC{E}{}{}{}{}{}\left({\blochOperator} \cdot \vec{\pOp}\right)}{2^{\nQubit }}=\frac{1}{2^{\nQubit }}\sum_{j=0}^{4^{\nQubit }-1}\left({\blochOperator}\right)_j \oC{E}{}{}{}{}{}\left(\pOpi{j}\right),
\end{align}
where we have used the linearity of the channel and the Bloch decomposition of the density operators (\eqAR{generalisedBlochForm}). Thus we can express the channel as a map between Bloch vectors:
\begin{align}
\left({\blochOperatorN}\right)_i=&\frac{1}{2^{\nQubit }}\traceOp{\sum_{j=0}^{4^{\nQubit }-1}\left({\blochOperatorN}\right)_j \pOpi{j}\pOpi{i}}=\traceOp{\pOpi{i}\rhoN }=\traceOp{\pOpi{i}\oC{E}{}{}{}{}{}\left(\rho\right)}\\
=&\frac{1}{2^{\nQubit }}\sum_{j=0}^{4^{\nQubit }-1} \traceOp{\pOpi{i}\oC{E}{}{}{}{}{}\left(\pOpi{j}\right)}\left({\blochOperator}\right)_j,
\end{align}
so we have:
\begin{align}
{\blochOperatorN}=&\transferMatrixi{\oC{E}{}{}{}{}{}} {\blochOperator},
\end{align}
were we define the Pauli transfer matrix as:
\begin{align}\eqAL{PauliTransferMatrix}
\left(\transferMatrixi{\oC{E}{}{}{}{}{}} \right)_{i,j}=&\frac{1}{2^{\nQubit }} \traceOp{\pOpi{i}\oC{E}{}{}{}{}{}\left(\pOpi{j}\right)}.
\end{align}
Since \({\blochOperatorN}\) must have all elements between \(-1\) and \(1\) (see \eqAR{iThElement}) \(\transferMatrixi{\oC{E}{}{}{}{}{}} \) must also have all elements between \(-1\) and \(1\), otherwise there would be some choices of \({\blochOperator}\) that would lead to unphysical \({\blochOperatorN}\).

\subsection{Circuit}
We consider a situation where we wish to implement the circuit:
\begin{align}\eqAL{circuitNoiseFree}
\gC{C}{}{}{}{}{Wireless}~=~\gG{U}{}{1}{}{}{Wireright}\gG{U}{}{2}{}{}{}\dots\gG{U}{}{\nGate}{}{}{WireleftL},
\end{align}
of \(\nGate\) gates where \(\oG{U}{}{n}{}{}{}\) is the \(n\)th noise-free gate. However, owing to noise affecting our gates, we actually implement the circuit:
\begin{align}\eqAL{circuitDephasingNoise}
\gCN{C}{}{}{}{}{Wireless}~=~\gGN{U}{}{1}{}{}{Wireright}\gGN{U}{}{2}{}{}{}\dots\gGN{U}{}{\nGate}{}{}{WireleftL},
\end{align}
where \(\oGN{U}{}{n}{}{}{}\) is the noisy gate given by:
\begin{align}\eqAL{dephasingNoisyGate}
\gGN{U}{}{n}{}{}{}~=~\gG{U}{}{n}{}{}{}\gC{E}{}{}{}{}{},
\end{align}
 and \(\oC{E}{}{}{}{}{}\) is its error channel.

\subsection{Assumptions and Requirements}
\begin{enumerate}
\item We assume a uniform noise model, which means every gate has an identical error channel. 
\item We assume that the noise-free components of all gates commute with their error channels, a requirement for most methods using identity insertion.
\item We assume that the noisy inverse gate has the same error channel as the noisy gate it is the (near) inverse of, a requirement for most methods using identity insertion.
\item We use a single qubit dephasing noise as our error channel for illustrative purposes, but clearly this would not be appropriate for two qubit gates. 
\item We assume there are no SPAM (state preparation and measurement) errors, as our mitigation methods are designed to mitigate gates, there are other procedures for mitigating measurements \cite{Bravyi2021}.
\end{enumerate}

\subsection{Noise Model Characterisation}
 We assume a dephasing-noise channel with noise amplitude (error probability) \(0\le \noiseAmplitude\le 1\):
\begin{align}
\oC{E}{}{}{}{}{}\!\left(\rho\right) = \left(1-\noiseAmplitude\right) \rho +\noiseAmplitude\zOp\rho\zOp,
\end{align}
where \(\zOp\) is a Pauli Z operator. Using circuit notation we have:
\begin{align}\eqAL{circuitNotationDephasingNoise}
\gC{E}{}{}{}{}{} = \begin{cases}
\gG{}{}{}{}{}{wire} & \text{ with probability }1-\noiseAmplitude,\\
\gO{Z}{}{}{}{}{}&\text{ with probability } \noiseAmplitude.
\end{cases}
\end{align}

\subsubsection{Transfer Matrix}
The transfer matrix for dephasing noise is given by (\eqAR{PauliTransferMatrix}):
\begin{align}\eqAL{dephasingTransferMatrix}
\transferMatrixi{\oC{E}{}{}{}{}{}} = \begin{pmatrix}
1 & 0				& 0				   & 0 \\
0  &1-2\noiseAmplitude & 0				   & 0 \\
0  &0				& 1-2\noiseAmplitude & 0 \\
0  &0				& 0				   & 1 
\end{pmatrix},
\end{align}
this is a diagonal matrix with the eigenvalues:
\begin{align}
\oV{\lambda}{}{0}{}{}{}=&1,&\oV{\lambda}{}{1}{}{}{}=&1-2\noiseAmplitude,
\end{align}
both of which are repeated.

\subsubsection{Canonical Decomposition}
We can decompose the channel into its canonical form by defining:
\begin{align}\eqAL{canonicalDecomposition}
\oC{E}{}{}{}{}{}= &\oV{f}{}{0}{}{}{}\!\left(\noiseAmplitude\right)\oC{N}{}{0}{}{}{}+\oV{f}{}{1}{}{}{}\!\left(\noiseAmplitude\right)\oC{N}{}{1}{}{}{},&\oV{f}{}{0}{}{}{}\!\left(\noiseAmplitude\right)=&1-\noiseAmplitude,&\oC{N}{}{0}{}{}{}=&\oC{I}{}{}{}{}{},\\
&&\oV{f}{}{1}{}{}{}\!\left(\noiseAmplitude\right)=&\noiseAmplitude,&\oC{N}{}{1}{}{}{}\!\left(\rho\right) =&\zOp\rho\zOp,
\end{align}
where \(\oC{I}{}{}{}{}{}\) is the identity channel (returns the input unchanged).  It is worth noting here that \(\oC{N}{}{n}{}{}{}\) is independent of the noise amplitude: \(\noiseAmplitude\); and that both \(\oC{I}{}{}{}{}{}\) and \(\oC{N}{}{1}{}{}{}\) are CPTP (completely-positive and trace-preserving) channels. This means that if they act on a physical state they will return a physical state.

Since the identity channel commutes with all other channels this decomposition is self-compatible (\ie its constituent noise channels commute). It is also closed under composition since the composition of two consecutive dephasing-noise channels is equivalent to a single dephasing-noise channel with a modified noise amplitude. That is, if we define:
\begin{align}
\oC{E}{}{\noiseAmplitude}{}{}{}=&\left(1-\noiseAmplitude\right)\oC{I}{}{}{}{}{}+\noiseAmplitude\oC{N}{}{1}{}{}{},&\oC{E}{}{\noiseAmplitudeY}{}{}{}=&\left(1-\noiseAmplitudeY\right)\oC{I}{}{}{}{}{}+\noiseAmplitudeY\oC{N}{}{1}{}{}{},
\end{align}
then:
\begin{align}
\oC{E}{}{\noiseAmplitude}{}{}{}\circ \oC{E}{}{\noiseAmplitudeY}{}{}{}=\oC{E}{}{\noiseAmplitudeY}{}{}{}\circ \oC{E}{}{\noiseAmplitude}{}{}{}=&\left(\left(1-\noiseAmplitude\right)\left(1-\noiseAmplitudeY\right)+\noiseAmplitude\noiseAmplitudeY\right)\oC{I}{}{}{}{}{}+\left(\noiseAmplitude\left(1-\noiseAmplitudeY\right)+\noiseAmplitudeY\left(1-\noiseAmplitude\right)\right)\oC{N}{}{1}{}{}{},
\end{align}
(since \(\oC{N}{}{1}{}{}{}^{2}=\oC{I}{}{}{}{}{}\)) which is just the dephasing-noise channel \(\oC{E}{}{\noiseAmplitudeZ}{}{}{}\) where:
\begin{align}\eqAL{addingNoiseDephasing}
\noiseAmplitudeZ = \noiseAmplitude+\noiseAmplitudeY\left(1-2\noiseAmplitude\right).
\end{align}

\subsubsection{Circuit Noise Model}
The channel that corresponds to the full circuit (see \eqAR{circuitDephasingNoise}) is:
\begin{align}\eqAL{circuitDephasingNoiseMaps}
\oCN{C}{}{}{}{}{}~=&~\oCN{U}{}{\nGate}{}{}{L}\circ \oCN{U}{}{\nGate-1}{}{}{}\circ \dots \oCN{U}{}{1}{}{}{}~=\oC{U}{}{\nGate}{}{}{L}\circ\oC{E}{}{}{}{}{L}\circ \oC{U}{}{\nGate-1}{}{}{}\circ\oC{E}{}{}{}{}{L}\circ \dots \oC{U}{}{1}{}{}{}\circ\oC{E}{}{}{}{}{L}\nonumber\\
~=&~\left(1-\noiseAmplitude\right)^{\nGate}\oC{U}{}{\nGate}{}{}{L}\circ \oC{U}{}{\nGate-1}{}{}{}\circ \dots \oC{U}{}{1}{}{}{}\nonumber\\
&+\noiseAmplitude\left(1-\noiseAmplitude\right)^{\nGate-1}\left(\oC{U}{}{\nGate}{}{}{L}\circ\oC{N}{}{1}{}{}{}\circ \oC{U}{}{\nGate-1}{}{}{}\circ \dots \oC{U}{}{1}{}{}{}+\dots+\oC{U}{}{\nGate}{}{}{L}\circ \oC{U}{}{\nGate-1}{}{}{}\circ \dots \oC{U}{}{1}{}{}{}\circ\oC{N}{}{1}{}{}{}\right)\nonumber\\
&+\dots+\noiseAmplitude^{\nGate}\oC{U}{}{\nGate}{}{}{L}\circ\oC{N}{}{1}{}{}{}\circ \oC{U}{}{\nGate-1}{}{}{}\circ\oC{N}{}{1}{}{}{}\circ \dots \oC{U}{}{1}{}{}{}\circ\oC{N}{}{1}{}{}{},
\end{align}
where we have used the canonical error channel decomposition of \eqAR{canonicalDecomposition}; \(\oCN{C}{}{}{}{}{}\) is the channel corresponding to the whole circuit; and \(\oCN{U}{}{n}{}{}{}\), \(\oC{U}{}{n}{}{}{}\), and \(\oCN{E}{}{}{}{}{}\) are the channels corresponding to the \(n\)th noisy-gate, noise-free component of the noisy gate, and error channel of the noisy gate, respectively. 

We can write the expansion in \eqAR{circuitDephasingNoiseMaps} more succinctly in the form:
\begin{align}\eqAL{circuitDephasingNoiseMapsSuccinct}
\oCN{C}{}{}{}{}{}~=&~\sum_{k=0}^{\nGate}\left(1-\noiseAmplitude\right)^{\nGate-k}\noiseAmplitude^{k} \chiTermi{\stochasticLabel,k},
\end{align}
where:
\begin{align}\eqAL{circuitDephasingNoiseMapsComponents}
\chiTermi{\stochasticLabel,0}=&\oC{U}{}{\nGate}{}{}{L}\circ \oC{U}{}{\nGate-1}{}{}{}\circ \dots \oC{U}{}{1}{}{}{},\\
\chiTermi{\stochasticLabel,1}=&\oC{U}{}{\nGate}{}{}{L}\circ\oC{N}{}{1}{}{}{}\circ \oC{U}{}{\nGate-1}{}{}{}\circ \dots \oC{U}{}{1}{}{}{}+\dots+\oC{U}{}{\nGate}{}{}{L}\circ \oC{U}{}{\nGate-1}{}{}{}\circ \dots \oC{U}{}{1}{}{}{}\circ\oC{N}{}{1}{}{}{},\\
\dots&\nonumber\\
\chiTermi{\stochasticLabel,\nGate}=&\oC{U}{}{\nGate}{}{}{L}\circ\oC{N}{}{1}{}{}{}\circ \oC{U}{}{\nGate-1}{}{}{}\circ\oC{N}{}{1}{}{}{}\circ \dots \oC{U}{}{1}{}{}{}\circ\oC{N}{}{1}{}{}{},
\end{align}
\ie \(\chiTermi{\stochasticLabel,k}\) is the term containing all the constituent maps of \(\oCN{C}{}{}{}{}{}\) that include exactly \(k\) occurrences of the channel \(\oC{N}{}{1}{}{}{}\). Clearly, \(\chiTermi{\stochasticLabel,0}\) is equivalent to the noise-free target circuit (\(\oC{C}{}{}{}{}{}\), see \eqAR{circuitNoiseFree}):
\begin{align}
\chiTermi{\stochasticLabel,0}=\oC{U}{}{\nGate}{}{}{L}\circ \oC{U}{}{\nGate-1}{}{}{}\circ \dots \oC{U}{}{1}{}{}{}=\oC{C}{}{}{}{}{}.
\end{align}

We christen the number of constituent channels of \(\chiTermi{\stochasticLabel,k}\) its multiplicity:
\begin{align}\eqAL{multiplicityDephasing}
\multiplicity{\chiTermi{\stochasticLabel,k}}=\frac{\nGate!}{k!\left(\nGate-k\right)!}.
\end{align}

\subsubsection{Proxy Bias and Noise Level}\appSL{proxyBias}
The bias (\(\eBiasO\)) due to noise in a circuit for an observable (\(\obsOp\), here assumed to be a Pauli word operator\footnote{We choose a Pauli word operator for our proxy bias observable to simplify the calculation but this is not really necessary. In fact, all we need is an operator with an expectation (on a physical state) bounded between \(-1\) and \(1\). Since rescaling and shifting an observable does not change its fractional error, we can obtain a bound for any observable simply by rescaling the proxy bias given here.}: \(\obsOp=\pOpi{i}\)) is given by:
\begin{align}\eqAL{biasDephasingExample}
\eBiasO=\left|\xObsN-\xObs\right|,
\end{align}
where:
\begin{align}\eqAL{observableCalcDephasing}
\xObs=&\traceOp{\obsOp\oC{C}{}{}{}{}{} \left(\initialState\right)},&
\xObsN=&\traceOp{\obsOp\oCN{C}{}{}{}{}{} \left(\initialState\right)},
\end{align}
where \(\initialState\) is the initial state of the qubits, \(\xObs\) is the noise-free observable expectation and \(\xObsN\) is the noisy observable expectation.  Using our notation in \eqAR{circuitDephasingNoiseMapsSuccinct} we can re-express the bias in terms of our chi-terms (see \eqAR{multiplicityDephasing}):
\begin{align}\eqAL{dephasingProxyBiasDefinition}
\eBiasO=&\left|\traceOp{\obsOp\left(\oCN{C}{}{}{}{}{} \left(\initialState\right)-\oC{C}{}{}{}{}{} \left(\initialState\right)\right)}\right|=\left|\traceOp{\obsOp\left(\sum_{k=0}^{\nGate}\left(1-\noiseAmplitude\right)^{\nGate-k}\noiseAmplitude^{k} \chiTermi{\stochasticLabel,k} \left(\initialState\right)-\chiTermi{\stochasticLabel,0} \left(\initialState\right)\right)}\right|\nonumber\\
= &\left|\left(\left(1-\noiseAmplitude\right)^{\nGate}-1\right)\traceOp{\obsOp\chiTermi{\stochasticLabel,0} \left(\initialState\right)}+\sum_{k=1}^{\nGate}\left(1-\noiseAmplitude\right)^{\nGate-k}\noiseAmplitude^{k}\traceOp{\obsOp \chiTermi{\stochasticLabel,k} \left(\initialState\right)}\right|\nonumber\\
\le &\left(1-\left(1-\noiseAmplitude\right)^{\nGate}\right)+\sum_{k=1}^{\nGate}\left(1-\noiseAmplitude\right)^{\nGate-k}\noiseAmplitude^{k}\frac{\nGate!}{k!\left(\nGate-k\right)!}=2\left(1-\left(1-\noiseAmplitude\right)^{\nGate}\right),
\end{align}
where we have inverted the binomial expansion in the final step and in the penultimate step used the triangle inequality and  the fact that (see \eqAR{multiplicityDephasing}):
\begin{align}\eqAL{multiplicityBoundDephasing}
-\multiplicity{\chiTermi{\stochasticLabel,k}}\le \traceOp{\obsOp \chiTermi{\stochasticLabel,k} \left(\initialState\right)}\le \multiplicity{\chiTermi{\stochasticLabel,k}}.
\end{align}
This holds because each of the sub-maps in the sum that makes up \(\chiTermi{\stochasticLabel,k} \left(\initialState\right)\) is a CPTP map (see \eqAR{circuitDephasingNoiseMapsComponents}), which means when it acts on \(\initialState\) (a physical state) it returns a physical state. Since the modulus of the expectation of a Pauli operator with respect to a physical state (\(\rho\)) is bounded by unity:
\begin{align}
-1\le \traceOp{\rho\pOpi{i} }\le 1,
\end{align}
the maximum magnitude of the expectation with respect to \(\chiTermi{\stochasticLabel,k} \left(\initialState\right)\) occurs when every sub-map expectation has the same sign. In this case we simply count the sub-maps (\ie use the multiplicity) to find the bounds for the expectation with respect to \(\chiTermi{\stochasticLabel,k} \left(\initialState\right)\). We define the upper bound to the bias in \eqAR{dephasingProxyBiasDefinition} as the proxy bias:
\begin{align}\eqAL{dephasingProxyBiasExample}
\eBiasProxyN =&2\left(1-\left(1-\noiseAmplitude\right)^{\nGate}\right),
\end{align}
which is independent of the structure of the gates (\(\oG{U}{}{n}{}{}{}\)) within the circuit, depending only on how many error channels there are and what their noise amplitude is.

We define the noise level (\(\noiseLevel\)) of the circuit as the first order expansion of the proxy bias in the small noise (\(\noiseAmplitude\nGate\ll 1\)) regime:
\begin{align}\eqAL{dephasingExampleNoiseLevel}
\noiseLevel =2 \noiseAmplitude\nGate.
\end{align}

\subsection{Mitigation: IISM:NA}
We will now work through the mitigation method evaluation procedure for IISM:NA (noise-aware identity-insertion-assisted synchronous mitigation). This method uses knowledge about the noise model to effect the mitigation. It makes use of identity insertions to amplify the errors for each gate in the circuit. It is a synchronous mitigation strategy so it amplifies the errors of each gate by the same amount for each circuit variant.

\subsubsection{Identity Insertion}\appSL{identityInsertionDephasing}
Identity insertion involves resolving the (noisy) identity before every noisy gate, in terms of the noisy gate and its inverse. So:
\begin{align}\eqAL{iiDephasingGate}
\gGN{U}{}{n}{}{i}{}~=~&\Big(\gGN{U}{}{n}{}{}{}\gGDN{U}{}{n}{}{}{}\Big)^{i}\gGN{U}{}{n}{}{}{}~=~\Big(\gG{U}{}{n}{}{}{}\gC{E}{}{}{}{}{}\gGD{U}{}{n}{}{}{}\gC{E}{}{}{}{}{}\Big)^{i}\gG{U}{}{n}{}{}{}\gC{E}{}{}{}{}{}\nonumber\\
~=~&\gG{U}{}{n}{}{}{}\Big(\gC{E}{}{}{}{}{}\Big)^{2i+1},
\end{align}
is the effective noisy gate obtained after \(i\) identity insertions for the \(n\)th noisy gate in the circuit. Here we have assumed that the noisy gate and its noisy inverse have the same error channel, which commutes with them both:
\begin{align}\eqAL{dephasingNoisyGateInverse}
\gGN{U}{}{n}{}{}{}~=&~\gG{U}{}{n}{}{}{}\gC{E}{}{}{}{}{}~=~\gC{E}{}{}{}{}{}\gG{U}{}{n}{}{}{},&\gGDN{U}{}{n}{}{}{}~=&~\gGD{U}{}{n}{}{}{}\gC{E}{}{}{}{}{}~=~\gC{E}{}{}{}{}{}\gGD{U}{}{n}{}{}{}.
\end{align}

So the effective error channel of the noise-amplified gate \(\oGN{U}{}{n}{}{i}{}\), after \(i\) identity insertions is \(\oC{E}{}{}{}{i}{}\), where (see \eqARB{circuitNotationDephasingNoise}{iiDephasingGate}):
\begin{align}
\gC{E}{}{}{}{i}{} = \Big(\gC{E}{}{}{}{}{}\Big)^{2i+1}  = \begin{cases}
\gG{}{}{}{}{}{wire} & \text{ with probability }1-\noiseAmplitudei{i},\\
\gO{Z}{}{}{}{}{}&\text{ with probability } \noiseAmplitudei{i},
\end{cases}
\end{align}
where (see \eqARB{dephasingTransferMatrix}{addingNoiseDephasing}):
\begin{align}\eqAL{noiseIIEnhancedDephasing}
\noiseAmplitudei{i}=\frac{1}{2}\left(1-\left(1-2\noiseAmplitude\right)^{2i+1}\right),
\end{align}
is the amplified noise amplitude, where \(\noiseAmplitudei{0}=\noiseAmplitude\) is the original noise amplitude. So, to leading order, the noise amplitude of the \(i\)th circuit is \(2i+1\) times the original noise amplitude:
\begin{align}\eqAL{noiseAmplitudeIIDephasing}
\noiseAmplitudei{i}=\left(2i+1\right)\noiseAmplitude+\orderZero \left(\noiseAmplitude^{2}\right).
\end{align}

\subsubsection{Zero-Noise Extrapolation}\appSL{ZNE}
Zero-noise extrapolation or Richardson extrapolation \cite{Temme2017} is a mitigation method that involves finding a weighted sum of noisy observables that is equivalent to the noise-free observable up to some power in the noise amplitude:
\begin{align}\eqAL{zeroNoiseMitigationDephasing}
\xObsM = \sum_{i=0}^{\nMit } \coefGlobali{i} \xObsNi{i}=\xObs+\orderZero \left(\noiseAmplitude^{\nMit +1}\right),
\end{align}
where (see \eqAR{observableCalcDephasing}):
\begin{align}
\xObsM=&\traceOp{\obsOp\sum_{i=0}^{\nMit } \coefGlobali{i}\oCN{C}{}{}{}{i}{} \left(\initialState\right)},&
\xObsNi{i}=&\traceOp{\obsOp\oCN{C}{}{}{}{i}{} \left(\initialState\right)},&
\xObs=&\traceOp{\obsOp\oC{C}{}{}{}{}{} \left(\initialState\right)},
\end{align}
are the mitigated observable, \(i\)th noisy observable, and noise-free observable, respectively; and \(\nMit\) is the order of the mitigation and the number of extra noisy circuit variants required for the mitigation. Each of the noisy circuit variants should be equivalent to the original circuit except for having an amplified noise amplitude.

We Taylor expand our noisy observables in terms of the noise amplitude:
\begin{align}
\xObsNi{i}=\xObsN\left(\noiseAmplitudei{i}\right) = \sum_{m=0}^{\infty} \dTaylori{m}\noiseAmplitudei{i}^{m}=\xObs+\orderZero\left(\noiseAmplitudei{i}\right),
\end{align}
where \(\dTaylori{m}\) are coefficients independent of \(\noiseAmplitudei{i}\). For dephasing noise we have (\eqAR{circuitDephasingNoiseMapsSuccinct}):
\begin{align}
\xObsNi{i}=\traceOp{\obsOp\oCN{C}{}{}{}{i}{} \left(\initialState\right)}=\sum_{k=0}^{\nGate}\left(1-\noiseAmplitudei{i}\right)^{\nGate-k}\noiseAmplitude^{k}\traceOp{\obsOp\chiTermi{\stochasticLabel,k} \left(\initialState\right)},
\end{align}
and therefore:
\begin{align}\eqAL{dDefinitionDephasing}
\dTaylori{m}=\begin{cases}
\sum_{k=0}^{m} \frac{\left(\nGate-k\right)!}{\left(\nGate-m\right)! \left(m-k\right)!}\left(-1\right)^{m-k}\traceOp{\obsOp\chiTermi{\stochasticLabel,k}\left(\initialState\right)}&\text{ for }m\le\nGate,\\
0&\text{ for }m> \nGate.
\end{cases}
\end{align}

This means \eqAR{zeroNoiseMitigationDephasing} is equivalent to:
\begin{align}\eqAL{mitigatedObservableZNEDephasing}
\xObsM = \sum_{i=0}^{\nMit } \coefGlobali{i} \xObsNi{i}= \sum_{m=0}^{\infty} \dTaylori{m}\sum_{i=0}^{\nMit } \coefGlobali{i}\noiseAmplitudei{i}^{m}=\xObs+\orderZero \left(\noiseAmplitude^{\nMit +1}\right),
\end{align}
or equivalently:
\begin{align}\eqAL{coefficientEquationsDephasing}
\sum_{i=0}^{\nMit }\coefGlobali{i}=&1, &\sum_{i=0}^{\nMit }\coefGlobali{i}\noiseAmplitudei{i}^k=\noiseAmplitude^k\sum_{i=0}^{\nMit }\coefGlobali{i}\left(\frac{\noiseAmplitudei{i}}{\noiseAmplitude}\right)^k=&0 \text{ for } 1\le k\le \nMit.
\end{align}
So the coefficients depend only on the relative quantity \(\noiseAmplitudei{i}/\noiseAmplitude\) and not on the absolute value of the noise amplitude. If we solve \eqAR{coefficientEquationsDephasing} we find (see \citeRef{He2020}):
\begin{align}\eqAL{mitigationCoefficientsDephasingIISM}
\coefGlobali{i}=&\prod_{m=0:m\ne i}^{\nMit }\frac{\noiseAmplitudei{m}}{\noiseAmplitudei{m}-\noiseAmplitudei{i}},&
\sum_{i=0}^{\nMit }\coefGlobali{i}\noiseAmplitudei{i}^{\nMit +1}=&\left(-1\right)^{\nMit }\prod_{n=0}^{\nMit }\noiseAmplitudei{n}.
\end{align}
Thus (\eqAR{mitigatedObservableZNEDephasing}):
\begin{align}
\xObsM = \xObs+ \left(-1\right)^{\nMit }\left(\prod_{n=0}^{\nMit }\noiseAmplitudei{n}  \right)\dTaylori{\nMit +1}+\orderZero \left(\noiseAmplitude^{\nMit +2}\right).
\end{align}

\subsubsection{Bias, Proxy Bias, and Noise Level after Mitigation}\appSL{mitigatedProxyBias}
The bias is given by (see \eqARD{multiplicityDephasing}{biasDephasingExample}{multiplicityBoundDephasing}{dDefinitionDephasing}):
\begin{align}\eqAL{biasBoundIISMNADephasing}
\eBiasO=&\left|\xObsM-\xObs\right|= \left(\prod_{n=0}^{\nMit }\noiseAmplitudei{n}  \right)\left|\dTaylori{\nMit +1}\right|+\orderZero \left(\noiseAmplitude^{\nMit +2}\right)\nonumber\\
=& \left(\prod_{n=0}^{\nMit }\noiseAmplitudei{n}  \right)\sum_{k=0}^{\nMit+1} \frac{\left(\nGate-k\right)!}{\left(\nGate-\nMit-1\right)! \left(\nMit+1-k\right)!}\left|\traceOp{\obsOp\chiTermi{\stochasticLabel,k}\left(\initialState\right)}\right|+\orderZero \left(\noiseAmplitude^{\nMit +2}\right)\nonumber\\
\le & \left(\prod_{n=0}^{\nMit }\noiseAmplitudei{n}  \right)\sum_{k=0}^{\nMit+1} \frac{\nGate!}{\left(\nGate-\nMit-1\right)! \left(\nMit+1-k\right)!}\frac{}{k!}+\orderZero \left(\noiseAmplitude^{\nMit +2}\right)\nonumber\\
 &= 2^{\nMit+1}\left(\prod_{n=0}^{\nMit }\noiseAmplitudei{n}  \right) \frac{\nGate!}{\left(\nGate-\nMit-1\right)! \left(\nMit+1\right)!}+\orderZero \left(\noiseAmplitude^{\nMit +2}\right),
\end{align}
where we have assumed that the observable is a Pauli word operator (\(\obsOp=\pOpi{i}\)), as in \appSR{proxyBias}, and therefore the modulus of its expectation for any physical operator is bounded by unity. We have also assumed that \(\nMit<\nGate\), if \(\nMit\ge\nGate\) then the mitigation is unbiased. So the mitigated proxy bias (an upper bound for the bias) can be given by:
\begin{align}\eqAL{proxyBiasDephasingSmallNoise}
\eBiasProxyM = 2^{\nMit+1}\left(\prod_{n=0}^{\nMit }\left(2n+1\right) \right) \frac{\nGate!}{\left(\nGate-\nMit-1\right)! \left(\nMit+1\right)!} x^{\nMit+1}+\orderZero \left(\noiseAmplitude^{\nMit +2}\right),
\end{align}
where we have used the expression for the noise amplitude of the \(n\)th circuit variant (\(\noiseAmplitudei{n}\)) given in \eqAR{noiseAmplitudeIIDephasing}. Since we have:
\begin{align}
\frac{\nGate!}{\left(\nMit+1\right)!\left(\nGate-\nMit-1\right)!}=\frac{\nGate^{\nMit+1}}{\left(\nMit+1\right)!}+\orderLarge\left(\nGate^{\nMit}\right),
\end{align}
in the large circuit (\(\nGate\gg 1\)) and small noise (\(\nGate \noiseAmplitude \ll 1\)) regime  we have:
\begin{align}\eqAL{proxyBiasDephasingFull}
\eBiasProxyM = 2^{\nMit+1}\left(\prod_{i=0}^{\nMit }\left(2i+1\right) \right)\frac{\left(\nGate\noiseAmplitude\right)^{\nMit+1}}{\left(\nMit+1\right)!} +\orderLargeSmall \left(\nGate^{\nMit}\noiseAmplitude^{\nMit +1},\nGate^{\nMit +2}\noiseAmplitude^{\nMit +2}\right),
\end{align}
where we have used \eqARB{dDefinitionDephasing}{mitigatedObservableZNEDephasing} in order to determine the correct scaling with \(\nGate\) and \(\noiseAmplitude\). This means the noise level (after mitigation) is given by (\eqAR{proxyBiasDephasingSmallNoise}):
\begin{align}\eqAL{proxyBiasDephasingIISM}
\noiseLevelM=\left(\prod_{i=0}^{\nMit }\left(2i+1\right) \right) \frac{\nGate!}{\nGate^{\nMit+1}\left(\nGate-\nMit-1\right)! \left(\nMit+1\right)!} \noiseLevel^{\nMit+1},
\end{align}
in the small noise regime (\(\nGate \noiseAmplitude \ll 1\)) and\footnote{The ratio of the value used to bound the mitigated noise level in \eqAR{noiseLevelDephasingLargeCircuit} to the true mitigated noise level is roughly \(\sqrt{\uppi\left(\nMit+1\right)}/2\)  (this value has been obtained numerically) for large \(\nMit\).}:
\begin{align}\eqAL{noiseLevelDephasingLargeCircuit}
\noiseLevelM=\frac{\left(2\nMit+1\right)!!}{\left(\nMit+1\right)!} \noiseLevel^{\nMit+1}<\frac{\left(2\nMit+2\right)!!}{2\left(\nMit+1\right)!} \noiseLevel^{\nMit+1}=\frac{\left(2\noiseLevel\right)^{\nMit+1}}{2},
\end{align}
in the small noise and large circuit regime (\(\noiseAmplitude\nGate ,\nGate^{-1} \ll 1\)). Where \(!!\) is the double factorial, defined by:
\begin{align}
n!!=&n\left(n-2\right)!!,&1!!=0!! =1,
\end{align}
and \(\noiseLevel=2\noiseAmplitude\nGate\) is defined in \eqAR{dephasingExampleNoiseLevel}.


\subsubsection{Sampling Cost, Length Scale Factor, and Runtime Scaling}
The most efficient way to implement our mitigation procedure (\eqAR{mitigatedObservableZNEDephasing}) is to use Monte-Carlo sampling \cite{Endo2018}. To implement this we define a probability (\(\probGlobali{i}\)) and sign (\(\signGlobali{i}\)) associated with each of circuit variants, so the \(i\)th circuit has:
\begin{align}\eqAL{probabilitiesCHILM}
\probGlobali{i}=&\frac{\left|\oV{c}{Gl}{}{}{i}{}\right|}{\oV{C}{Gl}{}{}{}{}},&\signGlobali{i}=&\frac{\oV{c}{Gl}{}{}{i}{}}{\left|\oV{c}{Gl}{}{}{i}{}\right|},&\oV{C}{Gl}{}{}{}{}=&\sum_{i=0}^{\nMit}\left|\oV{c}{Gl}{}{}{i}{}\right|,
\end{align}
where \(\oV{C}{Gl}{}{}{}{}\) is known as the sampling cost \cite{Cai2021}, for reasons that will become apparent. We simply make \(\nRun\)  runs of our circuit and each time we randomly choose one of our \(\nMit+1\) circuits to implement and multiply the resulting output by the sign \(\signGlobali{i}\) corresponding to the circuit \(i\) that we choose for that run. We sample circuit \(i\) with probability \(\probGlobali{i}\). The expectation of this procedure is \(\xObsM /\oV{C}{Gl}{}{}{}{}\) so we simply multiply our expectation by the sampling cost in order to obtain our mitigated value. The variance of the Monte Carlo sampling procedure is given by the shot noise:
\begin{align}
\sigma_{\mathrm{MC}}^2 \propto \frac{1}{\nRun}.
\end{align}
However, because we multiply our final result by \(\oV{C}{Gl}{}{}{}{}\) we will increase the variance by a factor of \(\oV{C}{Gl}{}{}{}{}^2\). This means the ratio of the number of runs needed for the mitigated circuit to the number of runs needed for the unmitigated circuit, to yield the same variance is approximately \cite{Endo2018}:
\begin{align}
\frac{\nRunM}{\nRun} \approx \oV{C}{Gl}{}{}{}{}^{2}.
\end{align}
So  \(\oV{C}{Gl}{}{}{}{}\) is the (square root of the) cost for sampling quasi-stochastic ensembles, if \(\oV{c}{Gl}{}{}{i}{}>0\) for all \(i\) then \(\oV{C}{Gl}{}{}{}{}=1\) (see \eqAR{coefficientEquationsDephasing}). For the coefficients given in \eqAR{mitigationCoefficientsDephasingIISM} (see \eqAR{noiseAmplitudeIIDephasing}):
\begin{align}\eqAL{coefficientExpansionDephasingIISM}
\coefGlobali{i}=&\prod_{m=0:m\ne i}^{\nMit }\frac{2m+1+\orderZero \left(\noiseAmplitude\right)}{2\left(m-i\right)+\orderZero \left(\noiseAmplitude\right)}=\left(-1\right)^{i}\frac{\left(2\nMit+1\right)!!}{2^{\nMit}\left(2i+1\right)\left(i!\right)\left(\left(\nMit-i\right)!\right)}+\orderZero \left(\noiseAmplitude\right).
\end{align}
So the sampling cost for using IISM:NA to mitigate dephasing noise is given by\footnote{The sampling cost sum (see \eqARB{coefficientExpansionDephasingIISM}{IISMCost}) was calculated using Mathematica \cite{Mathematica13}.}:
\begin{align}\eqAL{IISMCost}
\oV{C}{Gl}{}{}{}{}=&\sum_{i=0}^{\nMit}\left|\oV{c}{Gl}{}{}{i}{}\right|=\frac{\left(2\nMit+1\right)!!{}_2F_1\left(\frac{1}{2},-\nMit ;\frac{3}{2} ; -1\right)}{2^{\nMit}\left(\nMit!\right)}+\orderZero \left(\noiseAmplitude\right), 
\end{align}
where \(!!\) is the double factorial,
\begin{align}
{}_2F_1\left(a,b;c; z\right)=\sum_{k=0}^{\infty}\frac{\Gamma\left(a+k\right)}{\Gamma\left(a\right)}\frac{\Gamma\left(b+k\right)}{\Gamma\left(b\right)}\frac{\Gamma\left(c\right)}{\Gamma\left(c+k\right)}\frac{z^k}{k!},
\end{align}
is a Hypergeometric function and:
\begin{align}
\Gamma\left(z\right)=\int_0^{\infty}t^{z-1}\mathrm{e}^{-t}\mathrm{d} t,
\end{align}
is the Euler gamma function (see \cite{Mathematica13}).

As well as increasing the number of samples required, each run will take longer, on average, to implement because we are increasing the number of gates we use. The \(i\)th circuit has \(2i+1\) times the gates of the original noisy circuit and so will take \(2i+1\) times longer (neglecting initialisation and measurement time). Thus the average runtime of the mitigated circuit (\(\timeExecM\)) relative to the original noisy circuit (\(\timeExec\)) is:
\begin{align}\eqAL{lengthScaleINdependenceDephasing}
\oV{F}{Gl,L}{}{}{}{} = \frac{\timeExecM}{\timeExec}=\sum_{i=0}^{\nMit}\probGlobali{i}\left(2i+1\right)=\frac{\sum_{i=0}^{\nMit}\left|\oV{c}{Gl}{}{}{i}{}\right|\left(2i+1\right)}{\oV{C}{Gl}{}{}{}{}},
\end{align}
which we denote the length scale factor (again neglecting initialisation and measurement time). So the length scale factor for using IISM:NA to mitigate dephasing noise is given by:
\begin{align}\eqAL{IISMF}
\oV{F}{Gl,L}{}{}{}{}=&\frac{\left(2\nMit+1\right)!!}{2^{\nMit}\oV{C}{Gl}{}{}{}{}}\sum_{i=0}^{\nMit}\frac{1}{\left(i!\right)\left(\left(\nMit-i\right)!\right)}+\orderZero \left(\noiseAmplitude\right)=\frac{\left(2\nMit+1\right)!!}{\nMit!\oV{C}{Gl}{}{}{}{}}+\orderZero \left(\noiseAmplitude\right)\nonumber\\
=&\frac{2^{\nMit}}{{}_2F_1\left(\frac{1}{2},-\nMit ;\frac{3}{2} ; -1\right)}+\orderZero \left(\noiseAmplitude\right).
\end{align}

Thus the ratio of the total mitigation procedure runtime (\(\timeM\)) compared to that for the original noisy circuit (\(\timeN\)) to achieve the same variance is given by:
 \begin{align}\eqAL{scalabilityDephasingExample}
\oV{S}{Gl}{}{}{}{} = \frac{\timeM}{\timeN}\approx
\oV{C}{Gl}{}{}{}{}^{2}\oV{F}{Gl,L}{}{}{}{}=\frac{\left(\left(2\nMit+1\right)!!\right)^{2}{}_2F_1\left(\frac{1}{2},-\nMit ;\frac{3}{2} ; -1\right)}{2^{\nMit}\left(\nMit!\right)^{2}}+\orderZero \left(\noiseAmplitude\right),
 \end{align}
which we denote the runtime scaling (again we assume negligible initialisation and measurement times). The width scale factor doesn't feature here because we use the same number of qubits as the original noisy circuit.


\subsubsection{Noise Boundary}\appSL{noiseBoundaryWorkedExample}
The noise boundary is the maximum noise that can be mitigated for a fixed maximum runtime scaling (\(
\oV{S}{Gl}{}{}{}{} =\scalabilityTarget\)) and target bias (\(\eBiasO=\eBiasTarget\)). To ensure we do not exceed the bias we choose an order of mitigation (\(\nMit\)) that will reduce the proxy bias to below the target bias (\(\eBiasTarget\)). To make our analytical analysis more straightforward we will use the mitigated noise level bound derived in \eqAR{noiseLevelDephasingLargeCircuit} rather than the full proxy bias. So we set the target bias (\(\eBiasTarget\)) equal to the mitigated noise level bound and then invert the expression to find the required mitigation order (\(\nMit\)): 
\begin{align}\eqAL{mitigationOrderDephasingLargeCircuit}
\eBiasTarget=&\frac{\left(2\noiseLevel\right)^{\nMit+1}}{2}&\implies &&\nMit=&\frac{\ln\left(2\eBiasTarget\right)}{\ln\left(2\noiseLevel\right)}-1.
\end{align}
Now, we want to find the largest possible noise level (\(\noiseLevel\)) that we can mitigate so we will assume that \(\nMit\gg 1\), \ie our target bias is much smaller than our original noise level. In this case we can use the fact that:
 \begin{align}
{}_2F_1\left(\frac{1}{2},-\nMit ;\frac{3}{2} ; -1\right)=&\frac{2^{\nMit}}{\nMit}\left(1+\orderGeneral\left(\frac{1}{\nMit^2}\right)\right),\\
\frac{\left(2\nMit+1\right)!!}{2^{\nMit}\left(\nMit!\right)}=&2\sqrt{\frac{\nMit}{\uppi}}\left(1+\frac{3}{8\nMit}+\orderGeneral\left(\frac{1}{\nMit^2}\right)\right).
\end{align}
(found by numerical and analytical analysis) to simplify the form of our sampling cost and length scale factor (\eqARB{IISMCost}{IISMF}):
\begin{align}\eqAL{samplingCostExpansionIISMNADephasing}
\oV{C}{Gl}{}{}{}{}=&\frac{\left(2\nMit+1\right)!!{}_2F_1\left(\frac{1}{2},-\nMit ;\frac{3}{2} ; -1\right)}{2^{\nMit}\left(\nMit!\right)}+\orderZero \left(\noiseAmplitude\right)=\frac{2^{\nMit+1}}{\sqrt{\nMit\uppi}}\left(1+\orderGeneral\left(\frac{1}{\nMit}\right)\right)+\orderZero \left(\noiseAmplitude\right), \\
\oV{F}{Gl,L}{}{}{}{}=&\frac{2^{\nMit}}{{}_2F_1\left(\frac{1}{2},-\nMit ;\frac{3}{2} ; -1\right)}+\orderZero \left(\noiseAmplitude\right)=\nMit\left(1-\orderGeneral\left(\frac{1}{\nMit^2}\right)\right)+\orderZero \left(\noiseAmplitude\right),
\end{align}
which means the runtime scaling is given by (\eqAR{scalabilityDephasingExample}):
\begin{align}\eqAL{scalabilitySimplifiedExpressionDephasing}
\oV{S}{Gl}{}{}{}{} \approx
\oV{C}{Gl}{}{}{}{}^{2}\oV{F}{Gl,L}{}{}{}{}=\frac{4^{\nMit+1}}{\uppi}\left(1+\orderGeneral\left(\frac{1}{\nMit}\right)\right)+\orderZero\left(\noiseAmplitude\right).
\end{align}
to leading order. Now, we set our runtime scaling equal to our maximum runtime scaling (\(\scalabilityTarget\)) and substitute our mitigation order from \eqAR{mitigationOrderDephasingLargeCircuit}, then rearrange to determine the maximum noise level we can mitigate:
\begin{align}
\scalabilityTarget = &\frac{4^{\frac{\ln\left(2\eBiasTarget\right)}{\ln\left(2\noiseLevel\right)}}}{\uppi}&\implies &&\noiseLevel = 2^{2\frac{\ln\left(2\eBiasTarget\right)}{\ln\left(\uppi\scalabilityTarget\right)}-1}.
\end{align}
So the noise boundary is given by:
\begin{align}\eqAL{IISMDephasingBoundary}
\oV{e}{B}{}{}{}{}= 2^{2\frac{\ln\left(2\eBiasTarget\right)}{\ln\left(\uppi\scalabilityTarget\right)}-1}=\frac{\left(2\eBiasTarget\right)^{\frac{\ln\left(4\right)}{\ln\left(\uppi\scalabilityTarget\right)}}}{2}.
\end{align}
So we require a runtime scaling greater than \(4/\uppi\) to achieve any mitigation at all and the larger our tolerated runtime scaling the larger noise boundary we can tolerate. However large our tolerated runtime scaling is, the noise boundary will always be smaller than 0.5. 

\subsubsection{Robustness}\appL{IISMNARobustness}
A method is robust if the runtime scaling and the proxy bias are unchanged (to leading order) if there are some imperfections in the noise amplification procedure. For identity insertions (\appSR{identityInsertionDephasing}), the source of imperfections would be differences between the noise model of the inverse gate and the original noisy gate, \ie we have (\cf \eqAR{dephasingNoisyGateInverse}):
\begin{align}
\gGN{U}{}{n}{}{}{}~=&~\gG{U}{}{n}{}{}{}\gC{E}{}{}{}{}{}~=~\gC{E}{}{}{}{}{}\gG{U}{}{n}{}{}{},&\gGDN{U}{}{n}{}{}{}~=&~\gGD{U}{}{n}{}{}{}\gC{E}{}{\dagger}{}{}{}~=~\gC{E}{}{\dagger}{}{}{}\gGDN{U}{}{n}{}{}{}.
\end{align}
where:
\begin{align}
\gC{E}{}{}{}{}{L}=&\begin{cases}
\gG{}{}{}{}{}{Wire}~&\text{ with prob. }1-\noiseAmplitude\\
\gG{Z}{}{}{}{}{}~&\text{ with prob. }\noiseAmplitude
\end{cases},&
\gC{E}{}{\dagger}{}{}{L}=&\begin{cases}
\gG{}{}{}{}{}{Wire}~&\text{ with prob. }1-\oV{y}{}{\dagger}{}{}{}\noiseAmplitude,\\
\gG{Z}{}{}{}{}{}~&\text{ with prob. }\oV{y}{}{\dagger}{}{}{}\noiseAmplitude.
\end{cases}
\end{align}
In this case the amplified gate noise amplitudes are (see \eqARB{dephasingTransferMatrix}{addingNoiseDephasing}):
\begin{align}\eqAL{stochasticNoiseRobustIISMNA}
\noiseAmplitudei{i}= &\frac{1-\left(1-2\left(1+\oV{y}{}{\dagger}{}{}{}\right)\noiseAmplitude+4\oV{y}{}{\dagger}{}{}{}\noiseAmplitude^{2}\right)^{i}\left(1-2\noiseAmplitude\right)}{2}=\left(1+i\left(1+\oV{y}{}{\dagger}{}{}{}\right)\right)\noiseAmplitude+\orderZero\left(\noiseAmplitude^{2}\right).
\end{align}
This means the coefficients are given by (\eqAR{mitigationCoefficientsDephasingIISM}):
\begin{align}
\coefGlobali{i}=\prod_{m=0:m\ne i}^{\nMit }\frac{\noiseAmplitudei{m}}{\noiseAmplitudei{m}-\noiseAmplitudei{i}}=\left(-1\right)^{i}\frac{\prod_{m=0}^{\nMit }\left(1+m\left(1+\oV{y}{}{\dagger}{}{}{}\right)\right)}{\left(1+\oV{y}{}{\dagger}{}{}{}\right)^{\nMit}\left(1+i\left(1+\oV{y}{}{\dagger}{}{}{}\right)\right)i!\left(\nMit-i\right)!}+\orderZero\left(\noiseAmplitude\right).
\end{align}
This means the sampling cost is given by (\eqAR{IISMCost}):
\begin{align}
\oV{C}{Gl}{}{}{}{}=&\sum_{j=0}^{\nMit}\left|\oV{c}{Gl}{}{}{j}{}\right|=&\frac{\prod_{m=0}^{\nCircGlobal }\left(1+m\left(1+\oV{y}{}{\dagger}{}{}{}\right)\right){}_2F_1\left(\frac{1}{1+\oV{y}{}{\dagger}{}{}{}}, -\nMit;\frac{\oV{y}{}{\dagger}{}{}{}+2}{\oV{y}{}{\dagger}{}{}{}+1} ; -1\right)}{\left(1+\oV{y}{}{\dagger}{}{}{}\right)^{\nMit}\left(\nMit!\right)}+\orderZero\left(\noiseAmplitude\right), 
\end{align}
and the length scale factor by (\eqAR{lengthScaleINdependenceDephasing}):
\begin{align}
\oV{F}{Gl}{}{}{}{} =&\frac{\sum_{i=0}^{\nMit}\left|\oV{c}{Gl}{}{}{i}{}\right|\left(2i+1\right)}{\oV{C}{Gl}{}{}{}{}}=1+2\nMit\frac{{}_2F_1\left(\frac{2+\oV{y}{}{\dagger}{}{}{}}{1+\oV{y}{}{\dagger}{}{}{}}, 1-\nMit;\frac{2\oV{y}{}{\dagger}{}{}{}+3}{\oV{y}{}{\dagger}{}{}{}+1} ; -1\right)}{\left(\oV{y}{}{\dagger}{}{}{}+2\right){}_2F_1\left(\frac{1}{1+\oV{y}{}{\dagger}{}{}{}}, -\nMit;\frac{\oV{y}{}{\dagger}{}{}{}+2}{\oV{y}{}{\dagger}{}{}{}+1} ; -1\right)}+\orderZero\left(\noiseAmplitude\right).
\end{align}

The proxy bias is given by (see \eqARB{biasBoundIISMNADephasing}{stochasticNoiseRobustIISMNA}):
\begin{align}
\oV{\varepsilon}{\proxy}{}{\scriptscriptstyle M}{}{}=&2^{\nMit+1}\left(\prod_{n=0}^{\nMit }\noiseAmplitudei{n}  \right) \frac{\nGate!}{\left(\nGate-\nMit-1\right)! \left(\nMit+1\right)!}+\orderZero \left(\noiseAmplitude^{\nMit +2}\right)\nonumber\\
=&\frac{\prod_{n=0}^{\nMit }\left(2\nGate\noiseAmplitudei{n} \right) }{\left(\nMit+1\right)!}+\orderLargeSmall \left(\nGate^{\nMit}\noiseAmplitude^{\nMit +1},\nGate^{\nMit+2}\noiseAmplitude^{\nMit +2}\right)\nonumber\\
=&\left(2\nGate\noiseAmplitude \right)^{\nMit+1}\frac{\prod_{n=0}^{\nMit }\left(\left(1+n\left(1+\oV{y}{}{\dagger}{}{}{}\right)\right)\right) }{\left(\nMit+1\right)!}+\orderLargeSmall \left(\nGate^{\nMit}\noiseAmplitude^{\nMit +1},\nGate^{\nMit+2}\noiseAmplitude^{\nMit +2}\right).
\end{align}
So the mitigated noise level is given by:
\begin{align}
\noiseLevelM=\noiseLevel^{\nMit+1}\frac{\prod_{n=0}^{\nMit }\left(\left(1+n\left(1+\oV{y}{}{\dagger}{}{}{}\right)\right)\right) }{\left(\nMit+1\right)!}.
\end{align}

Thus we see that the imperfection in the inverse noisy gate's error channel will change the sampling cost and length scale factor, and therefore the runtime scaling; and also the proxy bias and therefore the mitigated noise level. This means that IISM:NA is not robust but it is quasi-robust. Since, though the metrics depend on the value of  \(\oV{y}{}{\dagger}{}{}{}\) to leading order the order of the leading order term is independent of \(\oV{y}{}{\dagger}{}{}{}\), \ie it is still zeroth order in the noise amplitude (\(\noiseAmplitude\)) for the sampling cost and the length scale factor and \(\nMit+1\)th order in the noise level (\(\noiseLevel\)) for the proxy bias. Here we have assumed that, despite the imperfect control, we are able to achieve perfect characterisation. So, we cannot generate noisy gate variants precisely but we know the impact of the errors that we make during the generation procedure.

\subsubsection{Qualitative Certification}
Now we have all the information we need (see \eqARC{proxyBiasDephasingIISM}{scalabilitySimplifiedExpressionDephasing}{IISMDephasingBoundary}) to provide a qualitative evaluation of IISM:NA. We will assess each of the certification criteria for mitigating dephasing noise with IISM:NA. We include the proxy bias (\(\eBiasProxyM \), \eqARB{proxyBiasDephasingFull}{noiseLevelDephasingLargeCircuit}), runtime scaling (\(\oV{S}{Gl}{}{}{}{} \), \eqARF{scalabilitySimplifiedExpressionDephasing}{scalabilitySimplifiedExpressionDephasing}{lengthScaleINdependenceDephasing}{IISMCost}{mitigationCoefficientsDephasingIISM}{noiseIIEnhancedDephasing}), and noise boundary (\(\oV{e}{B}{}{}{}{}\), \eqAR{IISMDephasingBoundary}) again, for illustrative purpose:
\begin{align}\eqAL{preciseDphIISMNA}
\eBiasProxyM <& \frac{\left(2\noiseLevel\right)^{\nMit+1}}{2} +\orderLargeSmall \left(\frac{\noiseLevel^{\nMit +1}}{\nGate},\noiseLevel^{\nMit +2}\right),
\end{align}
 \begin{align}\eqAL{scalabilityDPhIISMNA}
\oV{S}{Gl}{}{}{}{}\approx&
\oV{C}{Gl}{}{}{}{}^{2}\oV{F}{Gl,L}{}{}{}{}=\frac{4^{\nMit+1}}{\uppi}\left(1+\orderGeneral\left(\frac{1}{\nMit}\right)\right)+\orderZero\left(\noiseAmplitude\right),~~\oV{F}{Gl,L}{}{}{}{} =\frac{\sum_{i=0}^{\nMit}\left|\oV{c}{Gl}{}{}{i}{}\right|\left(2i+1\right)}{\oV{C}{Gl}{}{}{}{}},\nonumber\\
\oV{C}{Gl}{}{}{}{}=&\sum_{i=0}^{\nMit}\left|\oV{c}{Gl}{}{}{i}{}\right|,~~
\coefGlobali{i}=\prod_{m=0:m\ne i}^{\nMit }\frac{\frac{1}{2}\left(1-\left(1-2\noiseAmplitude\right)^{2m+1}\right)}{\frac{1}{2}\left(1-\left(1-2\noiseAmplitude\right)^{2m+1}\right)-\frac{1}{2}\left(1-\left(1-2\noiseAmplitude\right)^{2i+1}\right)},
\end{align}
\begin{align}\eqAL{unboundedDPhIISMNA}
\oV{e}{B}{}{}{}{}= 2^{2\frac{\ln\left(2\eBiasTarget\right)}{\ln\left(\uppi\scalabilityTarget\right)}-1}=\frac{\left(2\eBiasTarget\right)^{\frac{\ln\left(4\right)}{\ln\left(\uppi\scalabilityTarget\right)}}}{2}.
\end{align}

\begin{itemize}
\item \textbf{Scalable}: Yes (see \eqAR{scalabilityDPhIISMNA}); the coefficients of the mitigation are independent of the number of gates in the circuit (see \eqAR{mitigationCoefficientsDephasingIISM}), which means so is the sampling cost (\eqAR{IISMCost}). The length scale factor is also independent of the number of gates in the circuit as it depends only on the coefficients, the sampling cost, and the number of identity insertions (\eqAR{lengthScaleINdependenceDephasing}). Since the runtime scaling depends only on the length scale factor and the sampling cost, it is independent of the number of gates in the circuit and therefore the mitigation method is scalable.
\item \textbf{Unbounded}: No (see \eqAR{unboundedDPhIISMNA}); the maximum noise level we can mitigate is 0.5, no matter how large we allow the runtime scaling to be so we cannot mitigate arbitrarily large noise levels.
\item \textbf{Precise}: Yes (see \eqAR{preciseDphIISMNA}); we can make the proxy bias (and therefore the bias) as small as we like by increasing \(\nMit\) (the proxy bias is exponentially suppressed), provided that \(\noiseLevel< 0.5\). So we can achieve arbitrary accuracy with this method if we are willing to pay the price (of a long runtime).
\item \textbf{Efficient}: No (see \eqAR{scalabilityDPhIISMNA}); even if the noise level (and therefore the noise amplitude, \(\noiseAmplitude\)) tends to zero the runtime scaling will not be unity. Even a first order (\(\nMit=1\)) mitigation has \(\oV{S}{Gl}{}{}{}{}=6\) in the zero-noise limit (\(\noiseAmplitude\rightarrow 0\)) and all higher order mitigations would have a larger runtime scaling. So this method does not make good use of reductions in the noise level before mitigation.
\item \textbf{Robust}: Yes (well quasi-robust, see \appR{IISMNARobustness});  the runtime scaling and the proxy bias are modified to leading order if the inverse noisy gate has a different noise amplitude to the original noisy gate but the order of the leading order term is unchanged. This would not be an issue for Hermitian gates as the inverse noisy gate is the same as the original noisy gate.
\end{itemize}
So IISM:NA is not SUPER but it is SPqR.

\section{Mitigation Methods}\appSL{customChannels}
Here we will provide the parameters and usage directions for all the mitigation methods studied in this work. We will not provide the derivations of the results here but the full derivations can be found in \cite{MyLongWork} and a worked example of the derivation is given in \appSR{workedExample} for IISM:NA mitigating dephasing noise.

\subsection{Noise Models}\appSL{noiseModel}
We consider three types of noise:
\begin{itemize}
\item Stochastic noise (SN):
\begin{align}
\gC{E}{}{n}{}{}{}=\begin{cases}
\gG{}{}{}{}{}{Wire}~&\text{ with probability }1-\eStoch\\
\gC{N}{}{n}{}{}{}~&\text{ with probability }\eStoch
\end{cases}
\end{align}
where \(\oC{E}{}{n}{}{}{}\) is the error channel of the \(n\)th gate, \(\oC{N}{}{n}{}{}{}\) is a CPTP channel, and \(\eStoch\) is the noise amplitude. The noise level is given by:
\begin{align}\eqAL{stochNoiseLevel}
\noiseLevel = \oV{e}{SN}{}{}{}{} = 2\eStoch\nGate,
\end{align}
where \(\nGate\) is the number of gates in the circuit. \(\oC{E}{}{n}{}{}{}\) is assumed to be closed so:
\begin{align}\eqAL{SNnoisemodelA}
\oC{N}{}{n}{}{}{}^{2}=&\aParam\oC{I}{}{}{}{}{}+\left(1-\aParam\right)\oC{N}{}{n}{}{}{},&0\le& \aParam\le 1,
\end{align}
where \(\aParam\) is determined by the precise structure of \(\oC{N}{}{n}{}{}{}\) (our numerical results use \(\aParam=1\)).
\item Rotational errors (RE):
\begin{align}\eqAL{rotNoiseAmplitude}
\gC{E}{}{n}{}{}{}=\gG{~\rotG{\eAngle}{\hermitianOpi{n}}}{}{}{}{}{VVL}~,
\end{align}
where \(\oC{E}{}{n}{}{}{}\) is the error channel of the \(n\)th gate, \(\hermitianOpi{n}\) is a Hermitian operator, and \(\eAngle\) is the noise amplitude. The noise level is given by:
\begin{align}\eqAL{rotNoiseLevel}
\noiseLevel = \oV{e}{RE}{}{}{}{} = \left|\eAngle\right|\nGate.
\end{align}
\item (Stochastic-)over-rotation errors (ORE):
\begin{align}
\gC{E}{}{n}{}{}{}=\gG{~\rotG{\eAngle}{\hermitianOpi{n}}}{}{}{}{}{VVL}\begin{cases}
\gG{}{}{}{}{}{Wire}~&\text{ with probability }1-\eStoch\\
\gG{\hermitianOpi{n}}{}{}{}{}{}~&\text{ with probability }\eStoch
\end{cases}~,
\end{align}
where \(\oC{E}{}{n}{}{}{}\) is the error channel of the \(n\)th gate, \(\hermitianOpi{n}\) is the Hermitian operator that generates the \(n\)th gate, and \(\eAngle\) and \(\eStoch\) are the rotational-error and stochastic-noise amplitudes respectively. The full noise level is given by (see \eqARB{stochNoiseLevel}{rotNoiseLevel}):
\begin{align}
\noiseLevel = \oV{e}{RE}{}{}{}{} +\oV{e}{SN}{}{}{}{}=\left|\eAngle\right|\nGate+2\eStoch\nGate.
\end{align}
\end{itemize}

\subsection{Linear Mitigation}
Linear mitigation obtains a mitigated observable estimate from a linear combination of noisy observable estimates:
\begin{align}
\xObsM = \sum_{i=0}^{\nCircGlobal}\coefGlobali{i}\xObsNi{i}=\sum_{i=0}^{\nCircGlobal}\coefGlobali{i}\traceOp{\obsOp\oCN{C}{}{}{}{i}{}\left(\initialState\right)}=\traceOp{\obsOp\oCN{C}{}{}{M}{}{}\left(\initialState\right)},
\end{align} 
where \(\xObsM\) is the mitigated observable, \(\nCircGlobal\) is the number of extra circuit variants we use in the mitigation (generally \(\oCN{C}{}{0}{}{}{}\) is the original noisy circuit), \(\coefGlobali{i}\) is the coefficient used for the \(i\)th noisy observable estimate (or equivalently the \(i\)th noisy circuit variant), \(\obsOp\) is the observable we wish to measure, \(\oCN{C}{}{}{}{i}{}\) is the (channel/map corresponding to) the \(i\)th noisy circuit variant, \(\initialState\) is the initial state our circuit acts on, and \(\oCN{C}{}{}{M}{}{}\) is the effective mitigated circuit.

\subsubsection{Scope}\appSL{scope}
Mitigation methods can have different scopes:
\begin{itemize}
\item Synchronous methods use the same variant generation strategy for each gate in a given variant circuit:
\begin{align}
\gC{C}{}{}{SM}{}{Wireless}~=~\Big\{
\coefGlobali{j}: \gCN{C}{}{}{}{j}{Wireless}
\Big\}_{j=0}^{\nCircGlobal}~=~\Big\{
\coefGlobali{j}: \gGN{U}{}{1}{}{j}{Wireright} \gGN{U}{}{2}{}{j}{}\dots \gGN{U}{}{\nGate}{}{j}{WireleftL}
\Big\}_{j=0}^{\nCircGlobal}~,
\end{align}
where \(\oC{C}{}{}{SM}{}{}\)  is the (channel equivalent to the) synchronously mitigated circuit, \(\coefGlobali{j}\) is the coefficient of the \(j\)th noisy circuit variant, \(\oCN{C}{}{}{}{j}{}\) is the \(j\)th noisy circuit variant, \(\nCircGlobal\) is the number of circuit variants used, \(\oGN{U}{}{n}{}{j}{}\) is the \(j\)th variant of the \(n\)th gate.
\item Asynchronous methods can use different variant generation strategies for each gate in a given variant circuit:
\begin{align}
\gC{C}{}{}{AM}{}{Wireless}~=~\Big\{
\coefGlobali{j}: \gCN{C}{}{}{}{j}{Wireless}
\Big\}_{j=0}^{\nCircGlobal}~=~\Big\{
\coefGlobali{j}: \gGN{U}{}{1}{}{J_1\!\left(j\right)}{Wireright} \gG{}{}{}{}{}{WireVS}\gGN{U}{}{2}{}{J_2\!\left(j\right)}{}\gG{}{}{}{}{}{WireVS}\dots \gGN{U}{}{\nGate}{}{J_{\nGate}\!\left(j\right)}{WireleftL}
~~~~\Big\}_{j=0}^{\nCircGlobal}~,
\end{align}
where \(\oC{C}{}{}{AM}{}{}\)  is the (channel equivalent to the) asynchronously mitigated circuit, \(\coefGlobali{j}\) is the coefficient of the \(j\)th noisy circuit variant, \(\oCN{C}{}{}{}{j}{}\) is the \(j\)th noisy circuit variant, \(\nCircGlobal\) is the number of circuit variants used, \(\oGN{U}{}{n}{}{J_n\left(j\right)}{}\) is the \(J_n\left(j\right)\)th variant of the \(n\)th gate. In this case \(J_n\) is some function of \(j\) that depends on \(n\) and it need not be the case that \(J_n\left(j\right)=J_m\left(j\right)\) if \(n\ne m\).
\item Local methods effectively mitigate each gate independently and then these mitigated gates are concatenated in order to form circuit variants and their coefficients:
\begin{align}
\gC{C}{}{}{LM}{}{Wireless}~=&~\Big\{
\coefGlobali{j}: \gCN{C}{}{}{}{j}{Wireless}
\Big\}_{j=0}^{\nCircGlobal}~=~\Big\{
\coefGlobali{j}: \gGN{U}{}{1}{}{J_1\!\left(j\right)}{Wireright} \gG{}{}{}{}{}{WireVS}\gGN{U}{}{2}{}{J_2\!\left(j\right)}{}\gG{}{}{}{}{}{WireVS}\dots \gGN{U}{}{\nGate}{}{J_{\nGate}\!\left(j\right)}{WireleftL}
~~~~\Big\}_{j=0}^{\nCircGlobal}~,\nonumber\\
=&~\gGN{U}{}{1}{M}{}{Wireright}\gGN{U}{}{2}{M}{}{}\dots\gGN{U}{}{\nGate}{M}{}{WireleftL}~,~~~~\gGN{U}{}{n}{M}{}{}= \gG{}{}{}{}{}{WireVS}\Big\{
\coefLocali{n,i}: \gGN{U}{}{n}{}{i}{}\Big\}_{i=0}^{\nCircLocali{n}}\hspace{-24 pt}\gG{}{}{}{}{}{WireVS}~,
\end{align}
where \(\oC{C}{}{}{LM}{}{}\)  is the (channel equivalent to the) locally mitigated circuit; \(\coefGlobali{j}\) is the coefficient of the \(j\)th noisy circuit variant; \(\oCN{C}{}{}{}{j}{}\) is the \(j\)th noisy circuit variant; \(\nCircGlobal\) is the number of circuit variants used; \(\oGN{U}{}{n}{}{J_n\!\left(j\right)}{}\) is the \(J_n\left(j\right)\)th variant of the \(n\)th gate. For local mitigation:
\begin{align}
J_n\!\left(j\right) =& \left\lfloor\frac{j}{\oV{N}{C}{n-1}{}{}{}+1}\right\rfloor \bmod{\left(\oV{N}{L,C}{n}{}{}{}+1\right)}, &\oV{N}{C}{n}{}{}{}=&\left(\prod_{m=1}^{n}\left(\nCircLocali{m}+1\right)\right)-1,
\end{align}
where \(\oV{N}{C}{n}{}{}{}\) is the number of circuits required to mitigate \(n\) gates (\(\nCircGlobal=\oV{N}{C}{\nGate}{}{}{}\), where \(\nGate\) is the number of gates in the circuit), \(\oV{N}{L,C}{n}{}{}{}\) is the number of noisy variants we use to mitigate the \(n\)th noisy gate, \(\left\lfloor x/y\right\rfloor\) is the integer part of the division of \(x\) by \(y\), and \(x\bmod{y}\) is the remainder of \(x\) divided by \(y\). \(\oGN{U}{}{n}{M}{}{}\) is the mitigated version of the \(n\)th gate; \(\coefLocali{n,i}\) is the coefficient of the \(i\)th variant of the \(n\)th gate; and \(\oGN{U}{}{n}{}{i}{}\) is the \(i\)th noisy variant of the \(n\)th gate. The global circuit coefficients are defined by:
\begin{align}
\coefGlobali{j}=\prod_{n=1}^{\nGate}\coefLocali{n,J_n\left(j\right)}.
\end{align}
\end{itemize}

\subsubsection{Mitigation Method Implementation}\appSL{implementation}
We implement our mitigation methods using Monte Carlo sampling \cite{Piveteau2022}.

For a global mitigation method we have a mitigated circuit given by:
\begin{align}
\gCN{C}{}{}{M}{}{Wireless}~=&~\Big\{
\coefGlobali{j}: \gCN{C}{}{}{}{j}{Wireless}
\Big\}_{j=0}^{\nCircGlobal},&\costGlobal =&\sum_{i=0}^{\nCircGlobal }\left|\coefGlobali{i}\right|,&\probGlobali{j}=&\frac{\left|\coefGlobali{j}\right|}{\costGlobal},&\signGlobali{j}=&\frac{\coefGlobali{j}}{\left|\coefGlobali{j}\right|}.
\end{align}
where \(\oCN{C}{}{}{M}{}{Wireless}\) is the effective mitigated circuit, \(\coefGlobali{j}\) is the coefficient of the \(j\)th circuit variant (\( \oCN{C}{}{}{}{j}{Wireless}\)), \(\costGlobal\) is the sampling cost, \(\probGlobali{j}\) is the probability we select the \(j\)th circuit for a given run of the mitigated circuit, and \(\signGlobali{j}\) is the sign associated with the \(j\)th circuit. To implement this mitigated circuit we make \(\nRun\) runs of the circuit, for each run we select at random from the mitigation method's component circuit variants, selecting circuit \(j\) with probability \( \probGlobali{j}\). We multiply the output of each run by \(\signGlobali{j}\) (where \(j\) matches the circuit we implement for that run). We average all the \(\nRun\) (signed) outputs we obtain and multiply the result by \(\costGlobal\) to obtain our mitigated observable estimate.

For a local mitigation method we have a mitigated circuit given by:
\begin{align}
\gCN{C}{}{}{M}{}{Wireless}~=&~\Big(\gGN{U}{}{n}{M}{}{}
\Big)_{n=1}^{\nGate},&\gGN{U}{}{n}{M}{}{}=& \Big\{
\coefLocali{n,i}: \gGN{U}{}{n}{}{i}{}\Big\}_{i=0}^{\nCircLocali{n}},&\costLocali{n} =&\sum_{i=0}^{\nCircLocali{n}}\left|\coefLocali{n,i}\right|,\nonumber\\
\probLocali{n,j}=&\frac{\left|\coefLocali{n,j}\right|}{\costLocali{n}},&\signLocali{n,j}=&\frac{\coefLocali{n,j}}{\left|\coefLocali{n,j}\right|}, &\costGlobal=&\prod_{n=1}^{\nGate}\costLocali{n} .
\end{align}
where \(\oCN{C}{}{}{M}{}{Wireless}\) is the effective mitigated circuit, \(\oGN{U}{}{n}{M}{}{}\) is the \(n\)th effective mitigated gate in the mitigated circuit, \(\nGate\) is the number of gates in the circuit, \(\coefLocali{n,i}\) is the coefficient of the \(i\)th noisy gate variant of the \(n\)th gate (\( \oGN{U}{}{n}{}{j}{Wireless}\)), \(\nCircLocali{n}\) is the number of gate variants we use for the \(n\)th gate, \(\costLocali{n}\) is the local sampling cost of the \(n\)th gate, \(\probLocali{n,j}\) is the probability we select the \(j\)th variant for the \(n\)th gate, \(\signLocali{n,j}\) is the sign associated with the \(j\)th variant of the \(n\)th gate, and \(\costGlobal\) is the global sampling cost. To implement this mitigated circuit we make \(\nRun\) runs of the circuit. For each gate in each run we select at random from the mitigated gate's component gate variants, selecting variant \(j\) with probability \( \probLocali{n,j}\) (for the \(n\)th gate). We multiply the output of each run by the product of the signs associated to each gate used in that run, so for the \(m\)th run we have:
\begin{align}
\prod_{n=1}^{\nGate}\signLocali{n,j_{n,m}},
\end{align}
where \(j_{n,m}\) is the index of the gate variant selected for the \(n\)th gate for the \(m\)th run of the circuit. We take the average of the \(\nRun\) signed outputs and multiply the result by \(\costGlobal\) to obtain our mitigated observable estimate.

\subsection{CHILM}\appSL{CHILM}
Custom-channel-and-hidden-inverse-assisted local mitigation (CHILM) is a local method (see \appSR{scope}) that uses a mixture of custom channels and hidden inverses to amplify the noise in the noisy gate variants:
\begin{align}
\gCN{C}{}{}{M}{}{Wireless}~~~~=~&\Big(\gGN{U}{}{n}{M}{}{}\Big)_{n=1}^{\nGate}, &\gGN{U}{}{n}{M}{}{}~~=&~\Big\{
\coefLocali{n,j}: \gGN{U}{}{n}{}{j}{}
\Big\}_{j=0}^{\nCircLocali{n}}~,\\
\gGN{U}{}{n}{}{j<\nCircLocali{n}}{}~~~~=~&\gGN{U}{}{n}{}{}{}\gC{E}{}{n}{C}{j}{}~,&\gGN{U}{}{n}{}{\nCircLocali{n}}{}~~=&~\gGN{U}{}{n}{HI}{}{},
\end{align}
where \(\oCN{C}{}{}{M}{}{Wireless}\) is the mitigated circuit, \(\oGN{U}{}{n}{M}{}{Wireless}\) is \(n\)th mitigated gate, \(\nGate\) is the number of gates in the circuit, \(\coefLocali{n,j}\) is the mitigation coefficient for the \(j\)th gate variant of the \(n\)th gate, \(\oGN{U}{}{n}{}{j}{Wireless}\) is the \(j\)th gate variant of the \(n\)th gate, \(\nCircLocali{n}\) is the number of gate variants for the \(n\)th gate, \(\oGN{U}{}{n}{}{}{}\) is the \(n\)th noisy gate before mitigation, \(\oC{E}{}{n}{C}{j}{}\) is the custom error channel used to generate the \(j\)th variant of the \(n\)th noisy gate, and \(\oGN{U}{}{n}{HI}{}{}\) is the hidden inverse of the \(n\)th noisy gate (used for the final of the \(\nCircLocali{n}\) noisy gate variants, not counting the original noisy gate). 

CHILM is a special instantiation of probabilistic error cancellation (see \citeRef{Temme2017}). The use of custom channels for variant generation is discussed in \citeRef{Li2017} and hidden inverses are introduced in \citeRefs{Zhang2022, Leyton2022} (but they are not used therein for probabilistic error cancellation).

The execution time for the \(i\)th variant of the \(n\)th gate is assumed to be:
\begin{align}
\oV{\tau}{L}{n}{}{i}{} =&\begin{cases}
\oV{\tau}{L}{n}{}{}{},&\text{ for } i=0,\\
2\oV{\tau}{L}{n}{}{}{},&\text{ for } 0<i<\nCircLocali{n},\\
\oV{\tau}{L}{n}{}{}{},&\text{ for } i=\nCircLocali{n},
\end{cases}
\end{align}
where \(\oV{\tau}{L}{n}{}{}{}\) is the execution time for the original \(n\)th noisy gate (\(\oGN{U}{}{n}{}{0}{}=\oGN{U}{}{n}{}{}{}\)). That is, we assume that the execution time of every custom error channel is the same as the original gate and we assume the execution time of the hidden inverse is the same as the original noisy gate. 

We do not implement CHILM for stochastic noise.

For rotational errors we use \(\nCircLocali{n}=2\) where:
\begin{align}
\gGN{U}{}{n}{}{0}{} ~=~& \gGN{U}{}{n}{}{}{}~=~\gG{U}{}{n}{}{}{}\gG{~\rotG{\eAngle}{\hermitianOpi{n}}}{}{}{}{}{VVL},&\gGN{U}{}{n}{}{1}{} ~=~& \gGN{U}{}{n}{}{}{}\gG{~\rotG{\uppi}{\hermitianOpi{n}}}{}{}{}{}{VVL},\nonumber\\
\gGN{U}{}{n}{}{2}{} ~=~& \gGN{U}{}{n}{HI}{}{}~=~\gG{U}{}{n}{}{}{}\gG{~~~~~~~~~~\rotGD{\eAngle}{\hermitianOpi{n}}}{}{}{}{}{DoubleSize},
\end{align}
with the coefficients:
\begin{align}
\coefLocali{0}=&\frac{1}{2},&
\coefLocali{1}=&-\frac{1-\cos \eAngle}{2\cos \eAngle}=-\frac{\eAngle^2}{4}+\orderZero\left(\eAngle^{4}\right) ,&
\coefLocali{2}=& \frac{1}{2\cos \eAngle}=\frac{1}{2}+\frac{\eAngle^2}{4}+\orderZero\left(\eAngle^{4}\right),
\end{align}
where \(\eAngle\) is the noise amplitude of the original noisy gate (see \appSR{noiseModel}).  This method is unbiased so the proxy bias is zero.

For (stochastic) over-rotation errors we use the same variant generation strategy as for rotational errors (\(\nCircLocali{n}=2\)):
\begin{align}
\gGN{U}{}{n}{}{0}{}~ =~& \gGN{U}{}{n}{}{}{}~=~\gG{U}{}{n}{}{}{}\gG{~\rotG{\eAngle}{\hermitianOpi{n}}}{}{}{}{}{VVL}\begin{cases}
\gG{}{}{}{}{}{Wire}~&\text{ with prob. }1-\eStoch\\
\gG{\hermitianOpi{n}}{}{}{}{}{}~&\text{ with prob. }\eStoch
\end{cases},\nonumber\\
\gGN{U}{}{n}{}{1}{}~ =~& \gGN{U}{}{n}{}{}{}\gG{~\rotG{\uppi}{\hermitianOpi{n}}}{}{}{}{}{VVL},\nonumber\\
\gGN{U}{}{n}{}{2}{}~ =~& \gGN{U}{}{n}{HI}{}{}~=~\gG{U}{}{n}{}{}{}\gG{~~~~~~~~~~\rotGD{\eAngle}{\hermitianOpi{n}}}{}{}{}{}{DoubleSize}\begin{cases}
\gG{}{}{}{}{}{Wire}~&\text{ with prob. }1-\eStoch\\
\gG{\hermitianOpi{n}}{}{}{}{}{}~&\text{ with prob. }\eStoch
\end{cases},
\end{align}
and the coefficients:
\begin{align}
\coefLocali{0}=&\frac{1}{2}, ~~~
\coefLocali{1}=-\frac{1-\left(1-2\eStoch\right)\cos \eAngle}{2\left(1-2\eStoch\right)\cos \eAngle}=-\eStoch-\frac{\eAngle^2}{4}+\orderZero\left(\eStoch^2,\eAngle^{2}\eStoch,\eAngle^{4}\right) ,\\
\coefLocali{2}=& \frac{1}{2\left(1-2\eStoch\right)\cos \eAngle}=\frac{1}{2}+\eStoch+\frac{\eAngle^2}{4}+\orderZero\left(\eStoch^2,\eAngle^{2}\eStoch,\eAngle^{4}\right),
\end{align}
where \(\eStoch\) and \(\eAngle\) are the noise amplitudes of the stochastic and rotational components of the original noisy gate's error channel, respectively (see \appSR{noiseModel}).  This method is unbiased so the proxy bias is zero. It is worth noting that the hidden inverse used in the mitigation of stochastic over-rotation errors has an error channel that is not the inverse of the error channel of the original noisy gate. Only the coherent (rotational) part of the error channel is inverted. The stochastic component is assumed to be the same as the original gate.

\subsection{CHISM}\appSL{CHISM}
Custom-channel-and-hidden-inverse-assisted synchronous mitigation (CHISM) is a synchronous method (see \appSR{scope}) that uses a mixture of custom channels and hidden inverses to amplify the noise in the noisy gate variants:
\begin{align}
\gCN{C}{}{}{M}{}{Wireless}~=~&\Big\{
\coefGlobali{j}: \gCN{C}{}{}{}{j}{Wireless}
\Big\}_{j=0}^{\nCircGlobal}~=~\Big\{
\coefGlobali{j}: \Big(\gGN{U}{}{n}{}{j}{}\Big)_{n=1}^{\nGate}
\Big\}_{j=0}^{\nCircGlobal}~,\\
\gGN{U}{}{n}{}{j<\nCircGlobal}{}~~~=&~\gGN{U}{}{n}{}{}{}\gC{E}{}{n}{C}{j}{}~,~~~\gGN{U}{}{n}{}{\nCircGlobal}{}~=~\gGN{U}{}{n}{HI}{}{},
\end{align}
where \(\oCN{C}{}{}{M}{}{Wireless}\) is the mitigated circuit, \(\coefGlobali{j}\) is the mitigation coefficient for the \(j\)th circuit variant, \(\oCN{C}{}{}{}{j}{Wireless}\) is the \(j\)th circuit variant, \(\nCircGlobal\) is the number of circuit variants, \(\oGN{U}{}{n}{}{j}{}\) is the \(j\)th variant of the \(n\)th noisy gate, \(\nGate\) is the number of gates in the circuit,  \(\oGN{U}{}{n}{}{}{}\) is the \(n\)th noisy gate before mitigation, \(\oC{E}{}{n}{C}{j}{}\) is the custom error channel used to generate the \(j\)th variant of the \(n\)th noisy gate, and \(\oGN{U}{}{n}{HI}{}{}\) is the hidden inverse of the \(n\)th noisy gate (used for the final of the \(\nCircGlobal\) noisy gate variants, not counting the original noisy gate). 

CHISM requires all the mitigated noisy gates to have noise of the same type with the same noise amplitude. CHISM is a generalisation of the zero-noise extrapolation methods of \citeRefs{Temme2017,Li2017} to target rotational errors. The use of custom channels for variant generation is discussed in \citeRef{Li2017} and hidden inverses are introduced in \citeRefs{Zhang2022, Leyton2022}.

The execution time for the \(j\)th circuit is given by:
\begin{align}
\timeExeci{j} =&\begin{cases}
\timeExec,&\text{ for } i=0,\\
2\timeExec,&\text{ for } 0<i<\nCircGlobal,\\
\timeExec,&\text{ for } i=\nCircGlobal,
\end{cases}
\end{align}
where \(\timeExec\) is the execution time for the original noisy circuit (\(\oCN{C}{}{}{}{0}{Wireless}
=\oCN{C}{}{}{}{}{Wireless}
\)). We assume that the execution time of every custom error channel is the same as the original gate and that all the gates take the same time to implement. We also assume the execution time of the hidden inverse is the same as the original noisy gate. 

We do not implement CHISM for stochastic noise.

For rotational errors we use the gate variants:
\begin{align}
\gGN{U}{}{n}{}{0}{}~ =~& \gGN{U}{}{n}{}{}{}~=~\gG{U}{}{n}{}{}{}\gG{~\rotG{\eAngle}{\hermitianOpi{n}}}{}{}{}{}{VVL},&\gGN{U}{}{n}{}{0<i<\nCircGlobal}{} ~~~~~=~& \gGN{U}{}{n}{}{}{}\gG{~~~~~~~~~~\rotGNF{\frac{i\uppi}{\nCircGlobal}}{\hermitianOpi{n}}}{}{}{}{}{DoubleSize},\nonumber\\
\gGN{U}{}{n}{}{\nCircGlobal}{} ~=~& \gGN{U}{}{n}{HI}{}{}~=~\gG{U}{}{n}{}{}{}\gG{~~~~~~~~~~\rotGD{\eAngle}{\hermitianOpi{n}}}{}{}{}{}{DoubleSize},
\end{align}
which means we have the amplified noise amplitudes:
\begin{align}
\eAnglei{0}= &\eAngle,&\eAnglei{0<i<\nCircGlobal}= &\eAngle+\frac{2i}{\nCircGlobal}\uppi,&\eAnglei{\nCircGlobal}= &-\eAngle.
\end{align}
We use the coefficients:
\begin{align}
\coefGlobali{i}=&\prod_{j=0,j\ne i}^{\nCircGlobal} \frac{\sin \frac{\eAnglei{j}}{2}}{\sin\frac{ \eAnglei{j}- \eAnglei{i}}{2}},
\end{align}
where \(\eAngle\) is the noise amplitude of the original noisy gate (see \appSR{noiseModel}) and \(\nCircGlobal\) is the number of gates used for the mitigation. We do not have an analytic form for the proxy bias in general but if we set \(\nCircGlobal=2\nGate\) (as we do in the main text) the method is unbiased.

\subsection{CIILM}\appSL{CIILM}
Custom-channel-and-identity-insertion-assisted local mitigation (CIILM) is a local method (see \appSR{scope}) that uses custom channels and identity insertions to amplify the noise in the noisy gate variants:
\begin{align}
\gCN{C}{}{}{M}{}{Wireless}~=~&\Big(\gGN{U}{}{n}{M}{}{}\Big)_{n=1}^{\nGate}, &\gGN{U}{}{n}{M}{}{}~=&~\Big\{
\coefLocali{n,j}: \gGN{U}{}{n}{}{j}{}
\Big\}_{j=0}^{\nCircLocali{n}}~,\\
\gGN{U}{}{n}{}{j}{}~=~&\Big(\gGN{U}{}{n}{T}{}{}\gGDN{U}{}{n}{T}{}{}\Big)^{j}\gGN{U}{}{n}{T}{}{},&\gGN{U}{}{n}{T}{}{}~=&~\gGN{U}{}{n}{}{}{}\gC{E}{}{n}{C}{}{},~~\gGDN{U}{}{n}{T}{}{}~=~\gGDN{U}{}{n}{}{}{}\gC{E}{}{n}{C}{}{},
\end{align}
where \(\oCN{C}{}{}{M}{}{Wireless}\) is the mitigated circuit, \(\oGN{U}{}{n}{M}{}{Wireless}\) is \(n\)th mitigated gate, \(\nGate\) is the number of gates in the circuit, \(\coefLocali{n,j}\) is the mitigation coefficient for the \(j\)th gate variant of the \(n\)th gate, \(\oGN{U}{}{n}{}{j}{Wireless}\) is the \(j\)th gate variant of the \(n\)th gate, \(\nCircLocali{n}\) is the number of gate variants for the \(n\)th gate, \(\oGN{U}{}{n}{T}{}{}\) is the tailored version of the \(n\)th noisy gate, \(\oGDN{U}{}{n}{T}{}{}\) is the tailored version of the noisy inverse of the \(n\)th gate (assumed to have the same error channel as \(\oGN{U}{}{n}{T}{}{}\)), \(\oGN{U}{}{n}{}{}{}\) is the \(n\)th noisy gate before mitigation, \(\oC{E}{}{n}{C}{}{}\) is the custom error channel used to tailor the \(n\)th noisy gate, and \(\oGDN{U}{}{n}{}{}{}\) is the noisy inverse of the \(n\)th noisy gate. To save time in characterisation one need not ever characterise the gates \(\oGN{U}{}{n}{}{}{}\) and \(\oGDN{U}{}{n}{}{}{}\) since the control channel can be chosen independently of the noise amplitude. It suffices to characterise \(\oGN{U}{}{n}{T}{}{}\) and \(\oGDN{U}{}{n}{T}{}{}\).

The version of CIILM presented here requires the error channels associated with the noisy gate and its noisy inverse to be the same and that the noisy gate, the noisy inverse, their error channels, and the custom error channel all commute with each other. CIILM could be considered to be an example of NEPEC (see \citeRef{Mari2021}) a variant of probabilistic error cancellation (see \citeRef{Temme2017}). The use of custom channels for noise amplification is discussed in \citeRef{Li2017}. 

The execution time for the \(i\)th variant of the \(n\)th gate is assumed to be:
\begin{align}
\oV{\tau}{L}{n}{}{i}{} =&2\left(2i+1\right)\oV{\tau}{L}{n}{}{}{},
\end{align}
where \(\oV{\tau}{L}{n}{}{}{}\) is the execution time for the original \(n\)th noisy gate (\(\oGN{U}{}{n}{}{}{}\)). That is, we assume that the execution time of every custom error channel is the same as the original gate and that the forward and inverse gates have the same execution time.

We do not implement CIILM for stochastic noise.

For rotational errors we use \(\nCircLocali{n}=2\) and the custom channel is given by:
\begin{align}
\gC{E}{}{n}{C}{}{}~=&~\gG{~\rotGNF{\frac{\uppi}{3}}{\hermitianOpi{n}}}{}{}{}{}{VVL}, 
\end{align}
where \(\hermitianOpi{n}\) is the generator of the rotational error of the \(n\)th noisy gate. The coefficients are given by:
\begin{align}
\coefLocali{0}=&\frac{\sin\frac{\eAnglei{1}}{2}\sin\frac{\eAnglei{2}}{2}}{\sin\frac{\eAnglei{1}-\eAnglei{0}}{2}\sin\frac{\eAnglei{2}-\eAnglei{0}}{2}}=-\sqrt{3}\eAngle+\orderZero\left(\eAngle^{2}\right), \\
\coefLocali{1}=&\frac{\sin\frac{\eAnglei{0}}{2}\sin\frac{\eAnglei{2}}{2}}{\sin\frac{\eAnglei{0}-\eAnglei{1}}{2}\sin\frac{\eAnglei{2}-\eAnglei{1}}{2}}=
1+\orderZero\left(\eAngle^{2}\right),\\
\coefLocali{2}=&\frac{\sin\frac{\eAnglei{0}}{2}\sin\frac{\eAnglei{1}}{2}}{\sin\frac{\eAnglei{0}-\eAnglei{2}}{2}\sin\frac{\eAnglei{1}-\eAnglei{2}}{2}}=
\sqrt{3}\eAngle+\orderZero\left(\eAngle^{2}\right),
\end{align}
where the effective noise amplitudes are given by:
\begin{align}
\eAnglei{0}=&\frac{2\uppi}{3}+\eAngle,&\eAnglei{1}=&2\uppi+3\eAngle,&\eAnglei{2}=&\frac{10\uppi}{3}+5\eAngle,
\end{align}
and \(\eAngle\) is the noise amplitude of the original noisy gate (see \appSR{noiseModel}).  This method is unbiased so the proxy bias is zero.


\subsection{CLM}\appSL{CLM}
Custom-channel-assisted local mitigation (CLM) is a local method (see \appSR{scope}) that uses custom channels to amplify the noise in the noisy gate variants:
\begin{align}
\gCN{C}{}{}{M}{}{Wireless}~=~&\Big(\gGN{U}{}{n}{M}{}{}\Big)_{n=1}^{\nGate}, ~~~\gGN{U}{}{n}{M}{}{}~=~\Big\{
\coefLocali{n,j}: \gGN{U}{}{n}{}{j}{}
\Big\}_{j=0}^{\nCircLocali{n}}~,\\
\gGN{U}{}{n}{}{j}{}~=~&\gGN{U}{}{n}{}{}{}\gC{E}{}{n}{C}{j}{}~=~\gG{U}{}{n}{}{}{}\gC{E}{}{n}{}{}{}\gC{E}{}{n}{C}{j}{}~=~\gG{U}{}{n}{}{}{}\gC{E}{}{n}{}{j}{},
\end{align}
where \(\oCN{C}{}{}{M}{}{Wireless}\) is the mitigated circuit, \(\oGN{U}{}{n}{M}{}{Wireless}\) is \(n\)th mitigated gate, \(\nGate\) is the number of gates in the circuit, \(\coefLocali{n,j}\) is the mitigation coefficient for the \(j\)th gate variant of the \(n\)th gate, \(\oGN{U}{}{n}{}{j}{Wireless}\) is the \(j\)th gate variant of the \(n\)th gate, \(\nCircLocali{n}\) is the number of gate variants for the \(n\)th gate and the order of the mitigation used for that gate, \(\oGN{U}{}{n}{}{}{}\) is the \(n\)th noisy gate before mitigation, \(\oC{E}{}{n}{C}{j}{}\) is the custom error channel used to generate the \(j\)th variant of the \(n\)th noisy gate, \(\oC{E}{}{n}{}{}{}\) is the original error channel of the \(n\)th noisy gate, \(\oG{U}{}{n}{}{}{}\) is the noise-free component of the \(n\)th gate, and \(\oC{E}{}{n}{}{j}{}\) is the error channel of the \(j\)th variant of the \(n\)th gate. 

CLM could be considered to be an example of NEPEC (see \citeRef{Mari2021}). CLM:SN (used to mitigate stochastic noise) is functionally equivalent to the probabilistic error cancellation procedure that is declared optimal for dephasing noise in \citeRef{Takagi2022}. Though, in fact, it is possible to obtain a slightly lower sampling cost\textemdash as noted by \citeRef{Cai2023a}, but the difference is negligible. 

The execution time for the \(i\)th variant of the \(n\)th gate is assumed to be:
\begin{align}
\oV{\tau}{L}{n}{}{i}{} =&\begin{cases}
\oV{\tau}{L}{n}{}{}{},&\text{ for } i=0,\\
2\oV{\tau}{L}{n}{}{}{},&\text{ for } i>0,\\
\end{cases}
\end{align}
where \(\oV{\tau}{L}{n}{}{}{}\) is the execution time for the original \(n\)th noisy gate (\(\oGN{U}{}{n}{}{0}{}=\oGN{U}{}{n}{}{}{}\)). That is, we assume that the execution time of every custom error channel is the same as the original gate. An exception is the case of (stochastic) over-rotation errors where we break our custom channels into Clifford gates and assume that each Clifford gate has the same execution time (see below).

For stochastic noise we use \(\nCircLocali{n}=1\), with the noisy gate variants:
\begin{align}
\gGN{U}{}{n}{}{0}{}~=~&\gGN{U}{}{n}{}{}{},&\gGN{U}{}{n}{}{1}{}~=~&\gGN{U}{}{n}{}{}{}\begin{cases}
\gG{}{}{}{}{}{Wire}~&\text{ with probability }1-\sin^{2} \frac{\uppi}{2},\\
\gC{N}{}{n}{}{}{}~&\text{ with probability }\sin^{2} \frac{\uppi}{2},
\end{cases}~
\end{align}
and the corresponding coefficients:
\begin{align}
\coefLocali{0}=&\frac{\eStochi{1}}{\eStochi{1}-\eStochi{0}}=1+\eStoch +\orderZero\left(\eStoch^{2}\right),&\coefLocali{1}=&-\frac{\eStochi{0}}{\eStochi{1}-\eStochi{0}}=-\eStoch +\orderZero\left(\eStoch^{2}\right).
\end{align}
where the amplified noise amplitudes are given by:
\begin{align}
\eStochi{0}= &\eStoch,&\eStochi{1}= &\eStoch+\sin^{2} \frac{\uppi}{2}\left(1-\left(1+\aParam\right)\eStoch\right),
\end{align}
where \(\eStoch\) is the noise amplitude of the original noisy gate and \(0\le\aParam\le 1\) is determined by the nature of the error channel (see \eqAR{SNnoisemodelA}). This method is unbiased so the proxy bias is zero.

For rotational errors we use \(\nCircLocali{n}=2\) and the noisy gate variants:
\begin{align}\eqAL{CLMcustomChannelsRot}
\gGN{U}{}{n}{}{0}{}~ =~& \gGN{U}{}{n}{}{}{},&\gGN{U}{}{n}{}{1}{}~ =~& \gGN{U}{}{n}{}{}{}\gG{~\rotGNF{\frac{\uppi}{4}}{\hermitianOpi{n}}}{}{}{}{}{VVL},&
\gGN{U}{}{n}{}{2}{}~ =~& \gGN{U}{}{n}{}{}{}\gG{~\rotGNF{\frac{3\uppi}{4}}{\hermitianOpi{n}}}{}{}{}{}{VVL},
\end{align}
with the coefficients:
\begin{align}\eqAL{rotationalErrorsCoefsCLM}
\coefLocali{0}=&\cos \eAngle=1-\frac{\eAngle^{2}}{2}+\orderZero\left(\eAngle^{4}\right), &
\coefLocali{1}=&-\frac{\cos \eAngle-1+\sin\eAngle}{2}=-\frac{\eAngle}{2}+\frac{\eAngle^{2}}{4}+\orderZero\left(\eAngle^{4}\right) ,\\
&&\coefLocali{2}=& \frac{1-\cos \eAngle+\sin\eAngle}{2}=\frac{\eAngle}{2}+\frac{\eAngle^{2}}{4}+\orderZero\left(\eAngle^{4}\right),
\end{align}
where \(\eAngle\) is the noise amplitude of the original noisy gate (see \appSR{noiseModel}). This method is unbiased so the proxy bias is zero.

For (stochastic) over-rotation errors we use \(\nCircLocali{n}=2\) and the same custom channels as for rotational errors (\eqAR{CLMcustomChannelsRot}). When we make use of this method in the main text we decompose our (non-trivial) custom channels into Clifford gates:
\begin{align}
\gC{E}{}{n}{C}{1}{}&~=~\gG{~\rotGNF{\frac{\uppi}{4}}{\hermitianOpi{n}}}{}{}{}{}{VVL}~=~\gG{~\sOpi{\hermitianOpi{n}}}{}{}{}{}{L},&\gC{E}{}{n}{C}{2}{}&~=~\gG{~\sOpi{\hermitianOpi{n}}}{}{}{}{}{L}\gG{~\hermitianOpi{n}}{}{}{}{}{}, &\sOpi{\hermitianOpi{n}}^2=\hermitianOpi{n},
\end{align}
(\(\oC{E}{}{n}{C}{0}{}\) is the identity channel) and we assume every gate takes the same amount of time to implement. So:
\begin{align}
\oV{\tau}{L}{n}{}{i}{} =&\left(i+1\right)\oV{\tau}{L}{n}{}{}{}.
\end{align}
We use the coefficients:
\begin{align}
\coefLocali{0}=&\frac{\cos \eAngle}{1-2\eStoch}=1+2\eStoch-\frac{\eAngle^{2}}{2}+\orderZero\left(\eStoch^2,\eStoch\eAngle^2,\eAngle^{4}\right), \\
\coefLocali{1}=&-\frac{\cos \eAngle-1+2\eStoch+\sin\eAngle}{2\left(1-2\eStoch\right)}=-\eStoch-\frac{\eAngle}{2}+\frac{\eAngle^{2}}{4}+\orderZero\left(\eStoch^2,\eAngle\eStoch,\eAngle^{4}\right) ,\\
\coefLocali{2}=& \frac{1-2\eStoch-\cos \eAngle+\sin\eAngle}{2\left(1-2\eStoch\right)}=-\eStoch+\frac{\eAngle}{2}+\frac{\eAngle^{2}}{4}+\orderZero\left(\eStoch^2,\eAngle\eStoch,\eAngle^{4}\right),
\end{align}
where \(\eStoch\) and \(\eAngle\) are the noise amplitudes of the stochastic and rotational components of the original noisy gate's error channel, respectively (see \appSR{noiseModel}). This method is unbiased so the proxy bias is zero.

\subsection{CSM}\appSL{CSM}
Custom-channel-assisted synchronous mitigation (CSM) is a synchronous method (see \appSR{scope}) that uses custom channels to amplify the noise in the noisy gate variants:
\begin{align}
\gCN{C}{}{}{M}{}{Wireless}~=~&\Big\{
\coefGlobali{j}: \gCN{C}{}{}{}{j}{Wireless}
\Big\}_{j=0}^{\nCircGlobal}~=~\Big\{
\coefGlobali{j}: \Big(\gGN{U}{}{n}{}{j}{}\Big)_{n=1}^{\nGate}
\Big\}_{j=0}^{\nCircGlobal}~,\\
\gGN{U}{}{n}{}{j}{}~=~&\gGN{U}{}{n}{}{}{}\gC{E}{}{n}{C}{j}{}~=~\gG{U}{}{n}{}{}{}\gC{E}{}{n}{}{}{}\gC{E}{}{n}{C}{j}{}~=~\gG{U}{}{n}{}{}{}\gC{E}{}{n}{}{j}{},
\end{align}
where \(\oCN{C}{}{}{M}{}{Wireless}\) is the mitigated circuit, \(\coefGlobali{j}\) is the mitigation coefficient for the \(j\)th circuit variant, \(\oCN{C}{}{}{}{j}{Wireless}\) is the \(j\)th circuit variant, \(\nCircGlobal=\nMit\) is the number of circuit variants and the order of the mitigation, \(\oGN{U}{}{n}{}{j}{}\) is the \(j\)th variant of the \(n\)th noisy gate, \(\nGate\) is the number of gates in the circuit,  \(\oGN{U}{}{n}{}{}{}\) is the \(n\)th noisy gate before mitigation, \(\oC{E}{}{n}{C}{j}{}\) is the custom error channel used to generate the \(j\)th variant of the \(n\)th noisy gate, \(\oC{E}{}{n}{}{}{}\) is the original error channel of the \(n\)th noisy gate, \(\oG{U}{}{n}{}{}{}\) is the noise-free component of the \(n\)th gate, and \(\oC{E}{}{n}{}{j}{}\) is the error channel of the \(j\)th variant of the \(n\)th gate. 

CSM requires all error channels to have the same type (\eg they must all be stochastic or rotational) and have the same noise amplitude. All the custom channels have the same noise amplitude but may have different generators (their generator should match that of the error channel they are tailoring). CSM is a zero-noise extrapolation method akin to \citeRefs{Temme2017,Li2017}. Akin to \citeRef{Li2017} it uses custom error channels in order to amplify the noise of gates and produce circuit variants with differing noise levels to use for the mitigation. However, it generalises their approach by adding the ability to perform unbiased mitigation of rotational errors.

The execution time for the \(j\)th circuit is given by:
\begin{align}
\timeExeci{j} =&\begin{cases}
\timeExec,&\text{ for } i=0,\\
2\timeExec,&\text{ for } i>0,\\
\end{cases}
\end{align}
where \(\timeExec\) is the execution time for the original noisy circuit. We assume that the execution time of every custom error channel is the same as the original gate and that all the gates take the same time to implement. 

For stochastic noise we use the gate variants:
\begin{align}
\gGN{U}{}{n}{}{0}{}~=~&\gGN{U}{}{n}{}{}{},&\gGN{U}{}{n}{}{0<i}{}~=~&\gGN{U}{}{n}{}{}{}\begin{cases}
\gG{}{}{}{}{}{Wire}~&\text{ with probability }1-\sin^{2} \frac{i\uppi}{2\nMit},\\
\gC{N}{}{n}{}{}{}~&\text{ with probability }\sin^{2} \frac{i\uppi}{2\nMit},
\end{cases}~
\end{align}
with the coefficients:
\begin{align}
\coefGlobali{i}=&\prod_{m=0:m\ne i}^{\nMit }\frac{\eStochi{m}}{\eStochi{m}-\eStochi{i}},&\eStochi{m}=&\eStoch+\sin^{2} \frac{m\uppi}{2\nMit}\left(1-\left(1+\aParam\right)\eStoch\right),
\end{align}
where \(\eStoch\) is the noise amplitude of the original noisy gate and \(0\le\aParam\le 1\) is determined by the nature of the error channel (see \eqAR{SNnoisemodelA}). We do not have an analytic form for the proxy bias in general but if we set \(\nMit=\nGate\) (as we do in the main text) the method is unbiased.

For rotational errors we use the gate variants:
\begin{align}
\gGN{U}{}{n}{}{0}{} ~=&~ \gGN{U}{}{n}{}{}{},&\gGN{U}{}{n}{}{0<i}{} ~=~& \gGN{U}{}{n}{}{}{}\gG{~\rotG{\cAnglei{i}}{\hermitianOpi{n}}}{}{}{}{}{VVL},&\cAnglei{0<i}= &\frac{2i-1}{\nMit}\uppi.
\end{align}
with the coefficients:
\begin{align}
\coefGlobali{i}=&\prod_{j=0,j\ne i}^{\nMit} \frac{\sin \frac{\eAnglei{j}}{2}}{\sin\frac{ \eAnglei{j}- \eAnglei{i}}{2}},&\eAnglei{0}= &\eAngle ,&\eAnglei{0<i}= &\eAngle + \cAnglei{i},
\end{align}
where \(\eAngle\) is the noise amplitude of the original noisy gate (see \appSR{noiseModel}) and \(\nMit\) is the order of the mitigation. We do not have an analytic form for the proxy bias in general but if we set \(\nMit=2\nGate\) (as we do in the main text) the method is unbiased.


\subsection{IIAM}\appSL{IIAM}
Identity-insertion-assisted asynchronous mitigation (IIAM) is an asynchronous method (see \appSR{scope}) that uses identity insertions to amplify the noise in the noisy gate variants:
\begin{align}\eqAL{IIAMStandardForm}
\gCN{C}{}{}{M}{}{Wireless}~=~&\Big\{
\coefGlobali{i}: \gCN{C}{}{}{}{i}{Wireless}
\Big\}_{i=0}^{\nCircGlobal}~=~\Big\{
\coefGlobali{i}: \Big(\gGN{U}{}{n}{}{J_n\left(i\right)}{}~~\Big)_{n=1}^{\nGate}
\Big\}_{i=0}^{\nCircGlobal}~,\\
\gGN{U}{}{n}{}{J_n\left(i\right)}{}~=~&\Big(\gGN{U}{}{n}{}{}{}\gGDN{U}{}{n}{}{}{}\Big)^{J_n\!\left(i\right)}\gGN{U}{}{n}{}{}{}~=~\gG{U}{}{n}{}{}{}\gC{E}{}{n}{}{J_n\!\left(i\right)}{}~~,
\end{align}
where \(\oCN{C}{}{}{M}{}{Wireless}\) is the mitigated circuit, \(\coefGlobali{i}\) is the mitigation coefficient for the \(i\)th circuit variant, \(\oCN{C}{}{}{}{i}{Wireless}\) is the \(i\)th circuit variant, \(\nCircGlobal\) is the number of circuit variants, \(\oGN{U}{}{n}{}{J_n\!\left(i\right)}{}\) is the \(J_n\!\left(i\right)\)th variant of the \(n\)th noisy gate, \(\nGate\) is the number of gates in the circuit,  \(\oGN{U}{}{n}{}{}{}\) is the \(n\)th noisy gate before mitigation, \(\oGDN{U}{}{n}{}{}{}\) is the noisy inverse gate, \(\oG{U}{}{n}{}{}{}\) is the noise-free component of the \(n\)th gate, and \(\oC{E}{}{n}{}{J_n\!\left(i\right)}{}\) is the error channel of the \(J_n\!\left(i\right)\)th variant of the \(n\)th gate. 

IIAM requires all error channels to be identical and to commute with the gates they affect. IIAM is a generalisation of RIIM to more diverse noise models, RIIM is introduced in \citeRef{He2020} (see also \citeRef{Pascuzzi2022}) as a method for dealing with local depolarising noise. We only have a knowledge-free version of this method.

The definition of \(J_n\left(i\right)\) is rather complex so it is easier to represent the mitigation in another, equivalent way:
\begin{align}\eqAL{noisycircuitvariantExpansionIIAM}
\gCN{C}{}{}{M}{}{Wireless}~=&~\left\{1:\gCN{C}{A}{}{}{m}{Wireless}\right\}_{m=0}^{\nMit}~,&\gCN{C}{A}{}{}{m}{Wireless}~=&~\left\{\oV{c}{A}{m}{}{j}{}:\gQN{C}{A}{}{}{m,j}{Wireless}~~\right\}_{j=0}^{\nAMi{m}-1}~,
\end{align}
where \(\nMit\) is the order of the mitigation, \(\oCN{C}{A}{}{}{m}{Wireless}\) is a linear combination of all the circuit variants that can be produced by applying exactly \(m\) identity insertions anywhere in the circuit. This could be \(m\) identity insertions applied to any one gate, \(1\) identity insertion applied to any \(m\) gates, or anything in between.  \(\oQN{C}{A}{}{}{m,j}{Wireless}\) is a linear combination (with unit coefficients) of the \(j\)th class of ways to include \(m\) identity insertions and \(\nAMi{m}\) is the number of distinct configurations of \(m\) identity insertions that can be applied, \ie the number of integer partitions of \(m\). Each class has a list of instructions associated with it: \(\oV{L}{A}{}{}{m,j}{}\), where each element of this list denotes a number of identity insertions followed by a number of gates to apply that number of identity insertions to. For example, we have:
\begin{align}
\oV{L}{A}{}{}{0,0}{}=&\left\{\left\{0,0\right\}\right\},&\oV{L}{A}{}{}{1,0}{}=&\left\{\left\{1,1\right\}\right\},&\oV{L}{A}{}{}{4,3}{}=&\left\{\left\{2,1\right\},\left\{1,2\right\}\right\},&\oV{L}{A}{}{}{5,4}{}=&\left\{\left\{2,2\right\},\left\{1,1\right\}\right\},
\end{align}
the first (\(\oV{L}{A}{}{}{0,0}{}\)) indicates we use no identity insertions so \(\oQN{C}{A}{}{}{0,0}{Wireless}\) is just our original noisy circuit (\(\oCN{C}{}{}{}{}{}\)); the second (\(\oV{L}{A}{}{}{1,0}{}\)) indicates we use one identity insertion so \(\oQN{C}{A}{}{}{1,0}{Wireless}\) consists of all the noisy circuits that can be made using a single identity insertion, so it is made up of \(\nGate\) distinct circuits, each one applies the identity insertion to a different gate; the third (\(\oV{L}{A}{}{}{4,3}{}\)) indicates each circuit that makes up \(\oQN{C}{A}{}{}{4,3}{Wireless}\) has four identity insertions in total: one gate has two identity insertions and two gates have one identity insertion, we include all possible circuits that satisfy this requirement; and the fourth (\(\oV{L}{A}{}{}{5,4}{}\)) indicates each circuit that makes up \(\oQN{C}{A}{}{}{5,4}{Wireless}\) has five identity insertions in total: two gates have two identity insertions and one gate has one identity insertion, we include all possible circuits that satisfy this requirement. In \tabAR{IIAMCoefficients} we provide the configuration lists for \(m=0\) to \(m=5\); the multiplicity of \(\oQN{C}{A}{}{}{m,j}{Wireless}\) (\(\multiplicity{\oQN{C}{A}{}{}{m,j}{}}\), \ie the number of distinct circuits \(\oQN{C}{A}{}{}{m,j}{}\) is made up of); all the coefficients required for the mitigations; and the equivalent \(i\) index of each coefficient (see \eqAR{IIAMStandardForm}):
\begin{align}\eqAL{relationBetweenCoefficientsIIAM}
i =& j+\sum_{n=0}^{m-1}\nAMi{n},&
\oV{c}{Gl}{}{}{i}{}=&\multiplicity{\oQN{C}{A}{}{}{m,j}{}}\oV{c}{A}{m}{}{j}{},&\gCN{C}{}{}{}{i}{}~~=&~~\left\{\frac{1}{\multiplicity{\oQN{C}{A}{}{}{m,j}{}}}:\gQN{C}{A}{}{}{m,j}{}\,\right\}.
\end{align}
The implementation of IIAM is similar to the standard procedure (see \appSR{implementation}), the only difference is that \(\oCN{C}{}{}{}{i}{}\) is now a stochastic mixture of circuits. So each time we should sample \(\oCN{C}{}{}{}{i}{}\) we choose one of the constituent circuits of \(\oQN{C}{A}{}{}{m,j}{}\) uniformly at random and then divide the output by the multiplicity (\(\multiplicity{\oQN{C}{A}{}{}{m,j}{}}\)).

The execution time for every circuit contained within \(\oCN{C}{A}{}{}{m}{Wireless}\) is:
\begin{align}
\timeExeci{j} =&\left(1+\frac{2m}{\nGate}\right)\timeExec
\end{align}
where \(\timeExec\) is the execution time for the original noisy circuit and we have assumed that all gates take the same amount of time to implement.

Numerical and analytical calculations have indicated that the proxy bias is given by:
\begin{align}\eqAL{stochasticNoiseIIAMKFProxy}
\oV{\varepsilon}{\proxy}{}{\scriptscriptstyle M}{}{}=&\begin{cases}\frac{\left(2\nGate\eStoch\right)^{\nMit +1} }{\left(\nMit +1\right)!}&\text{ for stochastic noise}\\ 
\frac{\left(\nGate\left|\eAngle\right|\right)^{\nMit +1} }{\left(\nMit +1\right)!}&\text{ for rotational errors}\\ 
\end{cases}
\end{align}
to leading order.

It is worth noting that \tabAR{IIAMCoefficients} is slightly different from the corresponding table in \citeRef{He2020} (see Table I of \citeRef{He2020} or Table II of the Arxiv version: \citeRef{He2020Arxiv}). The reason for this difference is that there is a degree of freedom in the choice of coefficients when \(\nMit=4\) and one wishes to mitigate local depolarising noise. We choose coefficients that will also mitigate rotational errors and the authors of \citeRef{He2020} make a different choice. In addition it appears there may be a typographic error for the coefficient: \(a_{\left\{3,3\right\}}\) in the fourth order mitigation (corresponding to \(\oV{c}{A}{2}{}{1}{}\) in this work). We obtain consistent results using:
\begin{align}
a_{\left\{3,3\right\}}=\frac{N_{\mathrm{c}}^2+14 N_{\mathrm{c}}+\mathbf{58}}{32},
\end{align}
whereas in Table I of \citeRef{He2020} 59 is written in place of 58 in the numerator.

\begin{landscape}

\begin{table}[htbp]
\begin{subtable}{1\linewidth}
 \scalebox{0.9}{
\begin{tabular}{l|llllllllllll}
\(i\)							&0				&1				&2				&3							\\
\(\left(m,j\right)\)					&\(\left(0,0\right)\)		&\(\left(1,0\right)\)		&\(\left(2,0\right)\)		&\(\left(2,1\right)\)					\\
\hline
\(\oV{L}{A}{}{}{m,j}{}\)				&\(\left\{\{0,0\}\right\}\)	&\(\left\{\{1,1\}\right\}\)	&\(\left\{\{2,1\}\right\}\)	&\(\left\{\{1,2\}\right\}\)				\\
\(\multiplicity{\oQN{C}{A}{}{}{m,j}{}}\)	&\(1\)				&\(\nGate\)			&\(\nGate\)			&\(\frac{\nGate\left(\nGate-1\right)}{2}\)	\\[5pt]
\(\nMit: 0;\oV{c}{A}{m}{}{j}{}\)		&\(1\)\\[5pt]
\(\nMit: 1;\oV{c}{A}{m}{}{j}{}\)		&\(\frac{2+\nGate }{2}\)& \(-\frac{1}{2}\)\\[5pt]
\(\nMit: 2;\oV{c}{A}{m}{}{j}{}\)		& \(\frac{(2+\nGate ) (4+\nGate )}{8}\)& \(-\frac{4+\nGate }{4}\)& \(\frac{3}{8}\)& \(\frac{1}{4}\)\\[5pt]
\(\nMit: 3;\oV{c}{A}{m}{}{j}{}\)		& \(\frac{(2+\nGate ) (4+\nGate ) (6+\nGate )}{48}\)& \(-\frac{(4+\nGate ) (6+\nGate )}{16}\)& \(\frac{3 (6+\nGate )}{16}\)& \(\frac{6+\nGate }{8}\)\\[5pt]
\(\nMit: 4;\oV{c}{A}{m}{}{j}{}\)		&\(\frac{(2+\nGate ) (4+\nGate ) (6+\nGate ) (8+\nGate )}{384}\)& \(-\frac{(4+\nGate ) (6+\nGate ) (8+\nGate )}{96}\)& \(\frac{3 (6+\nGate ) (8+\nGate )}{64}\)& \(\frac{(6+\nGate ) (8+\nGate )}{32}\)\\[5pt]
\(\nMit: 5;\oV{c}{A}{m}{}{j}{}\)		&\(\frac{(2+\nGate ) (4+\nGate ) (6+\nGate ) (8+\nGate ) (10+\nGate )}{3840}\)& \(-\frac{(4+\nGate ) (6+\nGate ) (8+\nGate ) (10+\nGate )}{768}\)& \(\frac{(6+\nGate ) (8+\nGate ) (10+\nGate )}{128}\)& \(\frac{(6+\nGate ) (8+\nGate ) (10+\nGate )}{192}\)\\[5pt]
\hline
\end{tabular}
}
\caption{\(m=0\) to \(m=2\).}
\tabAL{IIAMCoefficients2}
\end{subtable}

\begin{subtable}{1\linewidth}
 \scalebox{0.9}{
\begin{tabular}{l|llllllllllll}
\(i\)							&4						&5						&6				&7				&8						&9				&10						&11\\
\(\left(m,j\right)\)					&\(\left(3,0\right)\)				&\(\left(3,1\right)\)				&\(\left(3,2\right)\)		&\(\left(4,0\right)\)		&\(\left(4,1\right)\)				&\(\left(4,2\right)\)		&\(\left(4,3\right)\)				&\(\left(4,4\right)\)\\
\hline
\(\oV{L}{A}{}{}{m,j}{}\)				&\(\left\{\{3,1\}\right\}\)			&\(\left\{\{2,1\},\{1,1\}\right\}\)	&\(\left\{\{1,3\}\right\}\)	&\(\left\{\{4,1\}\right\}\)	&\(\left\{\{3,1\},\{1,1\}\right\}\)	&\(\left\{\{2,2\}\right\}\)	&\(\left\{\{2,1\},\{1,2\}\right\}\)	&\(\left\{\{1,4\}\right\}\)\\
\(\multiplicity{\oQN{C}{A}{}{}{m,j}{}}\)	&\(\nGate\)					&\(\nGate\left(\nGate-1\right)\)		&\(\frac{\nGate\left(\nGate-1\right)\left(\nGate-2\right)}{3!}\)&\(\nGate\)&\(\nGate\left(\nGate-1\right)\)&\(\frac{\nGate\left(\nGate-1\right)}{2}\)&\(\frac{\nGate\left(\nGate-1\right)\left(\nGate-2\right)}{2}\)&\(\frac{\nGate!}{4!\left(\nGate-4\right)!}\)\\[5pt]
\(\nMit: 3;\oV{c}{A}{m}{}{j}{}\)		&\(-\frac{5}{16}\)&\(-\frac{3}{16}\)& \(-\frac{1}{8}\)\\[5pt]
\(\nMit: 4;\oV{c}{A}{m}{}{j}{}\)		& \(-\frac{5 (8+\nGate )}{32}\)&\(-\frac{3 (8+\nGate )}{32}\)& \(-\frac{8+\nGate }{16}\)& \(\frac{35}{128}\)& \(\frac{5}{32}\)& \(\frac{9}{64}\)& \(\frac{3}{32}\)& \(\frac{1}{16}\)\\[5pt]
\(\nMit: 5;\oV{c}{A}{m}{}{j}{}\)		&\(-\frac{5 (8+\nGate ) (10+\nGate )}{128}\)&\(-\frac{3 (8+\nGate ) (10+\nGate )}{128}\)& \(-\frac{(8+\nGate ) (10+\nGate )}{64}\)& \(\frac{35 (10+\nGate )}{256}\) & \(\frac{5 (10+\nGate )}{64}\)& \(\frac{9 (10+\nGate )}{128}\)& \(\frac{3 (10+\nGate )}{64}\)& \(\frac{10+\nGate }{32}\)\\[5pt]
\hline
\end{tabular}
}
\caption{\(m=3\) to \(m=4\).}
\tabAL{IIAMCoefficients4}
\end{subtable}

\begin{subtable}{1\linewidth}
 \scalebox{0.9}{
\begin{tabular}{l|llllllllllll}
\(i\)							&12						&13						&14						&15										&16											&17													&18\\
\(\left(m,j\right)\)					&\(\left(5,0\right)\)				&\(\left(5,1\right)\)				&\(\left(5,2\right)\)				&\(\left(5,3\right)\)								&\(\left(5,4\right)\)									&\(\left(5,5\right)\)											&\(\left(5,6\right)\)\\
\hline
\(\oV{L}{A}{}{}{m,j}{}\)				&\(\left\{\{5,1\}\right\}\)			&\(\left\{\{4,1\},\{1,1\}\right\}\)	&\(\left\{\{3,1\},\{2,1\}\right\}\)	&\(\left\{\{3,1\},\{1,2\}\right\}\)					&\(\left\{\{2,2\},\{1,1\}\right\}\)						&\(\left\{\{2,1\},\{1,3\}\right\}\)								&\(\left\{\{1,5\}\right\}\)\\
\(\multiplicity{\oQN{C}{A}{}{}{m,j}{}}\)	&\(\nGate\)					&\(\nGate\left(\nGate-1\right)\)		&\(\nGate\left(\nGate-1\right)\)		&\(\frac{\nGate\left(\nGate-1\right)\left(\nGate-2\right)}{2}\)	&\(\frac{\nGate\left(\nGate-1\right)\left(\nGate-2\right)}{2}\)		&\(\frac{\nGate\left(\nGate-1\right)\left(\nGate-2\right)\left(\nGate-3\right)}{3!}\)	&\(\frac{\nGate\left(\nGate-1\right)\left(\nGate-2\right)\left(\nGate-3\right)\left(\nGate-4\right)}{5!}\)\\[5pt]
\(\nMit: 5;\oV{c}{A}{m}{}{j}{}\)		& \(-\frac{63}{256}\)& \(-\frac{35}{256}\)& \(-\frac{15}{128}\)& \(-\frac{5}{64}\)& \(-\frac{9}{128}\)& \(-\frac{3}{64}\)& \(-\frac{1}{32}\)\\[5pt]
\hline
\end{tabular}
}
\caption{\(m=5\).}
\end{subtable}

    \caption{Table of IIAM coefficients and parameters (see \appSR{IIAM}).}
    \tabAL{IIAMCoefficients}
\end{table}

\end{landscape}

\subsection{IILM}\appSL{IILM}
Identity-insertion-assisted local mitigation (IILM) is a local method (see \appSR{scope}) that uses identity insertions to amplify the noise in the noisy gate variants:
\begin{align}
\gCN{C}{}{}{M}{}{Wireless}~=~&\Big(\gGN{U}{}{n}{M}{}{}\Big)_{n=1}^{\nGate}, ~~~\gGN{U}{}{n}{M}{}{}~=~\Big\{
\coefLocali{n,j}: \gGN{U}{}{n}{}{j}{}
\Big\}_{j=0}^{\nCircLocali{n}}~,\\
\gGN{U}{}{n}{}{j}{}~=~&\Big(\gGN{U}{}{n}{}{}{}\gGDN{U}{}{n}{}{}{}\Big)^{j}\gGN{U}{}{n}{}{}{}~=~\gG{U}{}{n}{}{}{}\gC{E}{}{n}{}{j}{},
\end{align}
where \(\oCN{C}{}{}{M}{}{Wireless}\) is the mitigated circuit, \(\oGN{U}{}{n}{M}{}{Wireless}\) is \(n\)th mitigated gate, \(\nGate\) is the number of gates in the circuit, \(\coefLocali{n,j}\) is the mitigation coefficient for the \(j\)th gate variant of the \(n\)th gate, \(\oGN{U}{}{n}{}{j}{Wireless}\) is the \(j\)th gate variant of the \(n\)th gate, \(\nCircLocali{n}=\nMit\) is the number of gate variants for the \(n\)th gate and the order of the mitigation,  \(\oGN{U}{}{n}{}{}{}\) is the \(n\)th noisy gate before mitigation, \(\oGDN{U}{}{n}{}{}{}\) is the noisy inverse of the \(n\)th gate, \(\oG{U}{}{n}{}{}{}\) is the noise-free component of the \(n\)th gate, and \(\oC{E}{}{n}{}{j}{}\) is the error channel of the \(j\)th variant of the \(n\)th gate. 

The version of IILM presented here requires the error channels associated with the noisy gate and its noisy inverse to be the same and that the noisy gate, the noisy inverse, and their error channels all commute with each other. IILM could be considered to be a special case of NEPEC (see \citeRef{Mari2021}). 

The execution time for the \(i\)th variant of the \(n\)th gate is given by:
\begin{align}
\oV{\tau}{L}{n}{}{i}{} =&\left(2i+1\right)\oV{\tau}{L}{n}{}{}{},
\end{align}
where \(\oV{\tau}{L}{n}{}{}{}\) is the execution time for the original \(n\)th noisy gate.

The knowledge-free version (IILM:KF) uses:
\begin{align}
\coefLocali{n,i}=\prod_{m=0:m\ne i}^{\nMit }\frac{2m+1}{2\left(m-i\right)}.
\end{align}
The proxy bias for stochastic noise (for IILM:KF) is given by:
\begin{align}\eqAL{stochasticNoiseIILMKFProxy}
\oV{\varepsilon}{\proxy}{}{\scriptscriptstyle M}{}{}=&2\nGate\eStoch^{\nMit+1}\left(1+\aParam\right)^{\nMit}\frac{\left(2\nMit+1\right)!!}{\left(\nMit+1\right)!}+\nGate\orderZero\left(\eStoch^{\nMit+2}\right),
\end{align}
where \(\eStoch\) is the noise amplitude of the original noisy gates (\(\oGN{U}{}{n}{}{0}{}=\oGN{U}{}{n}{}{}{}\)), and \(0\le\aParam\le 1\) is defined in \eqAR{SNnoisemodelA}. The proxy bias for rotational errors (for IILM:KF)  is given by:
\begin{align}\eqAL{rotationalErrorsIILMKFProxy}
\oV{\varepsilon}{\proxy}{}{\scriptscriptstyle M}{}{}=&\nGate\left|\eAngle\right|^{\nMit+1}\frac{\left(2\nMit+1\right)!!}{\left(\nMit+1\right)!}+\nGate\orderZero\left(\eAngle^{\nMit+2}\right).
\end{align}

The noise-aware version (IILM:NA) for stochastic noise uses \(\nMit=1\) and:
\begin{align}
\coefLocali{0}=&\frac{3-3\left(1+\aParam\right)\eStoch+\left(1+\aParam\right)^{2}\eStoch^{2}}{2-3\left(1+\aParam\right)\eStoch+\left(1+\aParam\right)^{2}\eStoch^{2}}=\frac{3}{2}+\frac{3}{4}\left(1+\aParam\right)\eStoch +\orderZero\left(\eStoch^{2}\right),\\
\coefLocali{1}=&-\frac{1}{2-3\left(1+\aParam\right)\eStoch+\left(1+\aParam\right)^{2}\eStoch^{2}}=-\frac{1}{2}-\frac{3}{4}\left(1+\aParam\right)\eStoch +\orderZero\left(\eStoch^{2}\right).
\end{align}
This method is unbiased so the proxy bias is zero.

The noise-aware version (IILM:NA) for rotational errors uses (\(\nMit=2\)):
\begin{align}
\coefLocali{0}=&\frac{\sin\frac{3\eAngle}{2}\sin\frac{5\eAngle}{2}}{\sin\eAngle\sin\left(2\eAngle\right)}=\frac{15}{8}-\frac{35}{32}\eAngle^{2}+\orderZero\left(\eAngle^{4}\right), \\
\coefLocali{1}=&-\frac{\sin\frac{\eAngle}{2}\sin\frac{5\eAngle}{2}}{\sin^{2}\eAngle}=-\frac{5}{4}+\frac{15}{16}\eAngle^{2}+\orderZero\left(\eAngle^{4}\right) ,\\
\coefLocali{2}=& \frac{\sin\frac{\eAngle}{2}\sin\frac{3\eAngle}{2}}{\sin\eAngle\sin\left(2\eAngle\right)}=\frac{3}{8}+\frac{5}{32}\eAngle^{2}+\orderZero\left(\eAngle^{4}\right).
\end{align}
This method is unbiased so the proxy bias is zero.

\subsection{IISM}\appSL{IISM}
Identity-insertion-assisted synchronous mitigation (IISM) is a synchronous method (see \appSR{scope}) that uses identity insertions to amplify the noise in the noisy gate variants:
\begin{align}
\gCN{C}{}{}{M}{}{Wireless}~=~&\Big\{
\coefGlobali{j}: \gCN{C}{}{}{}{j}{Wireless}
\Big\}_{j=0}^{\nCircGlobal}~=~\Big\{
\coefGlobali{j}: \Big(\gGN{U}{}{n}{}{j}{}\Big)_{n=1}^{\nGate}
\Big\}_{j=0}^{\nCircGlobal}~,\\
\gGN{U}{}{n}{}{j}{}~=~&\Big(\gGN{U}{}{n}{}{}{}\gGDN{U}{}{n}{}{}{}\Big)^{j}\gGN{U}{}{n}{}{}{}~=~\gG{U}{}{n}{}{}{}\gC{E}{}{n}{}{i}{},
\end{align}
where \(\oCN{C}{}{}{M}{}{Wireless}\) is the mitigated circuit, \(\coefGlobali{j}\) is the mitigation coefficient for the \(j\)th circuit variant, \(\oCN{C}{}{}{}{j}{Wireless}\) is the \(j\)th circuit variant, \(\nCircGlobal=\nMit\) is the number of circuit variants and the order of the mitigation, \(\oGN{U}{}{n}{}{j}{}\) is the \(j\)th variant of the \(n\)th noisy gate, \(\nGate\) is the number of gates in the circuit,  \(\oGN{U}{}{n}{}{}{}\) is the \(n\)th noisy gate before mitigation, \(\oGDN{U}{}{n}{}{}{}\) is the noisy inverse gate, \(\oG{U}{}{n}{}{}{}\) is the noise-free component of the \(n\)th gate, and \(\oC{E}{}{n}{}{i}{}\) is the error channel of the \(i\)th variant of the \(n\)th gate. 

IISM requires all the mitigated noisy gates and their noisy inverses to have noise of the same type with the same noise amplitude and the error channel of each gate must commute with the noisy gates and its noisy inverses. IISM is a special case of zero-noise extrapolation (see \citeRefs{Temme2017,Li2017}). 

The execution time for the \(j\)th circuit is given by:
\begin{align}
\timeExeci{j} =&\left(2j+1\right)\timeExec
\end{align}
where \(\timeExec\) is the execution time for the original noisy circuit.

The knowledge-free version (IISM:KF) uses:
\begin{align}\eqAL{IISMcoefficientsKnowledgeFree}
\coefGlobali{i}=\prod_{m=0:m\ne i}^{\nMit }\frac{2m+1}{2\left(m-i\right)}.
\end{align}
The proxy bias for stochastic noise (for IISM:KF) is given by:
\begin{align}\eqAL{stochasticNoiseIISMKFProxy}
\oV{\varepsilon}{\proxy}{}{\scriptscriptstyle M}{}{}=&\left(2\nGate\eStoch\right)^{\nMit +1}  \frac{\left(2\nMit+1\right)!!}{\left(\nMit +1\right)!}+\orderLargeSmall\left(\nGate^{\nMit}\eStoch^{\nMit+1},\nGate^{\nMit+2}\eStoch^{\nMit+2}\right),
\end{align}
where \(\eStoch\) is the noise amplitude of the original noisy gates (\(\oGN{U}{}{n}{}{0}{}=\oGN{U}{}{n}{}{}{}\)). The proxy bias for rotational errors (for IISM:KF)  is given by:
\begin{align}\eqAL{rotationalErrorsIISMKFProxy}
\oV{\varepsilon}{\proxy}{}{\scriptscriptstyle M}{}{}=&\left(\nGate\left|\eAngle\right|\right)^{\nMit+1} \frac{\left(2\nMit+1\right)!!}{\left(\nMit+1\right)!}+\orderLargeSmall\left(\nGate^{\nMit}\eAngle^{\nMit+1},\nGate^{\nMit+2}\eAngle^{\nMit+2}\right).
\end{align}

The noise-aware version (IISM:NA) for stochastic noise uses:
\begin{align}
\coefGlobali{i}=&\prod_{m=0:m\ne i}^{\nMit }\frac{\eStochi{m}}{\eStochi{m}-\eStochi{i}}=\prod_{m=0:m\ne i}^{\nMit }\frac{2m+1}{2\left(m-i\right)}+\orderZero\left(\eStoch\right), &\eStochi{i}= &\frac{1-\left(1-\eStoch\left(1+\aParam\right)\right)^{2i+1}}{1+\aParam},
\end{align}
where \(\eStochi{m}\) is the noise amplitude of the \(m\)th noisy variant, \(\nMit\) is the order of the mitigation, and \(0\le \aParam\le 1\) is a parameter of the noise model  (see \eqAR{SNnoisemodelA}). The proxy bias is the same to leading order as IISM:KF (\eqAR{stochasticNoiseIISMKFProxy}) for \(\nMit\ll \nGate\).

The noise-aware version (IISM:NA) for rotational errors uses:
\begin{align}
\coefGlobali{i}=&\prod_{j=0,j\ne i}^{\nMit} \frac{\sin \frac{\eAnglei{j}}{2}}{\sin\frac{ \eAnglei{j}- \eAnglei{i}}{2}}=\prod_{j=0,j\ne i}^{\nMit} \frac{ 1+2j}{2\left(j- i\right)}+\orderZero\left(\eAngle\right),&\eAnglei{i}= &\left(1+2i\right)\eAngle,
\end{align}
where \(\eAnglei{j}\) is the noise amplitude of the \(j\)th noisy variant and \(\nMit\) is the order of the mitigation. The proxy bias is the same to leading order as IISM:KF (\eqAR{rotationalErrorsIISMKFProxy}) for \(\nMit\ll 2\nGate\).



\subsection{LC}\appSL{LC} 
Local cancellation (LC) is a local tailoring method (see \appSR{scope}). It is similar to CLM but instead of producing unbiased gates it removes the coherent components of error channels so that the effective (locally cancelled) error channel is a Pauli channel (or some other stochastic channel if another basis is used):
\begin{align}
\gGN{U}{}{}{LC}{}{}~=~\Big\{
\coefLCi{j}: \gGN{U}{}{}{}{j}{}
\Big\}_{j=0}^{\nCircLCi{}}~=~\gG{U}{}{}{}{}{}
\Big\{
\coefLCi{j}: \gC{E}{}{}{}{j}{}
\Big\}_{j=0}^{\nCircLCi{}}~=~\gG{U}{}{}{}{}{}\gC{E}{SN}{}{}{}{},
\end{align}
where \(\oGN{U}{}{}{LC}{}{Wireless}\) is the locally cancelled version of the noisy gate: \(\oGN{U}{}{}{}{}{}\), \(\coefLCi{j}\) is the cancelling coefficient for the \(j\)th gate variant of the noisy gate, \(\oGN{U}{}{n}{}{j}{Wireless}\) is the \(j\)th gate variant of the noisy gate, \(\nCircLCi{}\) is the number of extra gate variants used to cancel the noisy gate, \(\oG{U}{}{}{}{}{}\) is the noise-free component of the noisy gate, \(\oC{E}{}{}{}{j}{}\) is the error channel of the \(j\)th variant of the noisy gate, and \(\oC{E}{SN}{}{}{}{}\) is the stochastic-noise channel of the locally cancelled noisy gate, which will generally be a Pauli error channel (\ie \(\oC{E}{SN}{}{}{}{}=\oC{E}{P}{}{}{}{}\)):
\begin{align}
\gC{E}{P}{}{}{}{}=&\Big\{\gProbi{k}: \gO{P}{}{k}{}{}{}\Big\}_{k=0}^{4^{\nQubit}-1},&\sum_{k=0}^{4^{\nQubit}-1}\gProbi{k}=&1,
\end{align}
where \(\gProbi{k}\ge 0\) is the probability an error channel equivalent to applying the \(k\)th Pauli operator (\(\oO{P}{}{k}{}{}{}\)) occurs if a gate with this error channel is implemented, and \(\nQubit\) is the number of qubits that the error channel affects.

LC is an alternative noise tailoring method to Pauli twirling (\citeRef{OGorman2016,Bennett1995,Bennett1996,Knill2004}), which can be generalised to randomised compiling (RC, \citeRef{Wallman2016}). Both techniques convert coherent error channels to Pauli-noise channels, when averaged over many runs. RC stochastically sandwiches noisy operators between Pauli and Clifford operators, which are redundant in the absence of noise. Whereas LC uses a stochastic mixture of the original gate and noisy variants of this gate to achieve the same ends. RC has the advantage that it can be implemented knowledge free, whereas LC can only be implemented knowledge free for rotational errors if the hidden inverse can be implemented perfectly. Otherwise it requires quantitative knowledge of the coherent error. The choice between the techniques may depend on whether noisy variants or Clifford operators are easier to implement. 

The execution time for the \(i\)th variant of the noisy gate is assumed to be:
\begin{align}
\oV{\tau}{L}{}{}{i}{} =\oV{\tau}{L}{}{}{}{},
\end{align}
where \(\oV{\tau}{L}{}{}{}{}\) is the execution time for the original noisy gate. That is, we assume that the execution time of every noisy gate variant is the same.

The coefficients are calculated using the transfer-matrix representation of the noisy gate variant\footnote{For the LC derivation we present we use the Pauli transfer matrix but we could use a transfer matrix in another basis and rid ourselves of coherences in that basis if we prefer, the derivation is analogous.} error channels (see \eqAR{PauliTransferMatrix} of \appSR{transferMatrices}):
\begin{align}
\left(\transferMatrixi{\oC{E}{}{}{}{j}{}} \right)_{k,l}=&\frac{1}{2^{\nQubit }} \traceOp{\pOpi{k}\oC{E}{}{}{}{j}{}\left(\pOpi{l}\right)},
\end{align}
where \(\transferMatrixi{\oC{E}{}{}{}{j}{}}\) is the transfer matrix corresponding to the error channel of the \(j\)th noisy gate variant, \(\pOpi{k}\) is the \(k\)th Pauli word operator (see \appSR{pauliWords}), and \(\nQubit\) is the number of qubits the error channel acts on.
The Pauli error channel has a diagonal Pauli transfer matrix, so the coefficients are chosen to cancel the off-diagonal elements. The coefficients are stochastic so we require:
\begin{align}\eqAL{normalisationConditionLC}
\sum_{j=0}^{\nCircLCi{}}\coefLCi{j}=\sum_{j=0}^{\nCircLCi{}}\left|\coefLCi{j}\right|=&1&\implies&\coefLCi{j}>0,
\end{align}
for all \(j\). So the resulting channel will be CPTP since it is a stochastic combination of CPTP channels. It is possible to relax the requirement that \(\coefLCi{j}>0\) but this will incur a sampling cost, whereas conventional LC has a sampling cost of unity, so it is cost free. To cancel the off-diagonal elements of the error channel we require:
\begin{align}
\sum_{j=0}^{\nCircLCi{}}\coefLCi{j}\left(\transferMatrixi{\oC{E}{}{}{}{j}{}}\right)_{k,l}=&0 \text{ for }k\ne l.
\end{align}
We place no restriction on the diagonal elements of the \(\transferMatrixi{\oC{E}{}{}{}{j}{}} \) except those that result from \eqAR{normalisationConditionLC}. 

We do not implement LC for stochastic noise.

For rotational errors generated by the Hermitian operator \(\hermitianOp\) we use \(\nCircLC=1\). We can produce a stochastic channel using local cancellation, which will be a Pauli channel if the Hermitian operator has a diagonal Pauli transfer matrix. If we have access to two noisy gate variants with rotational errors:
\begin{align}\eqAL{LCcustomChannelsRot}
\gGN{U}{}{}{}{0}{}~ =~& \gG{U}{}{}{}{}{}\gG{~\rotGNF{\frac{\eAnglei{0}}{2}}{\hermitianOp}}{}{}{}{}{VVL}
,&\gGN{U}{}{}{}{1}{}~ =~& \gG{U}{}{}{}{}{}\gG{~\rotGNF{\frac{\eAnglei{1}}{2}}{\hermitianOp}}{}{}{}{}{VVL},
\end{align}
where \(\eAnglei{0}\) and \(\eAnglei{1}\) are the noise amplitudes of the two noisy gate variants, then we use the coefficients:
\begin{align}\eqAL{rotationalErrorsCoefsLC}
\coefLocali{0}=&-\frac{\sin \eAnglei{1}}{\sin \eAnglei{0}-\sin \eAnglei{1}}, &\coefLocali{1}=&\frac{\sin \eAnglei{0}}{\sin \eAnglei{0}-\sin \eAnglei{1}},
\end{align}
to obtain the locally cancelled gate. These coefficients will be stochastic provided that \( \eAnglei{0} \eAnglei{1}<0\), \ie the rotational errors act in opposing directions. The resulting locally cancelled gate is given by:
\begin{align}
\gGN{U}{}{}{LC}{}{}=\Big\{
\coefLCi{j}: \gGN{U}{}{}{}{j}{}
\Big\}_{j=0}^{1}=\gG{U}{}{}{}{}{}\begin{cases}
\gG{}{}{}{}{}{Wire}~&\text{ with probability }1-\eStochLC\\
\gG{\hermitianOp}{}{}{}{}{}~&\text{ with probability }\eStochLC
\end{cases}
\end{align}
where:
\begin{align}
\eStochLC = \frac{1}{2}\left(1-\frac{\sin\left(\eAnglei{0}-\eAnglei{1}\right)}{\sin \eAnglei{0}-\sin \eAnglei{1}}\right).
\end{align}
So (if we had a circuit of \(\nGate\) gates with the same error channels, which we locally cancelled in this way), the proxy bias would be given by (see \appSR{noiseModel}):
\begin{align}\eqAL{rotationalErrorsLCProxy}
\oV{\varepsilon}{\proxy}{}{\scriptscriptstyle M}{}{}=&\nGate\left(1-\frac{\sin\left(\eAnglei{0}-\eAnglei{1}\right)}{\sin \eAnglei{0}-\sin \eAnglei{1}}\right).
\end{align}
If \(\eAngle=\eAnglei{0}=-\eAnglei{1}\) everything simplifies and we find:
\begin{align}
\coefLCi{0}=&\frac{1}{2}=\coefLCi{1},&\oV{\varepsilon}{\proxy}{}{\scriptscriptstyle M}{}{}=&\nGate\left(1-\cos\eAngle\right)=\nGate\left(\frac{\eAngle^2}{2}+\orderZero\left(\eAngle^4\right)\right) = \frac{\oV{e}{RE}{}{}{}{} ^{2}}{2\nGate}+\orderGeneral\left(\frac{\oV{e}{RE}{}{}{}{} ^{4}}{\nGate^{3}}\right),
\end{align}
where (see \eqAR{rotNoiseLevel}):
\begin{align}
\oV{e}{RE}{}{}{}{} = \left|\eAngle\right|\nGate,
\end{align}
is the original noise level. So, although this is designed as a tailoring rather than a mitigation method it can also mitigate the noise in the small noise (\(\oV{e}{RE}{}{}{}{} \ll 1\)) or large circuit (\(\nGate\gg 1\)) regime.


\subsection{TIILM} \appSL{TIILM}
Tuned identity-insertion-assisted local mitigation (TIILM) is a local (see \appSR{scope}) method that uses identity insertions to amplify the noise in the noisy gate variants. The number of identity insertions used for each gate variant is tuned to minimise the runtime scaling. The procedure is given by:
\begin{align}
\gCN{C}{}{}{M}{}{Wireless}~=~&\Big(\gGN{U}{}{n}{M}{}{}\Big)_{n=1}^{\nGate}, ~~~\gGN{U}{}{n}{M}{}{}~=~\Big\{
\coefLocali{n,j}: \gGN{U}{}{n}{}{j}{}
\Big\}_{j=0}^{\nCircLocali{n}}~,\\
\gGN{U}{}{n}{}{j}{}~=~&\Big(\gGN{U}{}{n}{}{}{}\gGDN{U}{}{n}{}{}{}\Big)^{m_{n,j}}\gGN{U}{}{n}{}{}{}~=~\gG{U}{}{n}{}{}{}\gC{E}{}{n}{}{j}{},
\end{align}
where \(\oCN{C}{}{}{M}{}{Wireless}\) is the mitigated circuit, \(\oGN{U}{}{n}{M}{}{Wireless}\) is \(n\)th mitigated gate, \(\nGate\) is the number of gates in the circuit, \(\coefLocali{n,j}\) is the mitigation coefficient for the \(j\)th gate variant of the \(n\)th gate, \(\oGN{U}{}{n}{}{j}{Wireless}\) is the \(j\)th gate variant of the \(n\)th gate, \(\nCircLocali{n}\) is the number of gate variants for the \(n\)th gate,  \(\oGN{U}{}{n}{}{}{}\) is the \(n\)th noisy gate before mitigation, \(\oGDN{U}{}{n}{}{}{}\) is the noisy inverse of the \(n\)th gate, \(m_{n,j}\) is the number of identity insertions used for the \(j\)th variant of the \(n\)th gate, \(\oG{U}{}{n}{}{}{}\) is the noise-free component of the \(n\)th gate, and \(\oC{E}{}{n}{}{j}{}\) is the error channel of the \(j\)th variant of the \(n\)th gate. 

Our presentation of TIILM requires all error channels to commute with the gates they affect and that the error channels are the same for each gate and its inverse. TIILM could be considered to be a special case of NEPEC (see \citeRef{Mari2021}). For our numerical results we numerically optimise the values of \(m_{n,j}\) but for our analytic results we give two example choices, the first for the small noise regime and the second for the large circuit regime.

The execution time for the \(i\)th variant of the \(n\)th gate is given by:
\begin{align}
\oV{\tau}{L}{n}{}{i}{} =&\left(2m_{n,i}+1\right)\oV{\tau}{L}{n}{}{}{},
\end{align}
where \(\oV{\tau}{L}{n}{}{}{}\) is the execution time for the original \(n\)th noisy gate and \(m_{n,i}\) is the number of identity insertions for the \(i\)th variant.

For stochastic noise we use \(\nCircLocali{n}=1\) and (for the analytic results):
\begin{align}\eqAL{nIITIILMSN}
\kappa=&\sqrt{8-\frac{2}{\nGate}},&m_{n,0} =&0,&m_{n,1}\approx & \begin{cases}
\frac{\kappa}{\sqrt{1+\aParam}} \frac{\nGate}{\sqrt{\oV{e}{SN}{}{}{}{}}}&\text{ for } \oV{e}{SN}{}{}{}{}\ll 1\\
\frac{\ln 2}{1+\aParam}\frac{\nGate}{\oV{e}{SN}{}{}{}{}}&\text{ for } \nGate\gg 1.
\end{cases}
\end{align}
where \(\oV{e}{SN}{}{}{}{}\) is the stochastic noise level (see \eqAR{stochNoiseLevel}), \(\nGate\) is the number of gates in the circuit, and \(0\le \aParam\le 1\) is determined by the error channel (see \eqAR{SNnoisemodelA}). The coefficients\footnote{The first equalities for the coefficients given in \eqAR{TIILMCoefSN} for TIILM are used for both numerical and analytic results. However the Taylor expansions are valid only for the analytic results where we use \eqAR{nIITIILMSN} to define the number of identity insertions we use. For numerical results we optimise \(m_{n,0}\) and \(m_{n,1}\) to give the lowest possible runtime scaling.} are given by:
\begin{align}\eqAL{TIILMCoefSN}
\oV{c}{L}{}{}{0}{}=&\frac{1-\oV{\lambda}{}{1}{}{}{}^{2m_{n,1}+1}}{\oV{\lambda}{}{1}{}{}{}^{2m_{n,0}+1}-\oV{\lambda}{}{1}{}{}{}^{2m_{n,1}+1}}=\begin{cases}
1+\frac{\sqrt{1+\aParam}}{2\kappa\nGate}\sqrt{\oV{e}{SN}{}{}{}{}}+\orderSmall\!\left(\oV{e}{SN}{}{}{}{}\right),&\text{ for } \oV{e}{SN}{}{}{}{}\ll 1\\
1+\left(1+\aParam\right)\frac{\oV{e}{SN}{}{}{}{}}{\nGate}+\orderLarge\left(\frac{1}{\nGate^{2}}\right)&\text{ for } \nGate\gg 1,
\end{cases}\\
\oV{c}{L}{}{}{1}{}=&-\frac{1-\oV{\lambda}{}{1}{}{}{}^{2m_{n,0}+1}}{\oV{\lambda}{}{1}{}{}{}^{2m_{n,0}+1}-\oV{\lambda}{}{1}{}{}{}^{2m_{n,1}+1}}=\begin{cases}
-\frac{\sqrt{1+\aParam}}{2\kappa\nGate}\sqrt{\oV{e}{SN}{}{}{}{}}\left(1+\orderSmall\!\left(\sqrt{\oV{e}{SN}{}{}{}{}}\right)\right)&\text{ for } \oV{e}{SN}{}{}{}{}\ll 1\\
\left(1+\aParam\right)\frac{\oV{e}{SN}{}{}{}{}}{\nGate}+\orderLarge\left(\frac{1}{\nGate^{2}}\right)&\text{ for } \nGate\gg 1,
\end{cases}
\end{align}
where:
\begin{align}
\oV{\lambda}{}{0}{}{}{}=&1,&\oV{\lambda}{}{1}{}{}{}=&1-\left(1+\aParam\right)\eStoch,
\end{align}
are the eigenvalues of the stochastic-noise channel. This method is unbiased so the proxy bias is zero.

For rotational errors we use \(\nMit=2\) and (for the analytic results):
\begin{align}\eqAL{nIITIILMRE}
m_{n,0} =&0,&m_{n,1}=& \begin{cases}
\kappa_1\frac{\nGate}{\sqrt{\oV{e}{RE}{}{}{}{}}}&\text{ for } \noiseLevel\ll 1\\
\frac{\uppi}{4}\frac{\nGate}{\oV{e}{RE}{}{}{}{}}&\text{ for } \nGate\gg 1.
\end{cases},&m_{n,2}=& \begin{cases}
\frac{\uppi}{2}\frac{\nGate}{\oV{e}{RE}{}{}{}{}}&\text{ for } \noiseLevel\ll 1\\
\frac{3\uppi}{4}\frac{\nGate}{\oV{e}{RE}{}{}{}{}}&\text{ for } \nGate\gg 1.
\end{cases}
\end{align}
where we define:
\begin{align}
\kappa_1=&\sqrt{\frac{2}{\uppi}}\sqrt{4-\frac{1}{\nGate}}.
\end{align}
Here \(\oV{e}{RE}{}{}{}{}\) is the rotational noise level (see \eqAR{rotNoiseLevel}) and \(\nGate\) is the number of gates in the circuit. The coefficients\footnote{The first equalities for the coefficients given in \eqAR{TIILMCoefRE} for TIILM are used for both numerical and analytic results. However, the Taylor expansions are valid only for the analytic results where we use \eqAR{nIITIILMRE} to define the number of identity insertions we use. For numerical results we optimise \(m_{n,0}\), \(m_{n,1}\), and \(m_{n,2}\) to give the lowest possible runtime scaling.} are given by:
\begin{align}\eqAL{TIILMCoefRE}
\oV{c}{L}{}{}{0}{}= &\frac{\sin\frac{\left(2m_{n,1}+1\right)\eAngle}{2}\sin\frac{\left(2m_{n,2}+1\right)\eAngle}{2}}{\sin\left(\left(m_{n,1}-m_{n,0}\right)\eAngle\right)\sin\left(\left(m_{n,2}-m_{n,0}\right)\eAngle\right)}\nonumber\\
=&\begin{cases}
1+\frac{1}{2\kappa_1}\frac{\sqrt{\oV{e}{RE}{}{}{}{}}}{\nGate}+\orderSmall\left(\oV{e}{RE}{}{}{}{}\right)&\text{ for } \oV{e}{RE}{}{}{}{}\ll 1,\\
1+\orderLarge\left(\frac{1}{\nGate^2}\right)&\text{ for } \nGate\gg 1,
\end{cases}\\
\oV{c}{L}{}{}{1}{}=&-\frac{\sin\frac{\left(2m_{n,0}+1\right)\eAngle}{2}\sin\frac{\left(2m_{n,2}+1\right)\eAngle}{2}}{\sin\left(\left(m_{n,1}-m_{n,0}\right)\eAngle\right)\sin\left(\left(m_{n,2}-m_{n,1}\right)\eAngle\right)}\nonumber\\
=&\begin{cases}
-\frac{1}{2\kappa_1}\frac{\sqrt{\oV{e}{RE}{}{}{}{}}}{\nGate}+\orderSmall\left(\oV{e}{RE}{}{}{}{}^{\frac{3}{2}}\right)&\text{ for } \oV{e}{RE}{}{}{}{}\ll 1,\\
-\frac{\oV{e}{RE}{}{}{}{}}{2\nGate}+\orderLarge\left(\frac{1}{\nGate^2}\right)&\text{ for } \nGate\gg 1,
\end{cases}\\
\oV{c}{L}{}{}{2}{}=& \frac{\sin\frac{\left(2m_{n,0}+1\right)\eAngle}{2}\sin\frac{\left(2m_{n,1}+1\right)\eAngle}{2}}{\sin\left(\left(m_{n,2}-m_{n,0}\right)\eAngle\right)\sin\left(\left(m_{n,2}-m_{n,1}\right)\eAngle\right)}\nonumber\\
=&\begin{cases}
\frac{\kappa_1}{2}\frac{\oV{e}{RE}{}{}{}{}^{\frac{3}{2}}}{\nGate}+\orderSmall\left(\oV{e}{RE}{}{}{}{}^{2}\right)&\text{ for } \oV{e}{RE}{}{}{}{}\ll 1\\
\frac{\oV{e}{RE}{}{}{}{}}{2\nGate}+\orderLarge\left(\frac{1}{\nGate^2}\right)&\text{ for } \nGate\gg 1.
\end{cases}
\end{align}
where \(\eAngle\) is the noise amplitude of the rotational-error channels (\eqAR{rotNoiseAmplitude}). This method is unbiased so the proxy bias is zero.

\section{Numerical}\appSL{numerical}
In this appendix we will discuss the details of the simulation procedure (\appSR{simulations}) used to generate the numerical plots in \secR{numericalAnalysis}, along with details of the precision (\appSR{precision}) and parameters (\appSR{parameters}) used.

\subsection{Simulation}\appSL{simulations}
All our numerical plots are the result of simulations rather than implementation on a quantum computer. We simulate the circuits using the Pauli-transfer-matrix formalism (see \appSR{transferMatrices}). First we produce the Bloch vector corresponding to the initial state of the qubits (\eqAR{blochVectorSimulation}). Then we multiply this Bloch vector by matrices corresponding to each of the gates and error channels in our circuit (see \appSR{transfermatricesSimulation}), in the order they appear in our circuit (\eqR{depolarisingFullCircuit} of \secR{benchCircuit}). Finally, we extract the observable expectation by combining the relevant elements of the final Bloch vector obtained after the matrix multiplications (see \appSR{blochVectorsSimulation}). For local mitigation procedures we simulate a circuit where every noisy gate is replaced by the equivalent mitigated gate, where the mitigated gate transfer matrices are linear combinations of noisy gates as specified by the mitigation procedure (see \appSR{localMitigationSimulation}). For non-local mitigation procedures we simulate every circuit that appears in the linear combination required for the mitigated circuit and then linearly combine the observable estimates to find the mitigated observable estimates (see \appSR{globalMitigationSimulation}).

\subsubsection{Initial States and Measurements}\appSL{blochVectorsSimulation}
The Bloch vector corresponding to our initial state is (see \appSR{pauliWords}):
\begin{align}\eqAL{blochVectorSimulation}
\blochOperator_0=\bigotimes_{n=1}^{\nQubit} \begin{pmatrix}
1\\
0\\
0\\
1
\end{pmatrix},
\end{align}
where \(\nQubit\) is the number of qubits in the circuit. This Bloch vector corresponds to the state: \(\bigotimes_{n=1}^{\nQubit}\xVac\) (\ie the tensor product of the \(+1\) eigenstates of the Pauli Z operators for each qubit). After we have multiplied this vector by all the transfer matrices corresponding to the gates in the circuit we obtain the output Bloch vector \(\blochOperatorN\). For our observable estimates we measure: 
\begin{align}\eqAL{measurementSim}
\langle\iOp_{\acton{1}}\rangle=&\left(\blochOperatorN\right)_0,
&\langle\xOp_{\acton{1}}\rangle=&\left(\blochOperatorN\right)_{1},
&\langle\yOp_{\acton{1}}\rangle=&\left(\blochOperatorN\right)_2,
&\langle\zOp_1\rangle=&\left(\blochOperatorN\right)_3.
\end{align}
Where \(\left(\blochOperatorN\right)_i\) is the \(i+1\)th element of the (0-indexed) Bloch vector. \(\langle\iOp_{\acton{1}}\rangle\) can be used to verify the state is properly normalised. \(\langle\xOp_{\acton{1}}\rangle\), \(\langle\yOp_{\acton{1}}\rangle\), and \(\langle\zOp_{\acton{1}}\rangle\) are the expectation of the X, Y, and Z Pauli operators for the first qubit.

\subsubsection{Transfer Matrices}\appSL{transfermatricesSimulation}
We use a basis set of single qubit Pauli transfer matrices:
\begin{align}\eqAL{definitionsTransferSIM}
\transferMatrixi{\rOpi{1}}\left(\cAngle,\eStoch\right)=&\begin{pmatrix}
1&0&0&0\\
0&1&0&0\\
0&0&\left(1-2\eStoch\right)\cos\cAngle&\left(1-2\eStoch\right)\sin\cAngle\\
0&0&-\left(1-2\eStoch\right)\sin\cAngle&\left(1-2\eStoch\right)\cos\cAngle\\
\end{pmatrix},\\
\transferMatrixi{\rOpi{2}}\left(\cAngle,\eStoch\right)=&\begin{pmatrix}
1&0&0&0\\
0&\left(1-2\eStoch\right)\cos\cAngle&0&-\left(1-2\eStoch\right)\sin\cAngle\\
0&0&1&0\\
0&\left(1-2\eStoch\right)\sin\cAngle&0&\left(1-2\eStoch\right)\cos\cAngle\\
\end{pmatrix},\\
\transferMatrixi{\rOpi{3}}\left(\cAngle,\eStoch\right)=&\begin{pmatrix}
1&0&0&0\\
0&\left(1-2\eStoch\right)\cos\cAngle&\left(1-2\eStoch\right)\sin\cAngle&0\\
0&-\left(1-2\eStoch\right)\sin\cAngle&\left(1-2\eStoch\right)\cos\cAngle&0\\
0&0&0&1
\end{pmatrix}.
\end{align}
These three are sufficient to define all the single qubit gates and operators that we use within this work. For example we have:
\begin{align}\eqAL{onequbitMatrixSim}
\transferMatrixi{\xOp}=&\transferMatrixi{\rOpi{1}}\left(\uppi,0\right),&\transferMatrixi{\yOp}=&\transferMatrixi{\rOpi{2}}\left(\uppi,0\right),&\transferMatrixi{\zOp}=&\transferMatrixi{\rOpi{3}}\left(\uppi,0\right),\\
\transferMatrixi{\hat{\sOp}_{1}}=&\transferMatrixi{\rOpi{1}}\left(\frac{\uppi}{2},0\right),&\transferMatrixi{\hat{\sOp}_{2}}=&\transferMatrixi{\rOpi{2}}\left(\frac{\uppi}{2},0\right),&\transferMatrixi{\hat{\sOp}_{3}}=&\transferMatrixi{\rOpi{3}}\left(\frac{\uppi}{2},0\right),\\
\transferMatrixi{\hat{\tOp}_{1}}=&\transferMatrixi{\rOpi{1}}\left(\frac{\uppi}{4},0\right),&\transferMatrixi{\hat{\tOp}_{2}}=&\transferMatrixi{\rOpi{2}}\left(\frac{\uppi}{4},0\right),&\transferMatrixi{\hat{\tOp}_{3}}=&\transferMatrixi{\rOpi{3}}\left(\frac{\uppi}{4},0\right),
\end{align}
where the top three are the single qubit Pauli operators and the second and third rows are the square and fourth roots of the Pauli operators, respectively. Inverse gates can be found by inverting the angle:
\begin{align}\eqAL{inverseOnequbitMatrixSim}
\transferMatrixi{\hat{\sOp}_{1}^\dagger}=&\transferMatrixi{\rOpi{1}}\left(-\frac{\uppi}{2},0\right),&\transferMatrixi{\hat{\sOp}_{2}^\dagger}=&\transferMatrixi{\rOpi{2}}\left(-\frac{\uppi}{2},0\right),&\transferMatrixi{\hat{\sOp}_{3}^\dagger}=&\transferMatrixi{\rOpi{3}}\left(-\frac{\uppi}{2},0\right),\\
\transferMatrixi{\hat{\tOp}_{1}^\dagger}=&\transferMatrixi{\rOpi{1}}\left(-\frac{\uppi}{4},0\right),&\transferMatrixi{\hat{\tOp}_{2}^\dagger}=&\transferMatrixi{\rOpi{2}}\left(-\frac{\uppi}{4},0\right),&\transferMatrixi{\hat{\tOp}_{3}^\dagger}=&\transferMatrixi{\rOpi{3}}\left(-\frac{\uppi}{4},0\right).
\end{align}

The only two-qubit transfer matrix that we make use of directly is \(\transferMatrixi{\rOpi{15}}\left(\cAngle,\eStoch\right)\), which corresponds to rotations of \(\cAngle\) radians followed by a stochastic-noise channel with noise amplitude \(\eStoch\), both generated by \(\pOpi{15}=\zOp\otimes\zOp\). This is too large to display in all its glory. So, instead, we shall note the elements that differ from those of the two qubit identity matrix (which is all zeros except for the diagonal elements, which are equal to unity):
\begin{align}
&\left(\transferMatrixi{\rOpi{15}}\left(\cAngle,\eStoch\right)\right)_{1,1}=\left(\transferMatrixi{\rOpi{15}}\left(\cAngle,\eStoch\right)\right)_{2,2}=\left(\transferMatrixi{\rOpi{15}}\left(\cAngle,\eStoch\right)\right)_{4,4}=\left(\transferMatrixi{\rOpi{15}}\left(\cAngle,\eStoch\right)\right)_{7,7}=\left(\transferMatrixi{\rOpi{15}}\left(\cAngle,\eStoch\right)\right)_{8,8}\nonumber\\
&=\left(\transferMatrixi{\rOpi{15}}\left(\cAngle,\eStoch\right)\right)_{11,11}=\left(\transferMatrixi{\rOpi{15}}\left(\cAngle,\eStoch\right)\right)_{13,13}=\left(\transferMatrixi{\rOpi{15}}\left(\cAngle,\eStoch\right)\right)_{14,14}=\left(1-2\eStoch\right)\cos\cAngle,\\
&\left(\transferMatrixi{\rOpi{15}}\left(\cAngle,\eStoch\right)\right)_{1,14}=\left(\transferMatrixi{\rOpi{15}}\left(\cAngle,\eStoch\right)\right)_{4,11}=\left(\transferMatrixi{\rOpi{15}}\left(\cAngle,\eStoch\right)\right)_{7,8}=\left(\transferMatrixi{\rOpi{15}}\left(\cAngle,\eStoch\right)\right)_{13,2}=\left(1-2\eStoch\right)\sin\cAngle\nonumber\\
&=-\left(\transferMatrixi{\rOpi{15}}\left(\cAngle,\eStoch\right)\right)_{14,1}=-\left(\transferMatrixi{\rOpi{15}}\left(\cAngle,\eStoch\right)\right)_{11,4}=-\left(\transferMatrixi{\rOpi{15}}\left(\cAngle,\eStoch\right)\right)_{8,7}=-\left(\transferMatrixi{\rOpi{15}}\left(\cAngle,\eStoch\right)\right)_{2,13},
\end{align}
where we use the zero-index notation so the indices run from \(0\) to \(15\). We have:
\begin{align}\eqAL{twoqubitMatrixSim}
\transferMatrixi{\hat{\sOp}_{15}}=&\transferMatrixi{\rOpi{15}}\left(\frac{\uppi}{2},0\right),&\transferMatrixi{\hat{\sOp}_{15}^\dagger}=&\transferMatrixi{\rOpi{15}}\left(-\frac{\uppi}{2},0\right),
\end{align}
which are the Pauli transfer matrices corresponding to \(\hat{\sOp}_{15}\) (where \(\hat{\sOp}_{15}^{2} =\pOpi{15}=\zOp\otimes\zOp \)) and \(\hat{\sOp}_{15}^\dagger\).

For error channels we use:
\begin{align}\eqAL{errorChannelsSim}
\transferMatrixi{\oC{E}{SN}{i}{}{}{}}=&\transferMatrixi{R_{i}}\left(0,\eStoch\right), &\transferMatrixi{\oC{E}{RE}{i}{}{}{}}=&\transferMatrixi{R_{i}}\left(\eAngle,0\right),&\transferMatrixi{\oC{E}{ORE}{i}{}{}{}}=&\transferMatrixi{R_{i}}\left(\eAngle,\eStoch\right),
\end{align}
which correspond to stochastic-noise, rotational-error, and (stochastic-)over-rotation-error channels generated by \(\pOpi{i}\), respectively. 

Before applying a gate to our state (by multiplying the state's Bloch vector (see \appSR{blochVectorsSimulation}) by our gate's Pauli transfer matrix) we must promote the gate to the correct dimensionality. In our case the state has \(4^\nQubit\) dimensions. We only use nearest neighbour gates in our circuit, which simplifies this procedure. We use the tensor product operator, to promote single and two qubit operators to the full Hilbert space, such that the first qubit is the furthest right and we move left for gates corresponding to higher numbered qubits. Thus, for example, a noisy T gate (\ie \(\hat{\tOp}_{3}\)) suffering from pure over-rotation errors (\(\oC{E}{}{}{RE}{}{}\)) that acts on the second qubit has a transfer matrix given by:
\begin{align}\eqAL{singleQUbitPrmo}
\iOp \otimes \iOp\otimes \transferMatrixi{\rOpi{3}}\left(\eAngle,0\right)\transferMatrixi{\hat{\tOp}_{3}}\otimes\iOp,
\end{align}
where \(\iOp\) is the two dimensional identity operator. Whereas the noisy two qubit gate: \(\hat{\sOp}_{15}\) that acts on the third and fourth qubits and suffers from stochastic over-rotation errors has a transfer matrix given by:
\begin{align}\eqAL{twoQUbitPrmo}
\transferMatrixi{\rOpi{15}}\left(\eAngle,\eStoch\right)\transferMatrixi{\hat{\sOp}_{15}}\otimes\iOp\otimes \iOp.
\end{align}
For our simulations the error channels follow the gates, but since the two commute the order is irrelevant.

\subsubsection{Non-Local Mitigation}\appSL{globalMitigationSimulation}
For the synchronous and (explicitly) asynchronous strategies, which mitigate the circuit as a whole rather than (effectively) mitigating each gate independently, we obtain an observable estimate (for each observable) for each circuit variant and then calculate the linear combination of the observable estimates for each circuit weighted by the mitigation coefficients, to obtain the mitigated observable estimates. The mitigated bias is given by the difference between this mitigated observable estimate and the observable expectation of the noise-free circuit:
\begin{align}\eqAL{mitigatedBlochVectorBiasSIM}
\eBiasi{\pOpi{j}}=\left|\traceOp{\pOpi{j}\left(\rho-\addnoise{\rho}_{\mitigationLabel}\right)}\right|=\left|\left(\blochOperator\right)_j-\sum_{i=0}^{\nCircGlobal}\coefGlobali{i}\left(\blochOperatorN_{;i}\right)_j\right|,
\end{align}
where \(\eBiasi{\pOpi{j}}\) is the mitigated bias for the \(j\)th Pauli observable (\(\pOpi{j}\)),  \(\rho\) is the noise-free state, \(\addnoise{\rho}_{\mitigationLabel}\) is the effective mitigated state, \(\left(\blochOperator\right)_j\) is the \(j\)th element of the noise-free Bloch vector (corresponding to the expectation of \(\rho\) with respect to \(\pOpi{j}\)), \(\left(\blochOperatorN_{;i}\right)_j\) is the \(j\)th element of the Bloch vector output by the \(i\)th noisy circuit variant, \(\nCircGlobal\) is the number of circuits used in the mitigation, and \(\coefGlobali{i}\) is the mitigation coefficient for the \(i\)th noisy circuit variant.

For example, we can consider first order (\(\nMit=\nCircGlobal=1\)) IISM (see \appSR{IISM}) for a circuit suffering from stochastic noise with noise amplitude \(\eStoch\). Here we have two circuit variants: the original noisy circuit and the circuit obtained by adding one identity insertion before each gate. The noise-free state is given by (see \eqARList{definitionsTransferSIM}{twoqubitMatrixSim}):
\begin{align}\eqAL{noiseFreeCircuitSIM}
\blochOperator=&\dots\left(\iOp \otimes \transferMatrixi{\hat{\tOp}_{1}}\otimes \iOp\otimes\iOp\right)\left(\iOp \otimes \transferMatrixi{\hat{\sOp}_{15}}\otimes\iOp\right)\left(\iOp \otimes \iOp\otimes \transferMatrixi{\hat{\tOp}_{1}}\otimes\iOp\right)\left(\iOp \otimes \iOp\otimes \transferMatrixi{\hat{\sOp}_{15}}\right)\nonumber\\
&\times\left(\iOp \otimes \iOp\otimes \iOp\otimes\transferMatrixi{\hat{\tOp}_{1}}\transferMatrixi{\yOp}\right)\blochOperator_0,
\end{align}
where we have explicitly noted the expression up to the \(6\)th gate in the circuit (see \eqR{depolarisingFullCircuit} of \secR{benchCircuit}). To obtain the state (Bloch vector) output by the original noisy circuit we perform the calculation (see \eqARList{definitionsTransferSIM}{errorChannelsSim}):
\begin{align}
\blochOperatorN_{;0}=&\dots\left(\iOp \otimes \transferMatrixi{\oC{E}{SN}{1}{}{}{}}\transferMatrixi{\hat{\tOp}_{1}}\otimes \iOp\otimes\iOp\right)\left(\iOp \otimes \transferMatrixi{\oC{E}{SN}{15}{}{}{}}\transferMatrixi{\hat{\sOp}_{15}}\otimes\iOp\right)\left(\iOp \otimes \iOp\otimes \transferMatrixi{\oC{E}{SN}{1}{}{}{}}\transferMatrixi{\hat{\tOp}_{1}}\otimes\iOp\right)\nonumber\\
&\times\left(\iOp \otimes \iOp\otimes \transferMatrixi{\oC{E}{SN}{15}{}{}{}}\transferMatrixi{\hat{\sOp}_{15}}\right)\left(\iOp \otimes \iOp\otimes \iOp\otimes\transferMatrixi{\oC{E}{SN}{1}{}{}{}}\transferMatrixi{\hat{\tOp}_{1}}\transferMatrixi{\oC{E}{SN}{2}{}{}{}}\transferMatrixi{\yOp}\right)\blochOperator_0.
\end{align}
To obtain the state output by the circuit obtained after a single identity insertion for each gate we perform the calculation (see \eqARList{definitionsTransferSIM}{errorChannelsSim}):
\begin{align}
\blochOperatorN_{;1}=&\dots\left(\iOp \otimes \transferMatrixi{\oC{E}{SN}{15}{}{}{}}\transferMatrixi{\hat{\sOp}_{15}}\transferMatrixi{\oC{E}{SN}{15}{}{}{}}\transferMatrixi{\hat{\sOp}_{15}^\dagger}\transferMatrixi{\oC{E}{SN}{15}{}{}{}}\transferMatrixi{\hat{\sOp}_{15}}\otimes\iOp\right)\nonumber\\
&\times\left(\iOp \otimes \iOp\otimes \transferMatrixi{\oC{E}{SN}{1}{}{}{}}\transferMatrixi{\hat{\tOp}_{1}}\transferMatrixi{\oC{E}{SN}{1}{}{}{}}\transferMatrixi{\hat{\tOp}_{1}^\dagger}\transferMatrixi{\oC{E}{SN}{1}{}{}{}}\transferMatrixi{\hat{\tOp}_{1}}\otimes\iOp\right)\nonumber\\&\times\left(\iOp \otimes \iOp\otimes \transferMatrixi{\oC{E}{SN}{15}{}{}{}}\transferMatrixi{\hat{\sOp}_{15}}\transferMatrixi{\oC{E}{SN}{15}{}{}{}}\transferMatrixi{\hat{\sOp}_{15}^\dagger}\transferMatrixi{\oC{E}{SN}{15}{}{}{}}\transferMatrixi{\hat{\sOp}_{15}}\right)\nonumber\\
&\times\left(\iOp \otimes \iOp\otimes \iOp\otimes\transferMatrixi{\oC{E}{SN}{1}{}{}{}}\transferMatrixi{\hat{\tOp}_{1}}\transferMatrixi{\oC{E}{SN}{1}{}{}{}}\transferMatrixi{\hat{\tOp}_{1}^\dagger}\transferMatrixi{\oC{E}{SN}{1}{}{}{}}\transferMatrixi{\hat{\tOp}_{1}}\left(\transferMatrixi{\oC{E}{SN}{2}{}{}{}}\transferMatrixi{\yOp}\right)^3\right)\blochOperator_0.
\end{align}
This means our biases after mitigation are given by (see \eqARB{IISMcoefficientsKnowledgeFree}{mitigatedBlochVectorBiasSIM}):
\begin{align}
\eBiasi{\xOp}=&\left|\left(\blochOperator\right)_1-\left(\frac{3}{2}\left(\blochOperatorN_{;0}\right)_1-\frac{1}{2}\left(\blochOperatorN_{;1}\right)_1\right)\right|,&\eBiasi{\yOp}=&\left|\left(\blochOperator\right)_2-\left(\frac{3}{2}\left(\blochOperatorN_{;0}\right)_2-\frac{1}{2}\left(\blochOperatorN_{;1}\right)_2\right)\right|,\nonumber\\\eBiasi{\zOp}=&\left|\left(\blochOperator\right)_3-\left(\frac{3}{2}\left(\blochOperatorN_{;0}\right)_3-\frac{1}{2}\left(\blochOperatorN_{;1}\right)_3\right)\right|.
\end{align}

\subsubsection{Local Mitigation}\appSL{localMitigationSimulation}
Since the exponential number of circuits required to implement a local mitigation is impractical to simulate, we instead simulate the mitigated circuit directly. We replace each noisy gate in the circuit with its mitigated equivalent in the original noisy circuit and then simulate this mitigated circuit to obtain the mitigated Bloch vector (\(\blochOperatorN_{\mitigationLabel}\)). The mitigated transfer matrix for the \(n\)th gate (used to simulate the \(n\)th mitigated gate) is given by:
\begin{align}\eqAL{mitigatedTransferMatricesSIM}
\transferMatrixi{n;\mitigationLabel} =\sum_{i=0}^{\nCircLocali{n}}\coefLocali{n,i} \transferMatrixi{n,i} 
\end{align}
where \(\transferMatrixi{n;\mitigationLabel}\) is the transfer matrix for the \(n\)th mitigated gate, \(\nCircLocali{n}\) is the number of noisy gate variants used to mitigate the \(n\)th gate, \(\coefLocali{n,i}\) is the mitigation coefficient for the \(i\)th noisy gate variant of the \(n\)th gate, and \(\transferMatrixi{n,i}\) is the transfer matrix of the \(i\)th noisy gate variant of the \(n\)th gate. The mitigated bias is given by the difference between the mitigated observable estimate (derived from \(\blochOperatorN_{\mitigationLabel}\)) and the observable expectation of the noise-free circuit:
\begin{align}\eqAL{mitigatedBlochVectorBiasLocalSIM}
\eBiasi{\pOpi{j}}=\left|\traceOp{\pOpi{j}\left(\rho-\addnoise{\rho}_{\mitigationLabel}\right)}\right|=\left|\left(\blochOperator\right)_j-\left(\blochOperatorN_{\mitigationLabel}\right)_j\right|,
\end{align}
where \(\eBiasi{\pOpi{j}}\) is the mitigated bias for the \(j\)th Pauli observable (\(\pOpi{j}\)),  \(\rho\) is the noise-free state, \(\addnoise{\rho}_{\mitigationLabel}\) is the effective mitigated state, \(\left(\blochOperator\right)_j\) is the \(j\)th element of the noise-free Bloch vector (corresponding to the expectation of \(\rho\) with respect to \(\pOpi{j}\)), and \(\left(\blochOperatorN_{\mitigationLabel}\right)_j\) is the \(j\)th element of the mitigated Bloch vector (corresponding to the expectation of \(\rhoM\) with respect to \(\pOpi{j}\)).

For example, we can consider CLM (\appSR{CLM}) for a circuit suffering from rotational errors with noise amplitude \(\eAngle\). The noise-free Bloch vector (\(\blochOperator\)) is calculated using the expression in \eqAR{noiseFreeCircuitSIM}. The mitigated Bloch vector (\(\blochOperatorN_{\mitigationLabel}\)) is calculated by replacing all the transfer matrices in \eqAR{noiseFreeCircuitSIM} by their mitigated equivalents: (see \eqARList{definitionsTransferSIM}{twoqubitMatrixSim}):
\begin{align}\eqAL{mitigatedCircuitSIM}
\blochOperatorN_{\mitigationLabel}=&\dots\left(\iOp \otimes \transferMatrixi{\hat{\tOp}_{1;\mitigationLabel}}\otimes \iOp\otimes\iOp\right)\left(\iOp \otimes \transferMatrixi{\hat{\sOp}_{15;\mitigationLabel}}\otimes\iOp\right)\left(\iOp \otimes \iOp\otimes \transferMatrixi{\hat{\tOp}_{1;\mitigationLabel}}\otimes\iOp\right)\left(\iOp \otimes \iOp\otimes \transferMatrixi{\hat{\sOp}_{15;\mitigationLabel}}\right)\nonumber\\
&\times\left(\iOp \otimes \iOp\otimes \iOp\otimes\transferMatrixi{\hat{\tOp}_{1;\mitigationLabel}}\transferMatrixi{\yOp_{;\mitigationLabel}}\right)\blochOperator_0,
\end{align}
where we have explicitly noted the expression up to the \(6\)th gate in the circuit (see \eqR{depolarisingFullCircuit} of \secR{benchCircuit}). The mitigated transfer matrix for the first gate is given by (see \eqAR{rotationalErrorsCoefsCLM}):
\begin{align}
\transferMatrixi{\yOp_{;\mitigationLabel}}=&\cos \eAngle\transferMatrixi{\oC{E}{RE}{2}{}{}{}}\transferMatrixi{\yOp}-\frac{\cos \eAngle-1+\sin\eAngle}{2}\transferMatrixi{\rOpi{2}}\left(\frac{\uppi}{2},0\right)\transferMatrixi{\oC{E}{RE}{2}{}{}{}}\transferMatrixi{\yOp}\nonumber\\
&+\frac{1-\cos \eAngle+\sin\eAngle}{2}\transferMatrixi{\rOpi{2}}\left(\frac{3\uppi}{2},0\right)\transferMatrixi{\oC{E}{RE}{2}{}{}{}}\transferMatrixi{\yOp},
\end{align}
where \(\transferMatrixi{\yOp_{;\mitigationLabel}}\) is the mitigated transfer matrix; \(\eAngle\) is the original noise amplitude of the rotational-error channel; \(\transferMatrixi{\oC{E}{RE}{2}{}{}{}}\) is the error channel of the original noisy gate; \(\transferMatrixi{\yOp}\) is the noise-free component of the original noisy gate (so the transfer matrix corresponding to the original noisy gate is \(\transferMatrixi{\oC{E}{RE}{2}{}{}{}}\transferMatrixi{\yOp}\)); and \(\transferMatrixi{\rOpi{2}}\left(\frac{\uppi}{2},0\right)\) and \(\transferMatrixi{\rOpi{2}}\left(\frac{3\uppi}{2},0\right)\) are the transfer matrices equivalent to the custom error channels used to generate the first and second noisy gate variants, respectively. The other mitigated gate transfer matrices are defined analogously, \ie the indices of the error and custom channels should be changed to match that of the gate, the coefficients are unchanged (if the noise amplitude is unchanged). Our biases after mitigation are given by (see \eqAR{mitigatedBlochVectorBiasLocalSIM}):
\begin{align}
\eBiasi{\xOp}=&\left|\left(\blochOperator\right)_1-\left(\blochOperatorN_{\mitigationLabel}\right)_1\right|,&\eBiasi{\yOp}=&\left|\left(\blochOperator\right)_2-\left(\blochOperatorN_{\mitigationLabel}\right)_2\right|,&\eBiasi{\zOp}=&\left|\left(\blochOperator\right)_3-\left(\blochOperatorN_{\mitigationLabel}\right)_3\right|.
\end{align}

\subsubsection{Local Cancellation}
Local cancellation is implemented in a similar fashion to the local mitigation strategies (see \appSR{localMitigationSimulation}). We replace each of the noisy gate transfer matrices in our simulation with the locally cancelled equivalents:
\begin{align}
\transferMatrixi{n;\localCancellationLabel} =\sum_{i=0}^{\nCircLCi{n}}\coefLCi{n,i} \transferMatrixi{n,i},
\end{align}
where \(\transferMatrixi{n;\localCancellationLabel}\) is the transfer matrix corresponding to the locally cancelled gate, \(\nCircLCi{n}\) is the number of noisy gate variants used for the cancellation, \(\coefLCi{n,i}\) is the coefficient for the \(i\)th variant of the \(n\)th gate, and \(\transferMatrixi{n,i}\) is the transfer matrix corresponding to the \(i\)th variant of the \(n\)th gate. We then treat the locally cancelled gates as if they were the original noisy gates for any further mitigation.

 For example, if we are locally cancelling a gate suffering from a pure over-rotation error of noise amplitude \(\eAngle\) then we can use the average of the original noisy gate and its hidden inverse. To be more explicit we can consider the \(\hat{\sOp}_{15}\) gate. The transfer matrices for the noisy gate and its hidden inverse are given by:
\begin{align}
\transferMatrixi{\oC{E}{RE}{15}{}{}{}}\transferMatrixi{\hat{\sOp}_{15}}=&\transferMatrixi{\rOpi{15}}\left(\eAngle,0\right)\transferMatrixi{\hat{\sOp}_{15}},&\transferMatrixi{\oC{E}{RE}{15}{}{}{}}^{\dagger}\transferMatrixi{\hat{\sOp}_{15}}=&\transferMatrixi{\rOpi{15}}\left(-\eAngle,0\right)\transferMatrixi{\hat{\sOp}_{15}},
\end{align}
respectively. Here \(\transferMatrixi{\hat{\sOp}_{15}}\) is the noise-free component of the noisy gate, \(\transferMatrixi{\oC{E}{RE}{15}{}{}{}}=\transferMatrixi{\rOpi{15}}\left(\eAngle,0\right)\) is the original error channel of the noisy gate, and \(\transferMatrixi{\oC{E}{RE}{15}{}{}{}}^{\dagger}=\transferMatrixi{\rOpi{15}}\left(-\eAngle,0\right)\) is the error channel of the hidden inverse of the original noisy gate. So the locally cancelled gate has a transfer matrix given by (\eqAR{definitionsTransferSIM}):
\begin{align}
\frac{1}{2}\transferMatrixi{\oC{E}{RE}{15}{}{}{}}\transferMatrixi{\hat{\sOp}_{15}}+\frac{1}{2}\transferMatrixi{\oC{E}{RE}{15}{}{}{}}^\dagger\transferMatrixi{\hat{\sOp}_{15}}=\transferMatrixi{\rOpi{15}}\left(0,\frac{1-\cos\eAngle}{2}\right)\transferMatrixi{\hat{\sOp}_{15}},
\end{align}
\ie the locally cancelled gate has an effective error channel that is a stochastic-noise channel with noise amplitude \(\frac{1-\cos\eAngle}{2}\). So we have successfully cancelled the coherent components of the channel.

\subsection{Precision and Code}\appSL{precision}
For all our noise-aware mitigation strategies we consider mitigation orders (\(\nMit\)) up to those sufficient to achieve unbiased mitigation. This serves to verify our analytic analyses and gives us confidence in our numerical results. However, to achieve unbiased mitigation for large circuits using synchronous mitigation methods we require coefficients and circuit simulations calculated to very high precision, to avoid rounding errors. We use up to 1500 decimal places for the calculations (see \tabAR{PlotParameters}).  We use Python for our simulations and make use of the high precision library mpmath and the symbolic library sympy (since the native operations such as tensor products and matrix-vector multiplications seem to run faster using sympy than mpmath matrices).

\subsection{Parameters}\appSL{parameters}
All the plots we use are characterised by two parameters: the noise level (\(\noiseLevel\)) and the number of gates (\(\nGate\)). For simplicity we split the noise level into two components, only one of which is non-zero for any one plot:
\begin{align}
\noiseLevel=\oV{e}{SN}{}{}{}{}+\oV{e}{RE}{}{}{}{},
\end{align}
where (\eqARB{stochNoiseLevel}{rotNoiseLevel} of \appSR{noiseModel}, respectively):
\begin{align}
\oV{e}{SN}{}{}{}{} = &2\nGate\eStoch,&\eExpectR = &\nGate\left|\eAngle\right|,
\end{align}
where \(\eStoch\) is the gate-wise error probability, and \(\eAngle\) is the angle of the rotational error. For simplicity, all the noise models we simulate are uniform. So, every noisy gate has the same noise amplitude (\(\eStoch\) and \(\eAngle\)). The parameters for the plots in \secR{numericalAnalysis} are given in the \tabAR{PlotParameters}. For all these plots we find the TIILM (\appSR{TIILM}) parameters: \(m_{n,0}=m_0\), \(m_{n,1}=m_1\), and \(m_{n,2}=m_2\) (\ie the numbers of identity insertions to use for each noisy variant gate) using an optimisation procedure. We use the values that lead to the smallest runtime scaling. These optimised values are also presented in \tabAR{PlotParameters}. We use 4 qubit (\(\nQubit=4\)) circuits. Since we are studying over-rotation errors, we have a Hermitian generator and therefore \(\aParam=1\) for our stochastic-noise channels (see \eqR{closureSN} of \secR{SN}).

\begin{table}[htbp]
    \centering
    \caption{Plot Parameters: Table of plot parameters used for the figures in \secR{numericalAnalysis}. We also include the optimised values used for  \(m_{n,0}=m_0\), \(m_{n,1}=m_1\), and/or \(m_{n,2}=m_2\) in TIILM (see \appSR{TIILM}). The final two columns give the precision used to calculate the mitigation coefficients and perform the simulation, respectively, in terms of the number of decimal places.}
    \tabAL{PlotParameters}
    \scalebox{1}{
    \begin{tabular}{c||rr|lll|rrr|cc}
        \hline
Plot Ref.	&\(\nGate\)&\(\nQubit\)& \(\oV{e}{SN}{}{}{}{} \)		&\(\oV{e}{RE}{}{}{}{} \)		& \(\oV{e}{GDN}{}{}{}{} \)&\(m_0\)&\(m_1\) &\(m_2\)&P. Coef & P. Sim.\\
        \hline
\figR{FCD001r1}&18&4&0.02 & 0 &0&0&246&N/A&500&1500\\
\figR{FCD001r10}&180&4&0.02 & 0 &0&0&2478&N/A&1500&1500\\
\figR{FCD1r1}&18&4&2 & 0 &0&0&16&N/A&500&1500\\
\figR{FCD1r10}&180&4&2 & 0 &0&0&169&N/A&1500&1500\\
\figR{FCO01r1}&18&4&0 & 0.2 &0&0&58&200&500&1500\\
\figR{FCO01r10}&180&4&0 & 0.2 &0&0&589&2002&1500&1500\\
\figR{FCO075r1}&18&4&0 & 1.5 &0&0&15&34&500&1500\\
\figR{FCO075r10}&180&4&0 & 1.5 &0&0&152&341&1500&1500\\
\figR{LCFCO01r1}&18&4&0 & 0.2 &0&0&1066&N/A&500&1500\\
\figR{LCFCO01r10}&180&4&0 & 0.2 &0&0&34068&N/A&1500&1500\\
\figR{LCFCO075r1}&18&4&0 & 1.5 &0&0&135&N/A&500&1500\\
\figR{LCFCO075r10}&180&4&0 & 1.5 &0&0&4487&N/A&1500&1500\\
        \hline		
    \end{tabular}
    }
\end{table}

\end{document}